\pdfoutput=1
\documentclass[11pt]{article}
\usepackage{jheppub}

\usepackage{customprelude}

\usepackage{amsmath}
\usepackage{amssymb}
\usepackage{graphicx,color}
\usepackage{amsthm}
\usepackage{mathrsfs}

\usepackage{tikz}
\usetikzlibrary{matrix,arrows}

\usepackage{ifpdf}

\usepackage{tabls}

\usepackage[boxsize=0.5em,aligntableaux=center]{ytableau}

\newcommand{\be}{\begin{equation}}
\newcommand{\ee}{\end{equation}}
\newcommand{\ba}{\begin{eqnarray}}
\newcommand{\ea}{\end{eqnarray}}

\renewcommand{\d}{\textrm{d}}

\newcommand\varpm{\mathbin{\vcenter{\hbox{  \oalign{\hfil$\scriptstyle+$\hfil\cr
          \noalign{\kern-.3ex}
          $\scriptscriptstyle({-})$\cr}}}}}
\newcommand\varmp{\mathbin{\vcenter{\hbox{  \oalign{$\scriptstyle({+})$\cr
          \noalign{\kern-.3ex}
          \hfil$\scriptscriptstyle-$\hfil\cr}}}}}

\title{\centering S-duality in $\mathcal{N}=1$ orientifold SCFTs}

\author[\spadesuit]{I\~naki Garc\'ia-Etxebarria}
\author[\clubsuit]{and Ben Heidenreich}

\affiliation[\spadesuit]{Max Planck Institute for Physics, F\"ohringer Ring 6, 80805 Munich, Germany}
\affiliation[\clubsuit]{Perimeter Institute for Theoretical Physics, Waterloo, Ontario, Canada N2L 2Y5}

\emailAdd{inaki@mpp.mpg.de}
\emailAdd{bheidenreich@perimeterinstitute.ca}

\abstract{We present a general solution to the problem of determining
  all S-dual descriptions for a specific (but very rich) class of
  $\cN=1$ SCFTs. These SCFTs are indexed by decorated toric diagrams,
  and can be engineered in string theory by probing orientifolds of isolated
   toric singularities with D3 branes. The S-dual phases are described by quiver gauge theories coupled to specific types of conformal matter which we describe explicitly. We \mbox{illustrate} our construction with many examples, including S-dualities in previously unknown SCFTs.}

\begin{document}

\makeatletter
\let\old@fpheader\@fpheader
\renewcommand{\@fpheader}{\old@fpheader\hfill
MPP-2016-338}
\makeatother

\maketitle
\newpage

\section{Introduction}


S-duality---where one quantum theory has multiple effective descriptions valid in different regions of the parameter space or moduli space of the theory---is a ubiquitous and rich phenomenon in superconformal field theories with more than four supercharges. Montonen-Olive duality~\cite{Montonen:1977sn,Goddard:1976qe,Witten:1978mh} of $\cN=4$ super-Yang-Mills (SYM) theory is the original and simplest example of this phenomenon. In this case, the strong coupling limit of an $\cN=4$ gauge theory with gauge group $G$ is equivalent to a weakly coupled $\cN=4$ theory whose gauge group is the Langlands dual $^LG$. The phenomenon extends to an $\SL(2,\bZ)$ duality whose elegant and intricate physics is best understood by considering the behavior of line operators in the theory~\cite{Aharony:2013hda}. More recently, S-duality has played a crucial role in the symbiotic development of four- and six-dimensional $\cN=2$ SCFTs, see, e.g.,~\cite{Gaiotto:2009we}.

Understanding S-duality in theories with less supersymmetry is a question of obvious interest, due to the increasingly rich dynamics that are possible in such theories and the wide-ranging consequences of S-duality in theories with extended supersymmetry.

Our paper will focus on S-duality in four-dimensional $\mathcal{N}=1$ SCFTs.\footnote{Seiberg-Witten theory could be considered as another instance of S-duality in four-dimensions, where S-dual low-energy effective descriptions are related to each other by motion along the Coulomb branch of moduli space. This phenomenon also occurs with $\cN=1$ supersymmetry~\cite{Intriligator:1994sm}, but is qualitative different than the SCFT S-dualities which we focus on in the following discussion, and in the rest of our paper.}
To gain an understanding of the phenomenon, one can construct examples of S-dualities in $\cN=1$ gauge theories by embedding them in a UV-complete string theory or field theory with a known S-duality. However, the resulting $\cN=1$ field theory is rarely (if ever) asymptotically free, and in many cases it is not superconformal either. In general, in the perturbative regime some of the couplings will be relevant and others will be irrelevant.

 This is not as bad a situation as might first seem. Consider, for instance, the $\mathcal{N}=1$ $\SU(N)$ gauge theory with two adjoint chiral multiplets and the superpotential
\be
W = h \Tr (\epsilon_{i j} \phi^i \phi^j)^2 \,,
\ee
see \cite{Argyres:1999xu}. This is a low-energy effective Lagrangian because $h$ is irrelevant at weak coupling. However, the infrared fixed point can be realized by flowing from a free theory, as follows. We start with the $h=0$ Lagrangian and flow to the infrared fixed point of the asymptotically free $\SU(N)$ gauge group, which is believed to be interacting~\cite{Intriligator:1994sm,Intriligator:2003mi}. Assuming it is, we can compute the exact dimension of the operator $\mathcal{O}=\Tr (\epsilon_{i j} \phi^i \phi^j)^2$ at the infrared fixed point using $a$-maximization~\cite{Intriligator:2003jj}. One finds that $\mathcal{O}$ is a marginal (dimension three) chiral operator. Since it is neutral under the $\SU(2)$ flavor symmetry, $\mathcal{O}$ is exactly marginal~\cite{Leigh:1995ep,Green:2010da}, and adding $h_\ast \int d^2\theta  \mathcal{O}+\mathrm{c.c.}$ to the Lagrangian generates a fixed line parameterized by $h_\ast$. Flows from the original effective field theory (EFT) will end somewhere on this line; 
 however, away from the ``cusp'' $h_\ast = 0$ there is no known UV complete flow connecting the fixed point to a free theory. Moreover, since parametric control of the effective theory requires $h \Lambda \ll 1$ for $\Lambda$ the $\SU(N)$ dynamical scale, all of the EFT flows will end near the cusp, and the dynamics of the fixed line become increasingly difficult to describe as we move farther away from it.

In this context, S-duality (at its most basic level) is the statement that by deforming the SCFT far out along the fixed line we eventually reach a dual description where there is another cusp associated to perturbative flows.\footnote{To be precise, since the fixed line has real dimension two (complex dimension one), not every path will take us to another cusp, but some paths will. The cusp we arrive at may be the same one we started at; in this case, the path still represents a non-trivial S-duality provided that it is non-trivial in the fundamental group of the fixed line, such as when it encircles another cusp or an orbifold singularity along the fixed line.} In the particular example we have chosen S-duality follows from Montonen-Olive duality~\cite{Argyres:1999xu}, as the effective theory considered above is nothing but a mass deformation of $\mathcal{N}=4$ super-Yang Mills (SYM). Thus, there are UV complete flows which end everywhere along the fixed line, but the cusps are distinguished as the endpoint of a flow or a series of flows from a free fixed point. Parametrically controlled effective field theories which flow to the fixed line will always end up near one of the cusps, which are the $\mathcal{N}=1$ analogs of weak coupling limits in S-dualities with extended supersymmetry.

This basic picture, where cusps along a fixed line are related by
exactly marginal deformations, can be thought of as a definition of
S-duality in SCFTs, and applies equally well with any number of
supercharges.  The aim of our paper is not to explore this field
theory picture directly, the main parts of which are
well-established~\cite{Leigh:1995ep, Argyres:1999xu} (if not widely
known). Rather, we seek to map out the realm of S-dualities in
four-dimensional $\mathcal{N}=1$ SCFTs, in order to establish how
common a phenomenon it is, what forms it takes, and especially which
cusps lie on the same conformal manifold and are therefore
S-duals. While a complete a classification is unfortunately beyond our
present abilities, we are able to make major progress by focusing on a
particular large but specific class of theories exhibiting S-duality,
continuing a series of recent
works~\cite{dualities1,Bianchi:2013gka,dualities2,dP1paper}. For
similar efforts in another class of theories see,
e.g.,~\cite{Gaiotto:2015usa,Franco:2015jna,Hanany:2015pfa}.\footnote{In
  addition (as in the above example), a large class of $\cN=1$
  S-dualities can be constructed by softly breaking supersymmetry in
  $\cN>1$ S-dual theories. However, some of the novel phenomena of
  $\cN=1$ theories, such as chirality, are absent in this case.}

Specifically, we will consider the class of SCFTs whose gravity duals arise from D3 branes probing an isolated toric Calabi-Yau orientifold singularity, i.e., an orientifold of a toric Calabi-Yau singularity such that the quotient space is both toric and smooth away from the singular point. Within this limited yet large class of theories, we are able to completely classify the S-dualities, giving a general prescription for how to construct the SCFTs from geometric data and determining which ones are related by the S-duality group $\SL(2,\bZ)$. (Note that our methods, which are based on T-duality, brane engineering, and topological data in the gravity dual, do not require large $N$.)

Our approach is based on the well-known fact that Montonen-Olive duality is related to the $\SL(2,\bZ)$ self-duality of type IIB string theory on $AdS_5\times S^5$, as realized by the near horizon limit $N$ D3 branes in a flat background. Different near-horizon geometries should generate similar string-derived S-dualities in other dual SCFTs provided that $\tau_{\rm IIB}$ remains a modulus, parameterizing the conformal manifold (or part of it). $\cN=1$ theories can be obtained by placing the D3 branes at a Calabi-Yau singularity; however, in this case the conformal manifold generically has dimension larger than one and the available SCFT duals are all related by Seiberg dualities, which can complicate efforts to distinguish S-duality phenomena from universality\footnote{Note, however, that Seiberg dualities are not always realized in string theory as universality, see, e.g.,~\cite{Strassler:2005qs}. For D3 branes at Calabi-Yau singularities they both play a role in the S-duality group as well as appearing as universalities in different gauge theory descriptions, and these two roles can be difficult to disentangle.} and to obtain a systematic understanding of the S-dualities.

As we will see, the S-dualities actually become simpler and easier to check when we include a particular class of orientifold planes that project out most of the moduli, leaving only $\tau_{\rm IIB}$ and a discrete set of parameters known as ``discrete torsion''. The prototypical examples of this are the $\cN=4$ $\SO(N)$ and $\Sp(N)$ gauge theories, which are realized by orientifolding AdS$_5 \times S^5$, corresponding to adding an O3 plane to the D3 brane stack. The connection between Montonen-Olive duality and the $\SL(2,\bZ)$ properties of discrete torsion was explained in~\cite{Witten:1998xy}. The present paper systematically generalizes this result to a large class of Calabi-Yau orientifold singularities.

Our analysis proceeds as follows. In~\S\ref{sec:geom} we review toric Calabi-Yau singularities and classify their orientifolds, identifying the broad class of isolated ``toric orientifolds'' that we are interested in. We then compute the cohomology groups which classify discrete torsion, along the lines of~\cite{Witten:1998xy}, and explain their properties and connection to the toric data. Appendices~\ref{app:toric} and \ref{app:Z2bilinear} provide supporting material on toric geometry and the $\bZ_2$ vector spaces that characterize discrete torsion, respectively. In~\S\ref{sec:NStorsion} we review the basics of how dual SCFTs can be engineered using NS5/D5 brane systems known as ``brane tilings'', focusing on the T-duality connection to the toric singularity. Using T-duality, we show how orientifolds appear in the brane tiling and identify the manifestation of NSNS discrete torsion as a discrete remnant of the moduli space of the unorientifolded tiling, providing compelling evidence that our prescription is correct. In~\S\ref{sec:RRtorsion} we discuss in detail the singular points in the moduli space of the brane tiling that the orientifold projection forces us to consider. We discuss how the multiple intersections of branes at these points lead to conformal matter in the quiver gauge theory, and engineer the SCFTs describing this matter using deconfinement, generalizing~\cite{dP1paper}. With a concrete understanding of these ``$\TO_k$'' CFTs in hand, we conjecture a natural dictionary relating the RR discrete torsion to their discrete parameters and show that it is self-consistent and mirrors the expected geometric properties of discrete torsion.

In the remainder of the paper we discuss examples---both those which exist in the literature as well as entirely new (and previously intractable) ones---to verify our claims and check the predicted S-dualities. In~\S\ref{sec:quad-CFTs} we briefly review and develop notation for the $\TO_2$ CFTs considered in~\cite{dP1paper}, which is sufficient for the examples we consider in~\S\ref{sec:examples}. By matching the superconformal index~\cite{Romelsberger:2005eg,Kinney:2005ej} (SCI) between elements of the same $\SL(2,\bZ)$ multiplet, we are able to check our S-duality predictions in detail,\footnote{One aspect of our predictions is matching the ranks of S-dual gauge groups by comparing their D3 charges. These can be obtained by elementary methods,
  see, e.g., \cite{Aspinwall:2004jr,Diaconescu:2006id} and the examples worked out in \cite{dualities2,dP1paper} using these techniques. We later propose a formula~\eqref{eqn:QD3} to obtain $Q_{\rm D3}$ directly from the brane tiling. The predicted rank relation is easily verified by 't Hooft anomaly matching.} verifying the discrete torsion dictionary that we have developed. Examples of these indices are given in appendix~\ref{app:SCI}, supplementing index computations in previous literature~\cite{dualities1,dP1paper}. We summarize our conclusions in~\S\ref{sec:conclusions}.

Note that, in the interest of saving space and limiting the broad scope of the present paper, our field theory analysis will not dwell on the some of the subtle but important distinctions highlighted above. Rather than constructing the cusps by a flow or series of flows from a free fixed point, we will work in the effective field theory description. This can be made precise by turning on the couplings of the effective field theory in a particular order to reach the cusp in a series of flows, as in the mass-deformed $\cN=4$ example above, but we will not do so, deferring such analyses to future work~\cite{SdualityFuture}. The effective field theory description is sufficient to compute the global symmetries, 't Hooft anomalies and the superconformal index of the infrared SCFT (all of which are unchanged by the exactly marginal operator) if we assume that there are no accidental symmetries along the flow and if chiral symmetry breaking does not occur. In at least some cases, this can be checked explicitly for sufficiently large $N$~\cite{SdualityFuture}. We also will not check explicitly that the SCFT has the expected fixed line (parameterized by $\tau_{\rm IIB}$) (see, e.g.,~\cite{Imamura:2007dc,Ardehali:2014zfa}). This can be checked in examples~\cite{dualities1, SdualityFuture}, and relates to the existence of a holomorphic coupling which is neutral under all the spurious flavor symmetries~\cite{Leigh:1995ep}.

Using more sophisticated field-theoretic tools, there are likely to be other, stronger checks of the proposed S-dualities that can be performed. For example, the S-duality group can be modified in an important way by the global structure of the gauge group (this is subtle for orientifolds, see~\cite{Garcia-Etxebarria:2015wns,Aharony:2016kai}) and the spectrum of line operators, as in~\cite{Aharony:2013hda}. Similarly, the $\mathcal{N}=1$ S-duality group may act non-trivially on half-BPS surface defects in a calculable way~\cite{Davide:2015}. In the present paper, we focus on local properties of the SCFTs, deferring a finer-grained study along these lines for the future.

\section{The geometry of toric orientifolds} \label{sec:geom}

In this section, we describe in detail a broad class of orientifold
geometries. In subsequent sections we will describe the theories
arising from branes probing these singularities.

\subsection{Toric Calabi-Yau singularities} \label{subsec:toricSing}

Consider space-time filling D3 branes probing $\bR^{3,1}\times Y_6$ at
point $p$ in $Y_6$.  In the absence of background flux, the infrared
behavior of the D3 brane world-volume gauge theory depends only on the
local geometry in a neighborhood $Y_p \subset Y_6$ of $p$. If $Y_6$ is
smooth at $p$, then the local geometry is $\bR^6$, and the gauge
theory flows to an $\mathcal{N}=4$ fixed point in the infrared. If
$Y_6$ is singular at $p$, however, the infrared behavior of the gauge
theory will depend on the nature of the singularity. We focus on the
case where $Y_p$ is toric and Calabi-Yau. In this case, the infrared
behavior of the world-volume gauge theory is that of a quiver gauge
theory---or more generally any one of a collection of Seiberg-dual
quiver gauge theories---which can be constructed explicitly using
brane-tiling techniques, as reviewed in~\S\ref{sec:NStorsion}.

\begin{figure}
  \centering
  \begin{subfigure}{0.31\textwidth}
    \centering
    \includegraphics{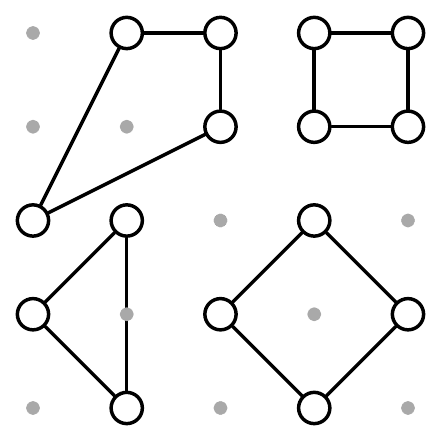}
    \caption{Examples of toric diagrams.}
    \label{sfig:toricDiags}
  \end{subfigure}
  \hspace{0.1\textwidth}
  \begin{subfigure}{0.31\textwidth}
    \centering
    \includegraphics{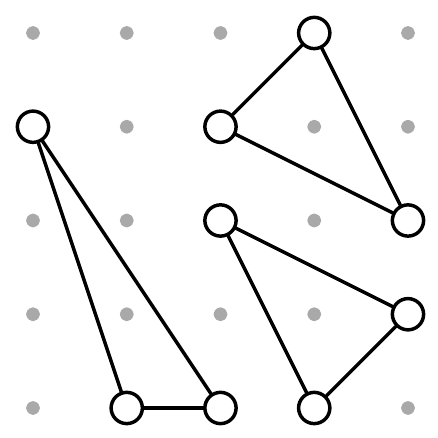}
    \caption{Equivalent toric diagrams.}
    \label{sfig:equivTDs}
  \end{subfigure}
  \caption{\subref{sfig:toricDiags}~Toric diagrams for (clockwise from
    upper left) the complex cone over $dP_1$, the conifold, the
    complex cone over $\bF_0$, and
    $\bC^2/\bZ_2\times\bC$. \subref{sfig:equivTDs}~Toric diagrams
    related by rotations and shears are equivalent. These all represent $\bC^3/\bZ_3$. 
    \label{fig:toricDiags}}
\end{figure}

As a Calabi-Yau toric variety,\footnote{See appendix~\ref{app:toric}
  for a brief review of toric geometry and notational conventions.}
$Y_p$ is described by a two-dimensional lattice polytope, the toric
diagram for $Y_p$. A few example toric diagrams are shown in
figure~\ref{fig:toricDiags}.
The toric geometry $Y_p$ can be recovered from the toric diagram as the vacuum manifold of a gauged linear sigma model (GLSM), where each of the $n$ corners $(u_i^1,u_i^2) \in \bZ^2$ of the toric diagram corresponds to a GLSM field $z_i$, and the gauge group is the subgroup $G \subseteq \U(1)^n$ which leaves the monomials
\be
\prod_i z_i\,,\;\; \prod_i z_i^{u_i^1}\,\mbox{, and}\;\; \prod_i z_i^{u_i^2}
\ee
invariant.

The simplest possible toric diagram, a minimal area triangle, corresponds to $\bC^3 \cong \bR^6$, whereas other triangular toric diagrams correspond to abelian orbifolds $\bC^3/\Gamma$, $\Gamma \subset \SU(3)$. In general, toric diagrams related by $\GL(2,\bZ)$ are equivalent (see figure~\ref{sfig:equivTDs}), whereas toric diagrams related by $\GL(2,\bR)$ transformations are orbifolds of the minimal area representative of their $\GL(2,\bR)$ equivalence class. For instance, parallelogram toric diagrams (e.g.\ the complex cone over $\bF_0$, pictured in figure~\ref{sfig:toricDiags}) correspond to orbifolds of the conifold.

A toric diagram corresponds to an \emph{isolated} singularity if its edges do not cross any lattice points. We will primarily be concerned with isolated singularities. A simple example of a non-isolated toric singularity is the orbifold $\bC^2/\bZ_2\times\bC$, see figure~\ref{sfig:toricDiags}.

\begin{figure}
  \centering
  \begin{subfigure}{0.31\textwidth}
    \centering
    \includegraphics{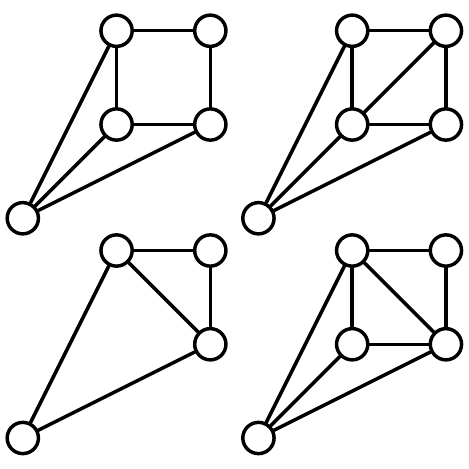}
    \caption{(Partial) resolutions of $dP_1$}
   \label{sfig:partialRes}
  \end{subfigure}
  \hfill
  \begin{subfigure}{0.31\textwidth}
    \centering
    \includegraphics{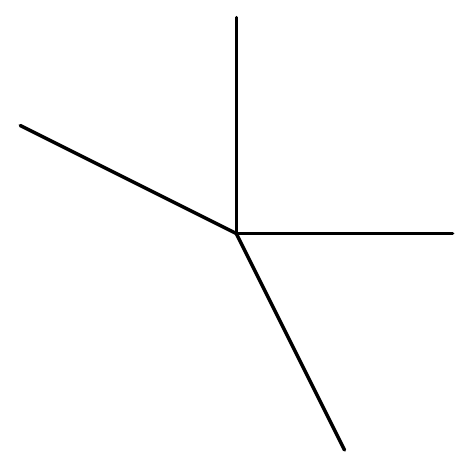}
    \caption{Web diagram for $dP_1$}
    \label{sfig:dP1web}
  \end{subfigure}
  \hfill
  \begin{subfigure}{0.31\textwidth}
    \centering
    \includegraphics{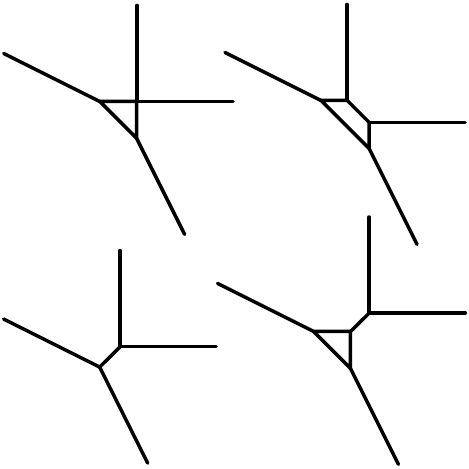}
    \caption{Resolved $dP_1$ webs}
    \label{sfig:dP1resolvedweb}
  \end{subfigure}
 \caption{\subref{sfig:partialRes}~Toric diagrams for some partial and complete resolutions of the $dP_1$ singularity. \subref{sfig:dP1web}~The web diagram for the affine $dP_1$ singularity. \subref{sfig:dP1resolvedweb}~Web diagrams for the resolutions of the $dP_1$ singularity shown in~\subref{sfig:partialRes}.\label{fig:webs}}
\end{figure}

Dividing the toric diagram into subpolytopes corresponds to partially resolving the toric singularity into component singularities represented by the subpolytopes.\footnote{See appendix~\ref{app:toric} for a more thorough discussion of these non-affine toric varieties.} If the subpolytopes share the same corners as the toric diagram, then this corresponds to introducing nonzero Fayet-Iliopoulos (FI) parameters into the GLSM. If the subpolytopes have additional corners (e.g.\ internal points in the toric diagram), then these must first be introduced as new GLSM fields, with a corresponding extension of the gauge group and the appearance of additional FI parameters.
For instance, there are two ways to divide a four-sided polytope into two triangles sharing the same corners. In the case of the conifold, these are the two possible resolutions, corresponding to different signs for the FI parameter. For larger four-sided toric diagrams, there are additional partial resolutions which introduce new GLSM fields. Some example resolutions of the $dP_1$ singularity are shown in figure~\ref{sfig:partialRes}.

Dividing the toric diagram into minimal area triangles corresponds to a \emph{complete} resolution of the singularity, resulting in a smooth toric variety. Each internal point of the toric diagram is now a compact toric divisor, whose fan (see~\S\ref{subapp:fans}) is built by extending the adjoining edges into rays separated by two dimensional cones.

Toric diagrams and their resolutions can alternately be described in terms of ``web diagrams'', as follows.\footnote{Web diagrams were introduced in \cite{Aharony:1997ju,Aharony:1997bh} with direct reference to networks of $(p,q)$ five-branes. Here we use them first abstractly and later to describe physical configurations of NS5 branes.} For each edge in the toric diagram, there is a dual edge in the web diagram orthogonal to it, with $k$ edges in the web diagram when the edge in the toric diagram crosses $k-1$ lattice points. The faces of the toric diagram are dual to vertices in the web diagram and vice versa, where the external edges (vertices) of the toric diagram are dual to edges (faces) in the web diagram which extend to infinity. Some examples of web diagrams are shown in figure~\ref{fig:webs}. Note that web diagrams are not drawn on a lattice, and the lengths of the internal edges are not fixed by the corresponding toric diagram.\footnote{In physical applications these lengths correspond to the K\"ahler moduli of the resolved toric singularity.} A web diagram is consistent if the lines have rational slopes $p/q$ and the sum of the vectors $(p,q)$ at each vertex is zero.\footnote{The overall sign of $(p,q)$ is not fixed by the slope of the edge $p/q$, but is chosen so that the vector $(p,q)$ points in the same direction as (rather than opposite to) the edge rooted at the vertex in question.} Web diagrams will prove useful when we discuss brane tilings below.

\subsection{Orientifolds of toric Calabi-Yau singularities} \label{subsec:toricOfolds}

We now consider D3 branes probing an orientifold geometry $X_6 = Y_6/\sigma$, where $\sigma: Y_6 \to Y_6$, $\sigma^2 = 1$ is an orientation-preserving involution and $Y_6$ ($X_6$) is the ``upstairs'' (``downstairs'') geometry. If the D3 branes sit at a point $p$ with $\sigma(p) \ne p$, then the geometry is locally unaffected by the orientifold, and the above description still applies, provided no background flux has been turned on. If $\sigma(p) = p$ but $p$ is a smooth point in $Y$, then the gauge theory flows to a $\mathcal{N}=4$, $\mathcal{N}=2$, or $\mathcal{N}=0$ fixed point, depending on whether the fixed locus of $\sigma$ is an O3 plane, O7 plane, or O5 plane, respectively. If, however, $p$ is a singular point in $Y$, the gauge theory flows to an infrared fixed point which depends both on $Y_p$ and on the action of $\sigma$ on $Y_p$.

As above, we focus on the case where $Y_p$ is toric and Calabi-Yau. For the orientifold to preserve $\mathcal{N}=1$ supersymmetry, we further require that $\sigma$ is holomophic with $\sigma^\star \Omega = -\Omega$,\footnote{We only consider O3/O7 orientifolds, since O5 planes are not mutually supersymmetric with D3 branes.} We classify the possible involutions, $z_i' = \sigma_i(z_j)$.
Taylor expanding about the singular point at the origin, we obtain:
\be
z_i' = S_i^j z_j +\frac{1}{2} S_i^{j k} z_j z_k+\ldots \;.
\ee
At leading order, the requirement $\sigma^2 = 1$ imposes $S_i^\ell S_\ell^j = \diag(\mu_1, \ldots, \mu_n) \in G$, so $S_i^j$ must be full rank. Thus, near $p$ the involution acts linearly on the homogenous coordinates $z_i$, and the higher-order terms will not affect the infrared behavior of the gauge theory. We therefore drop these terms, taking $\sigma$ to act linearly on the homogenous coordinates. 

\begin{figure}
  \centering
  \begin{subfigure}[t]{0.31\textwidth}
    \centering
    \includegraphics{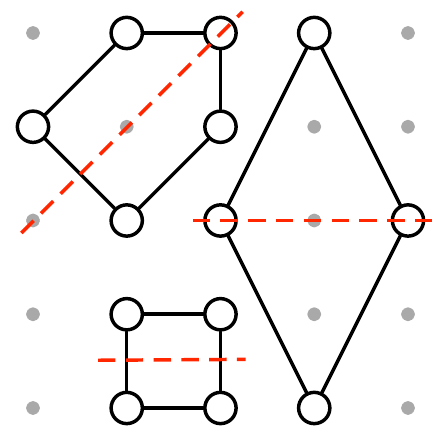}
    \caption{Non-toric involutions.}
    \label{sfig:nonToricInvol}
  \end{subfigure}
  \hfill
  \begin{subfigure}[t]{0.31\textwidth}
    \centering
    \includegraphics{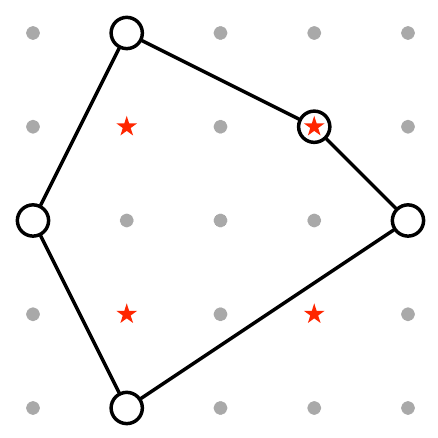}
    \caption{A toric involution.}
    \label{sfig:toricInvol}
  \end{subfigure}
  \hfill
  \begin{subfigure}[t]{0.31\textwidth}
    \centering
    \includegraphics{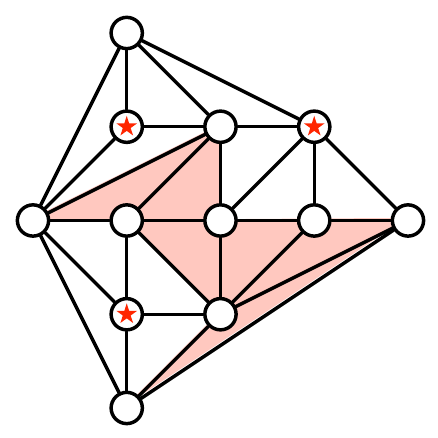}
    \caption{Complete resolution of a toric orientifold.}
    \label{sfig:toricInvolRes}
  \end{subfigure}
  \caption{\subref{sfig:nonToricInvol}~Examples of non-toric
    involutions for (clockwise from upper left) the $dP_2$
    singularity, $Y^{4,0}$, and the
    conifold. \subref{sfig:toricInvol}~A toric involution. The red
    stars indicate the even sublattice. \subref{sfig:toricInvolRes}~A
    complete resolution of the resulting toric orientifold. Each
    starred vertex represents an O7 plane, which wraps a compact
    (non-compact) divisor if the vertex lies in the interior (on the
    boundary) of the toric diagram. Each shaded triangle represents an
    O3 plane arising after complete resolution of the
    singularity.\label{fig:toricInvols}}
\end{figure}

Thus, $\sigma$ is a linear automorphism $\sigma \in \widehat{\Aut}(Y_p)$, as described in~\S\ref{subsec:automorph}. Since $\sigma$ has finite order, it is conjugate in $\widehat{\Aut}(Y_p)$ to 
\be
z_i' = P_i^j \tilde{\mu}_j z_j 
\ee
where the permutation matrix $P$ is induced by an automorphism of the toric diagram, $\Pi$, and $\tilde{\mu}_j$ is a phase factor. A generic toric diagram has no nontrivial automorphisms, in which case $P_i^j = \delta_i^j$ and $\sigma(T p) = T \sigma(p)$ for the natural torus action $p \to T p$, $T \in (\bC^{\star})^3$. Consequently, the quotient geometry $X_p \equiv Y_p/\sigma$ inherits a $(\bC^{\star})^3$ group action, and is therefore toric. We refer to this type of involution as a \emph{toric involution}, resulting in a \emph{toric orientifold} $X_p$. Non-toric involutions are possible for toric diagrams with nontrival $\widetilde{\Aut}(\Pi)$ (as defined in~\S\ref{subsec:automorph}), see for example figure~\ref{sfig:nonToricInvol}, but we consider only toric involutions henceforward.

Using~(\ref{eqn:Omega}), the conditions $\sigma^{\star} \Omega = - \Omega$ and $\sigma^2 = 1$ imply
\begin{align}
\prod_i \tilde{\mu}_i &= -1\,, & \prod_i \tilde{\mu}_i^{\vb{n}\cdot\vb{u}_i} &= \pm1\,.
\end{align}
for any $\vb{n} \in \bZ^2$. Since $\prod_i \tilde{\mu}_i = -1$, we can fix $\prod_i \tilde{\mu}_i^{\vb{n}\cdot\vb{u}_i} = 1$ by choosing the origin of $\bZ^2$ to lie on a particular even sublattice\footnote{In order to avoid confusion for the mathematically inclined reader, let us remark that we are slightly imprecise with language here, since the given subsets are not all subgroups of the original lattice. We will nevertheless keep on calling the relevant subsets ``sublattices'' henceforth for convenience.} $(2\bZ)^2 \subset \bZ^2$. Having done so, the points $u_i' \equiv (\vb{u}_i,2)$ generate the rays of the fan for $X_p$ (cf.~\S\ref{subapp:CYtoric}).

Thus, the toric involutions of $Y_p$ are in one-to-one correspondence with the even sublattices of the toric diagram, and there are four such involutions for any toric diagram.\footnote{Some of these involutions may be related by automorphisms of the toric diagram.} An example is shown in figure~\ref{sfig:toricInvol}.

For any corner of the toric diagram $\vec{u}_i$ that lies on the chosen even sublattice, the corresponding non-compact toric divisor $z_i = 0$ is wrapped by an O7 plane. When $Y_p$ is partially resolved, each subpolytope corresponds to a separate fixed point of $\sigma$, with an O7 plane on each toric divisor which lies on the even sublattice. In particular, a minimal area triangle corresponds to an O3 plane, unless any of its corners lies on the even sublattice, in which case it corresponds to a point on an O7 plane. An example resolution is shown in figure~\ref{sfig:toricInvolRes}.

\subsection{Discrete torsion and AdS/CFT}

So far, we have considered a class of toric Calabi-Yau geometries probed by $N$ D3 branes. The probe approximation is valid for $g_s N \ll 1$ for any value of $N$, and this is the regime in which the world-volume gauge theory is most-easily described. In the opposite limit, with $g_s N \gg 1$ but $g_s \ll 1$ so that $N \gg 1$, the physics is quite different. The backreaction of the D3 branes is strong in this limit, and the background Calabi-Yau is replaced by a warped Calabi-Yau metric and five-form flux emanating from a BPS horizon covering up the D3 branes. Near this horizon, the geometry is $AdS_5 \times Y_5$ for some compact five-manifold $Y_5$ describing the shape of the horizon.

By the AdS/CFT correspondence~\cite{Maldacena:1997re}, type IIB string theory in the near horizon region of the backreacted solution is dual to the infrared of the D3 brane world-volume gauge theory.\footnote{Strictly speaking, the near-horizon limit should correspond to the infrared superconformal fixed point of the field theory, but AdS/CFT has found many applications beyond exactly conformal field theories, see e.g.~\cite{Klebanov:2000hb}.} In the supergravity approximation, $g_s N \gg 1$ and $g_s \ll 1$, the low-energy excitations of this background are described by type IIB supergravity.

Because the near horizon geometry is dual to the world-volume gauge theory for $g_s N \gg 1$ and $g_s \ll 1$, different world-volume gauge theories must correspond to different near-horizon geometries. For D3 branes at Calabi-Yau singularities, the differences are rather obvious. By probing different singularities with D3 branes, we obtain distinct gauge theories. The local geometry near the singularity is also reflected in the near horizon region of the backreacted solution, which is the same up to warping and the introduction of five-form flux. The number of D3 branes (hence the rank of the gauge theory) is reflected in the number of five-form flux quanta which are turned on along $Y_5$.

The correspondence between the D-brane picture and the geometric picture is more subtle in the presence of orientifolds. The maximally supersymmetric case of D3 branes atop an O3 plane in a flat background was considered in~\cite{Witten:1998xy}.
 Placing $k$ D3 branes atop an O3$^+$ or O3$^-$, we obtain an $\mathcal{N}=4$ $\Sp(2k)$ or $\SO(2k)$ gauge group, respectively. In the latter case, we can add an extra ``half'' D3 brane to obtain an $\SO(2k+1)$ gauge group. In the worldsheet description, this extra brane is mapped to itself by the orientifold projection, and is therefore immobilized atop the O3$^-$; the combined object is commonly labeled the $\widetilde{\mathrm{O3}}^-$. The D3 charge of the O3$^{\pm}$ is $\pm\frac{1}{4}$, hence it is $-\frac{1}{4} + k$ for the $\SO(2k)$ stack and $-\frac{1}{4} + k + \frac{1}{2} = \frac{1}{4} + k$ for the $\SO(2k+1)$ and $\Sp(2k)$ stacks. The near-horizon geometry is $AdS_5 \times S^5/\bZ_2$ in all cases.
 
While the latter two cases appear to have the same gravity dual, in fact they differ by orientifold discrete torsion~\cite{Witten:1998xy,Hanany:2000fq}, i.e.\ flux in the non-trivial torsion component of the cohomology group which classifies the NSNS and RR two-form connections ($B_2$ and $C_2$, respectively). In the present example, this cohomology group is $H^3(S^5/\bZ_2,\tilde{\bZ})$, where $\tilde{\bZ}$ denotes the use of local coefficients (see e.g.~\cite{AT}) due to the orientifold action $B_2 \to -\sigma^\star B_2, C_2 \to -\sigma^\star C_2$.
We refer to (co)homology groups of this type as \emph{twisted} (co)homology groups.

A straightforward computation (see
e.g.~\cite{Witten:1998xy,Hanany:2000fq,dualities2}) gives
$H^3(S^5/\bZ_2,\tilde{\bZ}) \cong \bZ_2$, hence the classes $[F]$ and
$[H]$ --- for the connections $C_2$ and $B_2$, respectively --- each
assume two possible values, which we denote $0, 1 \in \bZ_2$ with
$1+1\equiv 0\pmod2$, for a total of four choices of discrete
torsion. It was argued in~\cite{Witten:1998xy} that the case
$([F],[H]) = (0,0)$ corresponds to the $\SO(2k)$ stack, whereas
$([F],[H]) = (1,0)$ and $([F],[H]) = (0,1)$ correspond to the
$\SO(2k+1)$ and $\Sp(2k)$ stacks, respectively. This explains the
Montonen-Olive duality between $\SO(2k+1)$ and $\Sp(2k)$ (as well as
the self-duality of $\SO(2k)$) as a consequence of S-duality in string
theory, which exchanges $[F]$ and $[H]$.

\begin{figure}
\begin{center}
\includegraphics{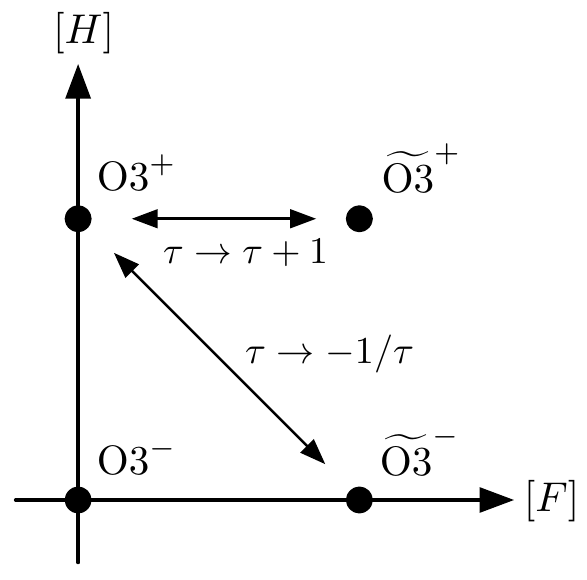}
\end{center}
\caption{The four types of O3 plane correspond to different choices of RR and NSNS discrete torsion ($[F]$ and $[H]$, respectively) in the gravity dual, where the action of $\SL(2,\bZ)$ on the O3 planes can be inferred from the known action on the RR and NSNS two form connections.
\label{fig:O3torsion}}
\end{figure}

The fourth case, $([F],[H]) = (1,1)$, is related to  $([F],[H]) = (0,1)$ by a shift $\tau \to \tau+1$ with $\tau\equiv C_0 + \frac{i}{g_s}$, which takes $[F]\to [F]+[H]$. Thus, it is perturbatively equivalent to the $\Sp(2k)$ theory, with the same gauge group but a different non-perturbative spectrum for any fixed $C_0$. The $([F],[H]) = (1,1)$ O3 plane is commonly labeled the $\widetilde{\mathrm{O3}}^+$. The $\widetilde{\mathrm{O3}}^+$ is self-dual under S-duality $(\tau\to-1/\tau)$ but forms a triplet with the O3$^+$ and $\widetilde{\mathrm{O3}}^-$ under the full $\SL(2,\bZ)$ of type IIB string theory, generated by $\tau \to \tau+1$ and $\tau \to-1/\tau$. The possible choices of discrete torsion, the corresponding O3 planes, and their transformation under $\SL(2,\bZ)$ are illustrated in figure~\ref{fig:O3torsion}.

Note that non-trivial discrete torsion does not affect the three-form fluxes $F_3$ and $H_3$. In particular, the de Rham cohomology class of $F_3$ ($H_3$) is the image of $[F]$ ($[H]$) under the natural map $H^3(X_5, \tilde{\bZ}) \to H^3(X_5, \tilde{\bR})$ which takes $\bZ \to \bR$ and $\bZ_n \to 0$ for each factor of the integral cohomology group. $H^3(X_5, \tilde{\bR})$ is trivial, so the three-form fluxes vanish in the absence of sources (such as wrapped five-branes).

As shown in~\cite{dualities2}, this discussion is readily generalized
to orientifolds of a large class of abelian orbifolds
$\mathbb{C}^3/\Gamma$, where $\Gamma \subset \SU(3)$ and we assume
that $\tilde{\Gamma} \equiv \Gamma \cup \sigma \Gamma$ acts freely on
$S^5$, so that the near-horizon geometry
$AdS_5 \times S^5/\tilde{\Gamma}$ is smooth and free of orientifold
fixed points. In these cases, we find
$H^3(S^5/\tilde{\Gamma}, \tilde{\bZ}) \cong \bZ_2$, as before, where
in general $\Gamma = \bZ_{2n+1}$ given our assumptions. The CFT dual
is an $\mathcal{N}=1$ quiver gauge theory of the form $\SO\times \SU^n$ or $\Sp\times \SU^n$, where the spectrum of chiral superfields and relative ranks of the gauge group factors depend on the orbifold in question, and the choice of $\SO(2k)\times \SU^n$, $\SO(2k+1)\times \SU^n$, or $\Sp(2k)\times \SU^n$ corresponds to the choice of discrete torsion, exactly as above.

S-duality covariance then implies a nontrivial duality between the
$\SO(2k+1)\times \SU^n$ and $\Sp(2\tilde{k})\times \SU^n$
theories,\footnote{Unlike the $\mathcal{N}=4$ case, in general
  $\tilde{k} \ne k$. Rather, the difference $k-\tilde{k}$ depends on
  the orbifold in question and can be fixed e.g.\ by matching the D3
  charge of the prospective dual theories.} analogous to
Montonen-Olive duality of $\mathcal{N}=4$ theories. This duality was
explored in~\cite{dualities1,dualities2,Bianchi:2013gka} using anomaly
matching for an infinite class of orbifolds and by matching the
superconformal index for the simple orbifold $\bC^3/\bZ_3$
in~\cite{dualities1}. The latter check was performed by expanding the
index in a fugacity $t$ corresponding to the superconformal R-charge
and computing the power series coefficients up to a fixed order using
a computer. Updated computations are presented in
\cite{dP1paper},\footnote{These updated results are not included in the text of~\cite{dP1paper}, but are available in the Mathematica file attached to the arXiv version.} showing perfect agreement up to a
high order in the series expansion, a very nontrivial check that the
expected S-duality is in fact present.

In~\cite{dP1paper}, Calabi-Yau cones over del Pezzo surfaces were considered. These are a relatively simple class of singularities which include the orbifold $\bC^3/\bZ_3$ which is a complex cone over $dP_0 = \bP^2$, the conifold orbifold $\mathcal{C}/\bZ_2$ which is a complex cone over $\bF_0 = \bP^1 \times \bP^1$, and three other toric singularities, as well as five non-toric singularities. Choosing an orientifold involution which fixes the del Pezzo surface, the twisted cohomology group is $H^3(X_5, \tbZ) \cong \bZ_2^{k+1}$ for the complex cone over $dP_k$, and $H^3(X_5, \tbZ) \cong \bZ_2^2$ for the cone over $\bF_0$.

\begin{figure}
  \centering
    \includegraphics{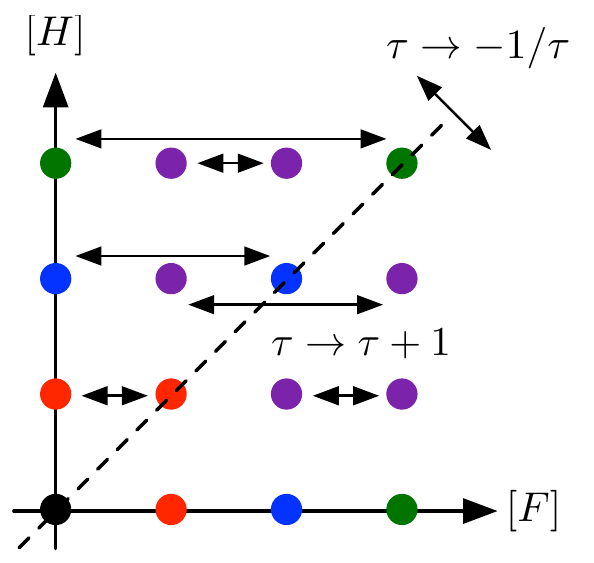}
    \caption{The action of $\SL(2,\bZ)$ on $[H], [F]$ in the case where $H^3(X_5,\tbZ) \cong \bZ_2^2$. The sixteen choices of discrete torsion fall into a singlet, three triplets, and a sextet under $\SL(2,\bZ)$, colored black, red/green/blue, and purple, respectively.} \label{fig:dP1torsion}
\end{figure}

The larger discrete torsion group gives rise to new phenomena. The case $H^3(X_5,\tbZ) \cong \bZ_2^2$, applicable to the complex cones over $dP_1$ and $\bF_0$, is illustrated in figure~\ref{fig:dP1torsion}. There are now five $\SL(2,\bZ)$ multiplets, including a sextet. The latter manifests as a triality in the CFT dual, since the six choices of torsion are related in pairs by the action of $\tau \to \tau+1$, which is a perturbative equivalence.
The same $\SL(2,\bZ)$ multiplets appear for yet larger discrete torsion groups. For $H^3(X_5, \tbZ) \cong \bZ_2^{p}$, there is one singlet $([H],[F])=(0,0)$ as well as $2^{p} - 1$ triplets $([H],[F]) \in \{(0,\alpha),(\alpha,0),(\alpha,\alpha)\}$ and $\frac{(2^p - 1)(2^{p-1} - 1)}{3}$ sextets $([H],[F]) \in \{(\alpha, \beta), (\alpha,\alpha+\beta), (\beta,\alpha), (\beta,\alpha+\beta), (\alpha+\beta,\alpha), (\alpha+\beta,\beta)\}$, where $\alpha \ne \beta$ and $\alpha, \beta \ne 0$.

The complex cone over $dP_1$ and its CFT duals were thoroughly
explored in~\cite{dP1paper}. Some of these duals
involve novel strongly coupled physics, which we review
in~\S\ref{sec:quad-CFTs}. The main purpose of our present work is to
generalize these results to arbitrary toric orientifolds, assuming
only that the horizon is smooth and free of orientifold planes. Just
as the size of the discrete torsion group grows with $k$ for the
complex cones over $dP_k$, the complexity of the analysis in the toric
case grows with the number of sides of the toric diagram. Our general
analysis in~\S\ref{sec:geom}--\S\ref{sec:RRtorsion} applies to toric
diagrams with an arbitrary number of sides, but in the present paper
we limit our explicit examples, in \S\ref{sec:examples}, to toric
diagrams with at most five sides, as some additional technical details
are required to describe the CFT duals for certain choices of discrete
torsion in toric diagrams with six or more sides. We give a systematic
prescription for analyzing such configurations in
\S\ref{sec:RRtorsion}, but we defer a full description of the more
technical details (which is straightforward but technically involved)
of these more complicated cases to a future work.

In the following sections we will give a complete characterization of the discrete torsion group for these toric orientifolds. To do so, it is most convenient to work with the Poincare dual $H_2(X_5,\tilde{\bZ}) \cong H^3(X_5,\tilde{\bZ})$. This is homology group classifies domain walls in AdS$_5$ arising from wrapped five branes. An torsion two-cycle $A \in H_2(X_5,\tilde{\bZ})$ corresponds to the discrete flux $\tilde{A} \in H^3(X_5,\tilde{\bZ})$ induced upon crossing a five-brane wrapping $A$, where to generate $[H]$ torsion we wrap NS5 branes and to generate $[F]$ torsion we wrap D5 branes. By dropping a five-brane wrapping $A$ through the horizon, we obtain the gravity dual with torsion $\tilde{A}$ from that with trivial torsion, as in~\cite{Hyakutake:2000mr}.

In all the examples we consider in this paper the de Rham cohomology
group $H^3(X_5,\tilde{\bR})$ is trivial. Thus, $F_3 = H_3 = 0$ far
from any sources---such as the wrapped five-branes discussed
above---and consequently the discrete torsion admits an alternate
description as a set of quantized Wilson lines. Consider a D1 brane
wrapped on a $\bZ_2$ torsion two-cycle $\Sigma$. The Chern-Simons
action $S_{CS} \propto \oint_\Sigma C_2$ admits an exact shift
symmetry $S_{CS} \to S_{CS} + 2\pi k$ corresponding to large gauge
transformations of $C_2$. Because $2\Sigma$ is homologically trivial,
whereas $S_{CS}$ is homologically invariant for $F_3 = H_3 = 0$ there
are two distinct possibilities, $S_{CS} = 0$ or $S_{CS} = \pi$.
Whether $e^{i S_{CS}} = \pm1$ depends on the $C_2$ gauge bundle, hence
on the torsion class $[F]$. Similar considerations apply to the
Chern-Simons term in the F-string worldsheet action and the torsion
class $[H]$.

Since there are exactly as many torsion two-cycles to wrap F and
D-strings on as there are choices of discrete torsion, one might
imagine that there is a one-to-one correspondence between these
quantized Wilson lines and the torsion classes $[F]$ and $[H]$. We
show that this is the case for the class of examples considered in our
paper in~\S\ref{sec:intersection}. As such, wrapped five branes and
wrapped strings provide complementary descriptions of discrete torsion. In the former case,
five-branes wrapping a torsion two-cycle correspond to a certain
torsion class: that generated by crossing the resulting domain wall
starting with trivial torsion. In the latter case, strings wrapping a
torsion two-cycle define a linear map on the torsion class, and for
each linear map there is a corresponding wrapped string. These dual
notions of discrete torsion will be useful in~\S\ref{sec:NStorsion} and~\S\ref{sec:RRtorsion}
when we discuss the dictionary between the geometry and the CFT dual.

\subsection{Discrete torsion for toric orientifolds} \label{subsec:torsioncalc}

In this subsection, we compute the twisted homology groups for toric
orientifolds---defined in~\S\ref{subsec:toricOfolds}---for which
the choice of involution $\sigma$ corresponds to the choice of one of
four even sublattices of the $\bZ^2$ lattice containing the toric
diagram, see figure~\ref{sfig:toricInvol}. We further assume a smooth
horizon free from orientifold fixed points. Equivalently, as explained
in~\S\ref{subsec:toricSing}--\ref{subsec:toricOfolds}, the sides of
the toric diagram must not cross lattice points, and the corners must
not lie on the designated even sublattice.

Despite these restrictions, this class of orientifolds contains a large (in fact, infinite) collection of interesting examples. For instance, it encompasses the special case of O3 planes in a flat background (an orientifold of $\bC^3$) as well as all the orbifold examples considered in~\cite{dualities1,dualities2,Bianchi:2013gka} and all of the nonorbifold examples considered in~\cite{dualities1}. Moreover, the infinite family of geometries $Y^{p,q}$~\cite{Gauntlett:2004yd,Martelli:2004wu} admit one such involution when either $p$ or $p-q$ is even, and two when both are even.

To compute the twisted homology groups, we employ the long exact sequence (see e.g.~\cite{AT}):
\begin{equation} \label{eqn:longexactTwisted}
\ldots \longrightarrow H_i(X,\tilde{\bZ}) \longrightarrow H_i(Y,\bZ) \;\overset{p_\ast^i}{\longrightarrow}\; H_i(X,\bZ) \longrightarrow H_{i-1}(X,\tilde{\bZ}) \longrightarrow \ldots
\end{equation}
where $X$ and $Y$ denote the downstairs and upstairs geometries, respectively, and $p_\ast^i$ is induced by the projection $p: Y \to X$. In the examples we encounter below, $p_\ast^i$ is always injective,\footnote{This is not true in general. For instance, the non-toric involution of the conifold $(z_1,z_2,z_3,z_4) \to (-z_1,-z_2,-z_3,z_4)$ (where $\sum_i z_i^2 = 0$)
gives $H_{\bullet}(\frac{S^3\times S^2}{\bZ_2},\bZ) \cong \{\bZ,\bZ_2,\bZ_2,\bZ_2, 0 ,\bZ \}$ ($\bullet = 0,1,2,\ldots$) whereas $H_{\bullet}(S^3\times S^2,\bZ) \cong \{\bZ,0,\bZ,\bZ, 0 ,\bZ \}$, so neither $p_\ast^2$ nor $p_\ast^3$ is injective.}
hence the long exact sequence breaks into short exact sequences
\begin{equation} \label{eqn:shortexactTwisted}
0 \longrightarrow H_i(Y,\bZ) \;\overset{p_\ast^i}{\longrightarrow}\; H_i(X,\bZ) \longrightarrow H_{i-1}(X,\tilde{\bZ}) \longrightarrow 0
\end{equation}
and we can compute the twisted homology groups given the homology groups of $X_5$ and $Y_5$ and the maps $p_\ast^i$.

\begin{figure}
  \centering
  \begin{subfigure}{0.31\textwidth}
    \centering
    \includegraphics{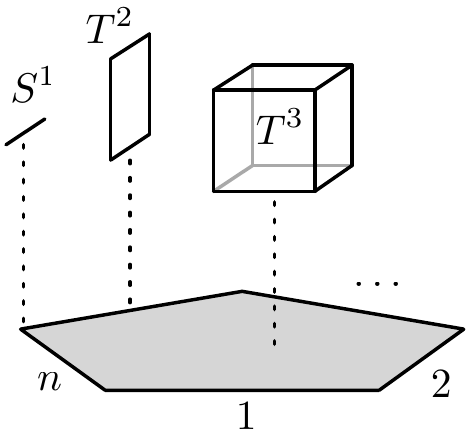}
    \caption{Horizon as a torus fibration.}
    \label{sfig:horizonfibration}
  \end{subfigure}
  \hfill
  \begin{subfigure}{0.31\textwidth}
    \centering
    \includegraphics{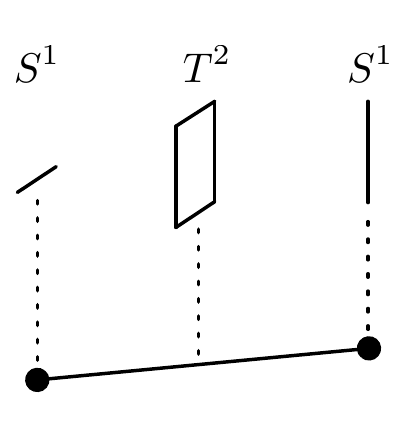}
    \caption{Torus fibration of $S^3$.}
    \label{sfig:S3fibration}
  \end{subfigure}
  \hfill
  \begin{subfigure}{0.31\textwidth}
    \centering
    \includegraphics{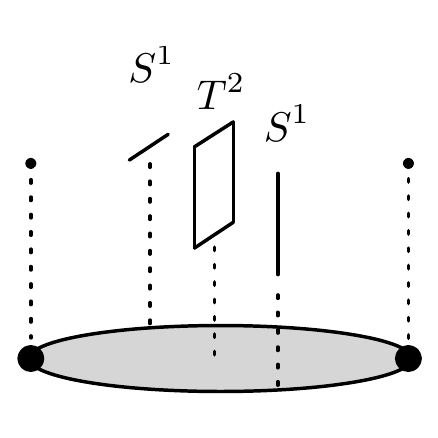}
    \caption{Torus fibration of $S^4$.}
    \label{sfig:S4fibration}
  \end{subfigure}
\caption{\subref{sfig:horizonfibration}~The horizon $X_5$ ($Y_5$) is a $T^3$ fibration over an $n$-sided polygon, where $|z_i| \to 0$ along the $i$th edge. The fiber, corresponding to the orbit of the $\U(1)^3$ isometry, degenerates to $T^2$ along each edge and to $S^1$ at each corner. \subref{sfig:S3fibration}~A $T^2$ fibration over an interval, homeomorphic to $S^3$. \subref{sfig:S4fibration}~A $T^2$ fibration over a disk, equivalent to a fibration of~\subref{sfig:S3fibration} over an interval contracted to a point at either end, hence homeomorphic to $S^4$.\label{fig:torusfibration}}
\end{figure}

Consider an $n$-sided toric diagram
with corners labeled $1,2,\ldots, n$ counterclockwise around the perimeter, and corresponding homogenous coordinates $z_i$, $i=1,\ldots, n$. The affine toric variety $X_p$ $(Y_p)$ is a real cone over the horizon $X_5$ $(Y_5)$.
As explained in~\S\ref{subsec:horizons}, $X_5$ $(Y_5)$ can be thought of as an orbifold of a $T^3$ fibration over an $n$-sided polygon with edges labelled $1,\ldots, n$, where $|z_i| \to 0$ along the $i$th edge. The $T^3$ fibration is parameterized by gauge-invariant combinations of the phases of the homogenous coordinates, whereas the magnitudes $|z_i|$ correspond to the position along the base. Thus, the $T^3$ degenerates to a $T^2$ along each edge, and to a $T^1$ at each vertex, see figure~\ref{sfig:horizonfibration}.

We consider the subspace $V_i \equiv \{z_i = 0\} \subset
X_5$ corresponding to the
$i$th edge of the polygon. Since this edge only intersects the edges
$i\pm 1$, we have $z_j \ne 0$ for $j\notin
\{i,i\pm1\}$. Moreover, the corners
$i-1,i,i+1$ are not colinear in the toric diagram (otherwise
$i$ is not a corner), so there exists a partial gauge fixing which
sets $\arg(z_j) = 0$ for $j\notin \{i,i\pm1\}$.

\begin{figure}
  \centering
  \begin{subfigure}{0.48\textwidth}
    \centering
    \includegraphics{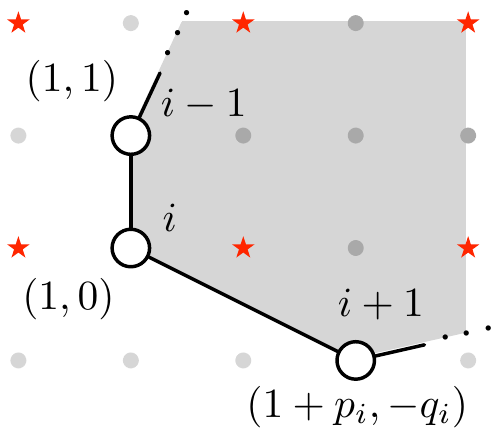}
    \caption{Toric diagram near $i$th corner.}
    \label{sfig:toricCornerI}
  \end{subfigure}
  \hfill
  \begin{subfigure}{0.48\textwidth}
    \centering
    \includegraphics{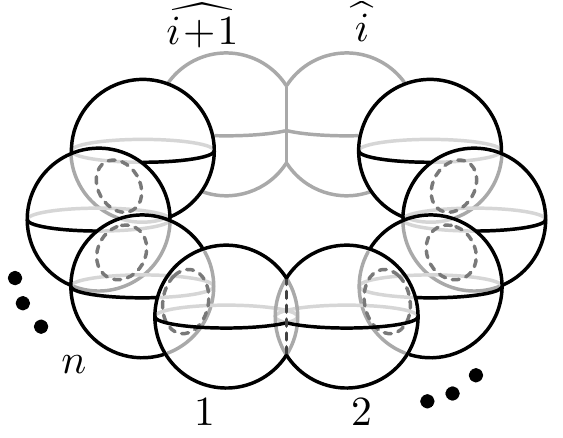}
    \caption{The space $X_3^{(i,i+1)}$.}
    \label{sfig:X3chain}
  \end{subfigure}
\caption{\subref{sfig:toricCornerI}~The form of the toric diagram near the $i$th corner, after an $\SL(2,\bZ)\ltimes (2\bZ)^2$ transformation. \subref{sfig:X3chain}~The space $X_3^{(i,i+1)}$ ($Y_3^{(i,i+1)}$) is a chain of lens spaces $V_j$ glued together along their torsion one-cycles with the spaces $V_i$, $V_{i+1}$ omitted.}
\end{figure}

To determine the residual gauge symmetry, we perform an $\SL(2,\bZ) \ltimes (2\bZ)^2$ transformation on the toric diagram---fixing the even sublattice specifying the involution to contain the origin---to express it in a convenient form. In general $\vb{u}_{i-1,i} \equiv \vb{u}_{i-1} - \vb{u}_i = (a,b)$, where $\gcd(a,b)=1$ for an isolated singularity. Thus, we can take $\vb{u}_{i-1,i} = (0,1)$ using $\SL(2,\bZ)$, so that $\vb{u}_i = (c,d)$ and $\vb{u}_{i-1} = (c,d+1)$ for odd $c$, since by assumption neither corner lies on the even sublattice. We are free to assume that $d$ is even, since otherwise we can map $d \to d+c$ by a shear transformation which leaves $\vb{u}_{i-1} - \vb{u}_i$ invariant. Fixing $\vb{u}_{i-1} = (1,1)$ and $\vb{u}_i = (1,0)$ after a $(2\bZ)^2$ translation, we have
$\vb{u}_{i+1} = (1+p_i,-q_i)$ for $p_i>0$, $\gcd(p_i,q_i)=1$, where $p_i$ is twice the area of the triangle formed between the corners $i,i\pm1$. Note that $q_i$ must be odd to satisfy $\gcd(p_i,q_i)=1$ with $p_i+1$, $q_i$ not both even, so $\gcd(2p_i,q_i)=1$, where we can fix $0< q_i < 2p_i$ by a shear transformation. The transformed toric diagram is shown in figure~\ref{sfig:toricCornerI}.

The residual gauge symmetry, $\hat{G}$, satisfies the constraints
\be
\mu_{i-1}^2 \mu_i^2 \mu_{i+1}^2 = \mu_{i-1}\mu_i \mu_{i+1}^{p_i+1} =
\mu_{i-1} \mu_{i+1}^{-q_i} = 1\, .
\ee
Thus, $\hat{G} \cong \bZ_{2p_i}$ is generated by $(\mu_{i-1},\mu_i,\mu_{i+1}) = (\omega_{2p_i}^{q_i}, \omega_{2p_i}^{p_i-q_i-1}, \omega_{2p_i})$, where $\omega_{k} \equiv e^{2\pi i/k}$. Therefore, the space $V_i$ is a $\bZ_{2p_i}$ orbifold of a $T^2$ fibration over an interval, where the fiber is parameterized by $(\arg(z_{i-1}), \arg(z_{i+1}))$ in the gauge fixing $\arg(z_j) = 0$, $j\notin \{i,i\pm1\}$, and the $A$ ($B$) cycle shrinks to zero size at the left (right) end of the interval. This fibration is homeomorphic to $S^3$, see figure~\ref{sfig:S3fibration}, so we obtain the lens space $V_i \cong L(2p_i,q_i)$. Neighboring subspaces $V_i$, $V_{i+1}$ intersect along their common torsion one-cycle, $V_i \cap V_{i+1} = \{z_i = z_{i+1} = 0\} \cong S^1$.

We now consider $X_5/X^{(i,i+1)}_3$, where $X_3^{(i, i+1)} \equiv \bigcup_{j\ne i,i+1} V_j$. Away from the locus $X_3^{(i,i+1)}$, $z_j \ne 0$ for $j \ne i,i+1$, so we can gauge fix $\arg(z_j)=0$, $j\ne i,i\pm1$, as before, where by the above argument the residual $\bZ_{2p_i}$ gauge symmetry acts transitively in the $z_{i-1}$ plane, so we can further fix $0 \le \arg(z_{i-1}) < \pi/p_i$. We view this as an $S^1$ fibration of $\arg(z_{i-1})$ over a base space, which is itself a torus fibration over a disk. The latter is homeomorphic to $S^4$, see figure~\ref{sfig:S4fibration}, so the total space is $(S^4\times S^1) / (\{0\}\times S^1)$.

We use the excision theorem to relate the homology of $X_3^{(i,i+1)}$ with that of $X_5$ and $X_5/X_3^{(i,i+1)}$. In general, there is a long exact sequence
\be
\ldots \longrightarrow \tilde{H}_n(A) \longrightarrow \tilde{H}_n(X) \longrightarrow \tilde{H}_n(X/A) \longrightarrow \tilde{H}_{n-1}(A) \longrightarrow \ldots
\ee
where $\tilde{H}_n$ are the reduced homology groups and $A \subset X$ is a sufficiently well-behaved closed subspace of $X$, see e.g.~\cite{AT}. A straightforward application with $X= S^4\times S^1$ and $A= \{0\}\times S^1\subset X$ gives $\tilde{H}_\bullet(X_5/X_3^{(i,i+1)}) \cong \tilde{H}_\bullet((S^4\times S^1) / (\{0\}\times S^1)) \cong \{0,0,0,0,\bZ,\bZ\}$ ($\bullet = 0,1,2,\ldots$). Thus, by excision $\tilde{H}_n(X_3^{(i,i+1)}) \cong \tilde{H}_n(X_5)$ for $n=0,1,2$. Since the higher homology groups are determined by Poincare duality combined with the universal coefficient theorem, the homology of $X_5$ is completely determined by that of $X_3^{(i,i+1)}$ for any $i$.

As shown above and illustrated in figure~\ref{sfig:X3chain}, the space $X_3^{(i,i+1)}$ is a chain of lens spaces $V_j \cong L(2p_j,q_j)$, with adjacent spaces glued together along their torsion one-cycles. By a similar argument, the homology of $Y_5$ is related to that of $Y_3^{(i,i+1)} \subset Y_5$, where $Y_3^{(i,i+1)}$ is a chain of lens spaces $U_j \cong L(p_j,q_j)$.

To compute the homology of $X_3^{(i,i+1)}$ and $Y_3^{(i,i+1)}$, we employ the Mayer-Vietoris sequence:
\be
\ldots \longrightarrow H_n(A \cap B) \;\overset{\Phi_n}{\longrightarrow}\; H_n(A) \oplus H_n(B) \longrightarrow H_n(X) \longrightarrow H_{n-1}(A \cap B) \longrightarrow \ldots
\ee
where $X$ is the union of the interiors of $A, B \subset X$ and $\Phi_n$ is induced by the chain map $x \to (x,-x)$. We show inductively that $H_\bullet(\cA_r) \cong \{\bZ,\bZ_{\gcd(p_1,\ldots,p_r)}, \bZ^{r-1}, \bZ^r\}$ for a chain $\cA_r$ of $r$ lens spaces $L(p_j,q_j)$, $j=1,\ldots,r$, glued together in this fashion.

The case $r=1$ follows from the the known homology $H_\bullet(L(p,q)) \cong \{\bZ,\bZ_{p},0,\bZ\}$. Suppose that $H_\bullet(\cA_{r-1}) \cong \{\bZ,\bZ_{\gcd(p_1,\ldots,p_{r-1})}, \bZ^{r-2}, \bZ^{r-1}\}$. We take $A = \cA_{r-1}$ and $B = L(p_r,q_r)$ in the Mayer-Vietoris sequence,\footnote{Technically, to fufill the requirement that $\cA_r$ is the union of the interiors of $A$ and $B$, we must include inside $A$ a thin ring of $L(p_r,q_r)$ surrounding the gluing point and likewise for $B$, but these modified subspaces deformation retract to $\cA_{r-1}$ and $L(p_r,q_r)$, respectively, so the result is unchanged.} so that $A\cap B \cong S^1$ is the torsion one-cycle. Since $H_{2,3}(A\cap B) = 0$, we find $H_3(\cA_r) \cong \bZ^{r-1} \oplus \bZ \cong \bZ^r$. Moreover, the map $\Phi_1: \bZ \to \bZ_{\gcd(p_1,\ldots,p_{r-1})} \oplus \bZ_{p_r}$ has kernel $\lcm(\gcd(p_1,\ldots,p_{r-1}),p_r)\cdot \bZ \cong \bZ$, hence there is a short exact sequence
\be
0 \longrightarrow \bZ^{r-2} \oplus 0 \longrightarrow H_2(\cA_r) \longrightarrow \bZ \longrightarrow 0 \,.
\ee
Since $\bZ$ is free, the sequence splits and $H_2(\cA_r) \cong \bZ^{r-1}$. The map $\Phi_0: \bZ \to \bZ \oplus \bZ$ is injective, hence there is a short exact sequence
\be
0 \longrightarrow \bZ_{\lcm(\gcd(p_1,\ldots,p_{r-1}),p_r)} \;\overset{\hat{\Phi}_1}{\longrightarrow}\; \bZ_{\gcd(p_1,\ldots,p_{r-1})} \oplus \bZ_{p_r} \longrightarrow H_1(\cA_r) \longrightarrow 0 \,.
\ee
Using the known form of $\hat{\Phi}_1$, we conclude that $H_1(\cA_r) \cong \bZ_{\gcd(\gcd(p_1,\ldots,p_{r-1}),p_r)} \cong \bZ_{\gcd(p_1,\ldots,p_r)}$, whereas $H_0(\cA_r) \cong \bZ$ follows trivially from the path-connectedness of $\cA_r$.

The generators of $H_3(\cA_r)$ are the lens spaces themselves, whereas $H_1(\cA_r)$ is generated by the torsion one-cycle of any one of them (all being equivalent), and elements of $H_2(\cA_r)$ are linear combinations $a_i \chi_i$ of two chains $\chi_i$ swept out by contracting a loop wrapped $p_i$ times around the torsion one-cycle to a point in $L(p_i,q_i)$, such that $\sum_i a_i p_i = 0$.\footnote{A minimal set of generators can be found by choosing an element $\Lambda^i_{\; j} \in \GL(r,\bZ)$ such that $\sum_j \Lambda^i_{\; j} p_j = (\gcd(p_1,\ldots,p_r),0,\ldots)$. The generators are then $\sum_j \Lambda^i_{\; j} \chi_j$ for $i=2,\ldots,r$.}

With a complete understanding of the homology of $H_\bullet(X_3^{(i,i+1)}) \cong \{ \bZ, \bZ_{2p}, \bZ^{n-3},\bZ^{n-2} \}$ and $H_\bullet(Y_3^{(i,i+1)}) \cong \{ \bZ, \bZ_{p}, \bZ^{n-3},\bZ^{n-2} \}$, we can reconstruct the homology of $X_5$ and $Y_5$ using Poincare duality and the universal coefficient theorem. We find:
\be
H_\bullet(X_5) = \{ \bZ, \bZ_{2p}, \bZ^{n-3}, \bZ_{2p}\oplus\bZ^{n-3}, 0,\bZ\} \;,\; H_\bullet(Y_5) = \{ \bZ, \bZ_{p}, \bZ^{n-3}, \bZ_{p}\oplus\bZ^{n-3}, 0,\bZ\} \,.
\ee
By excision, the generators of $H_{0,1,2}(X_5)$ are the same as those for $X_3^{(i,i+1)} \subset X_5$, whereas $H_3(X_5)$ is generated by the lens spaces $V_j$, $j\ne i, i+1$ subject to a single linear relation. 
Explicitly, the $V_j$ lift to torus-invariant divisors $z_j =0$ in $X_p$, which obey the linear relations:
\be \label{eqn:divisorclass}
\sum_j 2 [j] = 0 \;,\; \sum_j (\vb{u}_j\cdot\vb{n}) [j] = 0 \,,
\ee
for any $\vb{n} \in \bZ^2$, where $[j]$ denotes the divisor class of $z_j=0$. These relations are inherited by the three cycles, $V_j$, where a single relation remains after eliminating $[i]$, $[i+1]$ using the other conditions.

It is now straightforward to compute the twisted homology groups of $X_5$ using the long exact sequence~(\ref{eqn:longexactTwisted}). In particular, $p_\ast^3$ maps $U_i \to 2 V_i$, hence it takes $\bZ \to 2\bZ \subset \bZ$ and $\bZ_{p} \to \bZ_p \subset \bZ_{2p}$ for each factor in $H_3$. Likewise, $p_\ast^5$ takes $\bZ \to 2\bZ \subset \bZ$ and $p_\ast^1$ takes $\bZ_{p} \to \bZ_p \subset \bZ_{2p}$, whereas $p_\ast^{0,2}$ take $\bZ \to \bZ$ for each $\bZ$ factor. Thus, $p_\ast^i$ is always injective, and the long exact sequence breaks into short exact sequences~(\ref{eqn:shortexactTwisted}). We readily obtain
\be \label{eqn:twistedhomology}
H_\bullet(X_5,\tbZ) = \{ \bZ_2, 0, \bZ_2^{n-2}, 0, \bZ_2, 0\} \,,
\ee
where there is a $\bZ_2$ factor in $H_2(\tbZ)$ for each lens space in $X_3^{(i,i+1)}$. Notice that this agrees with the results of~\cite{dualities2} for the orbifold case, $n=3$, and with~\cite{dP1paper} for complex cones over $dP_k$, $k=0,1,2,3$, which correspond to $n=k+3$ sided toric diagrams.

\subsection{The divisor basis and intersection form}
\label{sec:divisor-basis}
\label{sec:intersection}

The relations~(\ref{eqn:divisorclass}) allow us to construct a basis for $H_2(X_5,\tbZ) \cong \bZ_2^{n-2}$ without fixing a subspace $X_3^{(i,i+1)}$. Recall that the induced maps $p_\ast^i$ in~(\ref{eqn:longexactTwisted}) are all injective for toric involutions, as shown above. Hence, we have the short exact sequence~(\ref{eqn:shortexactTwisted}) relating $H_2(X_5,\tbZ)$ to the third homology groups of $X_5$ and $Y_5$, and the relations~(\ref{eqn:divisorclass}) are inherited by $H_2(X_5,\tbZ)$. In particular, denoting the twisted two cycle in $V_i$ by $\langle i \rangle$, we have
\be \label{eqn:twistedclass}
\sum_i (\vb{u}_i\cdot\vb{n}) \langle i \rangle = 0
\ee
where the result only depends on the parity of $(\vb{u}_i\cdot\vb{n})$ since $2  \langle i \rangle = 0$. This can be restated as:
\be \label{eqn:twistedlattice}
\sum_{i \in L} \langle i \rangle = \sum_{j \in L'} \langle j \rangle
\ee
where $L$ and $L'$ denote any pair of even sublattices not containing the origin. We recover the old basis by eliminating $\langle i \rangle,  \langle i+1 \rangle$ using~(\ref{eqn:twistedlattice}). This is always possible, as adjacent corners must occupy distinct even sublattices for an isolated singularity.

There is a natural intersection parity $H_p(X_d,\tbZ) \times H_{d-p}(X_d,\bZ) \to \bZ_2$ for $d$-dimensional $X_d = Y_d/\sigma$, defined by the parity of the number of transverse intersections between generic representatives. For toric orientifolds, the intersection parity $H_2(X_5,\tbZ) \times H_{3}(X_5,\bZ) \to \bZ_2$ has physical significance as follows.
 Consider the change in the quantized Wilson line on a D1 brane wrapping $\langle i \rangle$ as it passes a D5 brane wrapping $\langle j \rangle$.\footnote{Note that the Wilson line is only defined where $F_3 \simeq 0$, far from the D5 brane.} Since $\langle i \rangle$ lies entirely within $V_i$, this Wilson line corresponds to the discrete torsion class $H^3(V_i,\tbZ) \cong \bZ_2$ which forms part of the overall discrete torsion group $H^3(X_5,\tbZ)  \cong \bZ_2^{n-2}$.
To compute the change in this torsion class, we consider the four-chain $W_i(r_1,r_2) = V_i \times [r_1, r_2]$ where the D5 brane is located at $r = r_0$ and $r_1 < r_0 < r_2$. As argued in more detail in~\cite{dualities2}, each point of intersection between $W_i(r_1,r_2)$ and $\langle j \rangle$ induces a relative change in the torsion classes $[F] \in H^3(V_i,\tbZ) \cong \bZ_2$ of the two ends. Thus, this change in torsion along $W_i(r_1,r_2)$ is equal to the intersection parity $W_i(r_1,r_2) \cap \langle j \rangle = V_i \cap \langle j \rangle$.

Therefore, the bilinear form
$\langle i \rangle \cdot \langle j \rangle \equiv V_i \cap \langle j
\rangle$
measures the change in torsion measured on $\langle i \rangle$ due to
a wrapped brane on $\langle j \rangle$. We now compute this bilinear
form. It is clear that $\langle i \rangle \cdot \langle j \rangle = 0$
for $i \notin \{j, j\pm 1\}$, because the torsion cycle
$\langle j \rangle$ lives entirely within $V_j$, which does not
intersect $V_i$ for $i \notin \{j, j \pm 1 \}$. Suppose that
$i = j \pm 1$; in this case $V_i$ and $V_j$ intersect transversely on
a one-cycle $\omega_{i j}$ which generates
$H_1(V_j,\bZ) \cong Z_{2 p_j}$, so
$\langle i \rangle \cdot \langle j \rangle = \omega_{i j} \cap \langle
j \rangle$.

To compute this intersection, we construct representatives of the
torsion two-cycle within each lens space.
A twisted two-cycle in $X \cong Y/\sigma$ is a closed chain in $Y$ such that $\sigma^\ast \Sigma = -\Sigma$. For $p$ odd, consider the twisted two-cycle in $L(2p,q)\cong L(p,q)/\bZ_2$ with the smooth immersion:
\begin{align} \label{eqn:torsionRepOddP}
x &= \cos \theta \,, & y &= e^{i\phi} \sin \theta \,,
\end{align}
for $\theta \in [0,\pi]$, $\phi \in [0,2\pi)$, where $\bigl(x,y\bigr) \cong \bigl(\omega_{2 p}\, x, \omega_{2 p}^q\, y\bigr)$ are the complex coordinates of $L(2p,q)$ and $\sigma$ maps $\{x, y\} \to - \{x,y\}$ up to the orbifold identification, hence $\theta \to \pi-\theta$, $\phi \to \phi+\pi$. This is an $\bR\bP^2$ embedded with a $p$-fold self-intersection along the equator $\theta = \pi/2$. By comparison, a representative one-cycle generating $\bZ_{2p}$ is
\begin{align} \label{eqn:oneCycleRep}
x &= e^{i \psi} \cos \theta_0\,, & y &= e^{i q \psi} \sin \theta_0 \,,
\end{align}
for fixed $0 \le \theta_0 \le \pi/2$ and $0 \le \psi \le \pi/p$ up to
the $\bZ_{2p}$ identification. Clearly~(\ref{eqn:torsionRepOddP})
and~(\ref{eqn:oneCycleRep}) intersect at a single point
$\theta = \theta_0$, $\phi =\psi =0$, up to the $\bZ_{2p}$
identification. Since the intersection parity is homologically
invariant, this proves that~(\ref{eqn:torsionRepOddP}) is a
non-trivial element of (and hence generates)
$H_2(V_i,\tbZ) \cong \bZ_2$, so the
intersection parity between the generators of
$H_2(V_i,\tbZ) \cong \bZ_2$ and $H_1(V_i,\bZ)\cong \bZ_{2p_i}$ is odd.

A similar argument applies for $p$ even, for which we consider the twisted two-cycle with the smooth embedding
\be
x = e^{i \phi} \cos \theta \;,\; y = e^{i (q-p) \phi} \sin \theta \,,
\ee
for $\theta \in [-\pi/2,\pi/2)$, $\phi \in [0,2\pi/p)$ up to the $\bZ_p \subset \bZ_{2p}$ identification, where $\sigma$ now maps $\phi \to \phi+ \pi/p$, $\theta \to - \theta$. This is a Klein bottle without self-intersections. It intersects the torsion one-cycle~(\ref{eqn:oneCycleRep}) at the point $\theta = \theta_0$, $\phi=\psi=0$, as before. Based on this, we conclude that $\langle i \rangle \cdot \langle i\pm 1 \rangle = 1$.

\begin{figure}
    \centering
    \includegraphics[width=8cm]{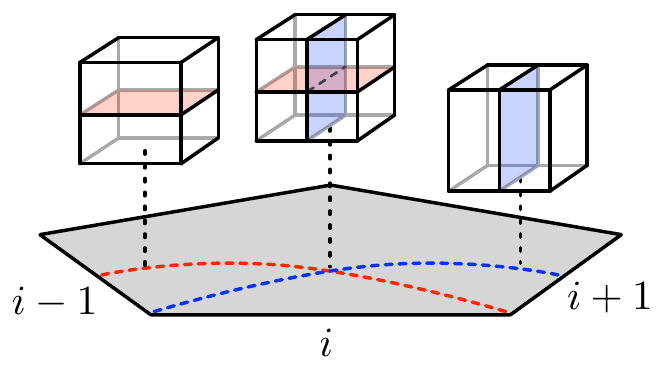}
\caption{Two three-cycles homologous to $V_i$ which intersect transversely along a one-cycle. As in figure~\ref{sfig:horizonfibration}, we represent the horizon as a $T^3$ fibration over an $n$-gon. $V_i$ corresponds to the $i$th edge, along which the fiber degenerates to $T^2$. The red and blue dashed curves represent different ways to deform $V_i$ into the bulk of the $n$-gon, consistent with different embeddings of $T^2 \subset T^3$, as described in the text.} \label{fig:selfintersection}
\end{figure}

It remains to compute $\langle i \rangle^2 = V_i \cap \langle i \rangle$. To do so, we first compute the transverse intersection between two different representatives of the $V_i$ homology class in $H_3(X,\bZ)$. This gives a one-cycle $\omega_{ii} \in H_1(V_i,\bZ)$ such that $V_i \cap \langle i \rangle = \omega_{i i} \cap \langle i \rangle$. We use the description of the horizon as a $T^3$ fibration over an $n$-gon, as shown in figure~\ref{sfig:horizonfibration}. Near $V_i$, the 
$T^3$ fiber can be parameterized by $(\arg z_{i-1},\arg z_{i},\arg z_{i+1})$, subject to the orbifold identification
\begin{equation} \label{eqn:T3orbifold}
(\arg z_{i-1},\arg z_{i},\arg z_{i+1}) \cong (\arg z_{i-1},\arg z_{i},\arg z_{i+1})+\frac{\pi}{p_i} (q_i,p_i-q_i-1,1) \,,
\end{equation}
with the fiber coordinate $\arg z_j$ degenerating on the $j$th edge of the $n$-gon. To lift $V_i$ off the $i$th edge of the $n$-gon, we need to specify $\arg z_i$. Two simple ways to do this, consistent with the identification~(\ref{eqn:T3orbifold}), are
\begin{equation}
\arg z_i = (p_i-q_i-1) \arg z_{i+1}\,, \qquad \mbox{or} \qquad \arg z_i = (p_i-q_i-1) \tilde{q}_i \arg z_{i-1}\,,
\end{equation}
where $\tilde{q}_i$ is chosen to satisfy $q_i \tilde{q}_i \cong 1 \pmod{2p_i}$. Either choice allows $V_i$ to be deformed to a $T^2$ fibration over a line segment connecting the edge $i-1$ with the edge $i+1$ through the bulk, except that in the former case one end must remain at the corner between edges $i$ and $i+1$, whereas in the latter the other end must remain at the corner between edges $i-1$ and $i$, so that the linked fibers degenerate simultaneously. These two choices intersect transversely in the interior of the $n$-gon, as shown in figure~\ref{fig:selfintersection}. For $q_i \ne p_i -1$, they intersect along the $p_i - q_i -1$ one-cycles
\begin{equation} \label{eqn:ViSelfIntersect}
\arg z_{i+1} = \tilde{q} \arg z_{i-1} + \frac{2\pi k}{p_i-q_i-1} \,,
\end{equation}
indexed by $0 \le k < p_i - q_i -1$. For $q_i = p_i -1$, $\arg z_i$ is
constant, and two representatives do not intersect (or intersect
non-transversely).

Each of the $p_i-q_i-1$ one-cycles in~(\ref{eqn:ViSelfIntersect}) composing $\omega_{ii}$ is a generator of $H_1(V_i,\bZ)$, hence its intersection parity with $\langle i \rangle$ is odd. Since $q_i$ is odd, we conclude that $V_i \cap \langle i \rangle = \omega_{i i} \cap \langle i \rangle = p_i \bmod 2$. Using figure~\ref{sfig:toricCornerI} to interpret the parity of $p_i$, we conclude that:
\begin{equation}
  \label{eq:torsion-product}
  \vev{i}\cdot \vev{j} = \begin{cases}
    1\quad \text{for } i=j\pm 1\\
    1\quad \text{for } i=j \quad \text{ if } L_{i+1}\neq L_{i-1}\\
    0\quad \text{otherwise}
    \end{cases}
\end{equation}
where by $L_k$ we denote the even sublattice on which the corner $k$
lives.

Several comments are in order. Firstly, the inner product is symmetric, although this is not obvious from the definition $\langle i \rangle \cdot \langle j \rangle \equiv V_i \cap \langle j \rangle$. Secondly, it is invariant under the equivalence relation~(\ref{eqn:twistedlattice}), as is required for it to be well-defined. In fact, invariance under~(\ref{eqn:twistedlattice}) can be used to derive the formula for $\langle i \rangle^2$ once $\langle i \rangle \cdot \langle j \rangle$ is known for $i \ne j$, as an alternative to computing the self-intersection as above. Thirdly, the inner product is non-degenerate, i.e. for any $A \in H_2(\tbZ)$, there exists $B \in H_2(\tbZ)$ such that $A\cdot B \ne 0$. (To show this, one can check that $A \cdot \langle i \rangle = 0$ for all $i$ implies $A$ is trivial up to the equivalence~(\ref{eqn:twistedlattice}).) Finally, the element
\begin{equation}
  \label{eq:norm-element}
  \eta \equiv \sum_i \langle i \rangle \,,
\end{equation}
plays a special role, in that $A\cdot (A+\eta) = 0$ for any
$A \in H_2(\tbZ)$; this is the ``norm'' element discussed in
appendix~\ref{app:Z2bilinear}.\footnote{Note that, in contrast to the
discussion in appendix~\ref{app:Z2bilinear}, we have not written
\eqref{eq:norm-element} as a sum over the elements of an
orthonormal basis, but one can check that $\sum_i \langle i \rangle$ is indeed the norm element, nonetheless.}

As a result, there are two distinct ways to specify a torsion class. We can either specify its components ``covariantly'' in the divisor basis:
\be
A = \sum_i A_i \langle i \rangle \,,
\ee
or we can specify the corresponding quantized Wilson lines ``contravariantly'' $A^i = A
\cdot \langle i \rangle$. The overcompleteness of the divisor basis implies different properties for the two cases.\footnote{Using the divisor basis to describe the space of discrete torsions is analogous to describing the (co)tangent space of a surface in terms of the (co)tangent space of the enveloping space. The pullback map is surjective and leads to equivalence classes of one-forms analogous to (\ref{eqn:covariantEquiv}). The push-forward map is injective, and leads to constraints on vectors analogous to (\ref{eqn:contravariantCons}).} In the covariant formulation there is an ambiguity due to the equivalence relation~(\ref{eqn:twistedlattice}):
\be \label{eqn:covariantEquiv}
A_i \cong A_i + \delta_{i \in L} + \delta_{i \in L'}  \,,
\ee
for any pair of even sublattices $L$, $L'$ not containing the origin. In the contravariant formulation, the $A^i$ are unambiguous but subject to the constraints
\be \label{eqn:contravariantCons}
\sum_{i \in L} A^i + \sum_{i \in L'} A^i = 0 \,.
\ee
The properties are complementary in that they ensure that the inner product $\langle A \rangle \cdot \langle B \rangle = A^i B_i = A_i B^i$ is well-defined.
Both co- and contravariant formulations will be useful in stating the AdS/CFT
dictionary for discrete torsion in the following sections.

\section{Brane tilings, orientifolds, and NSNS torsion}
\label{sec:NStorsion}

Having understood the geometry of toric orientifold singularities in
detail, we now review what is known about the $\cN=1$ SCFTs arising
from D3 branes probing these singularities. To do so, we use the
language of brane
tilings~\cite{Franco:2005rj,Feng:2005gw,Franco:2007ii}, focusing
especially on their construction as a five-brane system in the $g_s
\to \infty$ limit~\cite{Yamazaki:2008bt}. The recent progress
of~\cite{dP1paper} in constructing CFT duals for
intrinsically strongly coupled orientifold phases will be essential to
our analysis. In particular, as anticipated
in~\cite{dP1paper}, these strongly coupled phases
occur for every value of the discrete torsion when the toric diagram
has five or more sides, a category which includes orientifolds of some
simple and familiar geometries, such as the $dP_2$ and $dP_3$
singularities.\footnote{This explains the surprising absence of
  anomaly-free orientifolds for these geometries in earlier analyses
  such as~\cite{Franco:2010jv}.} We will review the results
of~\cite{dP1paper} in the next two sections,
generalizing them to the large class of toric orientifolds considered
in the previous section.

\subsection{Review of brane tilings} \label{subsec:tilings}

The $\cN=1$ gauge theory arising from $N$ D3 branes probing a toric
Calabi-Yau singularity can be engineered by $N$ D5 branes wrapping a
torus and intersecting a collection of NS5 branes in a manner which we
will describe.\footnote{Many of the results in this section are
  already well known, see \cite{Kennaway:2007tq,Yamazaki:2008bt} for
  reviews. We will nevertheless provide explicit derivations here,
  since the details are sometimes important for our later arguments.}
This ``brane tiling'', similar to brane box
models~\cite{Hanany:1997tb,Hanany:1998it}, is the result of
T-dualizing along a $T^2$ within the $T^3$ toric fiber of the
Calabi-Yau singularity. We begin by reviewing this T-duality for the
case without orientifolds before returning to their effects and the
correct description of discrete torsion in the brane tiling.

\subsubsection*{T-duality}

Consider a general toric singularity. There is a natural coordinate on the $T^3$
fiber defined by:
\begin{equation} \label{eqn:fiberpsi}
  \psi \equiv \sum_j \varphi_j \,,
\end{equation}
where $\varphi_j \equiv \arg z_j$. Surfaces of constant $\psi$ define
a subfiber $T^2 \subset T^3$ which covers the entire fiber along the
toric divisors $z_i = 0$ (where $\psi$ degenerates to a
point).

In order to T-dualize along this torus,
we choose coordinates along it. Suppose we are interested in the
vicinity of the divisor $z_i = 0$. The gauge-invariant
coordinates
\begin{equation} \label{eqn:fiberphi}
  \phi^a_{(i)} \equiv \sum_j u_{j i}^a \varphi_j \,, \qquad (u_{j i}^a \equiv u_j^a - u_i^a)\,,
\end{equation}
are well-defined in this region
because they do not depend on the degenerating coordinate $\varphi_i$. Different charts are appropriate near each toric divisor $z_i =0$, reflecting the non-trivial $T^3$ fibration. These charts are related by
\begin{equation}
  \phi^a_{(j)} = \phi^a_{(i)} + u_{i j}^a \psi,
\end{equation}
which becomes a gauge
transformation in the T-dual description, making $H = \dd B$ non-trivial
in de Rham cohomology and signaling the presence of NS5 branes. 

To describe the locations of these NS5 branes, we introduce coordinates along the base of the fibration, described by the dual cone as reviewed in~\S\ref{subsec:horizons}.
A general solution to the D-term conditions takes the form
\begin{equation}
  | z_j |^2 = \rho - u_j^a r_a \,,
\end{equation}
where $\rho$ and $r_a$ can be thought of as FI terms for the $U(1)^3$ global symmetry. In these coordinates, the dual cone $| z_j |^2 \geqslant 0$ is described by
\begin{equation}
  \rho \geqslant \max (u_1^a r_a, \ldots, u_n^a r_a) \,.
\end{equation}
Since $\psi$ degenerates on the faces $\rho = \rho_{\rm min}$, it is natural to think of $\rho$, or more precisely
\begin{equation}
R(\rho,r_a) \equiv \prod_j | z_j | = \prod_j \sqrt{\rho - u_j^a r_a} \,,
\end{equation}
as the radial coordinate paired with $\psi$, forming the complex combination $Z \equiv R e^{i \psi} = \prod_i z_i$.

The NS5 branes appear where the $T^2$ subfiber degenerates, namely along the curves $z_i = z_{i+1} = 0$ which correspond to the rays bounding the dual cone.
To locate these rays in the dual cone coordinates $r_a$, we consider
\begin{equation}
  | z_k |^2 - | z_i |^2 = u_{i k}^a r_a \geqslant 0 \,,\qquad (z_i = z_{i+1} = 0),
\end{equation}
with equality if and only if $k\in\{i,i+1\}$. This fixes $r_a$ to be proportional the outward normal of the side of the toric diagram connecting corners $i$ and $i+1$. Thus, the NS5 branes are rays in the $r_a$ plane forming the web diagram of the toric singularity in question, and sit at the origin of the $Z$ plane. Counting their worldvolume dimension, we conclude that they wrap one cycles (determined below) on the T-dual torus.

The Calabi-Yau metric takes the general form:
\begin{equation}
  \dd s^2 = \dd s_{\mathrm{base}}^2 + \tau_{a b} \dd \phi^a_{(i)}
  \dd \phi^b_{(i)} + 2 \tau_a^{(i)} \dd \phi^a_{(i)} \dd \psi +
  \tau^{(i)} \dd \psi^2 \,,
\end{equation}
where $\tau_{a b}, \tau_a^{(i)}$ and $\tau^{(i)}$ are functions of the base
coordinates $r_a$ and $\rho$. The absence of cross terms which mix the base and the fiber follows from the fact that the metric is K\"ahler and $\U(1)^3$ invariant~\cite{Abreu:2003}.\footnote{In particular, this follows from the $U(1)^3$ invariance of the K\"ahler potential. In general, one could imagine that $K \to K + f +\bar{f}$ under a $U(1)$ transformation, with $f$ holomorphic, but imposing that the metric is regular near $z=0$ for the associated GLSM field implies that $K$ can be made $U(1)$ invariant by a K\"ahler transformation.}
The metric components in different charts are related by
\begin{equation}
  \tau_b^{(j)} = \tau_b^{(i)} - u_{i j}^a \tau_{a b} \;, \qquad \tau^{(j)} =
  \tau^{(i)} - 2 \tau_a^{(i)} u_{i j}^a + u_{i j}^a \tau_{a b} u_{i j}^b \,,
\end{equation}
where $\tau_{a b}$ is invariant.

We apply Buscher's rules \cite{Buscher:1987sk,Buscher:1987qj}, which
can be written as
\begin{align}
  e^{2 \Phi'} &= \frac{e^{2 \Phi}}{G_{9 9}} \,, &
  G_{9 9}' &= \frac{1}{G_{9 9}} \,, &
  G_{9 i}' &= \frac{G_{9 i}}{G_{9 9}} \,, &
  G_{i 9}' &= - \frac{G_{i 9}}{G_{9 9}} \,, &
  G_{i j}' &= G_{i j} - \frac{G_{i 9} G_{9 j}}{G_{9 9}} \,,
\end{align}
where $G_{m n} = g_{m n} + B_{m n}$ combines the metric and $B$-field, and is in general neither symmetric nor antisymmetric. In this notation,
it is simple to write the multidimensional version of Buscher's rules:
\begin{align}
  e^{2 \Phi'} &=  |\mathcal{G}| e^{2 \Phi} \,, &
  G'_{\alpha \beta} &= \mathcal{G}^{\alpha \beta} \,, &
  G'_{\alpha i} &= \mathcal{G}^{\alpha \beta} G_{\beta i} \,, &
  G'_{i \beta} &= - G_{i \alpha} \mathcal{G}^{\alpha \beta} \,, &
  G'_{i j} &= G_{i j} - G_{i \alpha} \mathcal{G}^{\alpha \beta} G_{\beta j} \,,
\end{align}
where by $\alpha, \beta$ ($i, j$) we denote indices which are dualized
(not dualized),
$\mathcal{G}^{\alpha \beta} \equiv (G_{\alpha \beta})^{-1}$ and
$|\mathcal{G}| = \det \mathcal{G}$.

We find the T-dual background
\begin{align} \label{eqn:T-dualBkg}
\begin{split}
  \dd s^2 & = \dd s_{\mathrm{base}}^2 + \tau^{a b} \dd \tilde{\varphi}_a
  \dd \tilde{\varphi}_b + [\tau^{(i)} - \tau_a^{(i)} \tau^{a b} \tau_b^{(i)}] \dd \psi^2 \,,\\
  B_{(i)} & = \tau^{a b} \tau_a^{(i)} \dd \tilde{\varphi}_b \wedge \dd \psi \,,\\
  e^{2 \Phi} & = g_s^2 \det (\tau^{a b}) \,,
\end{split}
\end{align}
where $\tau^{a b} = (\tau_{a b})^{- 1}$ and the periodic coordinate
$\tilde{\varphi}_a \equiv \alpha' \tilde{\phi}_a$ has radius
$\tilde{R} = \alpha' / R = \alpha'$ (i.e. period
$2\pi\alpha'$). Notice that the combination
$\tau^{(i)} - \tau_a^{(i)} \tau^{a b} \tau_b^{(i)}$ is independent of
the choice of chart, so the metric element in~\eqref{eqn:T-dualBkg} is
globally defined, which implies that the T-dual coordinates
$\tilde{\phi}_a$ can be chosen globally and hence the fibration
is trivial.  Topologically, the space is now $T^2 \times \mathbb{R}^4$
with coordinates $(\tilde{\phi}_a, r_a, Z)$. The original non-trivial
topology is instead reflected in the gauge transformations
\begin{equation}
  B_{(i)} = B_{(j)} + \alpha' u_{i j}^a \dd \tilde{\phi}_a \wedge \dd \psi \,,
\end{equation}
relating different charts.
Near the toric divisor $z_i = 0$, we must have $\tau_a^{(i)}, \tau^{(i)} \rightarrow 0$ with $\tau_{a b}$
finite, in accordance with the fact that $\psi$ degenerates here with
$\phi^a_{(i)}$ well-defined. Thus $B_{(j)}$ ($j\ne i$) has a Dirac string along the divisor because $\psi$ degenerates whereas the prefactor $\tau^{a b} \tau_a^{(j)} \to - u^a_{i j}$ remains finite. These Dirac strings end along the curves $z_i = z_{i+1} = 0$, signaling the presence of NS5 branes.

We can measure the charge of an NS5 brane by integrating $H$ over the link of its worldvolume.
Consider the two-sphere formed by taking a path through the bulk of the dual
cone connecting points on the faces $i$ and $i + 1$ and fibering $\psi$ over
this interval. The size of this two-sphere can be varied by adjusting the size
of the arc and the locations of its endpoints. To construct the $S^3$ link of
an NS5 brane, we consider an interval crossing its worldvolume on the $T^2$
and fiber the above $S^2$ over this interval, shrinking the $S^2$ to zero size
on either end.

However, the T-dual background~(\ref{eqn:T-dualBkg}) produced by Buscher's rules is translationally invariant along the torus, and describes smeared NS5 branes. To capture the entire smeared charge, we deform the link, stretching the interval crossing the NS5 brane
worldvolume until it wraps around the torus and meets itself. At this point we
can deform the $S^3$ into $S^2 \times S^1$, where the $S^2$ is of fixed size
and the $S^1$ is the dual cycle on $T^2$ to that wrapped by the NS5 brane. More generally, when $S^1$ is an arbitrary cycle on the torus the integral of $H$ over $S^2\times S^1$ measures the intersection number between it and the NS5 brane worldvolume.

 Using Stokes' theorem, the integral of $H$ over the cycle described above reduces
to
\begin{equation}
  \frac{1}{(2 \pi)^2 \alpha'} \oint_{S^1 \times S^2} H = \frac{1}{(2 \pi)^2
  \alpha'}  \oint_{S^1 \times S^1_{\psi}} (B_{(i)} - B_{(i + 1)}) = u_{i, i +
  1}^a n_a \,,
\end{equation}
for winding $n_a = \frac{1}{2 \pi} \int_{S^1} \dd \tilde{\phi}_a$.
Thus, the NS5 brane has winding numbers
\begin{equation} \label{eqn:NS5winding}
  w_a^{i, i + 1} = \pm \varepsilon_{a b} u_{i, i + 1}^b \,,
\end{equation}
up to a convention-dependent overall sign. Since $r_a \propto w_a^{i, i + 1}$, the winding is fixed by the angle in the $r_a$ plane, as required for unbroken supersymmetry.

To recap, T-dualizing along a particular torus in the $T^3$ fiber of a toric Calabi-Yau singularity turns the geometry into NS5 branes on the flat background $T^2 \times \mathbb{R}^4$.\footnote{While~(\ref{eqn:T-dualBkg}) is not flat, this is due to the backreaction of the smeared NS5 branes.} The NS5 branes form the web diagram associated to the singularity in one plane of $\mathbb{R}^4$ and sit at the origin in the other plane. For each segment of slope $p/q$ in the web, the corresponding NS5 brane has winding numbers $(p,q)$ on the torus.\footnote{These statements were derived above in the affine case, but apply to partial and total resolutions as well.}

\subsubsection*{Weak and strong coupling limits}

\begin{figure}
  \centering
  \begin{subfigure}{0.49\textwidth}
    \centering
    \includegraphics[width=0.9\textwidth]{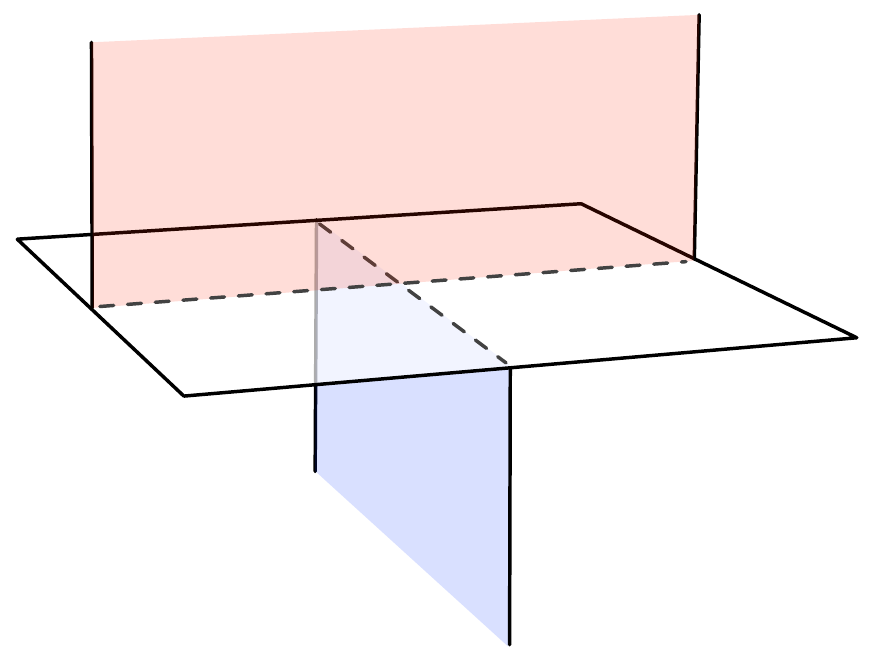}
    \caption{NS5 brane boundaries fail to meet}
    \label{sfig:NS5boundaries}
  \end{subfigure}
  \hfill
  \begin{subfigure}{0.49\textwidth}
    \centering
    \includegraphics[width=0.9\textwidth]{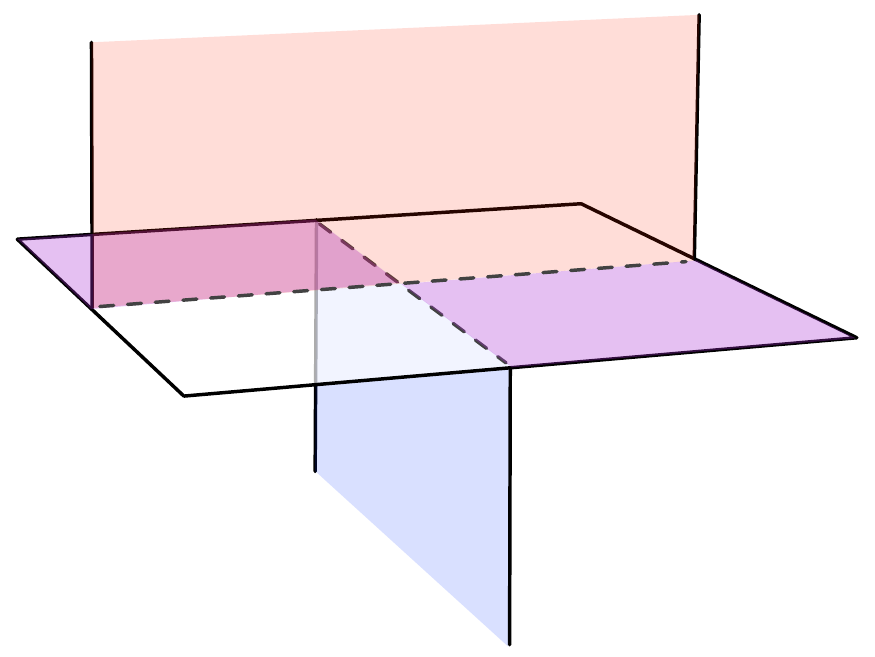}
    \caption{Connecting the boundaries along $T^2$}
    \label{sfig:NS5connected}
  \end{subfigure}
  \caption{\subref{sfig:NS5boundaries}~At junctions in the web diagram, the NS5 brane boundaries fail to meet along the $T^2$, despite adding to zero in homology. \subref{sfig:NS5connected}~This can fixed by adding NS5 brane segments along $T^2$. The resulting brane configuration is not BPS, and will relax into a holomorphic curve. However, the configuration pictured here correctly describes the topology of the NS5 branes, and is physically realized with D5 branes wrapping $T^2$ in the $g_s \to \infty$ limit.}
\end{figure}

So far, we have not addressed what happens at junctions in the web diagram, where several NS5 branes meet. It is easy to see that consistency of the web diagram implies that the NS5 brane tadpole vanishes at each junction, as required for consistency of the supergravity background (\ref{eqn:T-dualBkg}). However, if each NS5 brane wraps a fixed (localized rather than smeared) minimum length $S^1$ on the torus then their boundaries will not actually meet at the junctions, despite adding to zero in homology, see figure~\ref{sfig:NS5boundaries}. 
As a crutch, we can add NS5 brane segments along the torus connecting these edges, pictured in figure~\ref{sfig:NS5connected}. This joins the NS5 branes into a single brane on a piecewise-linear curve.

This description is heuristically correct, but in reality the NS5
brane will relax into a smooth shape, described by the holomorphic
curve \cite{Feng:2005gw}
\begin{equation} \label{eqn:NS5curve}
P(x,y) = 0\,,
\end{equation}
where $P(x,y)$ is a Laurent polynomial with Newton polygon (the convex hull of $(m,n)$ for all non-zero monomials $x^m y^n$ in $P(x,y)$) equal to the toric diagram. Here $x = e^{r_1/r_0+i \tilde{\phi}_1/\alpha'}$, $y = e^{r_2/r_0+i \tilde{\phi}_2/\alpha'}$ for some scale factor $r_0$.

We now consider the effect of adding $N$ D3 branes at the toric
singularity, focusing on the affine case for simplicity. After
T-dualizing, we find $N$ D5 branes wrapping the torus at the
intersection of the NS5 brane rays in the $r_a$ plane. To determine
the shape of the NS5 branes, we begin with the naive picture,
figure~\ref{sfig:NS5connected}. Unlike before, the NS5 branes parallel to the
$T^2$ will form bound states with the $N$ D5 branes, e.g.\ an
$(N,\pm 1)$ five-brane in a region with a single NS5 brane of either
orientation. For $g_s \ll 1$, the NS5 brane and $(N,\pm 1)$ five-brane
tensions are much larger than the D5 brane tension, and the NS5 brane---part of whose worldvolume consists of $(\pm N, 1)$ bound
states\footnote{Note that a $(-N,1)$ brane is the same as a $(N,-1)$
  of the opposite orientation. When discussing the wordvolume of the
  NS5 brane, the former description is more natural, whereas when
  discussing the worldvolume of the D5 branes, the latter is more
  natural. This change in viewpoint is the ``untwisting'' of
  \cite{Feng:2005gw}.}---again relaxes into the holomorphic
curve~(\ref{eqn:NS5curve}), with D5 brane disks ending along one
cycles on the curve.

While the exact shape of the D5 branes is hard to determine, to read off the resulting gauge theory it is sufficient to understand the topology of the D5 brane boundaries along the NS5 brane. In particular, each D5 brane disk gives rise to a $U(N)$ vector multiplet, whereas each intersection point between the boundaries of two disks along the NS5 brane gives a bifundamental chiral multiplet (or an adjoint chiral multiplet where a single boundary self-intersects), with the chirality fixed by the orientation of the intervening NS5 brane. In addition, for each $(\pm N, 1)$ brane there is a corresponding superpotential term linking the chiral multiplets which lie along its boundary.

Thus, the gauge theory can be determined by specifying the topology of the $(\pm N,1)$ brane configuration along the worldvolume of the NS5 brane, from which the locations of the D5 brane boundaries can be inferred by charge conservation. However, there is a simpler way to specify this topology. In the opposite limit $g_s \gg 1$, the NS5 brane tension is much less than the D5 brane and $(N,\pm1)$ five-brane tensions, and the D5 branes shrink onto the original torus. In this limit, the brane configuration is identical to the naive picture, figure~\ref{sfig:NS5connected}! The topology of the brane configuration, unchanged from the $g_s \ll 1$ limit, can now be specified by specifying the locations of the $(N,\pm 1)$ branes along the \emph{D5 brane} worldvolume. Since the D5 brane worldvolume is a torus, whereas the NS5 brane worldvolume is a punctured Riemann surface of arbitrarily high genus, the former is easier to draw and work with.

\begin{figure}
  \centering
  \begin{subfigure}{0.3\textwidth}
    \centering
    \includegraphics[height=3.5cm]{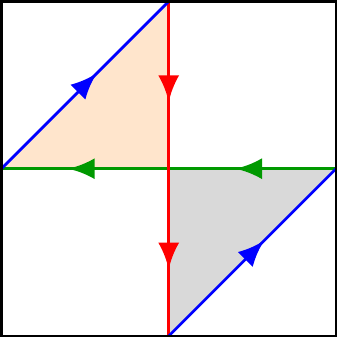}
    \caption{Example tiling.}
    \label{sfig:C3-tiling-example}
  \end{subfigure}
  \hfill
  \begin{subfigure}{0.31\textwidth}
    \centering
    \includegraphics[height=3.5cm]{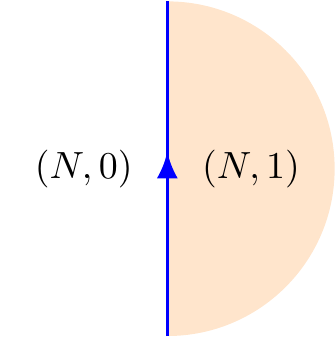}
    \caption{5-brane charge convention.}
    \label{sfig:5-brane-charge-conventions}
  \end{subfigure}
  \hfill
  \begin{subfigure}{0.3\textwidth}
    \centering
    \includegraphics[height=3.5cm]{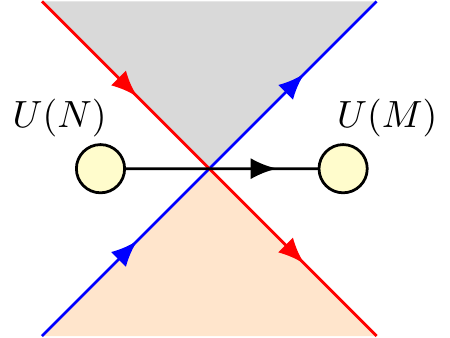}
    \caption{Bifundamental convention.}
    \label{sfig:tiling-bifundamental-conventions}
  \end{subfigure}
  \caption{\subref{sfig:C3-tiling-example} An example of a brane tiling, in this case T-dual to D3 branes in a flat background.
   The colored one-cycles are associated with NS5 branes ending on the
    tiling, whose slopes equal those of the external legs of the
    web diagram for the singularity. The white regions
    correspond to stacks of $N$ D5 branes, while the orange and gray shaded
    regions correspond to $(N,1)$ and $(N,-1)$ $(p,q)$-five brane bound states
    respectively. 
    In our conventions, \subref{sfig:5-brane-charge-conventions}~the NS5
    brane charge of the tiling increases as we cross an
    NS5 brane boundary oriented
    upwards from left to right, and \subref{sfig:tiling-bifundamental-conventions}~the arrows in the quiver diagram, corresponding to
    bifundamental chiral multiplets arising at the NS5 brane intersections in the tiling,
    follow the local orientation of the NS5 brane boundaries.}
  \label{fig:tiling-example}
\end{figure}

This five-brane system---as represented by the topology of the
$(N,\pm 1)$ branes along the D5 brane worldvolume---is commonly
referred to as a ``brane tiling'', see,
e.g., figure~\ref{fig:tiling-example}. We can recover the $g_s \ll 1$
picture by applying the ``untwisting'' procedure of~\cite{Feng:2005gw}
to the brane tiling, which is merely a formalization of the invariant
topology of the five-brane configuration. In particular, we can read
off the gauge theory without needing to untwist, following the rules
summarized in figure~\ref{fig:tiling-example}.

\begin{figure}
  \centering
  \includegraphics[width=\textwidth]{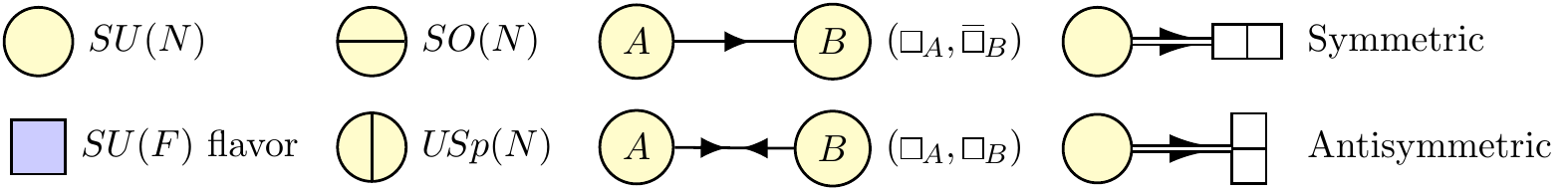}
  \caption{Summary of the quiver notation for weakly coupled fields
    (strongly coupled sectors are described by a notation to be
    introduced in figure~\ref{fig:quad-notation} below). Multiple
    arrowheads in the same direction along a given edge indicate a
    multiplicity of fields in the same representation, and reversing
    the direction of the arrowhead indicates taking the complex
    conjugate representation.}
  \label{fig:quiverdictionary}
\end{figure}

We will also often use quiver diagrams to encode the gauge group
and matter content of the resulting field theories. We adopt the same
quiver notation as in \cite{dP1paper}, summarized in
figure~\ref{fig:quiverdictionary}. (We will also introduce below an
extended quiver notation for representing the strongly coupled sectors
$\TO_k$ with $k\geq
2$.) As usual, nodes in the quiver diagram denote factors in a
semi-simple gauge group and arrows between nodes denote bifundamental
matter. Due to the presence of orientifold planes we also have
(anti)symmetric tensor matter, orthogonal and symplectic groups, and
quivers without a global orientation.

\subsubsection*{Moduli space}

The $g_s \gg 1$ limit also provides a simple picture of the supersymmetric moduli space~\cite{Imamura:2007dc,Yamazaki:2008bt}, which is $n-1$ dimensional for an $n$-sided toric diagram. The positions of the $n$ NS5 branes along $T^2$ represent $n-2$ degrees of freedom (accounting for translation invariance along the $T^2$). $n-3$ of these are T-dual to the holonomies $\oint B_2$ on the $n-3$ two cycles of the horizon $Y_5$. Their superpartners are Wilson lines on the NS5 branes, which T-dualize to the holonomies $\oint C_2$ on these same cycles. The final position modulus for the NS5 branes is tied to the axion $C_0$ by supersymmetry and is T-dual---together with the holonomy $\oint_{T^2} B_2$---to the beta deformation~\cite{Lunin:2005jy}. Together with the T-dual of the axio-dilaton (which is a combination of the metric and dilaton fields partnered with $\oint_{T^2} C_2$ in the five-brane description), these moduli match the $(n-1)$-dimensional conformal manifold of the dual SCFT.

\begin{figure}
  \centering
  \begin{subfigure}{0.4\textwidth}
    \centering
    \includegraphics[height=3cm]{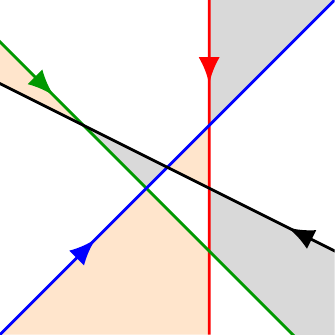}
    \caption{A brane tiling.}
    \label{sfig:non-Lagrangian-resolved}
  \end{subfigure}
  \hspace{1cm}
  \begin{subfigure}{0.4\textwidth}
    \centering
    \includegraphics[height=3cm]{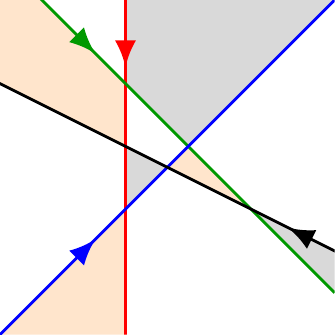}
    \caption{Its Seiberg dual.}
    \label{sfig:non-Lagrangian-resolved-b}
  \end{subfigure}
  \caption{Seiberg duality on a face of the tiling. As we move the NS5
    brane boundaries the central compact D5 shrinks to zero size, and
    then grows in a different topology.}
  \label{fig:non-Lagrangian-resolved}
\end{figure}

Moving in this moduli space changes area of the faces in the brane tiling and hence the corresponding gauge and superpotential couplings. At special points in moduli space Seiberg duality occurs naturally in the form of a D5 brane shrinking to zero size and regrowing in a different topology, see figure~\ref{fig:non-Lagrangian-resolved}. At other points, more singular behavior is possible, such as multiple simultaneous Seiberg dualities. In addition, in general there are regions of moduli space where $(N, k)$ branes appear for $k\ne 0,\pm 1$, and there is no (known) Lagrangian description of the five-brane system.

One can also consider brane tilings where the NS5 branes do not (and
can not, if we want a weak coupling description) form straight lines along the $T^2$, such as that in
figure~\ref{sfig:F07-III-tiling} below. These configurations do not
correspond to a BPS five-brane system, yet naively applying the
dictionary of figure~\ref{fig:tiling-example} leads to a gauge theory
in the same universality class as those constructed with BPS NS5
branes in the same $T^2$ holomology classes (as determined by the
toric diagram)~\cite{Beasley:2001zp}. Heuristically, one could think
of this as a deformation of the corresponding field theory which
preserves the same symmetries as the BPS theory.\footnote{Note that
  supersymmetry is imposed by hand on the gauge theories obtained from
  these non-BPS brane configurations.} In particular, conservation of
the R-symmetry implies that these deformations are marginal in the IR
SCFT. As a flavor-singlet marginal operator must be exactly
marginal~\cite{Green:2010da}, this suggests that the deformed theory
flows to somewhere on the same conformal manifold. Indeed, these
theories turn out to be Seiberg dual to the BPS theories, and the
early development of toric gauge theories (before,
e.g.,~\cite{Hanany:2005ss}) made no distinction between the two.

\subsection{Orientifolds and NSNS torsion}
\label{sec:NSNS-torsion}

We now describe the gauge theories arising from D3 branes at toric
orientifolds in the language of brane tilings. Acting on the local
fiber coordinates (\ref{eqn:fiberphi}) and $Z = \prod_i z_i$, the
involution described in \S\ref{subsec:toricOfolds} takes the form
\begin{align} \label{eqn:involFiber}
\phi_{(i)}^a &\to \phi_{(i)}^a + \pi u_i^a\,,  & Z &\to -Z\,.
\end{align}
Thus, in these local coordinates the involution combines inversion in the transverse $Z$ plane with a translation along $T^2$ determined by the even sublattice containing the corresponding corner of the toric diagram. After T-duality, the involution becomes $\tilde{\phi}_a \to -\tilde{\phi}_a$, $Z\to - Z$. There a four fixed planes located at $Z=0$ at four points on the $T^2$ and extending in the $r_a$ plane. Thus, toric orientifolds are T-dual to five-brane systems with four O5 planes.

To see the effect of the shift~(\ref{eqn:involFiber}), recall that the
T-dual of an orientifold without fixed points $\phi \to \phi + \pi$ is
the orientifold $\tilde{\phi} \to - \tilde{\phi}$ with two orientifold
planes $\tilde{\phi} = 0, \pi$ of opposite RR
charge~\cite{Dabholkar:1996pc,Witten:1997bs}. Since a basis of
one-cycles for the $T^2$ can always be chosen such that the shift acts
on a single generator of the basis, dualizing~(\ref{eqn:involFiber})
results in two O5$^+$ planes and two O5$^-$ planes, where their
relative positions on the $T^2$ are determined by the even sublattice
in question.

Since the $T^2$ is non-trivially fibered, this implies that the O5
plane charges change as we move between faces of the $(p,q)$ web in
the $r_a$ plane. This also follows from the well known fact that the RR-charge of an orientifold plane flips upon crossing an NS5 brane~\cite{Evans:1997hk}. To see how this works in detail, note that the RR charge of the O5 plane at $\tilde{\phi}_a \in \{0,\pi\}^2$ can be written as
\be \label{eqn:Qlocal}
Q_i(\tilde{\phi}) = 1-2 \biggl[\Bigl(h_i+\frac{\tilde{\phi}_a u^a_i}{\pi}\Bigr) \bmod 2\biggr] \,,
\ee
in the $i$th wedge of the web diagram,
for some choice of $h_i \in \{0,1\}^n$. Based on the RR charges of the O5 planes, the NS5 brane between the wedges $i$ and $i+1$ must intersect the O5 planes which satisfy
\be
h_{i+1}-h_{i} \equiv \frac{\tilde{\phi}_a u^a_{i+1,i}}{\pi} + 1 \pmod 2 \,.
\ee
Since $\gcd(u^1_{i+1,i},u^2_{i+1,i})=1$ for an isolated singularity,
there is always at least one solution. In fact, there are exactly two solutions: by~(\ref{eqn:NS5winding}) the
right hand side is invariant under $\tilde{\phi}_a \to \tilde{\phi}_a
+ \pi w_a^{(i,i+1)}$, hence each NS5 brane intersects exactly two O5
planes, determined by the winding numbers of the NS5.

\begin{figure}
  \centering
  \begin{subfigure}{0.54\textwidth}
    \centering
    \begin{subfigure}[t]{0.44\textwidth}
      \centering
      \includegraphics[width=0.8\textwidth]{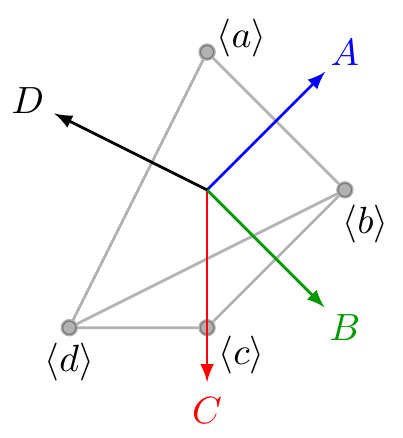}
      \caption{Nomenclature.}
      \label{sfig:dP1-example-conventions}
    \end{subfigure}
    \hfill
    \begin{subfigure}[t]{0.44\textwidth}
      \centering
      \includegraphics[width=0.8\textwidth]{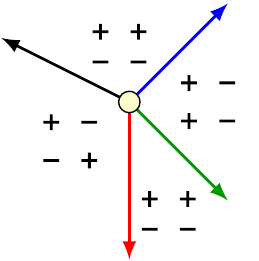}
      \caption{Condensed notation.}
      \label{sfig:local-charges-condensed}
    \end{subfigure}
  \end{subfigure}
  \hfill
  \begin{subfigure}{0.4\textwidth}
    \centering
    \begin{subfigure}{0.45\textwidth}
      \centering
      \includegraphics[height=2.5cm]{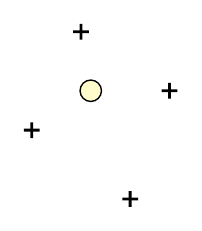}
      \caption{$(0,1)$}
      \label{sfig:local-charges-NW}
    \end{subfigure}
    \hfill
    \begin{subfigure}{0.45\textwidth}
      \centering
      \includegraphics[height=2.5cm]{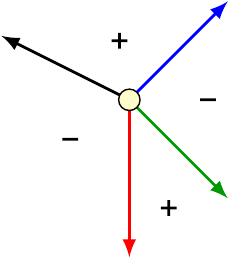}
      \caption{$(1,1)$}
      \label{sfig:local-charges-NE}
    \end{subfigure}\\
    \vspace{0.7cm}
    \begin{subfigure}{0.45\textwidth}
      \centering
      \includegraphics[height=2.5cm]{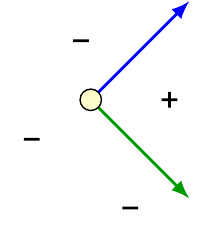}
      \caption{$(0,0)$}
      \label{sfig:local-charges-SW}
    \end{subfigure}
    \hfill
    \begin{subfigure}{0.45\textwidth}
      \centering
      \includegraphics[height=2.5cm]{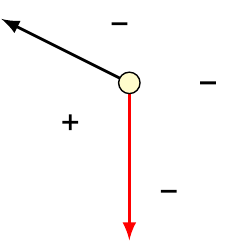}
      \caption{$(1,0)$}
      \label{sfig:local-charges-SE}
    \end{subfigure}
  \end{subfigure}
  \caption{Local charge assignment for the \IA\ phase of the complex
    cone over $dP_1$, analyzed in \S\ref{sec:dP1} below. For
    notational convenience throughout this paper we will use the
    condensed notation in~\subref{sfig:local-charges-condensed}, which
    simultaneously shows the orientifold charge of all four fixed
    points in every toric wedge. One can determine which NS5 branes
    intersect which fixed point, and the local charges of the O5, by
    taking the corresponding entry in the matrix of signs
    in~\subref{sfig:local-charges-condensed}. In
    \subref{sfig:local-charges-NW}-\subref{sfig:local-charges-SE} we
    show explicitly the local structure at each orientifold fixed
    point. By choosing any of the fixed points, and adding up the
    torsion generators associated to the wedges with non-zero
    (negative) local O5 charge, one obtains the $[H]$ NSNS torsion,
    via~\eqref{eqn:Htorsion}. In this particular case we have that
    $[H]=0$, up to~\eqref{eqn:twistedlattice}.}
  \label{fig:local-charges-example}
\end{figure}

The moduli space of NS5 brane positions has been reduced to a discrete set of choices $h_i \in \{0,1\}^n$. Redefining $\tilde{\phi}_a \to \tilde{\phi}_a + \pi n_a$, we see that
\be \label{eqn:hiequiv}
h_i \cong h_i + n_a u^a_i \,,
\ee
up to translations on the torus, so the discrete moduli space of NS5 brane positions is $\bZ_2^{n-2}$. Notice that~(\ref{eqn:hiequiv}) is identical to~(\ref{eqn:covariantEquiv}). Thus, $\sum_i h_i \langle i \rangle$ defines a class in $H^3(X_p, \tbZ)\cong \bZ_2^{n-2}$. Given the connection between the NS5 brane positions and the $B_2$ Wilson lines before orientifolding, it is very natural to conjecture
\be \label{eqn:Htorsion}
\boxed{[H] = \sum_i h_i \langle i \rangle} \,.
\ee
In other words, the \emph{local charges} of the O5 planes in each wedge of the web diagram determine both the involution and the NSNS discrete torsion $[H]$ via (\ref{eqn:Qlocal}). This prescription is summarized in figure~\ref{fig:local-charges-example}.

\medskip

A very strong check of~(\ref{eqn:Htorsion}) is as follows. D3 branes wrapped on three cycles $\Sigma$ of $X_5$ correspond to baryons in the dual gauge theory. A D3 brane wrapped once around $\Sigma$ is forbidden unless the RR and NSNS torsions restricted to this cycle vanish, $[H]_\Sigma = [F]_\Sigma = 0$~\cite{Witten:1998xy}; otherwise only an even wrapping number is permitted. For the three cycle $V_i$ this is the same as $\langle i \rangle \cdot [H] = \langle i \rangle \cdot [F] = 0$, as shown in~\S\ref{sec:intersection}.

In the coordinates $\phi^a_{(i)}, r_a,$ and $Z$, the three cycle $V_i$ spans the $T^2$ with $Z=0$, and extends along a line between the two neighboring NS5 branes in the $r_a$ plane. Therefore, in the five-brane description the baryon becomes a D1 brane stretched between adjacent NS5 branes in the web diagram. Since $V_i$ is mapped to itself by the orientifold involution, the D1 brane lies on top of one of the O5 planes. There are two obvious requirements for such a baryon to exist: (1) the O5 plane must intersect both of the NS5 branes in question, and (2) the RR charge of the O5 plane must be positive where it intersects the D1 brane. The latter requirement follows because $\Sp(1)$---the putative worldvolume gauge group for a single D1 brane coincident with an O5$^-$ plane---does not exist.

Let the O5 plane in question be located at $\tilde{\phi}_a = (0,0)$ without loss of generality. The above requirements become
\be \label{eqn:hconds}
h_i = 0\,, \qquad h_{i-1} = h_{i+1} = 1 \,.
\ee
For the ansatz $[H]_{\rm conj}$~(\ref{eqn:Htorsion}) this implies $\langle i \rangle \cdot [H]_{\rm conj} = 0$. Conversely, let $\langle i \rangle \cdot [H]_{\rm conj} = 0$. We can fix $h_i = 0$ and $h_{i+1}=1$ by the equivalence~(\ref{eqn:hiequiv}), hence by choosing an appropriate O5 plane. The constraint $\langle i \rangle \cdot [H]_{\rm conj} = 0$ then becomes $h_{i-1} = 1$, and we recover (\ref{eqn:hconds}). Thus, the above constraints on the NS5 brane positions and the O5 plane charges are equivalent to $\langle i \rangle \cdot [H]_{\rm conj} = 0$. Since two distinct torsion classes $[H]$ and $[H']$ cannot have the same contravariant components $H^i \equiv \langle i \rangle \cdot [H]$, this strongly supports the ansatz~(\ref{eqn:Htorsion}).\footnote{\label{fn:NSNS-proof} This is essentially a proof of (\ref{eqn:Htorsion}) if we put in the further assumption that the NS5 brane positions and O5 plane charges are T-dual to NSNS (rather than RR) degrees of freedom. Otherwise we have to contend with the additional constraint $\langle i \rangle \cdot [F] = 0$ which we have not discussed.}

\subsection{Constructing the orientifold gauge theory} \label{subsec:gaugetheoryconstruct}

Because the O5$^+$ and O5$^-$ carry opposite RR charge, there must be D5 branes ending on the lines of intersection between the O5 branes and NS5 branes. For instance, one can include D5 branes parallel to the O5$^-$ planes. However, these T-dualize to D7 branes wrapping toric divisors, which extend away from the toric singularity and modify the near horizon geometry. To avoid introducing additional D-branes in the T-dual, the D5 branes must be coincident with the NS5 branes, forming $(\pm k,1)$ five-brane bound states. In particular, the difference in charge between the O5$^+$ and O5$^-$ together with the orientifold projection require that the five-brane switches between a $(2,1)$ brane and a $(-2,1)$ brane as it crosses the O5 plane along the $T^2$.

There is a corresponding change in rank along the $T^2$ coming from
the edges of the $(\pm 2,1)$ branes, as follows. Unbroken
supersymmetry requires that the D5 branes in the $\tilde{\phi}_a$
plane and the O5 planes in the $r_a$ plane are oppositely oriented
with respect to a common volume-form $\Omega^{a b} = -\Omega^{b a}$,
$\Omega^{1 2} > 0$.\footnote{The NS5 branes satisfy $w_a \propto r_a$
  and their orientation is opposite viewed from the perspective of the
  two planes, hence the D5/O5 orientations must also be opposite along
  the two planes.} This implies that, relative to one of the two D5-O5
intersection points through which a given NS5 brane passes, the jump
in the RR charge has the same sign (and half the magnitude) as we
cross the NS5 brane along the D5 worldvolume as it does when we cross
the NS5 brane along the O5 worldvolume in the same angular direction
(e.g.\ in a clockwise sense about the D5/O5 intersection point).

The outcome of this rule is the same regardless of which of the two possible O5/D5 intersections we choose, because
\be
Q_i(\tilde{\phi}+\pi w^{(i,i+1)})=-Q_i(\tilde{\phi})\,,
\ee
which follows from
\be
\varepsilon_{a b} u^a_{i} u^b_{i+1} \in 2\bZ+1 \,,
\ee
which is equivalent to the requirement that adjacent corners lie on distinct even sublattices not containing the origin.

\begin{figure}
  \centering
  \begin{subfigure}{0.3\textwidth}
    \centering
    \includegraphics[height=4cm]{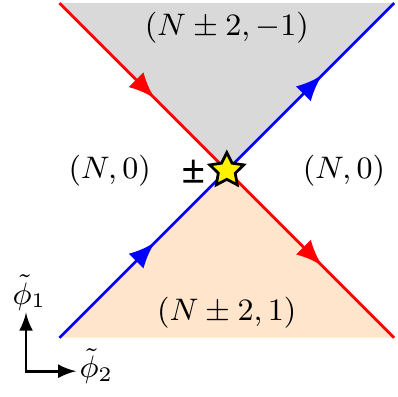}
    \caption{Tiling view.}
    \label{sfig:two-NS5s-tiling}
  \end{subfigure}
  \hspace{1cm}
  \begin{subfigure}{0.3\textwidth}
    \centering
    \includegraphics[height=4cm]{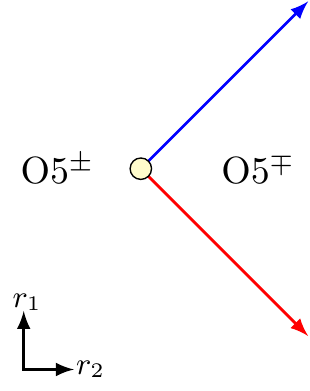}
    \caption{O5 view.}
    \label{sfig:two-NS5s-O5}
  \end{subfigure}
  \caption{Two NS5 branes intersecting on top of an O5 plane (the $\TO_1(N)$ configuration).
  \subref{sfig:two-NS5s-tiling}~The view from the brane tiling, where the point of intersection with the O5 plane is indicated by a star. At the intersection of the the two NS5 brane boundaries---coincident with the O5 plane---we obtain a two-index tensor of the $U(N)$ worldvolume gauge theory on the adjacent D5 brane stack. The ``T-parity'' of the fixed point~\cite{Imamura:2008fd} is indicated by a sign in the tiling, which is positive (negative) when the multiplet is a symmetric (antisymmetric) tensor. \subref{sfig:two-NS5s-O5}~The view from the O5 plane. The RR charge of the orientifold plane changes as we cross the NS5 branes. The T-parity is equal to the sign of the RR charge in the O5 segment which spans the largest angle in the $r_a$ plane.}
  \label{fig:O5-two-NS5s}
\end{figure}

The above rule determines the relative rank of any two faces of the
brane tiling by adding up the changes in RR charge along any path
connecting them. The result is independent of the chosen path because
the net change in RR charge upon circling one of the O5 planes is
zero, a direct consequence of the fact that the net change in RR
charge is zero upon circling the D5 brane stack along the O5 plane
worldvolume. We summarize the resulting changes in rank in
figure~\ref{fig:O5-two-NS5s}.

Once the ranks have been determined, the gauge theory can be read off using the same rules as before. Besides identifying elements that are mapped to each other by the orientifold involution, the only new feature is the potential for new degrees of freedom localized at the orientifold planes. We label the local configuration of branes where an O5 plane is crossed by $2k$ NS5 branes ``$\TO_k(N)$'',\footnote{Since the O5 plane charge changes across each NS5 brane, there must be an even number intersecting each O5 plane.} where \TO\ stands for ``toric orientifold'' and $N$ denotes the average rank of the $(N_i,0)$ D5 brane faces surrounding the O5 plane. For $\TO_0(N)$ there are no local degrees of freedom, but the enclosing face projects down to $\SO(N)$ ($\Sp(N)$) for an enclosed O5$^+$ (O5$^-$). For $\TO_1(N)$, there is a chiral multiplet in a tensor representation of $\SU(N)$ localized at the D5/O5 intersection point. Higgsing the multiplet recombines the NS5 branes in the web diagram and reduces to $\TO_0(N)$, hence for consistency the multiplet must be a symmetric (antisymmetric) tensor when the O5 plane enclosed by the corresponding NS5 branes in the web diagram has negative (positive) RR charge. This is the ``angle rule'' of~\cite{Imamura:2008fd}.

Another way to fix which type of tensor representation appears for $\TO_1$ is via anomaly cancellation. For a given $(N,0)$ face, the change in rank across a neighboring NS5 brane contributes to the $U(N)^3$ chiral anomaly of the two bifundamental chiral multiplets which cross it. Since these multiplets are in conjugate representations of $U(N)$, the contribution cancels. When one of these multiplets sits on an O5 plane (in the $\TO_1$ configuration), the $U(N)^3$ anomaly is determined by the type of tensor multiplet which appears, rather than by a change in rank. In order to cancel with the other contributions, the multiplet should be a symmetric (antisymmetric) tensor when the adjacent $(M,\pm1)$ five-branes have RR charge $M=N+2$ ($M=N-2$). This is sufficient to cancel all non-Abelian gauge anomalies, and is easily shown to be equivalent to the angle rule given above.

For the $\TO_0$ and $\TO_1$ configurations, it is convenient to specify the O5 charges by associating a sign known as the ``T-parity'' to the fixed point \cite{Imamura:2008fd}. For $\TO_0$, the T-parity is simply the O5 charge, whereas for $\TO_1$ it is the O5 charge in the wedge that covers an arc bigger than 180$^\circ$ in the $r_a$ plane, see figure~\ref{fig:O5-two-NS5s}. As we have seen, positive (negative) T-parity corresponds to $\SO$ ($\Sp$) gauge groups and symmetric (antisymmetric) tensor matter in the $\TO_0$ and $\TO_1$ cases, respectively.

\subsubsection*{Resolving singular cases}

The above rules are sufficient to construct the orientifold gauge theory when no more than two NS5 branes intersect a given O5 plane (i.e., when $\TO_k$ for $k\ge 2$ does not appear). However, there can be subtleties when $(N,p)$ five branes appear in the brane tiling for $p\notin \{0,\pm1\}$, or when two NS5 brane boundaries coincide. Unlike the case without orientifold planes, the NS5 brane moduli are fixed, and we cannot avoid these occurrences by moving to another part of moduli space.

We focus on the case of coincident NS5 brane boundaries, as occurs in
some examples later in the paper. Since the web diagram has no
parallel legs, this can only happen if there are antiparallel legs in
the web diagram. In this case overlapping NS5 boundaries inevitably
occur for at least one choice of $[H]$, see, e.g.,
figure~\ref{sfig:F07-III-tiling} below. One solution is to deform the NS5 brane boundaries slightly so that they only cross each other where they intersect the NS5 brane. This deformation breaks supersymmetry, but, as in~\S\ref{subsec:tilings}, there is reason to believe that imposing SUSY by hand on the resulting gauge theory gives a UV theory in the same universality class as the string theory we are interested in; this follows from Seiberg duality in the absence of orientifold planes.

The examples we provide later in the paper will further support this hypothesis. The case where $(N,p)$ five branes appear for $p\notin \{0,\pm1\}$ could be resolved analogously, except that a larger deformation of the branes is required to obtain a gauge theory description. We will not consider any examples of this latter type.

Note that the angle rule of~\cite{Imamura:2008fd} appears ambiguous
when there are anti-parallel legs in the web diagram. The type of tensor matter which actually appears depends on which way the
branes are deformed in the $\tilde{\phi}_a$ plane. 
To address this ambiguity, the angle rule can be restated in terms of the ``local web diagram'',\label{def:localwebdiagram} defined as the web diagram formed by the outgoing NS5 brane rays at the O5 fixed point in the $\tilde{\phi}_a$ plane. For BPS NS5 branes, the local web diagram is identical to the subset of the web diagram composed of the NS5 branes intersecting the O5 plane in question, see, e.g., figures~\ref{sfig:local-charges-NW}--\ref{sfig:local-charges-SE}. For non-BPS NS5 branes, it depends on the way the branes are deformed in the $\tilde{\phi}_a$ plane. Applying the angle rule to the local web diagram produces an anomaly-free spectrum that agrees with~\cite{Imamura:2008fd} in the BPS case and is consistent with Seiberg-duality and integrating in and out massive matter in the more general, non-BPS setting. Local web diagram will also prove useful in our discussion of the local degrees of freedom in $\TO_{k \ge 2}$ configurations below.

\section{\alt{$\TO_k$}{TOk} CFTs and RR torsion} \label{sec:RRtorsion}

So far we have not addressed the case where more than two NS5 branes
cross a single O5 plane. As argued in~\cite{dP1paper}
such configurations are intrinsically strongly coupled. They
correspond to infinite coupling points in the moduli space of the
unorientifolded parent theory that are halfway between Seiberg dual
gauge theories. Since the strongly coupled degrees of freedom are
localized at the multiple intersection, we expect that they are
described by a conformal field theory (CFT). Overloading notation, we label the class of
CFTs that occur in this way the $\TO_k(N)$ CFTs, $k \ge 2$, where $\TO_k(N)$ also labels the generating brane configuration ($2k$ NS5 branes crossing crossing atop an O5 plane), as above. We will see that, depending on additional discrete data, there are multiple $\TO_k(N)$ CFTs for each $k\ge 2$ and $N$ sufficiently large.

\subsection{Constructing $\TO_k$ CFTs using deconfinement} \label{subsec:TOkdec}

\begin{figure}
  \centering
  \includegraphics{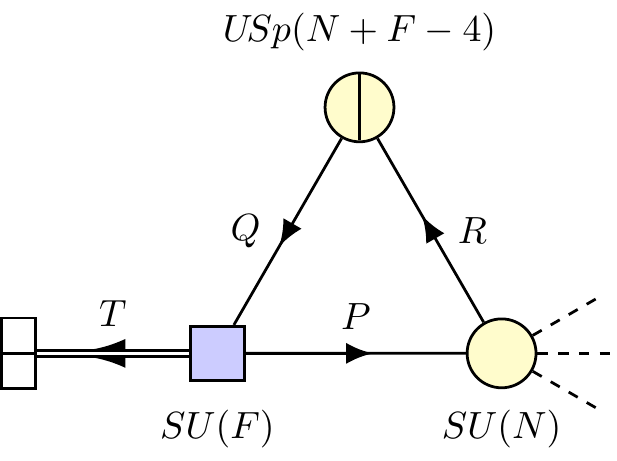}
  \caption{Deconfined description for an antisymmetric tensor, with the superpotential $W = PQR + TQ^2$.}
  \label{fig:antisymmetric-deconfinement-quiver}
\end{figure}
\begin{figure}
\centering
\begin{subfigure}[b]{0.41\textwidth}
  \centering
  \includegraphics[height=5cm]{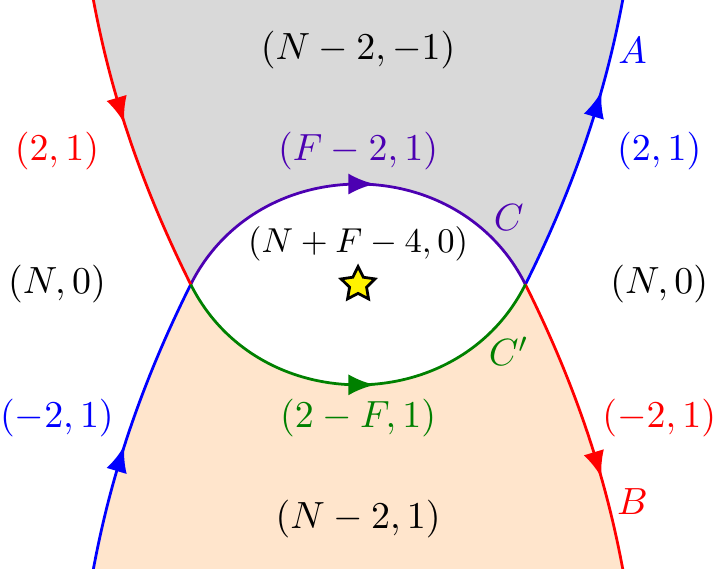}
  \caption{The D5 worldvolume}
  \label{subfig:postlift}
\end{subfigure}
\begin{subfigure}[b]{0.3\textwidth}
  \centering
  \includegraphics[height=5cm]{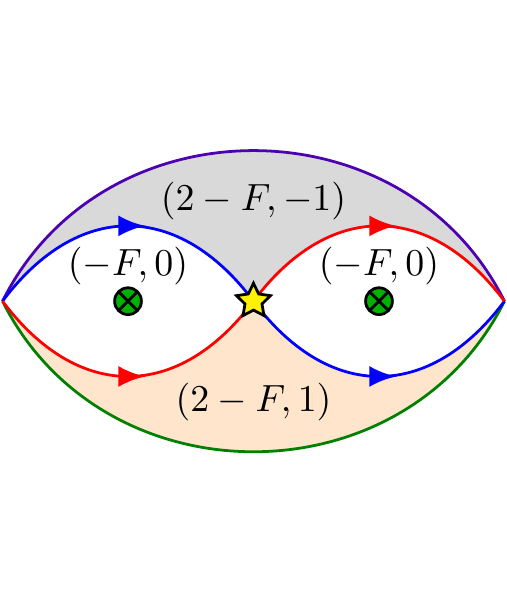}
  \caption{The flavor dome}
  \label{subfig:liftedeye}
\end{subfigure}
\begin{subfigure}[b]{0.27\textwidth}
  \centering
\includegraphics[height=5cm]{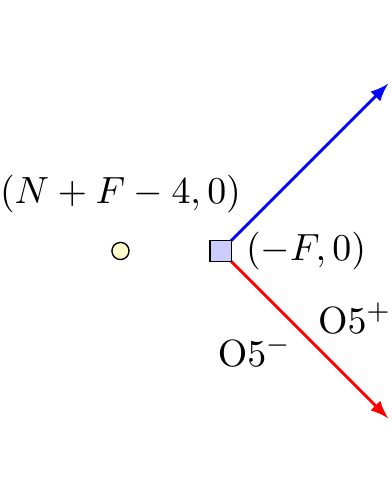}
\caption{The O5 worldvolume}
\label{subfig:O5deconfined}
\end{subfigure}
\caption{The brane configuration engineering the deconfined gauge theory in figure~\ref{fig:antisymmetric-deconfinement-quiver}. Here the brane configuration~\subref{subfig:liftedeye} is attached to the brane tiling~\subref{subfig:postlift} along the purple and green lines (the ``deconfinement ring''), with the interior displaced in the $r_a$ plane to form a dome over the $\Sp(N+F-4)$ face in the tiling. The spurious $\SU(F)$ flavor symmetry of figure~\ref{fig:antisymmetric-deconfinement-quiver} is generated by the $(-F,0)$ faces (the same as $(F,0)$ oppositely oriented) in the ``flavor dome''~\subref{subfig:liftedeye}, which are punctured (green crossed circles) to generate a global symmetry. The configuration of five branes which intersect the O5 plane is shown in~\subref{subfig:O5deconfined}. This differs from figure~\ref{sfig:two-NS5s-O5} in that the two NS5 branes intersect on top of the flavor dome, avoiding the brane tiling.
}
\label{fig:deconfinedbranes}
\end{figure}

As in~\cite{dP1paper}, we use deconfinement~\cite{Berkooz:1995km,Pouliot:1995me} to construct these CFTs. This is based on brane engineering the gauge theory shown in figure~\ref{fig:antisymmetric-deconfinement-quiver}, whose low energy limit (after confinement of the $\Sp(N+F-4)$ node) describes a single antisymmetric tensor of $\SU(N)$ with no superpotential, the $\TO_1(N)$ theory. We engineer this using the non-BPS brane configuration shown in figure~\ref{fig:deconfinedbranes}.
The net effect of this brane engineering is to transfer the O5 intersection point of the two NS5 branes from the brane tiling to a separate stack of $F$ flavor branes which is displaced from the tiling in the $r_a$ plane. This is done by blowing up a ``bubble'' in brane tiling, where the top of the bubble is a dome formed by punctured\footnote{Here by ``puncture'' we mean that a small disk is cut out of the D5 brane world-volume and a semi-infinite cylinder is attached. This makes the volume of the cycle wrapped by the brane infinite, hence the gauge coupling is zero.} flavor branes and recombinations of the NS5 branes and the bottom is a D5 brane face in the brane tiling. The recombined NS5 branes in this ``flavor dome'' generate the superpotential in figure~\ref{fig:antisymmetric-deconfinement-quiver} (see~\cite{dP1paper} for a more detailed explanation).

\begin{figure}
\centering
\begin{minipage}{0.25\textwidth}
\begin{subfigure}[b]{\textwidth}
  \centering
  \includegraphics[height=3cm]{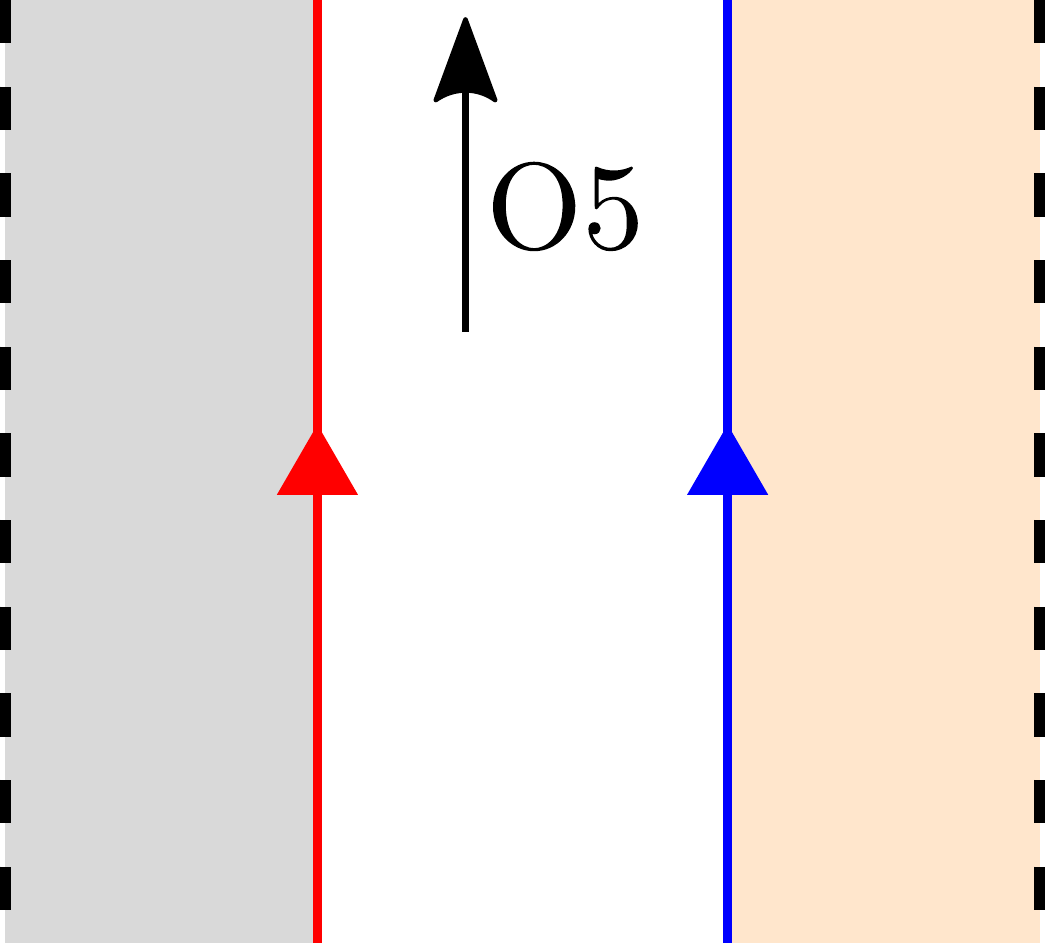}
  \caption{$\TO_1$ cylinder view}
  \label{subfig:TO1cylinder}
\end{subfigure} \\[0.5cm]
\begin{subfigure}[b]{\textwidth}
  \centering
  \includegraphics[height=3cm]{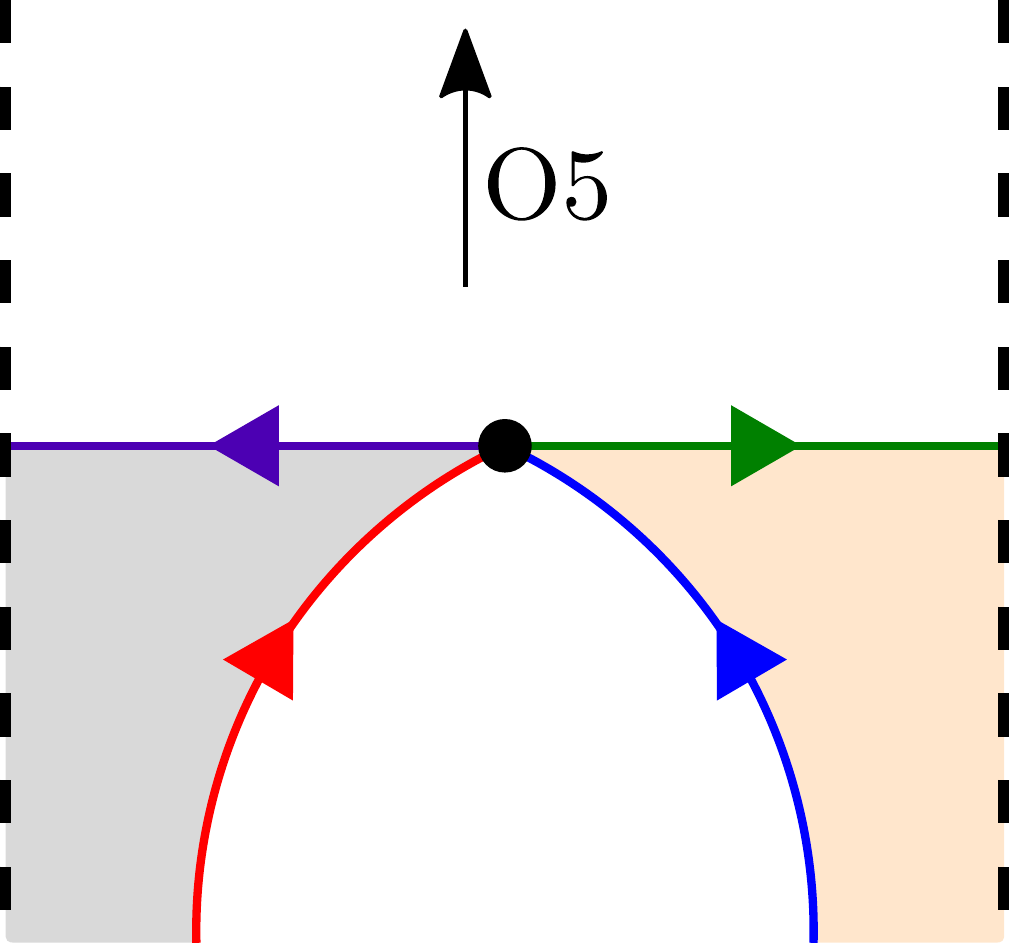}
  \caption{$\TO_1$ deconfined}
  \label{subfig:TO1cylinderDec}
\end{subfigure}
\end{minipage}
\begin{minipage}{0.74\textwidth}
\begin{subfigure}[b]{0.49\textwidth}
  \centering
\includegraphics[height=5cm]{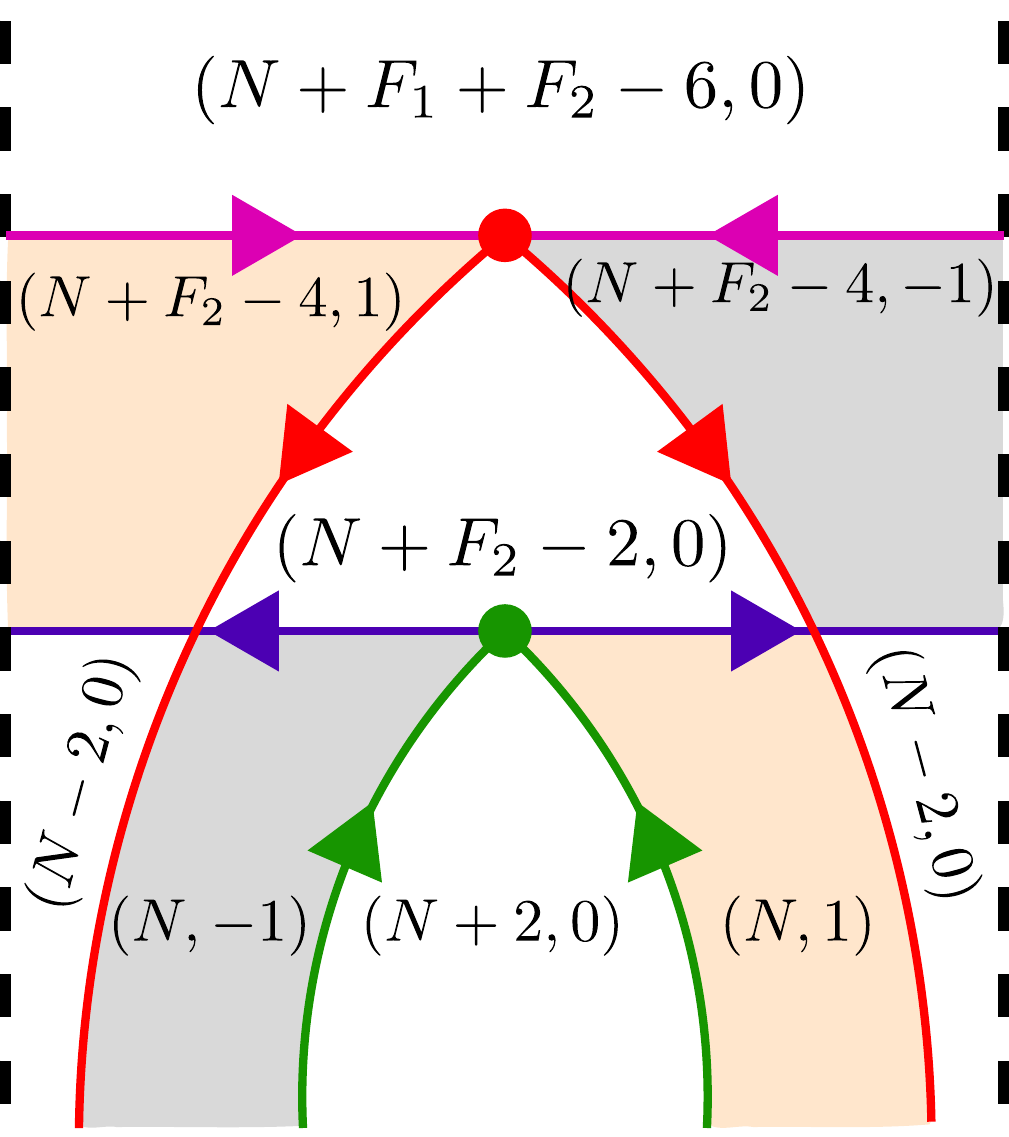}
\caption{$\TO_2$ deconfined}
\label{subfig:TO2cylinder}
\end{subfigure}
\begin{subfigure}[b]{0.49\textwidth}
  \centering
\includegraphics[height=5cm]{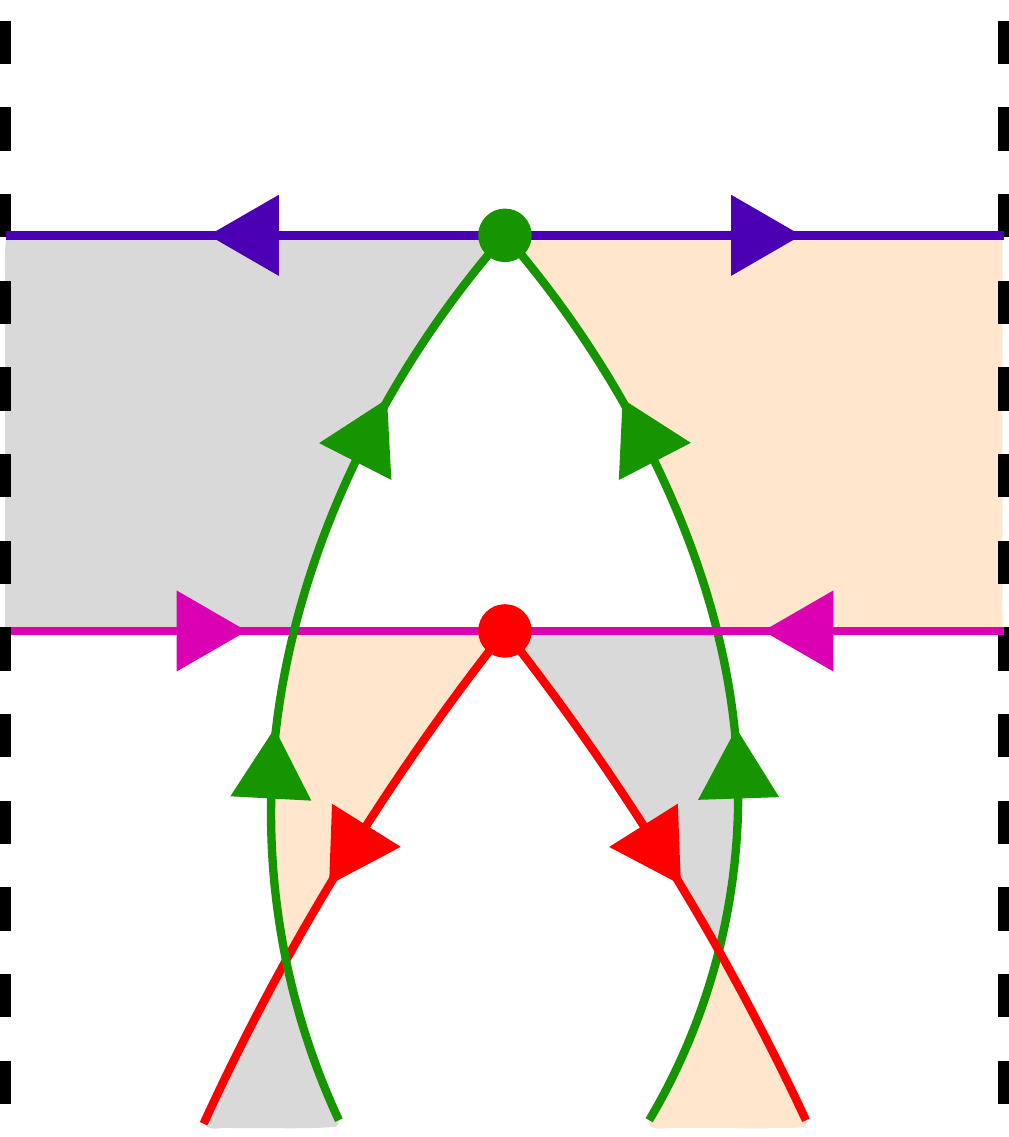}
\caption{Seiberg dual}
\label{subfig:TO2cylinderSD}
\end{subfigure}
\end{minipage}
\caption{\subref{subfig:TO1cylinder}~The $\TO_1$ configuration conformally mapped to a cylinder. The dashed lines---at angles $0$ and $\pi$ on the cylinder---are identified by the O5 involution. \subref{subfig:TO1cylinderDec}~Conformal mapping of the deconfined brane tiling corresponding to $\TO_1$, c.f.\ figure~\ref{subfig:postlift}. \subref{subfig:TO2cylinder}~Engineering a $\TO_2$ CFT using deconfinement. \subref{subfig:TO2cylinderSD}~A Seiberg-dual decription of~\subref{subfig:TO2cylinder}.}
\label{fig:TOkcylinder}
\end{figure}

Thus, deconfinement allows us to move the intersection point of pairs of NS5 branes from the brane tiling to a separate stack of flavor branes (one for each deconfined pair), and avoid higher multiplicity intersections. In this way, we can resolve the $\TO_{k \ge 2}$ configuration into a deconfined configuration with a Lagrangian description. The resulting brane configurations quickly become very complicated, and it is convenient to unroll the neighborhood of the tiling near the O5 plane into a cylinder with the O5 plane at infinity (as in the well-known conformal mapping $w = -i \log z$). Quotienting by the O5 involution, we obtain a cylinder of one-half the circumference which faithfully encodes the brane tiling near the O5 plane. For example, figure~\ref{sfig:two-NS5s-tiling} becomes figure~\ref{subfig:TO1cylinder} and figure~\ref{subfig:postlift} becomes figure~\ref{subfig:TO1cylinderDec}.

An example of deconfining a $\TO_2(N)$ CFT is shown in figure~\ref{subfig:TO2cylinder}. The edges of the deconfined bubbles now form two concentric ``deconfinement rings'' in the brane tiling, each with an associated flavor parity $F_i$ (with the index counting outward from the center), where $(-1)^{F_1 + F_2} = (-1)^N$ because the central $\Sp(N+F_1+F_2-6)$ face must have even rank.\footnote{We will not consider the case of symmetric tensor deconfinement. As argued in~\cite{dP1paper}, it should lead to a description where the RR torsion is encoded in a choice of branch on the moduli space rather than in the flavor parities. Due to the difficult of isolating the desired branch, we found this description to be of little practical use.} The inner ring can be reconfined, replacing $\Sp(N+F_1+F_2-6)$ with an antisymmetric tensor of the outer $\SU(N+F_2 - 2)$ gauge group. This shows that the spurious $\SU(F_1)$ flavor symmetry acts trivially in the infrared, as required. To show that $\SU(F_2)$ is likewise trivial in the infrared, we consider the Seiberg dual of the outer $\SU(N+F_2 - 2)$ gauge group, which exchanges the positions of the two concentric rings, see figure~\ref{subfig:TO2cylinderSD}. Upon confining the central face, $\SU(F_2)$ becomes manifestly trivial.\footnote{In the brane engineered picture, the infrared triviality of $\SU(F_i)$ corresponds to the fact that the $\SU(F_i)$ punctures detach when the deconfinement bubble is allowed to shrink to zero size (reconfine), rendering $\SU(F_i)$ charged states massive.}

The same argument shows that the infrared physics is independent of the choice of $F_i$ up to the parities $(-1)^{F_i}$, because the value of $F$ in figure~\ref{fig:antisymmetric-deconfinement-quiver} is arbitrary, but $(-1)^F$ is fixed by the requirement that the $\Sp(N+F-4)$ node has even rank. Since $(-1)^{F_1+F_2} = (-1)^N$, there are two a priori different flavor parities for any fixed $N$. As analyzed in detail in~\cite{dP1paper} and reviewed in~\S\ref{sec:quad-CFTs}, the two choices of flavor parity indeed give rise to different $\TO_2$ CFTs.

\begin{figure}
\centering
\begin{subfigure}[b]{0.8\textwidth}
  \centering
  \includegraphics[width=10cm]{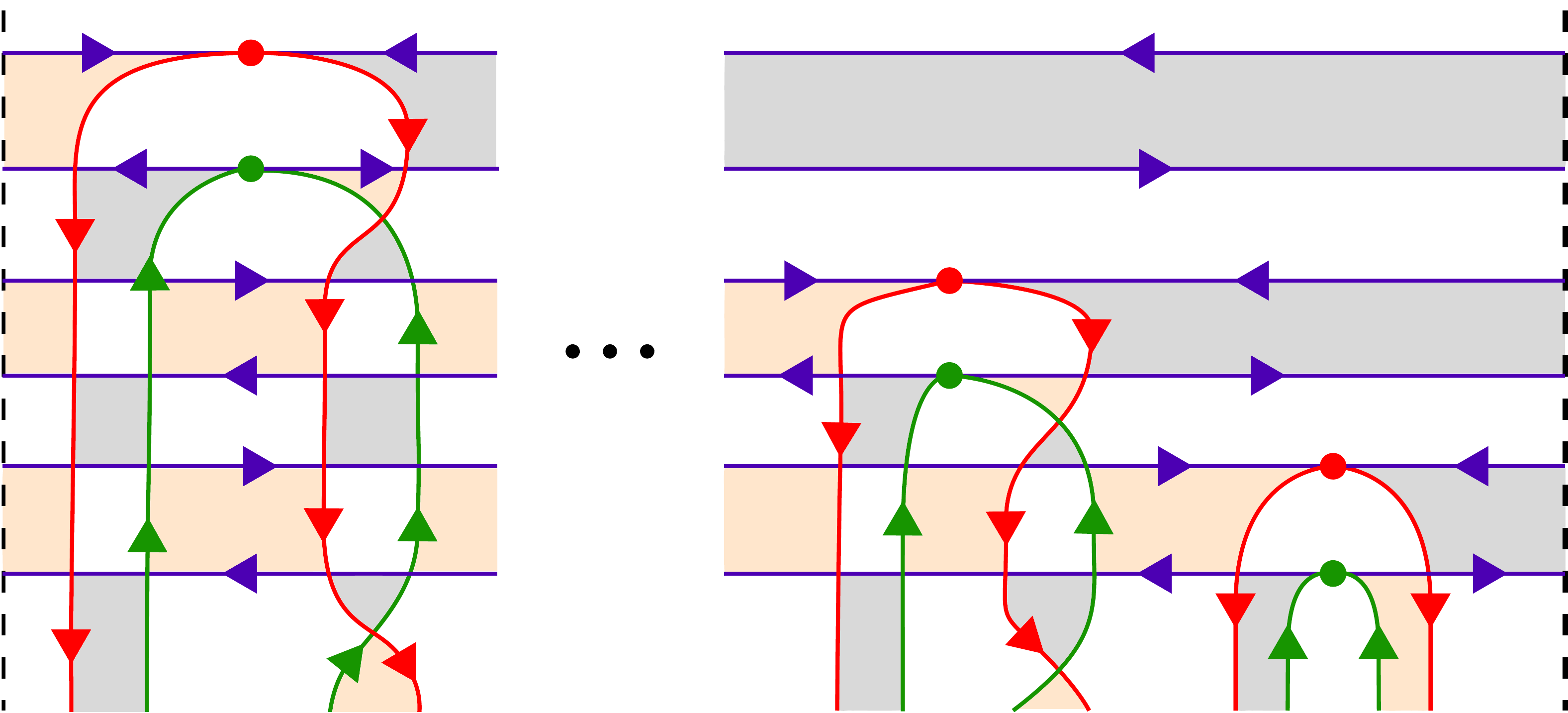}
  \caption{$\TO_{2p}$ deconfined}
  \label{subfig:TOevenk}
\end{subfigure} \\
\begin{subfigure}[b]{0.8\textwidth}
  \centering
  \includegraphics[width=10cm]{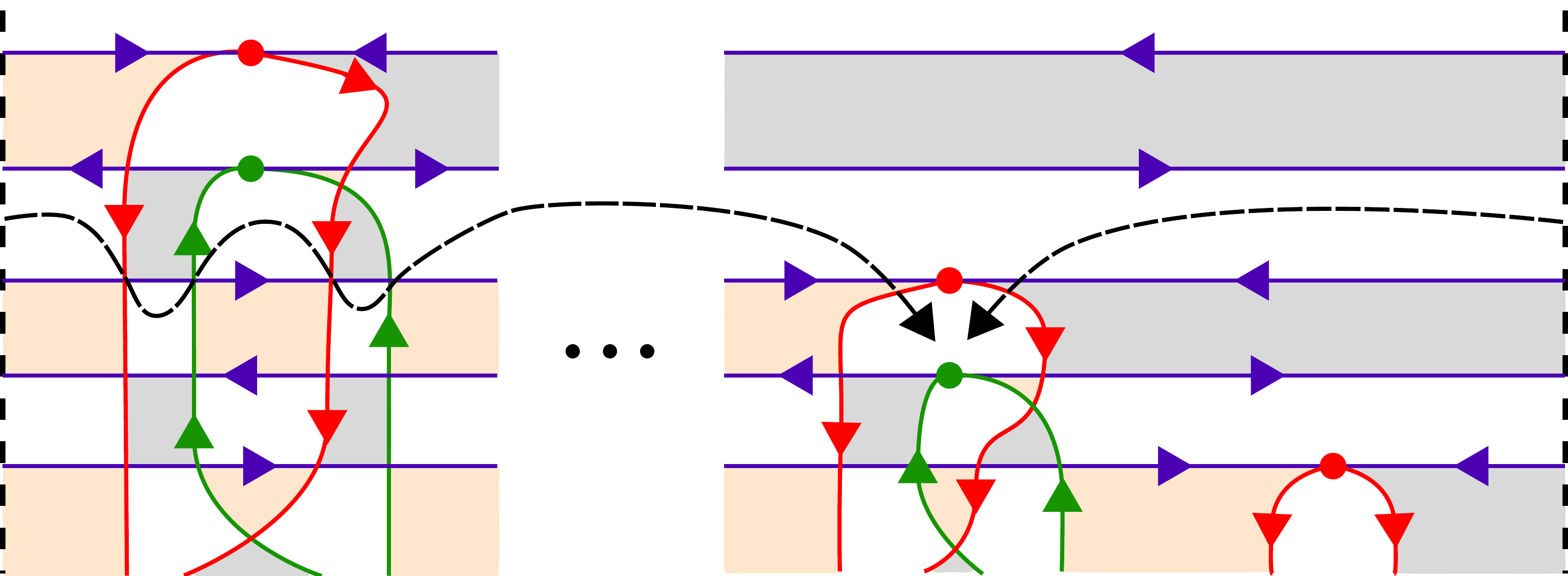}
  \caption{$\TO_{2p+1}$ deconfined}
  \label{subfig:TOoddk}
\end{subfigure}
\caption{Constructing a $\TO_k$ CFT via deconfinement for general~\subref{subfig:TOevenk} even and~\subref{subfig:TOoddk} odd $k \ge 2$. In~\subref{subfig:TOoddk} we also illustrate the path generating the contraction $\mathcal{M}_4$ which appears in the superpotential term $\mathcal{M}_4 Q_4^2 T_4$ generated by the fourth deconfinement ring (counting outwards).}
\label{fig:TOgenk}
\end{figure}

We can deconfine $\TO_k(N)$ for any $k$ in a similar fashion, see figure~\ref{fig:TOgenk}.
Now there are $k$ concentric rings associated to $k$ deconfinement bubbles.
As before, the innermost ring can be replaced with an antisymmetric tensor, and one can show that two adjacent rings can be exchanged by taking the Seiberg dual of each of the D5 brane faces between them.\footnote{It turns out that, so long as we apply Seiberg duality to four-sided faces between the two rings one at a time, the order in which we do so does not matter, and any face with more than four sides will reduce to a four-sided face by the time all other four-sided faces have been dualized. (Seiberg duality for a face with more than four sides does not have a simple brane interpretation.)} Note that each deconfinement ring is attached to flavor branes in the configuration shown in figure~\ref{subfig:liftedeye}, which generate the same collection of flavored fields as in figure~\ref{fig:antisymmetric-deconfinement-quiver}. The only difference is the superpotential, now of the form $P Q R + T Q^2 \mathcal{M}$ where $\mathcal{M}$ is the contraction of the fields along the path that skirts the inside of the bubble through $(N_a, 0)$ faces, see, e.g., figure~\ref{subfig:TOoddk}.

Close cousins of the CFTs engineered by the brane configurations in figure~\ref{fig:TOgenk} can be obtained by ``flipping'' certain mesons $\Phi$~\cite{Gaiotto:2015usa}, i.e., adding a fundamental field $\phi$ in the conjugate representation to $\Phi$ along with a mass deformation $\delta W = m^{2-\Delta_\Phi} \Phi \phi$. For the simplest class of mesons in figure~\ref{fig:TOgenk}, the net effect is to cross or uncross adjacent NS5 branes bordering shaded faces in the bottom row of the figure. 

Notice that the theories constructed in this way all have the property that only adjacent legs in the local web diagram are deconfined. Since we restrict to anti-symmetric (versus symmetric) tensor deconfinement, this implies that each deconfined pair encloses an O5$^+$ wedge, hence the pairs are uniquely determined by the local web diagram and O5 charges. Given these data, we can always construct a deconfined description from figure~\ref{fig:TOgenk} or a flipped variant of it. In fact, there are multiple ways to do this, but they only differ by permuting the deconfinement rings, hence they describe the same infrared fixed point.

It is interesting to consider whether there are additional deconfined brane configurations which are not Seiberg dual to those in figure~\ref{fig:TOgenk} or flipped versions thereof. We comment on this question briefly, though we have been unable to definitively answer it.

\begin{figure}
\centering
\begin{subfigure}[b]{0.3\textwidth}
  \centering
  \includegraphics[height=4cm]{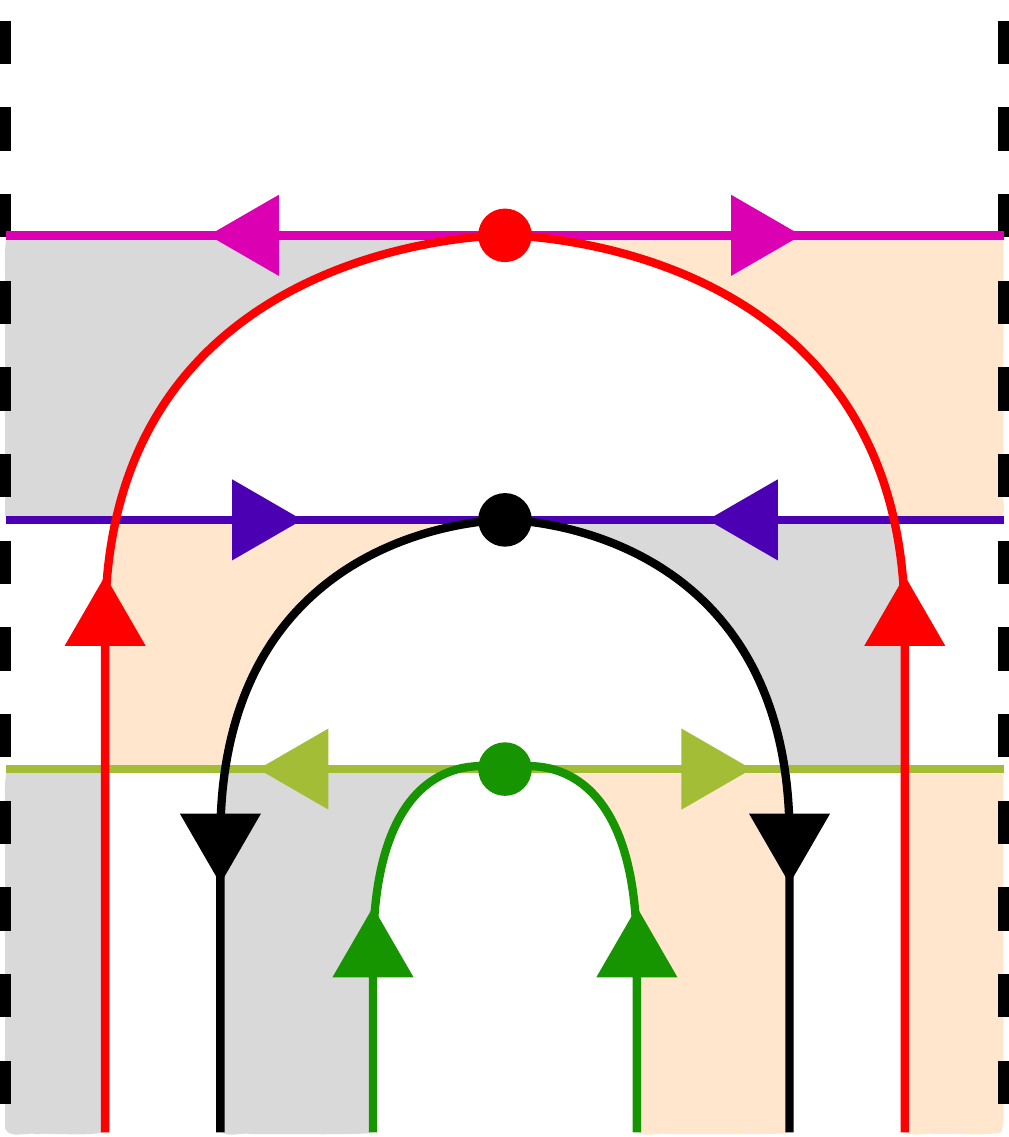}
  \caption{Inconsistent deconfinement}
  \label{subfig:inconsistentDec}
\end{subfigure}
\begin{subfigure}[b]{0.31\textwidth}
  \centering
  \includegraphics[height=3cm]{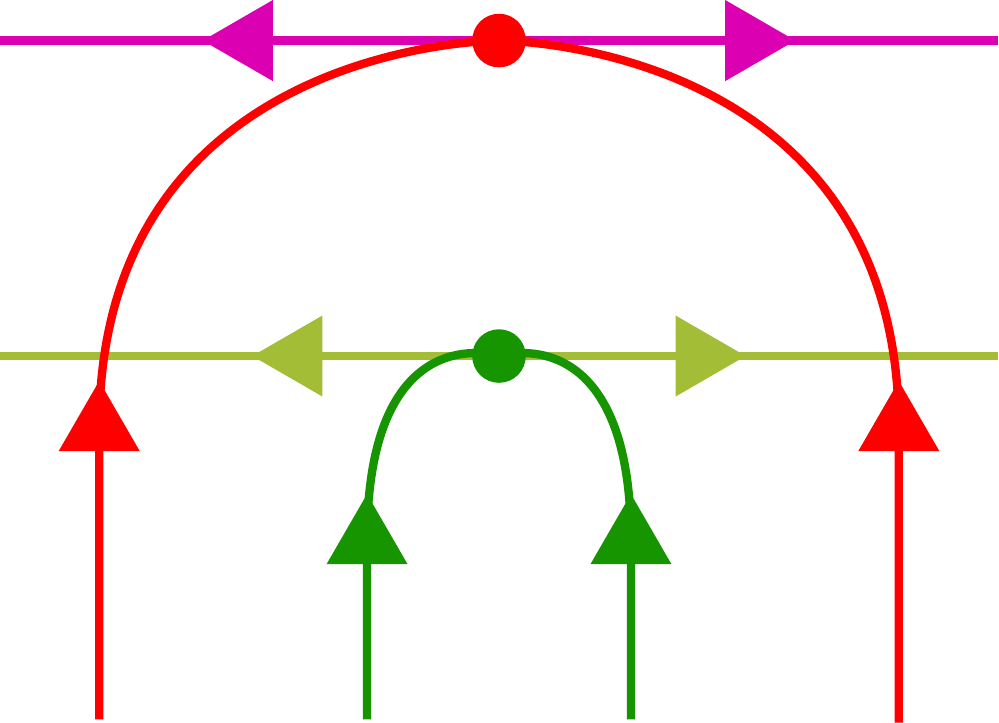}
  \caption{Locked rings}
  \label{subfig:badBubbles}
\end{subfigure}
\begin{subfigure}[b]{0.37\textwidth}
  \centering
  \includegraphics[height=3cm]{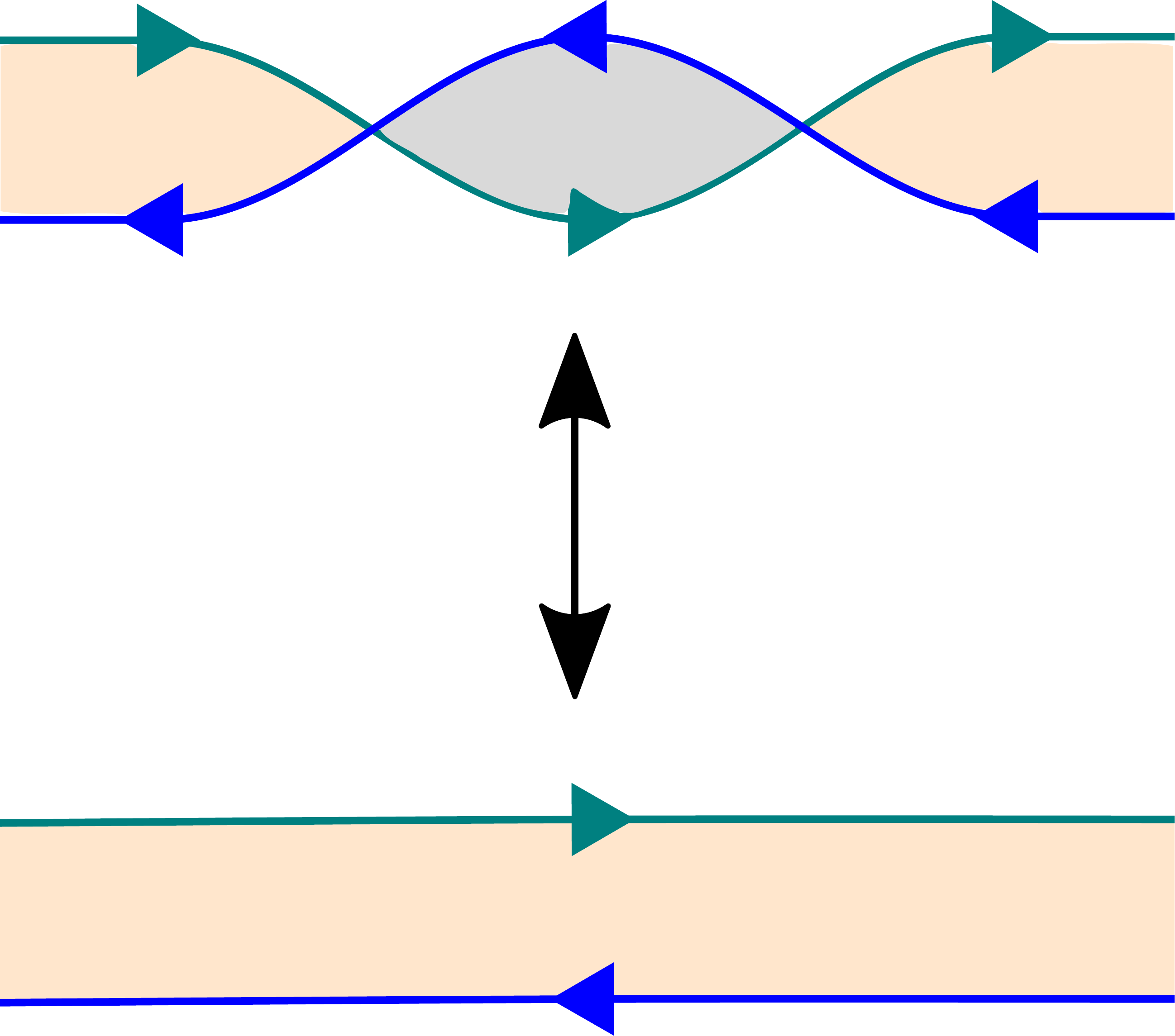}
  \caption{Antiparallel NS5 branes can cross}
  \label{subfig:massCross}
\end{subfigure}
\caption{\subref{subfig:inconsistentDec}~An inconsistent attempt at constructing a $\TO_3$ CFT. The outer deconfinement ring cannot be reconfined, and the associated flavor symmetry $\SU(F_3)$ fails to be trivial in the infrared. \subref{subfig:badBubbles}~The inconsistency of \subref{subfig:inconsistentDec} can be traced to the NS5 brane configuration shown here, which prevents the two deconfinement rings from crossing, because Seiberg duality (figure~\ref{fig:non-Lagrangian-resolved}) and integrating out massive matter~\subref{subfig:massCross} never cross parallel (unlike antiparallel) NS5 branes.}
\end{figure}

We first remark that not all deconfined brane configurations are consistent. For instance, in the configuration shown in figure~\ref{subfig:inconsistentDec} it is not possible to move the outer bubble inside the other two using Seiberg duality.
 This can be traced to the configuration of NS5 branes pictured in figure~\ref{subfig:badBubbles}, where a deconfined pair of NS5 branes encloses another deconfined pair with the same orientation. The parallel NS5 branes prevent two bubbles from crossing, as Seiberg duality never crosses parallel NS5 branes, see figures~\ref{fig:non-Lagrangian-resolved} and~\ref{subfig:massCross}. As a consequence, $\SU(F_3)$ is not trivial in the infrared. (In particular, the anomaly $\SU(F_3)^2 \U(1)_R$ is non-vanishing.) Since the $\SU(F_i)$ were introduced as spurious symmetries needed to construct the deconfined description, this is inconsistent.

A sufficient---and almost certainly necessary---condition for the $\SU(F_i)$ flavor symmetries to be trivial in the infrared is that we can move any given deconfinement ring so that it lies innermost around the O5 plane, and thus can be reconfined into an antisymmetric tensor. As we have seen, the configuration in figure~\ref{subfig:badBubbles} prevents this. Through experimentation, we further hypothesize that no pair of NS5 branes that are both inwards (or both outwards) directed can cross in the brane tiling; this is similar to the consistency conditions on brane tilings without orientifold planes \cite{Hanany:2005ss,Hanany:2006nm,Broomhead:2008an}. Although not obvious sufficient to ensure consistency, the two conditions mentioned above imply that only adjacent legs of the local web diagram can be deconfined, as was true in our construction in figure~\ref{fig:TOgenk}.

On the other hand, if we assume that the deconfinement rings are arranged concentrically without intersecting each other, then the above hypothesized constraints do restrict us to the construction in figure~\ref{fig:TOgenk}, and its Seiberg dual and flipped versions. This assumption is not fully general; for instance, other Seiberg dualities can be used to intertwine the rings in figure~\ref{fig:TOgenk}. However, the concentric arrangement is always possible if for any allowed $\TO_k$ deconfinement we can arrange for one ring to enclose the others, such that the enclosed tiling is an allowed $\TO_{k-1}$ deconfinement. This is physically motivated, in the sense that it should describe how a $\TO_k$ CFT can be partially resolved to a $\TO_{k-1}$ CFT. However, the details are somewhat subtle, and we will not pursue this reasoning any further in the present work.

For the purposes of our paper, it is sufficient to note that the consistent $\TO_k$ configurations constructed above are specified uniquely by the local web diagram, the O5 charges, and flavor parities associated to each O5$^+$ wedge. As we argue in the next section, this gives exactly the right set of CFTs to reproduce the RR torsion. This strongly suggests that other consistent deconfined brane configurations (if any exist) do not describe the $\TO_k$ configuration, but rather have a different physical significance.

\begin{figure}
\centering
\begin{subfigure}[b]{0.49\textwidth}
  \centering
  \includegraphics[height=5cm]{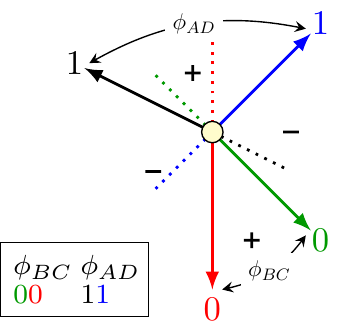}
  \caption{Binary sequence for a $\TO_2$ CFT}
\end{subfigure}
\begin{subfigure}[b]{0.49\textwidth}
  \centering
  \includegraphics[height=5cm]{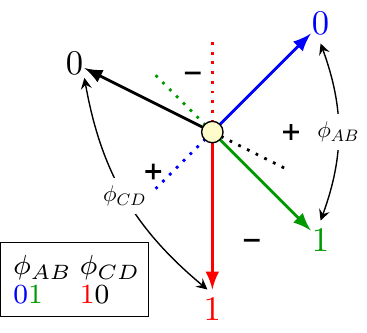}
  \caption{Effect of reversing the O5 charges}
\end{subfigure}
\caption{The web diagram and O5 charges determine a binary sequence which can be decorated with the flavor parities $\phi_\alpha = (-1)^{F_\alpha}$ to label a specific $\TO_k$ CFT.}
\label{fig:binarysequence}
\end{figure}

Before proceeding, we comment briefly on the taxonomy of the $\TO_k(N)$ CFTs constructed above. Due to flavor parities and the choice of the local web diagram (with different choices related by flipping mesons), there are multiple $\TO_k(N)$ CFTs for any fixed $k$ and $N$.\footnote{To be precise, we fix $\floor{N/2}$, since the flavor parities encode $(-1)^N$ and qualitative properties of the CFT depend on it.} Given a choice of O5 charges, the topology of the local web diagram can be fixed by associating a bit 0 (1) to each NS5 brane ray in the web diagram whose antiparallel ray lies within an O5$^+$ (O5$^-$) plane, see figure~\ref{fig:binarysequence}.\footnote{To allow deconfinement, the antiparallel ray to each ray in the local web diagram must lie in the interior of one of the two O5$^{\pm}$ wedges adjacent to the ray $k$ spaces away in the diagram. Obtaining a suitable local web diagram sometimes requires a deformation away from the BPS configuration, as discussed in~\S\ref{subsec:gaugetheoryconstruct}.}
 The resulting binary sequence can be shown to be antiperiodic (periodic) under a shift by $k$ places when $k$ is even (odd), hence the local web diagram encodes $k$ bits of information. Attaching the flavor parities to this sequence in the locations of the associated O5$^+$ wedges, we obtain a chain of $k$ tuples (e.g., $0^+10^-1\ldots1^+0$, abbreviated as \raisebox{4pt}{\scriptsize $\underset{01}{+}\underset{01}{-}\ldots\underset{10}{+}$}
) on which there is a natural $D_k$ action generated by cyclic permutation of the tuples and by reversing the sequence (e.g., $\raisebox{4pt}{\scriptsize $\underset{01}{+}\underset{01}{-}\ldots\underset{10}{+}$} \longrightarrow \raisebox{4pt}{\scriptsize $\underset{01}{+}\ldots\underset{10}{-}\underset{10}{+}$}$). The set of distinct $\TO_k(N)$ CFTs is therefore $\frac{2^k\times2^k}{D_k}$.

Specific $\TO_k$ CFTs can be specified using a notation
$\TO\raisebox{4pt}{\scriptsize $\underset{01}{+}\underset{01}{-}\ldots\underset{10}{+}$}(N)$
which incorporates the sequence of tuples. As an example, a symmetric (antisymmetric) tensor of $\SU(N)$ corresponds to $\TO\raisebox{4pt}{\scriptsize$\underset{00}{n}$}(N)$ ($\TO\raisebox{4pt}{\scriptsize$\underset{11}{n}$}(N)$), where $n=(-1)^N$.

\subsection{RR torsion}
\label{sec:RR-torsion}

In the previous section, we found that for each O5$^+$ wedge, the associated $\TO_k$ CFT has a flavor parity $F_\alpha$, subject to the constraint 
\be \label{eqn:colorcons}
\sum_{\alpha \in P} F_\alpha \equiv N \pmod 2 \,,
\ee
where the sum is taken over the flavor parities of the $\TO_k$ CFT associated to the fixed point~$P$.

Heuristically, we can think of the flavor parities as torsions associated to a single D5 brane transversely intersecting an O5$^+$ plane. Such a D5 brane is its own orientifold image, and consequently cannot be moved away in the $Z$ plane. There is no analogous pinned brane for an O5$^-$ because the putative worldvolume gauge group, $\Sp(1)$, does not exist. Consequently, for an O5 plane divided into O5$^\pm$ regions by NS5 branes the pinned branes cannot cross the O5$^-$ regions, and there is a separate ``flavor parity'' associated to each O5$^+$ region.

As in the well known case of $Op^-$ and $\widetilde{Op}^-$ planes (see e.g.~\cite{Witten:1998xy,Hanany:2000fq,Hyakutake:2000mr}), these flavor parities are likely associated to discrete fluxes of RR form fields. Although the complicated arrangement of NS5 branes makes an explicit analysis difficult, this suggests that that the flavor parities correspond to RR torsion in the T-dual, as in~\cite{dP1paper}.

As a preliminary check of this conjecture, we count the number of
flavor parity bits and compare with the RR torsion. The details depend
on the number of ``complete'' O5 planes (those which intersect no NS5
branes). There cannot be more than two of these, and when there are
two their RR charges must be opposite, since none of the corners lie
on the even sublattice containing the origin. In this case, for an
$n$-sided toric diagram there are $n$ flavor parities subject to two
constraints of the form~(\ref{eqn:colorcons}), with the color parity
fixed to be even by the $\Sp(N)$ face, for a total of $n-2$ bits. If
there is one complete O5 plane with positive RR charge then there are $n$
flavor parities subject to three constraints as well as the color
parity, for a total of $n-2$ bits. If the complete O5 plane has negative
RR charge then the color parity is fixed to be even, and there are
$n-3$ bits. Finally, if there are no complete O5 planes then there are $n$
flavor parities subject to four constraints as well as the color
parity, for a total of $n-3$ bits.

Thus, there are $n-2$ flavor parity bits when there is a complete O5 plane of positive charge and $n-3$ otherwise. The RR torsion $[F]$ contains $n-2$ bits, but when $[H] \ne 0$ the choices $[F]$ and $[F]+[H]$ are related by $\tau \to \tau+1$, and one bit is absorbed by doubling the period of the theta angle. By~(\ref{eqn:Htorsion}) $[H]$ is trivial if and only if there is a complete O5 plane of positive charge, so the two counts match.

This counting and the interpretation of odd flavor parity as non-trivial torsion suggests that $F_\alpha$ are the components of $[F]$ in some basis, i.e.,
\be \label{eqn:Falpha}
F_\alpha \equiv [F]\cdot Y_\alpha \pmod 2\,,
\ee
for $Y_\alpha$ independent of $[F]$. Because $[F]$ and $[F]+[H]$ give equivalent flavor parities, we require $Y_\alpha \cdot [H]=0$, whereas the $Y_\alpha$ should span the orthogonal complement of $[H]$ to encode the remaining bits of $[F]$ in $F_\alpha$.
To reproduce~(\ref{eqn:colorcons}), we require that
\be \label{eqn:Ycolor}
Y \equiv \sum_{\alpha \in P} Y_\alpha \,,
\ee
is independent of the choice of fixed point, $P$. Moreover, $Y$ should
vanish if and only if there is a complete O5 plane with negative charge, i.e., when $[H]=\eta$ where $\eta = \sum_i \langle i \rangle$ is the norm element satisfying $A \cdot \eta = A^2$. (This reflects the color parity constraint from a $\Sp(N)$ gauge group factor.)

We now construct a basis with these properties. An elementary solution to the color parity constraint is $Y=[H]+\eta$, which satisfies $Y\cdot [H]=0$ and implies a simple formula for the color parity
\be \label{eqn:colorparity}
N \equiv [F]\cdot([F]+[H]) \pmod 2 \,,
\ee
using the properties of $\eta$. Thus,
\be
\sum_{\alpha \in P} Y_\alpha = \sum_{i \in V_{(P)}} \langle i \rangle \,,
\ee
where $V_{(P)}$ denotes the set of corners of the toric diagram which lie within O5$^+$ wedges in the local web diagram associated to the fixed point $P$. Since $V_{(P)} = \bigcup_{\alpha \in P} V_\alpha$ where $V_\alpha$ is the set of corners within a given O5$^+$ wedge, we make the natural guess
\be \label{eqn:Yalpha}
Y_\alpha = \sum_{i \in V_\alpha} \langle i \rangle \,,
\ee
which determines $[F]$ using~(\ref{eqn:Falpha}). This is consistent with the intuition that the flavor parities encode torsions associated to the O5$^+$ wedges.

It is straightforward to check that the ansatz~(\ref{eqn:Yalpha}) satisfies $Y_\alpha \cdot [H]=0$, whereas most other consistency conditions are true by construction. It remains to be checked that the $Y_\alpha$ span the orthogonal complement of $[H]$. To do so, it is helpful to work in the $n$-dimensional vector space $\bZ_2^n$ before imposing the relations~(\ref{eqn:twistedlattice}). Now $Y$ defined by~(\ref{eqn:Ycolor}) depends on $P$, with $Y_{(P)} - Y_{(P')} = \sum_{i \in L} \langle i \rangle + \sum_{i \in L'} \langle i \rangle$ for a pair of even sublattices $L,L'$.
A single linear relation remains
\be \label{eqn:Ysumreln}
\sum_\alpha Y_\alpha = 0 \,,
\ee
where the sum is taken over all fixed points. When $[H] \ne 0$, there are $n$ O5$^+$ planes, each bounded by a pair of NS5 branes. For $[H] = 0$, there is a further complete O5$^+$ plane. Thus, the $Y_\alpha$ span the orthogonal complement of $[H]$ iff~(\ref{eqn:Ysumreln}) is the only linear dependence within $\bZ_2^n$. This follows from the assumption that adjacent corners occupy distinct even sublattices not containing the origin.\footnote{In particular, each corner $\langle i \rangle$ appears in exactly two O5$^+$ planes and adjacent corners $\langle i \rangle$, $\langle i+1 \rangle$ share at least one O5$^+$ plane between them. This establishes~(\ref{eqn:Ysumreln}) and shows that there are no further linear relations.}

To summarize, we have argued that
\be \label{eqn:RRtorsionprescription}
\boxed{F_\alpha \equiv [F]\cdot \sum_{i \in V_{\alpha}} \langle i \rangle \pmod 2}\,,
\ee
fixes the relation between the flavor parities and the RR torsion, where $V_\alpha$ denotes the set of corners of the toric diagram falling within the O5$^+$ plane $\alpha$ in the local web diagram. Along with~(\ref{eqn:Htorsion}), this is one of our principal results, and completes the AdS/CFT dictionary for the class of toric orientifolds considered in our paper.

The relation~(\ref{eqn:RRtorsionprescription}) can be inverted by choosing a dual basis $Y^\alpha$, such that
\be
Y_\alpha \cdot Y^\beta_{(I)} = \delta_\alpha^\beta \,,\qquad \alpha \in I\,.
\ee
Here the indexing set $I$ of O5$^+$ planes is any maximal choice such that the $Y_\alpha$, $\alpha \in I$ are linearly independent.
For $[H]=0$, this is a standard linear algebra problem with a unique solution for any given maximal $I$;\footnote{For a more thorough treatment of bases and dual bases in $\bZ_2$ vector spaces, see Appendix~\ref{app:Z2bilinear}.} for $[H] \ne 0$, the solution is ambiguous up to $Y^\alpha_{(I)} \to Y^\alpha_{(I)} + n^\alpha [H]$. In either case, we have
\be
[F] = \sum_{\alpha \in I} F_\alpha Y^\alpha \,,
\ee
which fixes $[F]$ up to $[F] \to [F] + [H]$.

\medskip

\begin{figure}
  \centering
  \begin{subfigure}{0.3\textwidth}
    \centering
    \includegraphics[height=3.5cm]{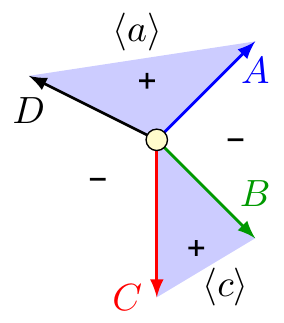}
    \caption{$(1,1)$}
    \label{sfig:local-charges-NE-shaded}
  \end{subfigure}
  \hspace{1cm}
  \begin{subfigure}{0.3\textwidth}
    \centering
    \includegraphics[height=3.5cm]{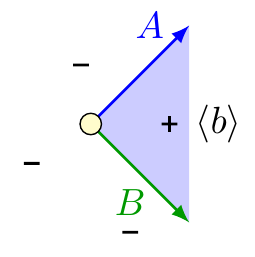}
    \caption{$(0,0)$}
    \label{sfig:local-charges-SW-shaded}
  \end{subfigure}
  \caption{Wedges with positive charge for the \IA\ phase of the
    complex cone over $dP_1$, analyzed in \S\ref{sec:dP1} below, at
    the fixed points $(0,0)$ and $(1,1)$ in the conventions of
    figure~\ref{fig:local-charges-example}. We have indicated the
    torsion generators associated with the shaded cones.}
  \label{fig:dP1-example-flavor-charges}
\end{figure}

Let us illustrate this discussion in the same example
discussed in \S\ref{sec:NSNS-torsion}, see in particular
figure~\ref{fig:local-charges-example}. Let us focus for instance on
the fixed point at $(1,1)$, shown in
figure~\ref{sfig:local-charges-NE}. The wedges with positive charge
are those in figure~\ref{sfig:local-charges-NE-shaded}. The wedges
$AD$ and $BC$ enclose the toric torsion generators $\vev{a}$ and
$\vev{c}$ respectively (in the conventions of
figure~\ref{sfig:dP1-example-conventions}), so according
to~\eqref{eqn:RRtorsionprescription} we have for the two flavor
parities of the quad-CFT at $(1,1)$
\begin{equation}
  F_{AD}^{(1,1)} \equiv [F]\cdot \vev{a} \pmod 2
\end{equation}
and
\begin{equation}
  F_{BC}^{(1,1)} \equiv [F]\cdot \vev{c} \pmod 2\, .
\end{equation}
It is an easy exercise to verify that choosing any other fixed point
in figure~\ref{fig:local-charges-example} leads to results consistent
with these. For instance, if we choose the point at $(0,0)$ we obtain
the positive wedge show in
figure~\ref{sfig:local-charges-SW-shaded}. The $AB$ wedge encloses
the $\vev{b}$ generator, so for the global parity of the tiling
we have
\begin{equation}
  N \equiv [F]\cdot\vev{b}\pmod 2
\end{equation}
which indeed satisfies \eqref{eqn:colorcons} once we take into
account~\eqref{eqn:twistedlattice}.

As above, we can invert the relationship between $[F]$ and the color
and flavor parities using a dual basis $Y^\alpha$. We choose the
generators $Y_1=Y_{AD}=\vev{a}$ and $Y_2=Y_{AB}=\vev{b}$. An
elementary computation then shows that the dual basis is given by
\begin{equation}
  Y^1 = \vev{b}\quad ; \quad Y^2 = \vev{c} \, .
\end{equation}
Therefore
\begin{equation}
  [F] = F_{AD}^{(1,1)} \vev{b} + N \vev{c} \,.
\end{equation}

\section{Details of \alt{$\TO_2$}{TO2} CFTs}
\label{sec:quad-CFTs}

\begin{figure}
  \centering
  \includegraphics[width=5cm]{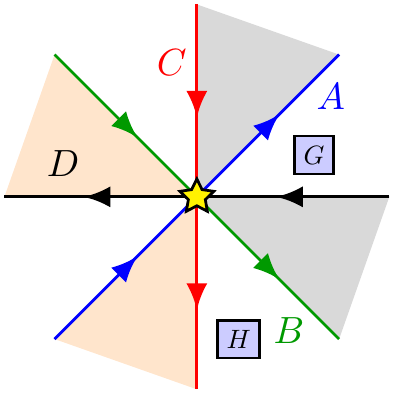}
  \caption{The $\TO_2$ configuration. We have named $A,\ldots,D$ the NS5
    branes intersecting at the fixed point, and $G\times H$ the
    manifest non-Abelian global symmetry group.}
  \label{fig:local-four-NS5s}
\end{figure}
Having established the AdS/CFT dictionary for toric orientifolds, the remainder of our paper is devoted to constructing examples and checking that their S-duality properties are as expected. For simplicity, we restrict our attention to toric diagrams with at most five sides, so it is sufficient to understand the $\TO_2$ CFTs---coming from the $\TO_2$ brane configuration, figure~\ref{fig:local-four-NS5s}---whose field theoretic details we now review (see also~\cite{dP1paper}).

\subsection{Mesons, baryons, global symmetries, and anomalies}

There are two classes of $\TO_2$ CFTs, depending on the local web diagram and O5 charges. We label them as\footnote{By a $D_2$ reflection $\quadSO^+(2p+1) \cong \quadSO^-(2p+1)$. It can be shown that $\quadSO^+(2p) \cong \quadSO^-(2p)$ as well~\cite{dP1paper}, hence the flavor parity label can be dropped in this case.}
\begin{align}
&\begin{aligned}
\quadSp^\phi(M) &= \TO\raisebox{4pt}{\scriptsize $\underset{00}{\phi}\,\underset{11}{m\phi}$}(M+2) \,,\\
\quadSO^\phi(M) &= \TO\raisebox{4pt}{\scriptsize $\underset{01}{\phi}\,\underset{10}{m\phi}$}(M+2) \,,
\end{aligned} &  m &= (-1)^M \,,
\end{align}
in relation to the notation defined in~\S\ref{subsec:TOkdec}. In either case, the global symmetry manifest in the brane tiling description is $\SU(M)\times\SU(M+4)\times\U(1)^3\times\U(1)_R$, where the anomaly-free $\U(1)$ symmetries correspond to the $\U(1)$ gauge fields on the four NS5 branes (with an overall decoupled $\U(1)$). It turns out that this is enhanced to $\Sp(2M)\times\SU(M+4)\times\U(1)^2\times\U(1)_R$ ($\SU(M)\times\SO(2(M+4))\times\U(1)^2\times\U(1)_R$) in the \quadSp\ (\quadSO) CFTs~\cite{dP1paper}, justifying their names.\footnote{The letter $\mathfrak{q}$ is in reference to the label ``quad CFT'' used in~\cite{dP1paper}.} Similar rank-preserving enhancements play a role in some of the toric orientifold CFTs constructed later in the paper.

\subsubsection*{The \alt{\quadSp}{qSp} theories}

\begin{figure}
  \centering
  \begin{subfigure}[b]{0.4\textwidth}
    \centering
    \includegraphics[height=6cm]{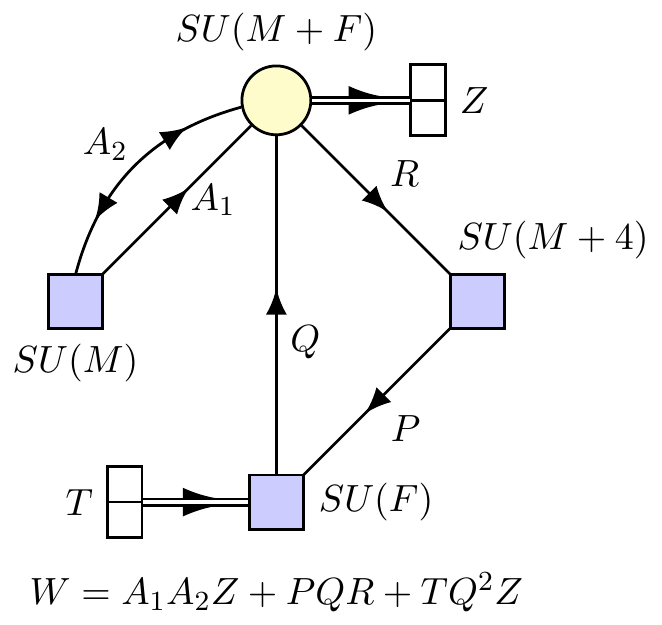}
    \caption{Quiver and superpotential for \LSp{A}.}
    \label{sfig:qSp-A-quiver}
  \end{subfigure}
  \hfill
  \begin{subfigure}[b]{0.55\textwidth}
    \centering
    \includegraphics[height=6cm]{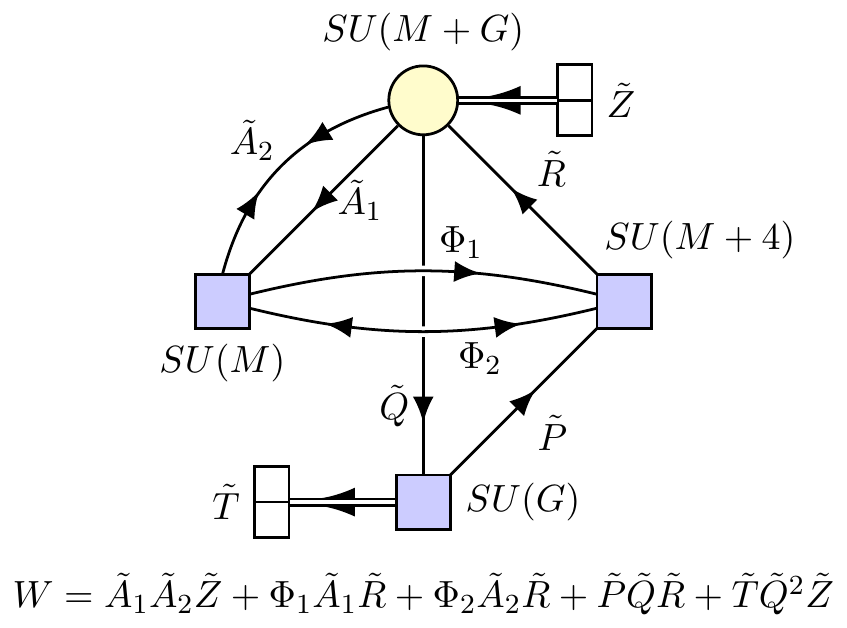}
    \caption{Quiver and superpotential for \LSp{B}.}
    \label{sfig:qSp-B-quiver}
  \end{subfigure}
  \caption{The two choices for the deconfined Lagrangian description
    of $\quadSp$. Both gauge theories flow to the same fixed point when $F+G\equiv M \pmod 2$.}
  \label{fig:qSp-deconfinement}
\end{figure}

The brane tiling for a deconfined description of $\quadSp^\phi(M)$ is shown in figure~\ref{subfig:TO2cylinder}. Reconfining the center face to produce an antisymmetric tensor and reading off the gauge group and superpotential using the methods described in sections~\ref{sec:NStorsion} and~\ref{sec:RRtorsion}, we obtain the quiver gauge theory in figure~\ref{sfig:qSp-A-quiver}.
An alternate description can be obtained by reconfining the center face of the Seiberg dual brane tiling in figure~\ref{subfig:TO2cylinderSD}, which gives the quiver gauge theory in figure~\ref{sfig:qSp-B-quiver}. For fixed $M$ and flavor parity $\phi=(-1)^F = (-1)^{G+M}$, these two gauge theories, which we label \LSp{A} and \LSp{B}, lie in the same universality class as the $\TO_2$ theory $\quadSp^\phi(M)$ that we are interested in. Notice that the enhanced symmetry $\Sp(2M) \supset \U(M)$ is manifest in the quiver diagrams, though it is hidden in the brane tiling description.

We focus on the description \LSp{A} for definiteness. The charge table for this quiver gauge theory is
{\small
\begin{equation}
  \label{eq:qSpA-charges}
        \begin{array}{c|c|ccccccc}
    & \SU(M+F) & \SU(M) & SU(M+4) & SU(F) & \U(1)_B & U(1)_X & U(1)_Y & U(1)_R\\
    \hline
    A_1 & \ov\fund & \fund & \singlet & \singlet & -\frac{1}{M+F}  & 1 & - \frac{M+4}{2(M+F)} & 1- \frac{M+4}{4(M+F)}\\
    A_2 & \ov\fund & \ov\fund & \singlet & \singlet & -\frac{1}{M+F}  & -1 & - \frac{M+4}{2(M+F)} & 1- \frac{M+4}{4(M+F)} \\
    R & \fund & \singlet & \ov\fund & \singlet & \frac{1}{M+F} & 0 & -1+\frac{M+4}{2(M+F)} & \frac{M+4}{4(M+F)} \\
    Z & \asymm & \singlet & \singlet & \singlet & \frac{2}{M+F} & 0 & \frac{M+4}{M+F} & \frac{M+4}{2(M+F)} \\
    P & \singlet & \singlet & \fund & \ov\fund & \frac{1}{F} & 0 & 1-\frac{M+4}{2F} & 2+\frac{M-4}{4 F} \\
    Q & \ov\fund & \singlet & \singlet & \fund & -\frac{1}{M+F} -\frac{1}{F} & 0 & \frac{M+4}{2F}- \frac{M+4}{2(M+F)} & -\frac{M-4}{4F}- \frac{M+4}{4(M+F)} \\
    T & \singlet & \singlet & \singlet & \ov\asymm & \frac{2}{F} & 0 & -\frac{M+4}{F} & 2+\frac{M-4}{2F}
  \end{array} 
\end{equation}}\\[-5pt]
where we have chosen a slightly different set of conventions
compared to that in \cite{dP1paper}:
\begin{align}
U(1)_Y^{\text{here}} &= U(1)_Y^{\text{there}} - \frac{M+4}{2} U(1)_B\,, & U(1)_R^{\text{here}} &= U(1)_R^{\text{there}} + \frac{M-4}{4} U(1)_B \,.
\end{align}

After imposing the $F$-term conditions, two chiral meson operators remain,\footnote{We ignore operators that are not $\SU(F)$ or $\SU(G)$ invariant. These are lifted in the infrared by quantum corrections, and are artifacts of the deconfined description.} $\Phi_1$ in the $(\fund, \ov\fund)$ representation of
$SU(M)\times SU(M+4)$ and $\Phi_2$ in the $(\ov\fund,\ov\fund)$ representation. 
In the \LSp{A} description both mesons are composite ($\Phi_i = A_i R$ in the notation of figure~\ref{fig:qSp-deconfinement}), whereas in the $\LSp{B}$ description both are elementary. In addition there are chiral baryon operators, of the form
\begin{equation}
\begin{aligned}
  \label{eqn:quadCFTbaryons}
  \mathcal{A}_k &= A_1^k A_2^{M-k} Q^F \,, & 0&\le k \le M \,,\\
  \mathcal{S}_k &= Z^{\frac{F+k-4}{2}} R^{M+4-k} \,, & 0&\le k \le M+4\,, & (-1)^k &= (-1)^F \,,
\end{aligned}
\end{equation}
in the \LSp{A}\ description.\footnote{To be precise, the formula for $\widetilde{\mathcal{O}}_{0,1}$ applies for $F>2$. When $F=1$ ($F=2$) we have $\widetilde{\mathcal{O}}_{1}=P$ ($\widetilde{\mathcal{O}}_{0}=T$).} Note that the baryons $\mathcal{A}_k$ combine to form the $M$-index antisymmetric tensor representation of $\Sp(2M)$, whereas the mesons $\Phi_i$ combine to form the $(\fund,\ov\fund)$ representation of $\Sp(2M)\times\SU(M+4)$. The baryons $\mathcal{S}_k$ are $\Sp(2M)$ singlets; their $\SU(M+4)$ representations, which depend on the flavor parity, are discussed in the next section.

The anomalies of the \quadSp\ theory can be divided into two classes:
\begin{align}
A&:
\begin{array}{|c|c|}
\hline
\SU(M+4)^3 & -M\\
\SU(M)^2 U(1)_Y & -(M+4)\\
\SU(M+4)^2 U(1)_Y & -M \\
\U(1)_Y^3 & -M (M+4) \\
\U(1)_X^2 \U(1)_Y & -M (M+4)\\
\U(1)_Y & - M (M+4) \\ \hline
\end{array}
& B&:
\begin{array}{|c|c|}
\hline
\SU(M)^2 U(1)_B & -2 \\
\SU(M)^2 U(1)_R & -\frac{M+4}{2} \\
\SU(M+4)^2 U(1)_B & 2\\
\SU(M+4)^2 U(1)_R & -\frac{M}{2} \\
\hline
U(1)_X^2 U(1)_B & -2 M\\
U(1)_X^2 U(1)_R & -\frac{M(M+4)}{2} \\
U(1)_Y^2 U(1)_B & 2 (M+4)\\
U(1)_Y^2 U(1)_R &  -\frac{M(M+4)}{2} \\
\hline
U(1)_B^2 U(1)_R & -4 \\
U(1)_B U(1)_R^2 & -2 \\
U(1)_R^3 & \frac{M(M+4)}{4}-1 \\
U(1)_B & 6\\
U(1)_R & -\frac{M(M+4)}{2}-1 \\ \hline
\end{array}
\label{eqn:quadSpAnomalies}
\end{align}
where the first class represents the contribution of the mesons
$\Phi_i$ if they are counted as ``half'' chiral multiplets and the second class exhibits an underlying enhanced symmetry, as discussed below. In
addition to the above anomalies, there is a Witten anomaly $(-1)^{M+F}$ for the $\Sp(2M)$ global symmetry, which is a simple illustration of the fact that $\quadSp^+(M) \ne \quadSp^-(M)$.

\subsubsection*{The \alt{\quadSO}{qSO} theory}

\begin{figure}
  \centering
  \begin{subfigure}[t]{0.4\textwidth}
    \centering
    \includegraphics[height=6cm]{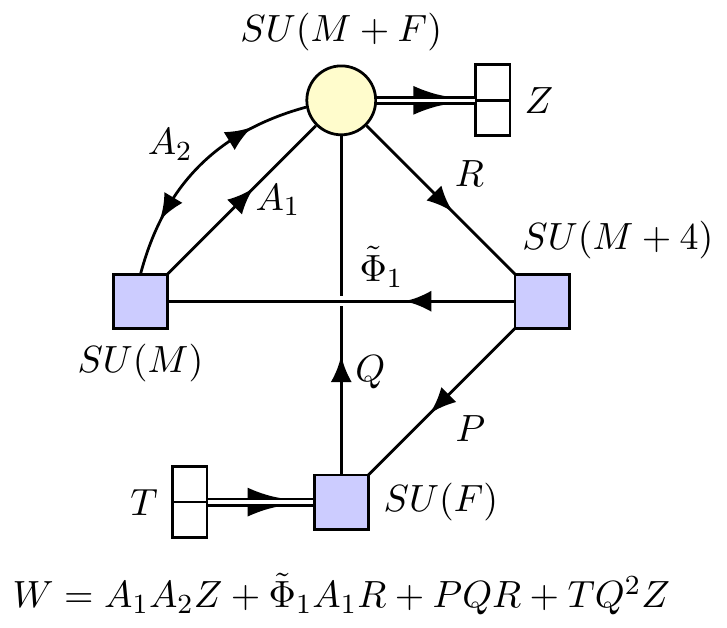}
    \caption{Quiver and superpotential for \LSO{A}.}
    \label{sfig:qSO-A-quiver}
  \end{subfigure}
  \hspace{2cm}
  \begin{subfigure}[t]{0.4\textwidth}
    \centering
    \includegraphics[height=6cm]{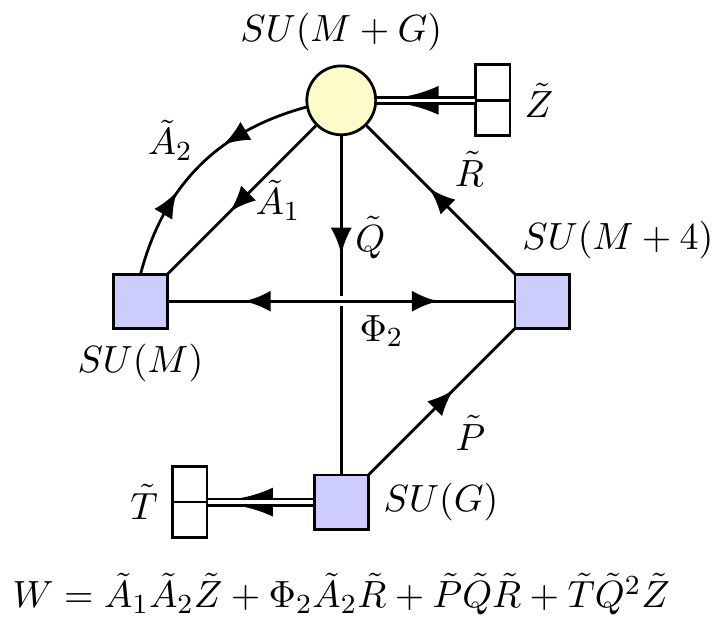}
    \caption{Quiver and superpotential for \LSO{B}.}
    \label{sfig:qSO-B-quiver}
  \end{subfigure}
  \caption{The two deconfined quivers for $\quadSO$, related by
    deconfinement duality when the ranks satisfy $F+G\equiv M \pmod 2$. They
    are isomorphic up to complex conjugation of the $SU(M)$
    representations and relabeling of the fields.}
  \label{fig:qSO-deconfinement}
\end{figure}

To obtain the brane tiling for $\quadSO^\phi(M)$, we flip one of the two mesons $\Phi_{1,2}$ in figure~\ref{subfig:TO2cylinder} (flipping both gives back figure~\ref{subfig:TO2cylinderSD}). The resulting quiver gauge theory is either figure~\ref{sfig:qSO-A-quiver} or figure~\ref{sfig:qSO-B-quiver}, depending on whether we start with the description in figure~\ref{sfig:qSp-A-quiver} or figure~\ref{sfig:qSp-B-quiver}.
As a result, the meson $\Phi_1$ is lifted and replaced with $\tilde{\Phi}_1$ in the $(\ov\fund,\fund)$ representation of $\SU(M)\times\SU(M+4)$. In the \LSO{A} description (figure~\ref{sfig:qSO-A-quiver}) $\tilde{\Phi}_1$ is composite and $\Phi_2$ is elementary whereas in the \LSO{B} description (figure~\ref{sfig:qSO-B-quiver}) the opposite is true. As above the two descriptions are equivalent for fixed $M$ and $\phi = (-1)^F = (-1)^{G+M}$.

Flipping $\Phi_1$ breaks the enhanced symmetry $\Sp(2M)$ back to $\SU(M)\times\U(1)_X$. However, there is now an hidden $\SO(2(M+4)) \supset \SU(M+4) \times \U(1)_Y$ symmetry in the infrared. To see how this works, note that $\tilde{\Phi}_1$ and $\Phi_2$ fill out the $(\ov\fund,\fund)$ representation of $\SU(M)\times\SO(2(M+4))$. Moreover,
the baryon $\mathcal{S}_k$ in (\ref{eqn:quadCFTbaryons}) transforms as a $k$-index antisymmetric tensor of charge $k - \frac{M+4}{2}$ under $\SU(M+4) \times \U(1)_Y$ with the constraint $(-1)^k = (-1)^F$. Thus, the baryon operators $\mathcal{S}_k$ fill out the spinor $S$ (conjugate spinor $\bar S$) representation of $\SO(2(M+4))$ for even (odd) $F$. A very strong argument for the enhancement $\SU(M+4) \times \U(1)_Y \longrightarrow \SO(2(M+4))$ (which amounts to a proof at the level of the superconformal index) is given in \cite{dP1paper}.

Since $S$ and $\bar{S}$ are related by an outer automorphism of $\SO(2(M+4))$, the spectrum of mesons and baryons is identical for $\quadSO^+(M)$ and $\quadSO^-(M)$. In fact, these are the same CFT. For odd $M$, this is a simple consequence of the fact that the quiver diagrams for \LSO{A} and \LSO{B} are isomorphic up to charge conjugation of some of the nodes, whereas $(-1)^F = - (-1)^G$. For even $M$, a proof using deconfinement is given in~\cite{dP1paper}.

Nonetheless, the isomorphism between $\quadSO^+(M)$ and $\quadSO^-(M)$ changes how $\SU(M+4) \times \U(1)_Y$ is embedded into the enhanced symmetry group $\SO(2(M+4))$ (as can be seen from the different $\SU(M+4) \times \U(1)_Y$ spectrum of baryons $\mathcal{S}_k$). When embedded into a larger brane tiling, the $\SO(2(M+4))$ symmetry is broken, and the flavor parity must be specified to identify the unbroken $\SU(M+4)\times\U(1)$ subgroup. To do so unambiguously, we associate to each $\SU(M)\times\SU(M+4)$ bifundamental meson $\Phi$ a parity equal to the flavor parity in the deconfined description where $\Phi$ is composite. The parities of the two mesons are related by $(-1)^M$, so either can be specified to fix the flavor parity. The same convention can be applied to $\quadSp^\phi(M)$, where now the two mesons have the same parity, equal to $\phi$.\footnote{A more intrinsic definition of the parity of $\Phi$ is $(-1)^{\hat{F}}$ where $\hat{F}$ is chosen such that $\Phi^{\hat{F}} \mathcal{S}$ contains an $\SU(M+4)$ singlet for some $\SU(M+4)$ baryon $\mathcal{S}$.}

The charge table for $\quadSO^\phi(M)$ is the same as~(\ref{eq:qSpA-charges}) with the flipped meson $\tilde{\Phi}_1$ added. The class $A$ anomalies become
\begin{align}
A&:\begin{array}{|c|c|}
\hline
\SU(M)^3 & -(M+4)\\
\SU(M)^2 U(1)_X & -(M+4)\\
\SU(M+4)^2 U(1)_X & -M \\
\U(1)_X^3 & -M (M+4) \\
\U(1)_Y^2 \U(1)_X & -M (M+4)\\
\U(1)_X & - M (M+4) \\ \hline
\end{array}
\end{align}
whereas the class $B$ anomalies are the same as in~(\ref{eqn:quadSpAnomalies}). As before, the class $A$ anomalies are equal to the contribution of the mesons if they are counted as ``half'' chiral multiplets. The class $B$ anomalies, shared in common between the $\quadSO$ and $\quadSp$ CFTs, exhibit a fictitious $\Sp(2M)\times\SO(2(M+4))\times\U(1)_B\times\U(1)_R$ symmetry, which is formally present in the baryonic sector but broken by the mesons (and by the superconformal $R$-charge at the infrared fixed point).

While the enhanced symmetries discussed above are broken when the $\TO_2$ CFT is embedded into a larger brane tiling, in some cases a remnant still contributes to the non-Abelian symmetries of a toric orientifold CFT. For instance, the subgroup $\Sp(M+4)\times\SU(2)\subset\SO(2(M+4))$ contributes to the $\SU(2)$ global symmetry of certain $Y^{p,q}$ orientifolds, despite not being manifest in any deconfined description.\footnote{For future reference we note that, as $\SO(2(M+4))$ has no chiral anomalies, the $\SO(2(M+4))$ symmetric sector associated to $\quadSO$ contributes nothing to the Witten anomalies of either $\Sp(M+4)$ or $\SU(2)$.}

\subsection{Abstract quivers and charge tables}
\label{sec:charge-table-conventions}

Rather than giving an explicit deconfined description for every theory built using $\quadSp$ and $\quadSO$, we now develop an abstract notation for charge tables and quiver diagrams from which it is straightforward to recover the deconfined description given above.

To notate the $\TO_2$ CFTs in a charge table, we use the following conventions.
  Consider a gauge theory coupled to $\quadSp$ with an Abelian global symmetry $\U(1)_A$ and an R-symmetry $\U(1)_R$. 
We describe the embedding of the flavor symmetries of \quadSp\ into the gauge theory using a charge table
\begin{equation} \label{eqn:abstractSp}
\begin{array}{c|cc|cccc}
& \SU(M) & \SU(M+4) & \U(1)_A & \U(1)_R \\ \hline
\quadSp & \ast & \ast & a_B & r_B  \\
\Phi_1 &  \ydiagram{1} & \overline{\ydiagram{1}} & a_1 & r_1 \\
\Phi_2 &  \overline{\ydiagram{1}} & \overline{\ydiagram{1}} & a_2 & r_2 
\end{array}
\end{equation}
This table indicates how the Abelian and non-Abelian symmetries of the quiver gauge theory couple to $\TO_2$ CFT. The second and third line indicate the charges of the mesons $\Phi_{1,2}$ under the symmetries of the quiver gauge theory. This fixes the relationship between the non-Abelian symmetries and those of the deconfined theory~(\ref{eq:qSpA-charges}). It also partially fixes the relationship between $\U(1)_{A,R}$ and the Abelian symmetries of the deconfined theory---henceforth denoted $Q_B^{(0)}$, $Q_X^{(0)}$, $Q_Y^{(0)}$, $Q_R^{(0)}$ in the basis defined by~(\ref{eq:qSpA-charges}). However, the $\U(1)$ charges on the first line are needed to fix the admixture of $Q_B^{(0)}$. In particular, these charges are the charges of a baryon with $Q_B^{(0)}=1$ and $Q_X^{(0)} = Q_Y^{(0)} = Q_R^{(0)} = 0$, for instance
\begin{equation}
  \mathcal{S}_{\frac{M+4}{4}} \biggm/ \biggl(\Phi_1 \Phi_2 \biggr)^{\frac{M+4}{8}} \,.
\end{equation}
This fixes
\begin{align}
\begin{split}
Q_A &= a_B Q_B^{(0)} + \frac{a_1 - a_2}{2} Q_X^{(0)} - \frac{a_1 + a_2}{2} Q_Y^{(0)}\,,\\
Q_R &= r_B Q_B^{(0)} + \frac{r_1 - r_2}{2} Q_X^{(0)} - \frac{r_1 + r_2-2}{2} Q_Y^{(0)} + Q_R^{(0)}\,.
\end{split}
\end{align}

The \quadSO\ case is very similar, except that the flipping of one of the mesons needs to be accounted for. For instance, the charge table
\begin{equation} \label{eqn:abstractSO}
\begin{array}{c|cc|cccc}
& \SU(M) & \SU(M+4) & \U(1)_A & \U(1)_R \\ \hline
\quadSO & \ast & \ast & a_B & r_B  \\
\Phi_1 &  \ov\fund & \fund & a_1 & r_1 \\
\Phi_2 &  \ov\fund & \ov\fund & a_2 & r_2 
\end{array}
\end{equation}
corresponds to the charge assignment
\begin{align}
\begin{split}
Q_A &= a_B Q_B^{(0)} - \frac{a_1 + a_2}{2} Q_X^{(0)} + \frac{a_1 - a_2}{2} Q_Y^{(0)}\,,\\
Q_R &= r_B Q_B^{(0)} - \frac{r_1 + r_2-2}{2} Q_X^{(0)} + \frac{r_1 - r_2}{2} Q_Y^{(0)} + Q_R^{(0)}\,.
\end{split}
\end{align}
To complete the specification of the $\TO_2$ CFT, the flavor parity is needed. When desired, we specify this by attaching a parity label to one of the mesons $\Phi \to \Phi[\phi]$ for $\phi\in\{+,-\}$, which fixes the flavor parity by the conventions discussed in the previous section.

\begin{figure}
  \centering
  \begin{subfigure}{0.45\textwidth}
    \centering
    \includegraphics[width=\textwidth]{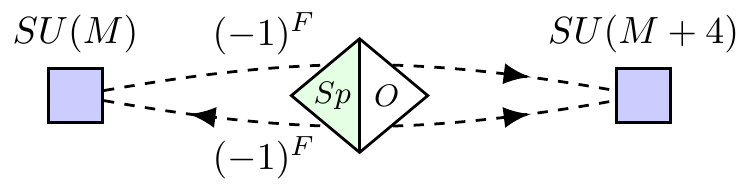}
    \caption{Notation for $\quadSp$ nodes.}
    \label{sfig:quad-notation-Sp}
  \end{subfigure}
  \hspace{1cm}
  \begin{subfigure}{0.45\textwidth}
    \centering
    \includegraphics[width=\textwidth]{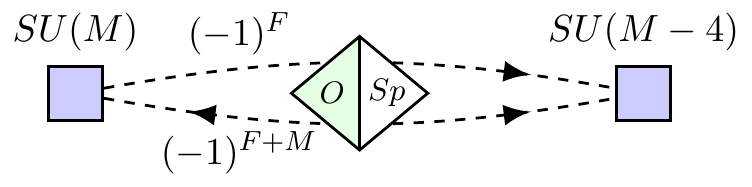}
    \caption{Notation for $\quadSO$ nodes.}
    \label{sfig:quad-notation-SO}
  \end{subfigure}
  \caption{Abstract quiver notation for $\TO_2$ CFTs, similar to~(\ref{eqn:abstractSp}), (\ref{eqn:abstractSO}). The dashed lines indicate the mesons and the attached labels are their associated parities. The shaded half of the diamond distinguishes between \quadSp\ and \quadSO. (This information can be inferred from the directions of the meson arrows here, but this is no longer true if both flavor nodes are replaced with $\SO$ or $\Sp$ groups.)
    }
  \label{fig:quad-notation}
\end{figure}

\begin{figure}
  \centering
  \includegraphics[width=\textwidth]{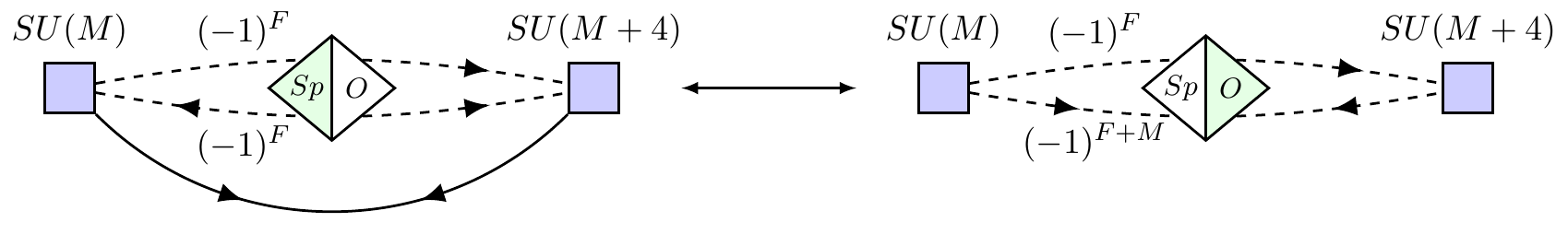}
  \caption{Flipping a meson in the language of the abstract quiver. This can be thought of as splitting the fundamental field (solid line) into two half-chirals (dashed lines) and then integrating out a vector-like pair of half-chirals. In the process, the flavor parity of the flipped meson changes by $(-1)^M$ and \quadSO\ and \quadSp\ are exchanged.}
  \label{fig:quad-flipping}
\end{figure}

\begin{figure}
  \centering
  \begin{subfigure}[b]{0.45\textwidth}
    \centering
    \includegraphics[width=\textwidth]{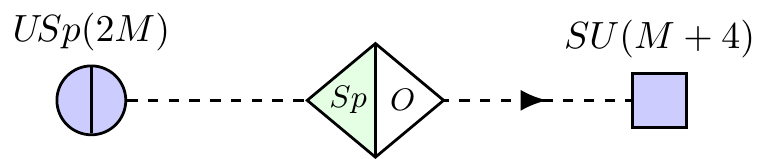}
    \caption{Manifest $\Sp(2M)$ symmetry.}
    \label{sfig:quad-manifest-Sp}
  \end{subfigure}
  \hspace{1cm}
  \begin{subfigure}[b]{0.45\textwidth}
    \centering
    \includegraphics[width=\textwidth]{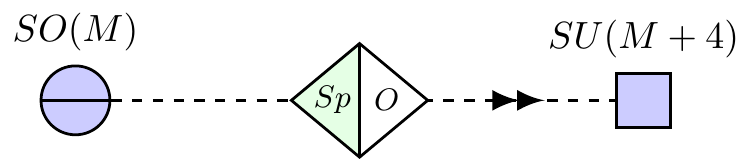}
    \caption{$SO(M)\times SU(2)\subset \Sp(2M)$.}
    \label{sfig:quad-manifest-SO}
  \end{subfigure}
  \caption{The quiver notation is easily adapted to emphasize
    different non-Abelian subgroups (besides $\SU(M)\times\SU(M+4)$) of the full
    flavor symmetry. \subref{sfig:quad-manifest-Sp}~Showing the full
    $\Sp(2M)\times\SU(M+4)$ flavor
    symmetry. \subref{sfig:quad-manifest-SO}~Showing the
    $\SO(M)\times \SU(M+4)\times\SU(2)$ subgroup.}
  \label{fig:quad-manifest}
\end{figure}

A similar notation can be used to draw quiver diagrams containing $\TO_2$ CFTs, see figure~\ref{fig:quad-notation}.
One advantage of this notation is that the idea of the mesons as ``half'' chiral multiplets can often be taken literally. For instance, the process of flipping a meson is heuristically that of integrating out a half-chiral multiplet, see figure~\ref{fig:quad-flipping}.
The class A anomalies discussed above, including all non-Abelian gauge anomalies, can also be computed using this heuristic, so anomaly cancellation is straightforward to enforce in the abstract quiver. The same notation can also be adapted to make various subgroups of the enhanced symmetries of the $\TO_2$ CFTs manifest, see figure~\ref{fig:quad-manifest}.

\begin{figure}
\centering
\begin{subfigure}[b]{0.305\textwidth}
  \centering
  \includegraphics[height=7cm]{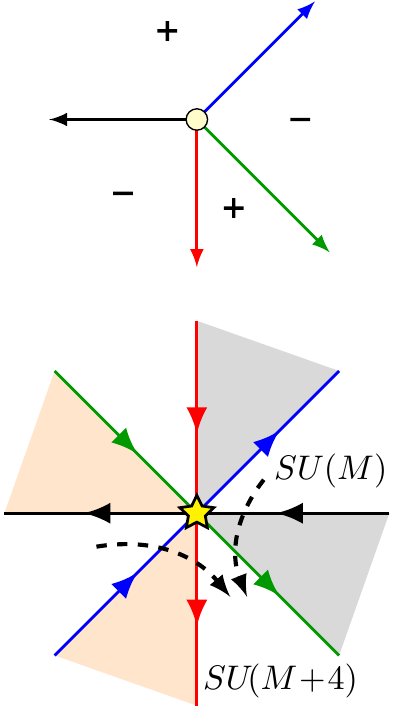}
  \caption{\quadSp\ from the tiling}
  \label{subfig:qSpDictionary}
\end{subfigure}
\begin{subfigure}[b]{0.305\textwidth}
  \centering
  \includegraphics[height=7cm]{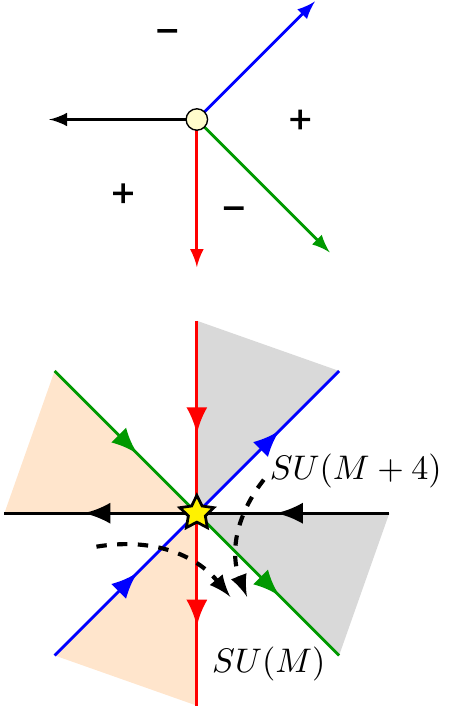}
  \caption{\quadSO\ from the tiling}
  \label{subfig:qSODictionary}
\end{subfigure}
\begin{subfigure}[b]{0.37\textwidth}
  \centering
  \includegraphics[height=5.5cm]{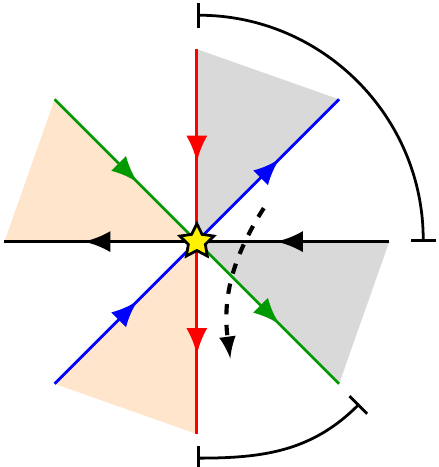}
 \caption{Relating meson and flavor parities}
 \label{subfig:MesonParity}
\end{subfigure}
\caption{Reading off \subref{subfig:qSpDictionary}~\quadSp\ and \subref{subfig:qSODictionary}~\quadSO\ abstract quivers from the brane tiling. As there is only one possible topology for the local web diagram, the distinction between \quadSp\ and \quadSO\ rests on the O5 charges. The $(N_i,0)$ faces of the tiling generate the non-Abelian flavor symmetries of the $\TO_2$ CFT, which are gauged when the local tiling is embedded in a larger whole. Likewise, there is a meson (indicated by a dashed arrow) connecting each adjacent pair of $(N_i,0)$ faces, which couples to a superpotential term in the larger brane tiling. \subref{subfig:MesonParity}~For a given meson, the two outgoing (incoming) NS5 branes adjacent to the head (tail) of the meson specify a pairing. One of these two pairs encloses an O5$^+$ plane in the local web diagram, and the meson parity is equal to the flavor parity associated to that pair.
  }
\label{fig:quaddictionary}
\end{figure}

To completely bypass the details of deconfinement in $\TO_2$ CFTs, we describe how the abstract quiver diagrams described above can be read off from the brane tiling directly. Once this has been done once using the full deconfinement machinery, doing so again is simply a matter of pattern recognition without the need for repeated deconfinement of the $\TO_2$ configuration. We summarize the dictionary in figure~\ref{fig:quaddictionary}.

\section{Examples}
\label{sec:examples}

In the preceeding sections, we have given a proposal that relates the geometric data
defining the discrete torsion to the structure of the field theory
living on the singularity. We have proven that it makes sense as a
definition, but does it give the correct physical predictions? In this
section we will answer this question in the affirmative in all examples previously considered in the literature as well as a few new ones, giving very strong evidence that our proposal is correct. These examples include $\cN=4$ Montonen-Olive duality and all
previously conjectured $\cN=1$ Montonen-Olive duality analogs coming from
branes at orientifolded singularities. Our methods can easily be applied to
construct SCFTs and obtain S-duality predictions in cases which were previously intractable,
such as the complex cone over $dP_2$, discussed in \S\ref{sec:dP2}.

\subsection{Flat space and orbifolds}
\label{sec:C3}

Our first example is the worldvolume theory on $k$ D3 branes on top of an O3 plane in
a flat background, producing an $\mathcal{N}=4$ gauge thery. As discussed in \cite{Witten:1998xy}, the
Montonen-Olive duality between $\SO(2k+1)$ and $\Sp(2k)$ and the
self-duality of $\SO(2k)$ follow from the $\SL(2,\bZ)$
self duality of type IIB string theory. We now
reproduce these results using our formalism.

\begin{figure}[t]
  \centering
  \begin{subfigure}[t]{0.31\textwidth}
    \centering
    \includegraphics[width=0.7\textwidth]{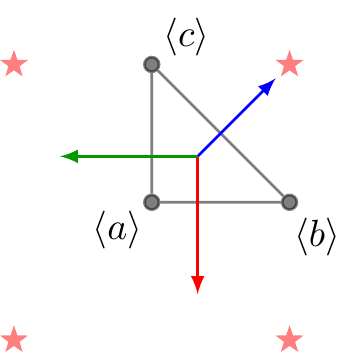}
    \caption{Toric structure for $\bC^3$.}
    \label{sfig:C3-web}
  \end{subfigure}
  \hfill
  \begin{subfigure}[t]{0.31\textwidth}
    \centering
    \includegraphics[width=0.8\textwidth]{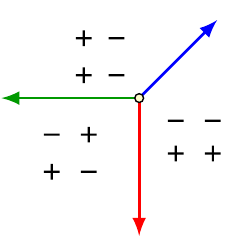}
    \caption{Local charges for $h=0$.}
    \label{sfig:C3-positive-charge}
  \end{subfigure}
  \hfill
  \begin{subfigure}[t]{0.31\textwidth}
    \centering
    \includegraphics[width=0.8\textwidth]{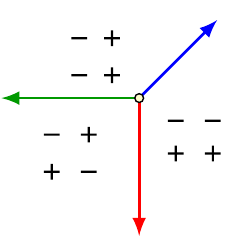}
    \caption{Local charges for $h=1$.}
    \label{sfig:C3-negative-charge}
  \end{subfigure}
  \caption{\subref{sfig:C3-web} Web and toric diagram for flat
    space. The stars denote the even sublattice associated with the
    toric involution leaving a fixed point. We have also named the
    generators of torsion in the divisor basis described in
    \S\ref{sec:divisor-basis}. \subref{sfig:C3-positive-charge} and
    \subref{sfig:C3-negative-charge} show the local charges for the
    two choices of NSNS torsion in the parameterization
    $[H]=h\vev{c}$.}
  \label{fig:C3-orientifold}
\end{figure}

\begin{figure}
\centering
\includegraphics[width=0.9\textwidth]{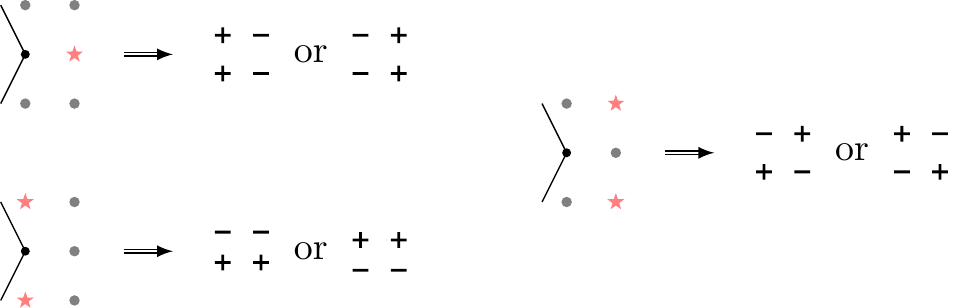}
\caption{The local charges of the four O5 planes within each wedge of
  the web diagram are constrained by the choice of involution.}
\label{fig:localchargesmnemonic}
\end{figure}

We start with the toric and web diagrams for $\bC^3$, shown in figure~\ref{sfig:C3-web}.
As shown in the figure, there is a unique choice of toric
involution that has an isolated fixed point---i.e., such that there is no corner
of the toric diagram on the even sublattice encoding the involution,
as explained in~\S\ref{subsec:toricOfolds}. Using the involution and NSNS discrete torsion $[H]$, we can construct the local charges of the tiling. In particular, if $[H] = \sum_i h_i \vev{i}$ is a particular divisor basis representation of $[H]$, then the local charge at the fixed point $n_a \in \{ (0,0), (0,1), (1,0), (1,1) \}$ in the $i$th wedge of the web diagram is given by $(-1)^{h_i + n_a u^a_i }$. A useful mnemonic for constructing these charges by hand is as follows: we first fill in the lower left fixed point with the charges $(-1)^{h_i}$. To determine the local charges for the remaining fixed points, we examine each wedge of the web diagram and impose one of the patterns shown in figure~\ref{fig:localchargesmnemonic}, depending on where the associated corner of the toric diagram lies in relation to the even sublattice specifying the involution.

In the present example, the divisor basis $\vev{a}, \vev{b}, \vev{c}$ satisfies $\vev{a}=\vev{b}=\vev{c}$ by~(\ref{eqn:twistedlattice}), so we can write the NSNS torsion as $[H]=h\vev{c}$ in full generality. Using this representation, we obtain the local charges shown in figure~\ref{sfig:C3-positive-charge} for $h=0$ and those in figure~\ref{sfig:C3-negative-charge} for $h=1$. Had we started with a different but equivalent divisor-basis representation for $[H]$, we would obtain the same local charges up to half-period translations on $T^2$.

\begin{figure}
  \centering
  \begin{subfigure}{0.45\textwidth}
    \centering
    \includegraphics[width=0.5\textwidth]{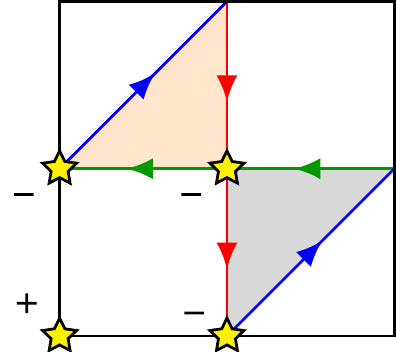}
    \caption{Tiling for $h=0$.}
    \label{sfig:C3-positive-tiling}
  \end{subfigure}
  \hspace{0.5cm}
  \begin{subfigure}{0.45\textwidth}
    \centering
    \includegraphics[width=0.5\textwidth]{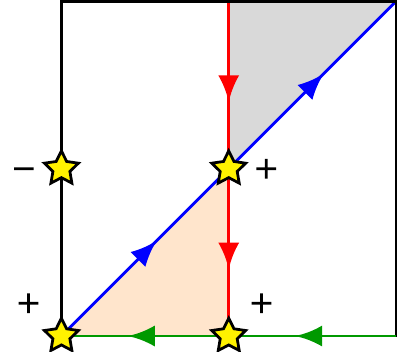}
    \caption{Tiling for $h=1$.}
    \label{sfig:C3-negative-tiling}
  \end{subfigure}
  \caption{\subref{sfig:C3-positive-tiling} Brane tiling corresponding
    to the local orientifold charges in
    figure~\ref{sfig:C3-positive-charge}. We have indicated the
    parities associated to each fixed point, in the conventions of
    \cite{Franco:2007ii,Imamura:2008fd}. The resulting theory is the
    $\cN=4$ theory with $SO$
    projection. \subref{sfig:C3-negative-tiling} Brane tiling
    corresponding to figure~\ref{sfig:C3-negative-charge},
    corresponding to the $\cN=4$ theory with $\Sp$ projection.}
  \label{fig:C3-tilings}
\end{figure}

Once the local charges are fixed, the brane tiling can be constructed by recalling that the local charge of a fixed point changes across a leg of the web diagram if and only if the corresponding NS5 brane intersects that fixed point. This allows us to locate all the NS5 branes on $T^2$ and determine the brane tiling, with the result shown in figure~\ref{fig:C3-tilings}.
In agreement with~\cite{Witten:1998xy}, we find that trivial NSNS torsion
corresponds to the $\cN=4$ theory with gauge group $\SO(N)$, while
non-trivial NSNS torsion corresponds to that with gauge group $\Sp(N)$.

Note that the tilings in figures~\ref{sfig:C3-positive-tiling} and \ref{sfig:C3-negative-tiling} are isomorphic up to an overall sign flip of the T-parities. This can be made manifest by translating figure~\ref{sfig:C3-negative-tiling} upwards by half a period, which then corresponds to the local charges generated by the divisor basis representation $[H] = \vev{a}+\vev{b}+\vev{c}$, equivalent to $[H] = \vev{c}$ by the relations $\vev{a} = \vev{b} = \vev{c}$. In future examples, we will make free use of half-period translations to present the brane tilings in whatever form is most convenient.

To fix the RR torsion, we take the ansatz $[F] = f \vev{c}$ as above. For $h=0$, we consider, e.g., the upper-right fixed point in figure~\ref{sfig:C3-positive-tiling}. By (\ref{eqn:Falpha}) the associated flavor parity is
\be
F = \vev{a} \cdot [F] = f \,,
\ee
since $Y_\alpha = \vev{a}$ for the red-green NS5 brane pair. For $\TO_1$ we have $F \equiv N \pmod 2$, so the $[F]$ torsion is
\be
[F] = N \vev{c} \,,
\ee
for an $\SO(N)$ gauge group. The same result can be obtained by choosing any other fixed point, or by using the constraint~(\ref{eqn:colorparity}).

By contrast, for $h=1$ we obtain $Y_\alpha = \vev{a}+\vev{c} \cong 0$ for the red-blue NS5 brane pair at the upper-right fixed point in figure~\ref{sfig:C3-negative-tiling}. This implies that $F$, hence $N$, must be even, in perfect agreement with the constraint on $\Sp(N)$, which is only defined for even $N$. It is straightforward to check that all the flavor and color parities are even, implying that we can set $[F] = 0$. This is consistent with the equivalence $[F] \to [F]+[H]$, $\tau\to \tau+1$, which combines the $[F]$ torsion with the theta angle when $[H] \ne 0$. Indeed, in string-theory derived normalization, the $\Sp$ holomorphic gauge coupling has period $\tau \cong \tau+2$.

As above, these torsion assignments exactly match those of \cite{Witten:1998xy}. Because S-duality of type IIB string theory takes $\tau \to -1/\tau$, $[F] \to [H]$ and $[H] \to -[F]$, we recover the well-known result that $\Sp(2k)$ is S-dual to $\SO(2k+1)$ and $\SO(2k)$ is self dual, as already illustrated from the O3 plane perspective in figure~\ref{fig:O3torsion}.
Note that the relation between the ranks of the S-dual theories is a consequence of the invariance of the D3 charge $Q_{\rm D3}$ under $\SL(2,\bZ)$. In this example we have $Q_{\rm D3} = -\frac{1}{4} + N/2$ for $\SO(N)$ and $Q_{\rm D3} = \frac{1}{4} + N/2$ for $\Sp(N)$ in the normalization where $Q_{\rm D3} = 1$ for a single mobile D3 brane. This can be computed using the known D3 charges of the O3$^-$ and O3$^+$ planes, and fixes $k=k'$ in the duality between $\SO(2k+1)$ and $\Sp(2k')$.

For future examples, it will be useful to be able to read off $Q_{\rm D3}$ directly from the brane tiling. T-duality smears the D5 charge across $T^2$, which suggests the simple formula\footnote{Note that the color-parity constraint~(\ref{eqn:colorparity}) combined with the $\SL(2,\bZ)$ invariance of $Q_{\rm D3}$ implies that the fractional part of $Q_{\rm D3}$ is fixed by the discrete torsion, $2Q_{\rm D3} \equiv 2Q_{\rm D3}^{(0)} + [F]\cdot[F]+[H]\cdot[H]+[F]\cdot[H] \pmod 2$ where $Q_{\rm D3}^{(0)}$ is the D3 charge for the $[F]=[H]=0$ $\SL(2,
\bZ)$ singlet phase. For consistency, this formula should follow directly from~(\ref{eqn:QD3}), without imposing that $Q_{\rm D3}$ is $\SL(2,\bZ)$ invariant. It would be interesting to prove this. (We observe in passing---also without giving a proof---that $2Q_{\rm D3}^{(0)}\equiv -A \pmod 2$, where $A$ is the area of the toric diagram in units where the $\bC^3$ toric diagram has area $1/2$.)}
\be \label{eqn:QD3}
2 Q_{\rm D3} = \frac{1}{\Vol T^2} \int_{T^2} Q_{\rm D5}(\tilde{\phi}_1, \tilde{\phi}_2)\, d^2 \tilde{\phi} \,,
\ee
where the factor of two accounts for the orientifold identification. Indeed, this formula reproduces the above result. We will use it in more complicated examples below, where it is often the most straightforward way to compute $Q_{\rm D3}$.

\subsubsection*{The \alt{$\bC^3/\bZ_3$}{C3/Z3} orbifold} 
\label{sec:C3/Z3}

\begin{figure}[t]
  \centering
  \begin{subfigure}[t]{0.31\textwidth}
    \centering
    \includegraphics[width=\textwidth]{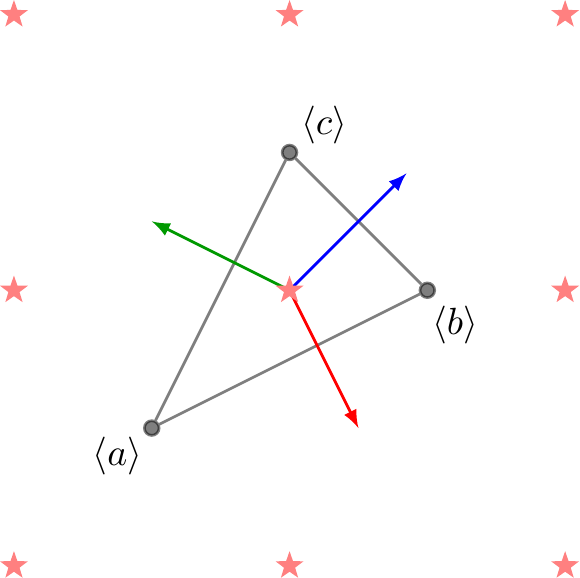}
    \caption{Toric structure for $\bC^3/\bZ_3$.}
    \label{sfig:Z3-web}
  \end{subfigure}
  \hfill
  \begin{subfigure}[t]{0.31\textwidth}
    \centering
    \includegraphics[width=0.8\textwidth]{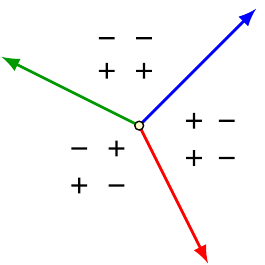}
    \caption{Local charges for $h=0$.}
    \label{sfig:Z3-trivial-H}
  \end{subfigure}
  \hfill
  \begin{subfigure}[t]{0.31\textwidth}
    \centering
    \includegraphics[width=0.8\textwidth]{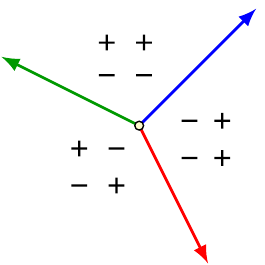}
    \caption{Local charges for $h=1$.}
    \label{sfig:Z3-nontrivial-H}
  \end{subfigure}
  \caption{\subref{sfig:Z3-web} Web and toric diagram for
    $\bC^3/\bZ_3$ inside the even sublattice associated with an
    isolated orientifold. \subref{sfig:Z3-trivial-H} and
    \subref{sfig:Z3-nontrivial-H} show the local charges for the two
    choices of NSNS torsion, in the conventions where the local
    charges of the bottom left fixed point in the tiling are given by
    the toric torsion $[H]=h(\vev{a}+\vev{b}+\vev{c})$.}
  \label{fig:Z3-orientifold}
\end{figure}

\begin{figure}
  \centering
  \begin{subfigure}{0.45\textwidth}
    \centering
    \includegraphics[width=0.5\textwidth]{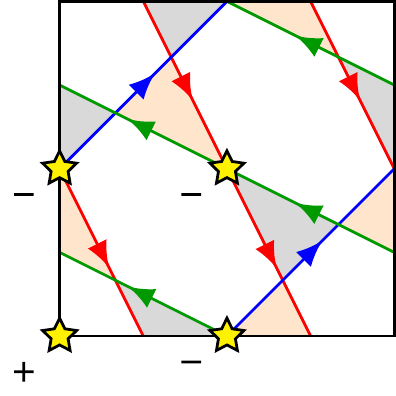}
    \caption{Tiling for $h=0$.}
    \label{sfig:Z3-trivial-H-tiling}
  \end{subfigure}
  \hspace{0.5cm}
  \begin{subfigure}{0.45\textwidth}
    \centering
    \includegraphics[width=0.5\textwidth]{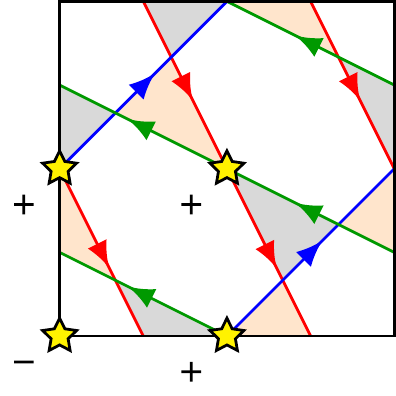}
    \caption{Tiling for $h=1$.}
    \label{sfig:Z3-nontrivial-H-tiling}
  \end{subfigure}
  \caption{Brane tilings for the two possible choices of NSNS
    torsion for $\bC^3/\bZ_3$.}
  \label{fig:Z3-tilings}
\end{figure}

We now move on to one of the simplest non-trivial examples of a toric orientifold, the
isolated orientifold of the $\bC^3/\bZ_3$ orbifold
\cite{dualities1}. The discussion proceeds very similarly to the
$\bC^3$ case just considered. The toric data are reviewed in
figure~\ref{sfig:Z3-web}.
As a minor variation, we now take the ansatz $[H] = h (\vev{a}+\vev{b}+\vev{c})$,\footnote{Since $\vev{a} = \vev{b}$, this is the same as the ansatz $[H]=h \vev{c}$. However, it produces local charges which differ by a half-period translation from the latter ansatz, which makes the similarity between figures~\ref{sfig:Z3-trivial-H-tiling} and~\ref{sfig:Z3-nontrivial-H-tiling} easier to see.}
which gives the local charges shown in figures~\ref{sfig:Z3-trivial-H} and~\ref{sfig:Z3-nontrivial-H}. From these we obtain the brane tilings in figure~\ref{fig:Z3-tilings}.

\begin{figure}
  \centering
  \begin{subfigure}{0.4\textwidth}
    \centering
    \includegraphics[height=3cm]{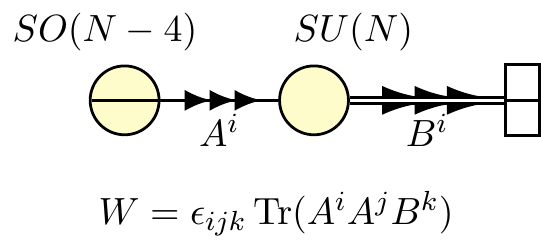}
    \caption{Quiver for $h=0$.}
    \label{sfig:Z3-SO}
  \end{subfigure}
  \hspace{1cm}
  \begin{subfigure}{0.4\textwidth}
    \centering
    \includegraphics[height=3cm]{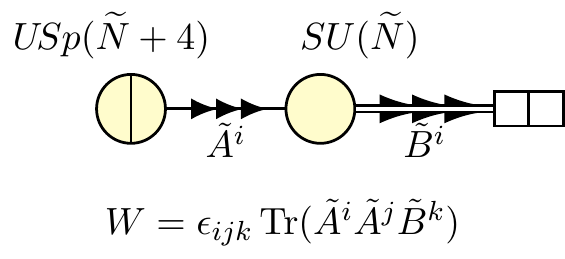}
    \caption{Quiver for $h=1$.}
    \label{sfig:Z3-USp}
  \end{subfigure}
  \caption{Quivers for the two possible choices of NSNS torsion for
    $\bC^3/\bZ_3$.}
  \label{fig:Z3-quivers}
\end{figure}

The tilings in figure~\ref{sfig:Z3-trivial-H-tiling} and figure~\ref{sfig:Z3-nontrivial-H-tiling} are again isomorphic up to an overall sign change for the T-parities, which is the origin of the ``negative rank duality'' observed in~\cite{dualities1}. Moreover, the cases $h=0$ and $h=1$ correspond to gauge groups with $\SO$ and $\Sp$ factors, respectively, as predicted by~\cite{dualities2}. We show the resulting quivers in figure~\ref{fig:Z3-quivers}.

To read off $[F]$, we follow exactly the same steps as before, with the result that $[F] = N \vev{c}$ for the case $h=0$, and $[F] = 0$ for $h=1$ (up to the $[F] \to [F]+[H]$, $\tau \to \tau+1$ equivalence) with $\tN$ constrained to be even. This, too, agrees with the predictions of~\cite{dualities2} and explains the S-duality between the $\SO(2k-1)\times\SU(2k+3)$ and $\Sp(2k+4)\times\SU(2k)$ theories observed in~\cite{dualities1}. As above, the relation between the ranks can be fixed by computing the D3 charge
\be
Q_{\rm D3} = \begin{cases}\frac{N}{2}-\frac{3}{4}\,, & h=0\,, \\  \frac{\tN}{2}+\frac{3}{4}\,,  & h=1\,, \end{cases}
\ee
which fixes $N = \tN+3$, and can be found using~(\ref{eqn:QD3}), or using exceptional collections as in~\cite{dualities2}.

\subsubsection*{General orbifolds}

\begin{figure}
  \centering
  \begin{subfigure}[b]{0.5\textwidth}
    \centering
    \includegraphics[height=4cm]{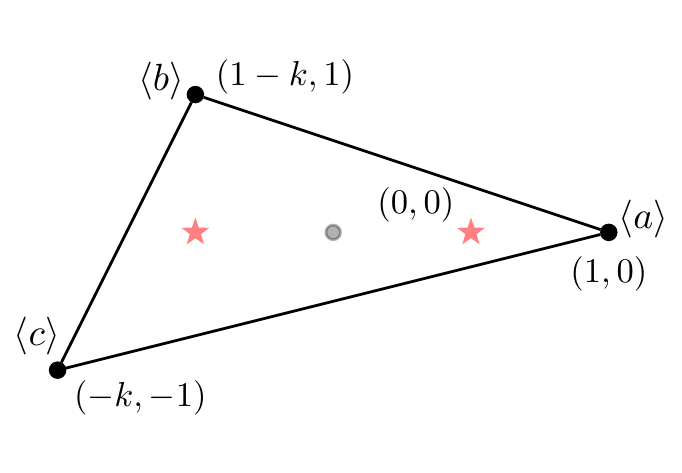}
    \caption{Toric data.}
    \label{sfig:Zk-toric}
  \end{subfigure}
  \hspace{1cm}
  \begin{subfigure}[b]{0.3\textwidth}
    \centering
    \includegraphics[height=4cm]{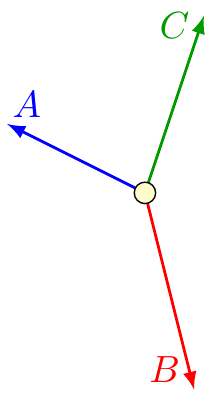}
    \caption{Web diagram.}
    \label{sfig:Zk-web}
  \end{subfigure}
  \caption{\subref{sfig:Zk-toric} Toric data for the toric
    orientifold $(z_1,z_2,z_3)\to (-z_1,-z_2,-z_3)$ of the $\bC^3/\bZ_{2k+1}$ orbifold $(z_1,z_2,z_3) \to (\omega_{2k+1} z_1, \omega_{2k+1} z_2, \omega_{2k+1}^{-2} z_3)$, where $\omega_n=\exp(2\pi i/n)$. We show the $k=3$ case for concreteness. \subref{sfig:Zk-web}~The corresponding web diagram.}
  \label{fig:Zk-toric}
\end{figure}

The above examples are easily generalized to any isolated toric orientifold of an orbifold singularity, such as the infinite family pictured in figure~\ref{fig:Zk-toric} and considered in~\cite{Bianchi:2013gka,dualities2} (of which both $\bC^3$ and $\bC^3/\bZ_3$ are members) as well as many other orbifolds, such as the $\bC^3/\bZ_7$ orbifold $(z_1, z_2, z_3) \to (\omega_7 z_1, \omega_7^2 z_2, \omega_7^4 z_3)$ considered in~\cite{dualities2}.

In particular, for an isolated orientifold singularity, the divisor basis satisfies $\vev{a} = \vev{b} = \vev{c}$, as the corresponding corners must lie on distinct even sublattices. Taking the ansatz $[H]=h(\vev{a}+\vev{b}+\vev{c})$, computing the local charges as above, and constructing the tiling, it is straightforward to check that the O5 plane $n_a = (0,0)$ does not intersect any NS5 branes, whereas the other O5 planes all do. We therefore obtain a gauge theory of the form $\SO(N) \times \prod_i \SU(N_i)$ or $\Sp(N) \times \prod_i \SU(N_i)$ for $h=0$ and $h=1$, respectively, where the ranks $N_i = N + 4 k_i$ are determined by the method discussed in~\S\ref{subsec:gaugetheoryconstruct}. The color parity is given by~(\ref{eqn:colorparity}), repeated below
\be
N \equiv [F] \cdot ([F]+[H]) \pmod 2\,.
\ee
Thus, $N$ is even and $[F]$ is undetermined for $h=1$ and $[F] = N (\vev{a}+\vev{b}+\vev{c})$ for $h=0$. This exactly reproduces the pattern of dualities hypothesized in~\cite{dualities2}, predicting S-dualities between the $\SO(2k+1)\times \SU^p$ theories and the $\Sp(2k') \times \SU(p)$ theories (for some difference $k-k'$ which can be fixed by computing $Q_{\rm D3}$). In particular, it explains all the examples discussed in~\cite{Bianchi:2013gka,dualities2}.

\subsection{Complex cone over \alt{$dP_1$}{dP1}}
\label{sec:dP1}

\begin{figure}
  \begin{subfigure}{0.21\textwidth}
    \centering
    \includegraphics[width=\textwidth]{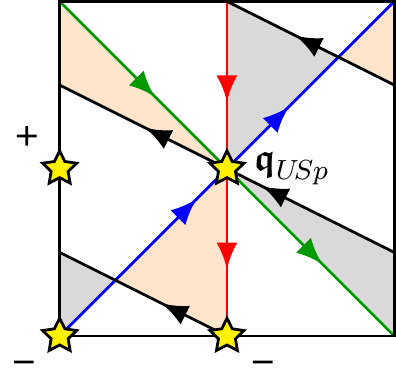}
    \caption{Phase \IA.}
    \label{sfig:dP1-IA-confined-tiling}
  \end{subfigure}
  \hfill
  \begin{subfigure}{0.21\textwidth}
    \centering
    \includegraphics[width=\textwidth]{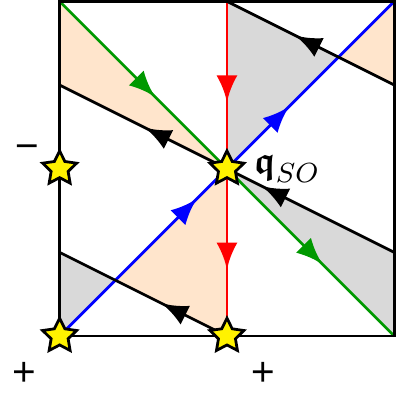}
    \caption{Phase \IB.}
    \label{sfig:dP1-IB-confined-tiling}
  \end{subfigure}
  \hfill
  \begin{subfigure}{0.21\textwidth}
    \centering
    \includegraphics[width=\textwidth]{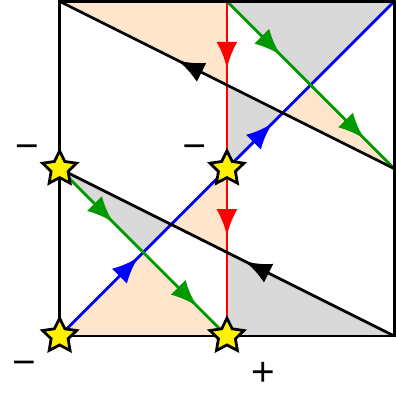}
    \caption{Phase \IIA.}
    \label{sfig:dP1-IIA-tiling}
  \end{subfigure}
  \hfill
  \begin{subfigure}{0.21\textwidth}
    \centering
    \includegraphics[width=\textwidth]{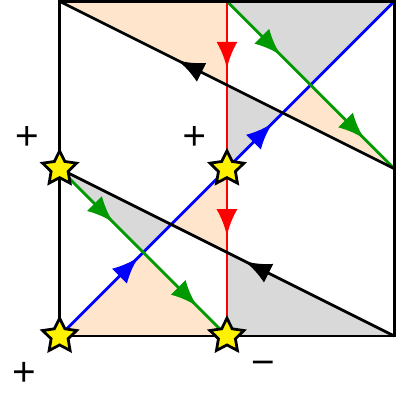}
    \caption{Phase \IIB.}
    \label{sfig:dP1-IIB-tiling}
  \end{subfigure}
  \caption{Brane tilings for the four phases of the $dP_1$ orientifold, as deduced in~\cite{dP1paper}. For the sake of uniform conventions in the present paper, we label the four phases differently than in~\cite{dP1paper}. In particular $\IA^{\rm (here)} = \II^{\rm (there)}$, $\IB^{\rm (here)} = \III^{\rm (there)}$, $\IIA^{\rm (here)} = \IA^{\rm (there)}$ and $\IIB^{\rm (here)} = \IB^{\rm (there)}$.
    }
    \label{fig:dP1-tilings}
\end{figure}

As our first non-orbifold example, we consider the orientifold of the complex cone over the first del Pezzo surface ($dP_1$), whose S-duality properties were recently understood~\cite{dP1paper}.
In that paper four phases of the
worldvolume field theory were identified: two ``classical'' (gauge theory) phases and two in
which an intrinsically strongly coupled sector appeared. A variety of
methods were used for matching these theories to discrete fluxes, thus
obtaining predictions for $\cN=1$ S-dualities of the associated infrared SCFTs. We now
illustrate how the methods presented in this paper significantly
simplify the discussion, and allow us to straightforwardly rederive
the results of~\cite{dP1paper}. For the sake of variety, we take the opposite approach to the orbifold examples above: we begin with the brane tilings for the four phases constructed in~\cite{dP1paper}, figure~\ref{fig:dP1-tilings}, and read off the geometry and discrete torsion.\footnote{In other words, we apply a ``forward algorithm''~\cite{Feng:2000mi,Franco:2005rj} for toric orientifolds, whereas in the previous section we applied an ``inverse algorithm''. Both are straightforward, but we illustrate the forward algorithm here for completeness.}

\begin{figure}
  \centering
  \begin{subfigure}{0.35\textwidth}
    \centering
    \includegraphics[width=\textwidth]{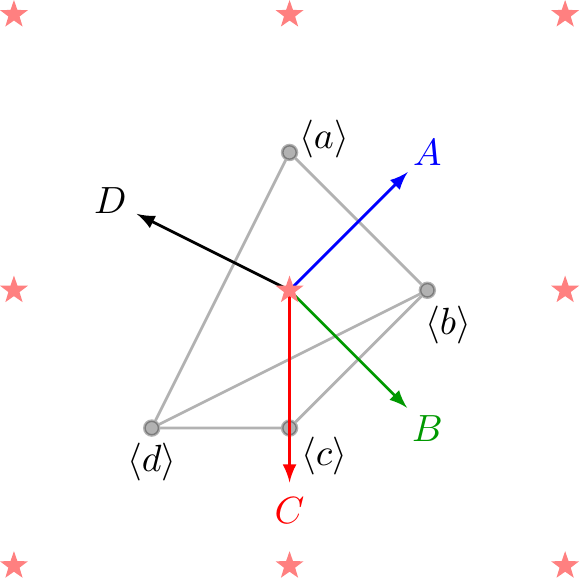}
    \caption{Toric structure.}
    \label{sfig:dP1-web}
  \end{subfigure}
  \hfill
  \begin{subfigure}{0.5\textwidth}
    \begin{subfigure}{0.45\textwidth}
      \centering
      \includegraphics[height=2.5cm]{dP1-IA-charges}
      \caption{Phase \IA.}
      \label{sfig:dP1-IA-charges}
    \end{subfigure}
    \hfill
    \begin{subfigure}{0.45\textwidth}
      \centering
      \includegraphics[height=2.5cm]{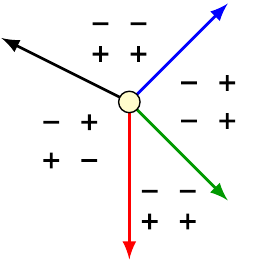}
      \caption{Phase \IB.}
      \label{sfig:dP1-IB-charges}
    \end{subfigure}
    \vspace{1cm}
    \\
    \begin{subfigure}{0.45\textwidth}
      \centering
      \includegraphics[height=2.5cm]{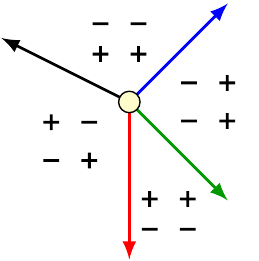}
      \caption{Phase \IIA.}
      \label{sfig:dP1-IIA-charges}
    \end{subfigure}
    \hfill
    \begin{subfigure}{0.45\textwidth}
      \centering
      \includegraphics[height=2.5cm]{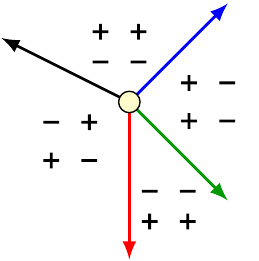}
      \caption{Phase \IIB.}
      \label{sfig:dP1-IIB-charges}
    \end{subfigure}
  \end{subfigure}
  \caption{The toric data associated to the brane tilings in figure~\ref{sfig:dP1-IIB-tiling}. The legs of the web diagram are fixed by the NS5 brane winding numbers, which in turn fix the toric diagram. The local charges \subref{sfig:dP1-IA-charges}--\subref{sfig:dP1-IIB-charges} are fixed by the T-parity and $\mathfrak{q}_{\SO/\Sp}$ assignments of the fixed points by cross referencing with figures~\ref{fig:O5-two-NS5s} and \ref{fig:quaddictionary}. To fix the starred even sublattice in~\subref{sfig:dP1-web} specifying the involution, we cross reference the local charges with figure~\ref{fig:localchargesmnemonic}.
The geometry specified by~\subref{sfig:dP1-web} is a toric orientifold of the complex cone over $dP_1$ with an isolated fixed point.
We indicate the partial resolution to the $\bC^3/\bZ_3$ orientifold singularity plus an O3 plane and name the external legs for later reference. 
    }
    \label{fig:dP1-web-local-charges}
\end{figure}

Applying the same steps as above in reverse, we obtain the web diagram, toric diagram, and local charges shown in figure~\ref{fig:dP1-web-local-charges}.
It is straightforward to check that the decorated toric diagram in figure~\ref{sfig:dP1-web} indeed corresponds to the $dP_1$ orientifold geometry considered in~\cite{dP1paper}.
From it, we read off the divisor basis relations $\vev{a}+\vev{c} = \vev{b} = \vev{d}$. We find it convenient to eliminate $\vev{b}$ and $\vev{d}$ using these relations, so that $[H] = \alpha\vev{a}+\gamma\vev{c}$. This basis is particular natural when considering the partial resolution to the $\bC^3/\bZ_3$ orientifold plus an O3 plane shown in figure~\ref{sfig:dP1-web}, as was done in~\cite{dP1paper}. In this case, the generator $\vev{a}$ corresponds to the discrete torsion of the $\bC^3/\bZ_3$ component, whereas the generator $\vev{c}$ corresponds to the O3 discrete torsion.

To read off $[H]$ for any given phase, we find the local charges $(-1)^{h_i}$ for any given fixed point and write $[H] = \sum_i h_i \vev{i}$, as in~(\ref{eqn:Htorsion}). We then eliminate $\vev{b}$ and $\vev{d}$ to express the result in our chosen basis. For instance, in phase \IIA, we obtain $[H] = \vev{a} + \vev{b}$ by selecting the upper-left fixed point. Using the relation $\vev{b} = \vev{a}+\vev{c}$, this is equivalent to $[H]=\vev{c}$, which is the result we would have read off directly if we had chosen the lower-left fixed point. Proceeding analogously, we obtain the $[H]$ torsion assignments shown in table~\ref{table:dP1-phases}, in complete agreement with~\cite{dP1paper}.
\begin{table}
  \centering
  \begin{tabular}{r|cccc}
    Phase: & \IA & \IB & \IIA & \IIB \\
    \hline
    $H_3$ torsion & (00) & (11) & (01) & (10)
  \end{tabular}
  \caption{The NSNS torsions for the four phases of the $dP_1$ orientifold, in the form $(\alpha \gamma)$, where $[H] = \alpha \vev{a} + \gamma \vev{c}$. This agrees with~\cite{dP1paper}, accounting for the different labels $\IA^{\rm (here)} = \II^{\rm (there)}$, $\IB^{\rm (here)} = \III^{\rm (there)}$, $\IIA^{\rm (here)} = \IA^{\rm (there)}$ and $\IIB^{\rm (here)} = \IB^{\rm (there)}$.}
  \label{table:dP1-phases}
\end{table}

\begin{figure}
  \centering
  \begin{subfigure}{0.41\textwidth}
    \centering
    \includegraphics[width=\textwidth]{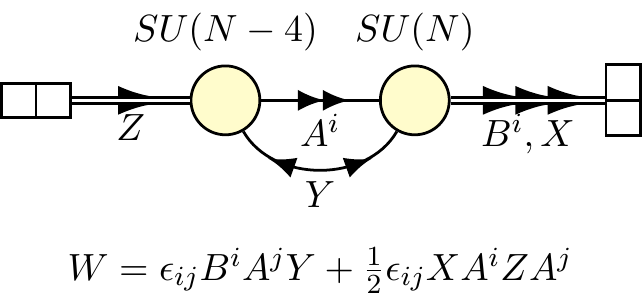}
    \caption{Phase \IIA.}
    \label{sfig:dP1-IIA-quiver}
  \end{subfigure}
  \hspace{2cm}
  \begin{subfigure}{0.41\textwidth}
    \centering
    \includegraphics[width=\textwidth]{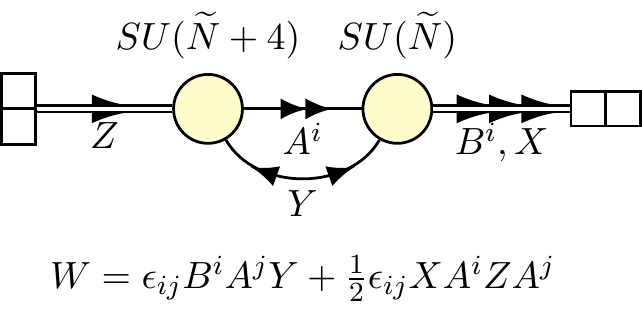}
    \caption{Phase \IIB.}
    \label{sfig:dP1-IIB-quiver}
  \end{subfigure}
  \caption{Quiver and superpotential for the classical phases of the
    complex cone over $dP_1$. (These phases were denoted as $\IA$ and
    $\IB$ in \cite{dualities1,dP1paper}.)}
  \label{fig:dP1-orientifold-quiver}
\end{figure}

Next, we read off the RR torsion $[F]$ and compare with the results of~\cite{dP1paper}, which were derived by consistency under partials resolutions and the matching of discrete symmetries between putative S-duals. We take the ansatz $[F] = \alpha_F \vev{a} + \gamma_F \vev{c}$ and start with the classical phases \IIA\ and \IIB, corresponding to the quiver diagrams shown in figure~\ref{fig:dP1-orientifold-quiver}.
The color parity constraint~(\ref{eqn:colorparity}) fixes 
\be \label{eqn:II-Ftorsion}
\begin{split}
\IIA&: N \equiv [F] \cdot ([F]+\vev{c}) = \alpha_F \pmod 2 \,,\\
\IIB&: \tN \equiv [F] \cdot ([F]+\vev{a}) = \gamma_F \pmod 2 \,,
\end{split}
\ee
where in the former (latter) case $\gamma_F$ ($\alpha_F$) is unfixed because of the $[F] \to [F] + [H]$, $\tau \to \tau+1$ equivalence. The same result, in agreement with~\cite{dP1paper}, can be recovered by considering the flavor parity associated to any of the four $\TO_1$ configurations at the fixed points. We label these phases as $\IIA^n$ and $\IIB^{\tilde{n}}$ for future reference, where $n = (-1)^N$ and $\tilde{n} = (-1)^{\tilde{N}}$.

\begin{figure}
  \centering
  \begin{subfigure}[t]{0.48\textwidth}
    \centering
    \includegraphics[width=\textwidth]{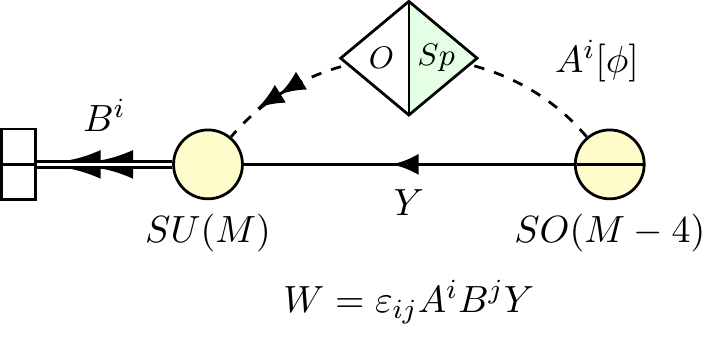}
    \caption{Phase \IA.}
    \label{sfig:dP1-IA-confined-quiver}
  \end{subfigure}
  \hfill
  \begin{subfigure}[t]{0.48\textwidth}
    \centering
    \includegraphics[width=\textwidth]{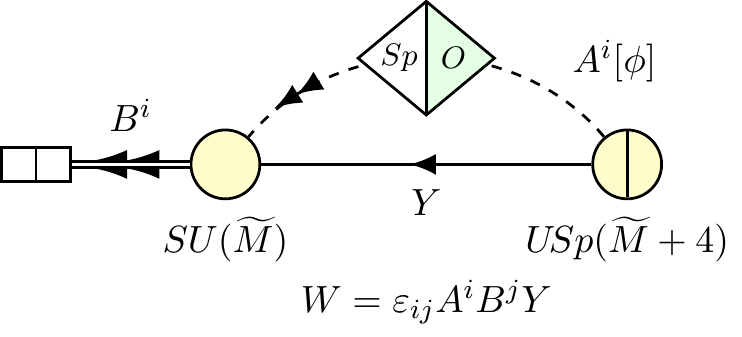}
    \caption{Phase \IB.}
    \label{sfig:dP1-IB-confined-quiver}
  \end{subfigure}
  \caption{Quivers for phases \IA\ and \IB\ of $dP_1$, where $\phi = (-1)^F$ is the meson parity associated to either member of the $\SU(2)$ doublet of $\TO_2$ mesons $A^i$.}
  \label{fig:IA-IB-quivers}
\end{figure}

Using the notation developed in~\S\ref{sec:charge-table-conventions}, the quiver diagram for the non-classical phase \IA\ is shown in figure~\ref{sfig:dP1-IA-confined-quiver},
 and the corresponding charge table is
\begin{equation} \label{eqn:dP1-IA-charge-table}
    \begin{array}{c|cc|cccc}
    & \SO(M-4) & \SU(M) & \SU(2) & \U(1)_B & \U(1)_Y & \U(1)_R \\ \hline
    Y & \ydiagram{1} & \overline{\ydiagram{1}} & \mathbf{1} & \frac{1}{M} & -1+\frac{1}{M} & 1 + \frac{1}{M}\\
    B^i & \mathbf{1} & \ydiagram{1,1} & \ydiagram{1} & -\frac{2}{M} & 1-\frac{2}{M} & - \frac{2}{M} \\ \hline
    \quadSp & \ast & \ast & \ast & 1 & -\frac{M-2}{2} & \frac{M-8}{4}  \\
    A^i[\phi] & \ydiagram{1} & \overline{\ydiagram{1}} & \ydiagram{1} & \frac{1}{M} & \frac{1}{M} & 1 + \frac{1}{M} 
  \end{array} 
\end{equation}
using the same basis for the global symmetries as~\cite{dP1paper}, where gauging $\SO(M-4)$ breaks the $\Sp(2(M-4))$ flavor symmetry of \quadSp\ down to $\SO(M-4)\times\SU(2)$.
Applying (\ref{eqn:colorparity}) and~(\ref{eqn:RRtorsionprescription}), we read off
\be \label{eqn:IA-Ftorsion}
\begin{split}
M &\equiv [F]\cdot [F] = \alpha_F + \gamma_F \pmod 2 \,,\\ 
F &\equiv [F] \cdot \vev{c} = \gamma_F \pmod 2 \,.
\end{split}
\ee
Here $\phi=(-1)^F$ is the meson parity associated to either of the mesons in the \quadSp\ theory, hence it is the flavor parity associated to the BC pair at the upper-right fixed point ($F \equiv [F] \cdot Y_{BC} \pmod 2$ where $Y_{BC} = \vev{c}$) by the prescription shown in figure~\ref{subfig:MesonParity}. We conclude that $[F] = (M+F) \vev{a} + F \vev{c}$, as was found in~\cite{dP1paper} by partial resolution. We label this phase as $\IA^{m;\phi}$ for future reference, where $m = (-1)^M$ and $\phi=(-1)^F$.

The non-classical phase \IB\ is similar. The quiver diagram is shown in figure~\ref{sfig:dP1-IB-confined-quiver}, and the charge table is
\begin{equation} \label{eqn:dP1-IB-charge-table}
  \begin{array}{c|cc|cccc}
    & \Sp(\tM+4) & \SU(\tM) & \SU(2) & \U(1)_B & \U(1)_Y & \U(1)_R \\ \hline
    Y & \ydiagram{1} & \overline{\ydiagram{1}} & \mathbf{1} & \frac{1}{\tM} & -1-\frac{1}{\tM} & 1 - \frac{1}{\tM}\\
    B^i & \mathbf{1} & \ydiagram{2} & \ydiagram{1} & -\frac{2}{\tM} & 1+\frac{2}{\tM} & \frac{2}{\tM} \\  \hline 
    \quadSO & \ast & \ast & \ast & -1 & \frac{\tM+2}{2} & -\frac{\tM+8}{4}  \\
    A^i[\phi] & \ydiagram{1} & \overline{\ydiagram{1}} & \ydiagram{1} & \frac{1}{\tM} & -\frac{1}{\tM} & 1 - \frac{1}{\tM}
  \end{array}
\end{equation}
As above, the gauging of $\Sp(\tM+4)$ breaks the $\SO(2(\tM+4))$ down to $\Sp(\tM+4) \times \SU(2)$. Unlike before, however, descriptions of this theory using antisymmetric tensor deconfinement futher break $\SU(2)$ down to $\U(1)_X$, with the non-Abelian enhancement occuring accidentally in the infrared. One advantage of the abstract notation developed in~\S\ref{sec:charge-table-conventions} is that we can keep the $\SU(2)$ global symmetry manifest.

Applying (\ref{eqn:colorparity}) and~(\ref{eqn:RRtorsionprescription}) as before, we now obtain
\be \label{eqn:IB-Ftorsion}
\begin{split}
\tM &\equiv [F]\cdot ([F]+\vev{a}+\vev{c}) = 0 \pmod 2 \,,\\ 
F &\equiv [F] \cdot \vev{b} = [F]\cdot \vev{d} = \alpha_F + \gamma_F \pmod 2 \,.
\end{split}
\ee
Here $\phi=(-1)^F$ is again the meson parity associated to either of the mesons in the \quadSO\ theory (which are equal since $\tM$ is even), hence it is the flavor parity associated to either the AB pair or the CD pair at the upper-right fixed point (where $Y_{AB} = \vev{b}$ and $Y_{CD} = \vev{d}$). As usual, $[F]$ is partially unfixed, due to the equivalence $[F]\to [F]+[H]$, $\tau \to \tau+1$, and the constraint that $\tM$ is even directly corresponds to the presence of a $\Sp(\tM+4)$ gauge group factor. We label this phase as $\IB^\phi$ for future reference, where $\phi = (-1)^F$.

The result~(\ref{eqn:IB-Ftorsion}) once again matches~\cite{dP1paper}, but this case is particularly interesting because it cannot be obtained by partial resolution to $\bC^3/\bZ_3$. Instead~\cite{dP1paper} resorted to matching discrete global symmetries between putative S-dual theories. This is unnecessary using the results of the current paper; the correct torsion assignments are fixed \emph{a priori}, without the need to match up the properties of S-dual theories.

Using the torsion assignments in table~\ref{table:dP1-phases}, (\ref{eqn:II-Ftorsion}), (\ref{eqn:IA-Ftorsion}), and (\ref{eqn:IB-Ftorsion}), as well as
the generators $S: \tau\to -1/\tau, [F]\to [H], [H] \to - [F]$ and $T: \tau \to \tau+1, [F]\to [F]+[H]$ of $\SL(2,\bZ)$,
we find the following $\SL(2,\bZ)$ multiplets
\begin{align} \label{eqn:dP1multiplets}
&(\IA^{+;+}), & &(\IA^{+;-}, \IB^{+}), & &(\IA^{-;+}, \IIB^+), & &(\IA^{-;-}, \IIA^+), & &(\IB^-, \IIA^-, \IIB^-)\,,
\end{align}
as in~\cite{dP1paper}. Accounting for the fact that each of the phases $\IB^{\pm}$, $\IIA^{\pm}$ and $\IIB^{\pm}$ represent two $[F]$ torsions related by $T$, we label this multiplets as ``singlets'', such as $(\IA^{+;+})$, ``triplets'', such as $(\IA^{+;-}, \IB^{+})$, and ``sextets'', such as $(\IIA^-, \IIB^-, \IB^-)$.

\begin{figure}
\begin{center}
  \includegraphics[height=7cm] {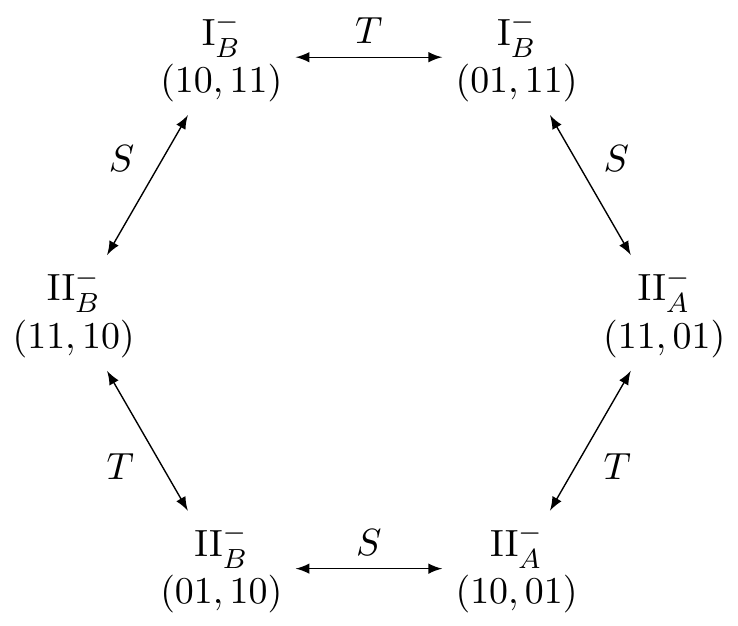}
  \caption{The action of $\SL(2,\bZ)$ on the sextet $(\IB^-, \IIA^-, \IIB^-)$, where $(\alpha_F\gamma_F,\alpha\gamma)$ denotes $([F],[H])$. The S-duality between $\IIA^-$ and $\IIB^-$ was originally observed in~\cite{dualities1} and $\IB^-$ was discovered---filling out the multiplet---in~\cite{dP1paper}.}
  \label{fig:sextet}
\end{center}
\end{figure}

All of the $\SL(2,\bZ)$ multiplets in our paper fall into one of these classes, as in any case where $[F]$ and $[H]$ are valued in a $\bZ_2$ vector space. We have already seen singlets $(\SO(2k))$ and triplets $(\SO(2k+1),\Sp(2k))$ in $\mathcal{N}=4$ theories and their orbifold cousins, discussed in the previous section. Sextets are a new phenomenon; we illustrate the action of $\SL(2,\bZ)$ on the sextet $(\IIA^-, \IIB^-, \IB^-)$ in figure~\ref{fig:sextet}.

The D3 charges
\be
\begin{aligned}
\IA&: Q_{\rm D3} = \frac{M}{2}-1\,, &\IB&: Q_{\rm D3} = \frac{\tM}{2}+1\,, \\
\IIA&: Q_{\rm D3} = \frac{N}{2}-\frac{1}{2}\,, &\IIB&: Q_{\rm D3} = \frac{\tN}{2}+\frac{1}{2}\,, \\
\end{aligned}
\ee
can be computed using~(\ref{eqn:QD3}), or by partial resolution to $\bC^3/\bZ^3$ plus an O3 plane, as in~\cite{dP1paper}. These fix the rank relations between the duals to be $M-2=N-1=\tN+1=\tM+2$. Notice that these relations constrain the relative color parities between S-dual theories. As demanded by self-consistency, this is the same constraint imposed by the torsion argument given above.

The S-dualities discussed above, (\ref{eqn:dP1multiplets}), were checked in~\cite{dP1paper} by matching 't~Hooft anomalies, discrete symmetries, and the superconformal indices between putative dual theories. We will perform the same highly non-trivial checks in the new examples discussed below.

\subsubsection*{Comment on ``negative rank duality''}

Notice that, similar to a phenomenon we have already observed in orbifold examples, the brane tilings in figures~\ref{sfig:dP1-IA-confined-tiling} and~\ref{sfig:dP1-IB-confined-tiling} are related by changing the sign of all the local charges, as are those in figures~\ref{sfig:dP1-IIA-tiling} and~\ref{sfig:dP1-IIB-tiling}. This is reflected in the labels $(\ldots)_A$ and $(\ldots)_B$, and is related to the formal replacement $N \to -N$ in the corresponding quiver gauge theory, following the rules of ``negative rank duality'' (see, e.g., \cite{Cvitanovic:2008zz} and references therein) reviewed in~\cite{dualities1}. Contrary to the connection between S-duality and negative rank duality hypothesized in~\cite{dualities1}, we have already seen that there are many S-dualities which related theories which are not negative rank duals, such as the $\SL(2,\bZ)$ triplets $(\IA^{-;+}, \IIB^+)$ and $(\IA^{-;-}, \IIA^+)$. In fact, the only S-dualities involving negative rank duals among the $dP_1$ orientifolds considered above are the $\IIA^- \longleftrightarrow \IIB^-$ part of the sextet (the bottom arrow in figure~\ref{fig:sextet}, seen in~\cite{dualities1}) and the triplet $(\IA^{+;-}, \IB^{+})$. Furthermore, the $\bF_0$ orientifold theories considered in the next section are invariant under negative rank duality, but have non-trivial S-duals.

In general, negative rank duality takes $[H] \to [H]+\eta$ where $\eta = \sum_i \vev{i}$. Only for specific choices of $[F]$ are the two phases related by $[H] \to [H] + \eta$ actually S-dual. Instead of being a predictor of S-dualities, we find negative rank duality to be a useful way of organizing a large collection of theories into ones with similar superficial properties, such as in the charge tables (\ref{eqn:dP1-IA-charge-table}) and (\ref{eqn:dP1-IB-charge-table}), which are formally negative rank duals even though, e.g., $\IB^-$ is not in the same $\SL(2,\bZ)$ multiplet as any phase of $\IA$. 

\subsection{Complex cone over \alt{$\bF_0$}{F0}: Isolated O7 plane} \label{sec:F0-O7}

\begin{figure}
  \centering
  \begin{subfigure}{0.35\textwidth}
    \centering
    \includegraphics[width=\textwidth]{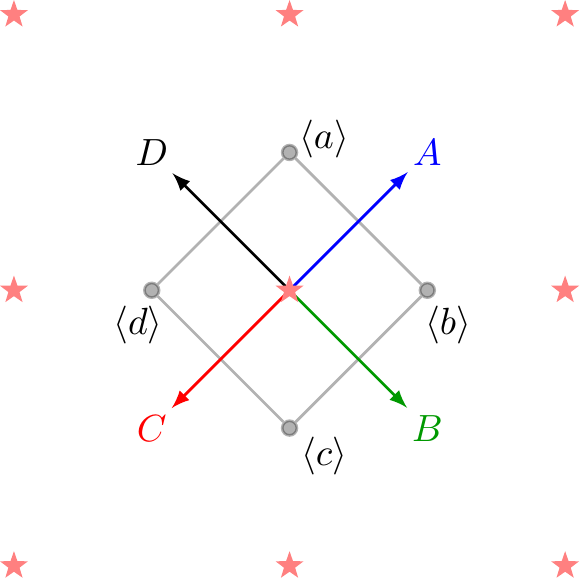}
    \caption{Toric structure.}
    \label{sfig:F07-web}
  \end{subfigure}
  \hfill
  \begin{subfigure}{0.62\textwidth}
    \begin{subfigure}{0.45\textwidth}
      \centering
      \includegraphics[height=3cm]{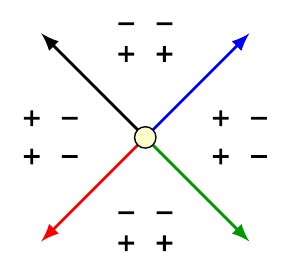}
      \caption{Phase $\I$.}
      \label{sfig:F07-I-charges}
    \end{subfigure}
    \hfill
    \begin{subfigure}{0.45\textwidth}
      \centering
      \includegraphics[height=3cm]{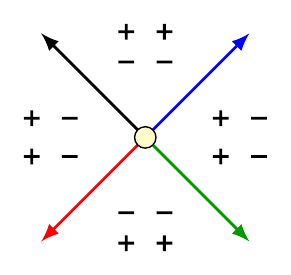}
      \caption{Phase $\II$.}
      \label{sfig:F07-II-charges}
    \end{subfigure}
    \vspace{0.5cm}
    \\
    \begin{subfigure}{0.45\textwidth}
      \centering
      \includegraphics[height=3cm]{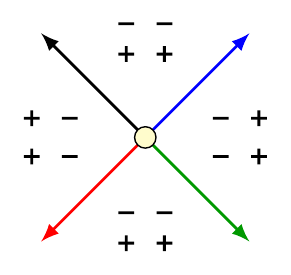}
      \caption{Phase \tII.}
      \label{sfig:F07-tII-charges}
    \end{subfigure}
    \hfill
    \begin{subfigure}{0.45\textwidth}
      \centering
      \includegraphics[height=3cm]{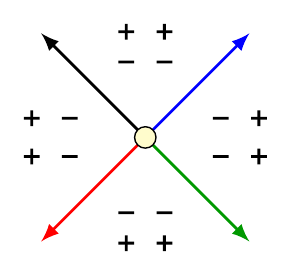}
      \caption{Phase \III.}
      \label{sfig:F07-III-charges}
    \end{subfigure}
  \end{subfigure}
  \caption{\subref{sfig:F07-web} Toric data for the isolated O7
    orientifold of the complex cone over $\bF_0$. We have named the
    external legs and torsion generators for future reference.}
  \label{fig:F07-toric}
\end{figure}

Our next example goes slightly beyond the cases that were understood in the previous literature. We consider the orientifold of the complex cone of $\bF_0$ defined by the toric data in figure~\ref{sfig:F07-web}.
This is the same orientifold geometry considered briefly in~\cite{dualities1,dP1paper}, and the corresponding world-volume gauge theories include a dual pair that was detected in~\cite{dualities1} using 't~Hooft anomaly matching and matching of discrete symmetries. However, a complete account of dualities in this class of theories has not yet been given, and in particular, there was some question in~\cite{dualities1} as to the string-theory nature of the duality found in that paper. We will see that it is an S-duality inherited from the $\SL(2,\bZ)$ self-duality of type IIB string theory, just as in the above examples. We further provide a full classification of the orientifold phases and check the predicted S-dualities using the superconformal index.

Note that for the toric orientifold considered in this section the $\bF_0$ exceptional divisor is wrapped by an O7 plane, as can be seen by resolving the singularity. In~\S\ref{sec:F0-O3} , we consider a different toric involution which instead resolves to four O3 planes.

The divisor basis relations are $\vev{a}=\vev{c}$ and $\vev{b}=\vev{d}$, so we will write the torsions as $[H] = \alpha \vev{a}+\beta \vev{b}$ and $[F] = \alpha_F \vev{a}+\beta_F \vev{b}$. We label the four possible choices of $[H]$ with the phase labels shown in table~\ref{table:F07-phases}, with the corresponding local charges shown in figures \ref{sfig:F07-I-charges}--\subref{sfig:F07-III-charges}.
\begin{table}
  \centering
  \begin{tabular}{r|cccc}
    Phase: & \I & \II & \tII & \III \\
    \hline
    $H_3$ torsion & (00) & (10) & (01) & (11)
  \end{tabular}
  \caption{The four different choices of NSNS torsion in the O7
    isolated orientifold of the complex cone over $\bF_0$, in the form $(\alpha\beta)$ for $[H] = \alpha \vev{a} + \beta \vev{b}$.}
  \label{table:F07-phases}
\end{table}
Note that the toric diagram has the large symmetry group $D_4$. This includes a $\bZ_2 \times \bZ_2$ subgroup of the $\SU(2)\times\SU(2)$ isometry group of $\bF_0 \cong \bP_1 \times \bP_1$, but also a $\bZ_2$ which exchanges the two $\bP_1$s and maps $\vev{a} \leftrightarrow \vev{b}$. As a consequence, phases \II\ and \tII\ have isomorphic properties up to a relabeling of the global symmetries, and we will not discuss \tII\ further until we consider the S-duality properties of these theories.

\subsubsection*{Phase \II}

\begin{figure}
  \centering
  \begin{subfigure}[t]{0.3\textwidth}
    \centering
    \includegraphics[height=3cm]{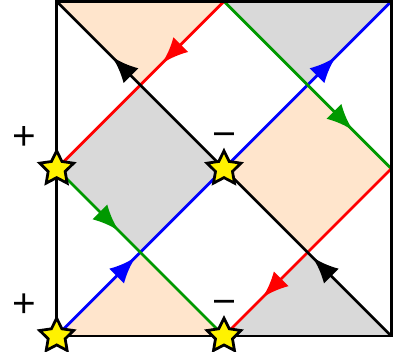}
    \caption{Brane tiling.}
    \label{sfig:F07-II-tiling}
  \end{subfigure}
  \hspace{1.5cm}
  \begin{subfigure}[t]{0.5\textwidth}
    \centering
    \includegraphics[width=\textwidth]{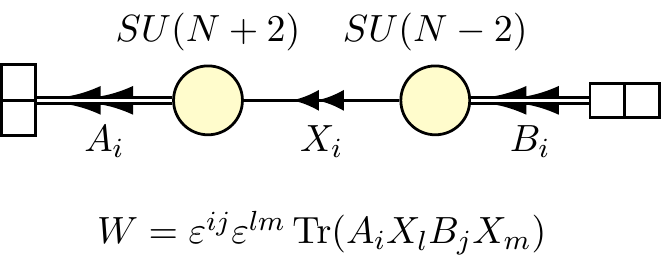}
    \caption{Quiver and superpotential.}
    \label{sfig:F07-II-quiver}
  \end{subfigure}
  \caption{Tiling and field theory data corresponding to the local
    charges for phase \II, in figure~\ref{sfig:F07-II-charges}.}
  \label{fig:F07-II}
\end{figure}

We begin with phase \II, which is the most straightforward to describe. Using the local charges in figure~\ref{sfig:F07-II-charges}, we obtain the brane tiling shown in figure~\ref{sfig:F07-II-tiling}, corresponding to the quiver gauge theory in figure~\ref{sfig:F07-II-quiver}.\footnote{This gauge theory was labelled as ``phase I'' of the $\bF_0$ orientifold in~\cite{dualities1}.}
The global symmetries are
\begin{equation}
  \label{eq:F07-II-charges}
    \begin{array}{c|cc|cccc}
    & SU(N-2) & SU(N+2) & SU(2)_1 & SU(2)_2 & U(1)_B & U(1)_R\\
    \hline
    B_i & \ov\symm & {\bf 1} & {\bf 1} & \fund & -\frac{1}{N-2} &
                    \frac{1}{2} + \frac{3}{N-2} \\
    X_i & \fund & \ov\fund & \fund & {\bf 1} & \frac{N}{N^2-4} & \frac{N^2-16}{2(N^2-4)} \\
    A_i & {\bf 1} & \asymm & {\bf 1} & \fund & -\frac{1}{N+2} &
                    \frac{1}{2} - \frac{3}{N+2}
  \end{array}
\end{equation}
in the same basis as~\cite{dualities1}. Reading off $[F]$ from the brane tiling is a straightforward application of the rules applied in the previous examples. We find
\be \label{eqn:F0-II-Ftorsion}
N \equiv \beta_F \pmod 2 \,,
\ee
with $\alpha_F$ undetermined, as expected from the $\tau \to \tau+1$, $[F] \to [F]+[H]$ equivalence. We denote this phase as $\II^n$ with $n = (-1)^N$ for future reference.

\subsubsection*{Phase \III}
\label{sec:F07-III}

\begin{figure}
  \centering
  \begin{subfigure}[t]{0.3\textwidth}
    \centering
    \includegraphics[height=4cm]{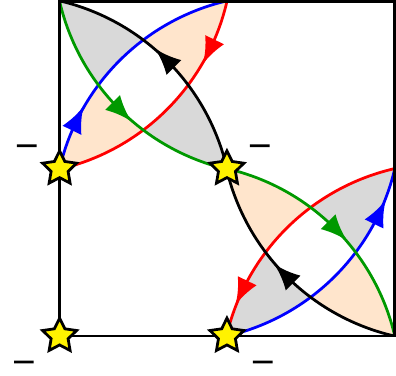}
    \caption{Brane tiling.}
    \label{sfig:F07-III-tiling}
  \end{subfigure}
  \hspace{1.5cm}
  \begin{subfigure}[t]{0.5\textwidth}
    \centering
    \includegraphics[height=3.5cm]{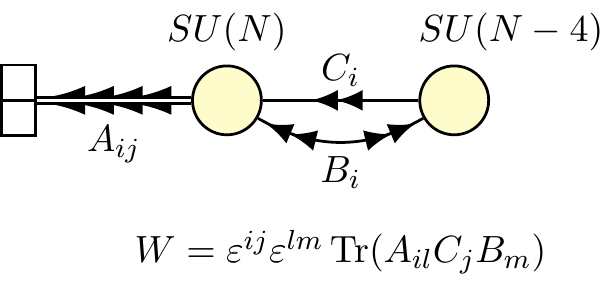}
    \caption{Quiver and superpotential.}
    \label{sfig:F07-III-quiver}
  \end{subfigure}
  \caption{Tiling and field theory data corresponding to the local
    charges for phase \III, figure~\ref{sfig:F07-III-charges}.}
  \label{fig:F07-III}
\end{figure}

Phase \III\ is slightly more subtle, because the brane tiling contains coincident anti-parallel NS5 brane boundaries. To resolve it into a gauge theory, we bend these boundaries slightly to obtain the tiling shown in figure~\ref{sfig:F07-III-tiling}, corresponding to the quiver gauge theory in figure~\ref{sfig:F07-III-quiver},\footnote{Here we apply the angle rule to the local web diagram (p.~\pageref{def:localwebdiagram}). This gauge theory and its Seiberg dual were labelled ``phase II'' in~\cite{dualities1}.}
with the global symmetries
\begin{equation}
  \label{eq:F07-III-charges}
   \begin{array}{c|cc|cccc}
    & SU(N-4) & SU(N) & SU(2)_1 & SU(2)_2 & U(1)_B & U(1)_R\\
    \hline
    C_i & \fund & \ov\fund & \fund & {\bf 1} & \frac{1}{N-4} & \frac{1}{2} + \frac{2}{N} \\
    B_i & \ov\fund & \ov\fund & {\bf 1} & \fund & -\frac{1}{N-4} & \frac{1}{2} + \frac{2}{N} \\
    A_{ij} & {\bf 1} & \asymm & \fund & \fund & 0 & 1-\frac{4}{N}
  \end{array}
\end{equation}
in the same basis as above.

\begin{figure}
  \centering
  \begin{subfigure}[t]{0.3\textwidth}
    \centering
    \includegraphics[height=4cm]{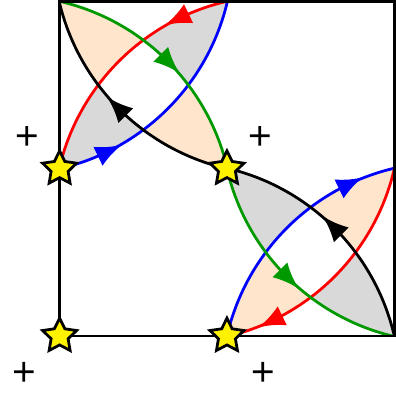}
    \caption{Brane tiling.}
    \label{sfig:F07-IIIp-tiling}
  \end{subfigure}
  \hspace{1.5cm}
  \begin{subfigure}[t]{0.5\textwidth}
    \centering
    \includegraphics[height=3.5cm]{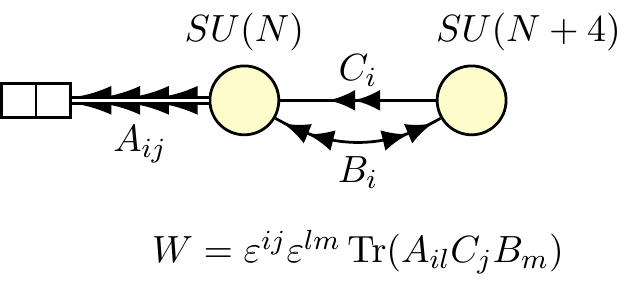}
    \caption{Quiver and superpotential.}
    \label{sfig:F07-IIIp-quiver}
  \end{subfigure}
  \caption{Tiling and field theory data corresponding to the local
    charges for phase \III, in figure~\ref{sfig:F07-III-charges},
    choosing an alternative way of resolving the overlapping NS5
    branes.}
  \label{fig:F07-IIIp}
\end{figure}

In the true brane configuration, the D5 brane face corresponding to the $\SU(N-4)$ gauge group factor shrinks to zero size and the corresponding gauge coupling blows up. The resulting physics is strongly coupled, but nevertheless it is reasonable to guess that the infrared fixed point is in the same universality class as the quiver gauge theory in figure~\ref{sfig:F07-III-quiver}, in the same spirit as our construction of the $\TO_k$ CFTs via deconfinement. Alternately, we can apply Seiberg duality to the strongly coupled $\SU(N-4)$ factor, giving the brane tiling in figure~\ref{sfig:F07-IIIp-tiling}, corresponding to the quiver gauge theory in figure~\ref{sfig:F07-IIIp-quiver}.
This is merely a different way of deforming the NS5 branes to obtain a gauge theory description, and provides no better handle on the strongly coupled physics of the coincident brane boundaries. Instead, the brane picture suggests that the correct description is ``half-way'' between the Seiberg-dual gauge theories.

Despite these subtleties, it is straightforward to compute $[F]$, and the answer is the same regardless of which of the Seiberg-dual brane tilings we work with. In a pattern that should by now be familiar from other ``classical'' phases of four-sided toric diagrams, such as in (\ref{eqn:II-Ftorsion}) and (\ref{eqn:F0-II-Ftorsion}), we find
\be
N \equiv \alpha_F + \beta_F \pmod 2 \,,
\ee
with $\alpha_F$ and $\beta_F$ each individually undetermined. We denote this phase as $\III^n$ with $n = (-1)^N$ for future reference.

\subsubsection*{Phase \I}

Phase \I\ offers the twin subtleties of coincident antiparallel NS5 brane boundaries along with quadruple intersections of these boundaries atop O5 planes ($\TO_2$ configurations). Nonetheless, the basic approach is the same as in phase \III: we deform the boundaries slightly to obtain a brane tiling with a gauge theory interpretation (apart from the $\TO_2$ configurations) and use this tiling to obtain a gauge theory description of the infrared fixed point and to read off the $[F]$ torsion, where we use the local web diagram (p.~\pageref{def:localwebdiagram}) to resolve ambiguities as needed.

\begin{figure}
  \centering
  \begin{subfigure}[c]{0.3\textwidth}
  \begin{subfigure}[b]{\textwidth}
    \centering
    \includegraphics[height=3.75cm]{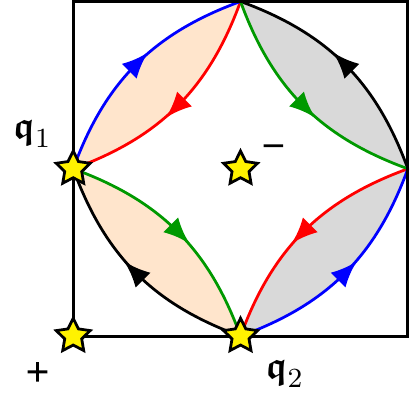}
    \caption{Deformed brane tiling}
    \label{sfig:F07-I-tiling}
  \end{subfigure} \\[0.25cm]
   \begin{subfigure}[b]{\textwidth}
    \centering
    \includegraphics[height=3.75cm]{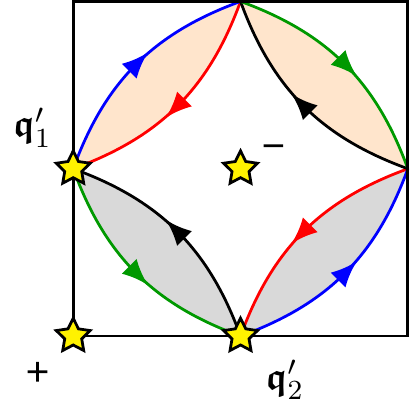}
    \caption{Another deformation}
    \label{sfig:F07-I-tiling-var}
  \end{subfigure}
  \end{subfigure}
  \begin{subfigure}[c]{0.6\textwidth}
    \centering
    \includegraphics[height=5.5cm]{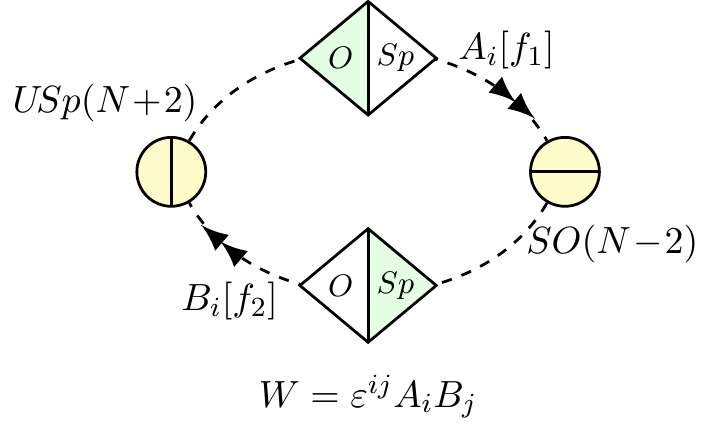}
    \caption{Quiver and superpotential}
    \label{sfig:F07-I-quiver}
  \end{subfigure}
  \caption{Tiling and field theory data corresponding to the local charges
    for phase \I\ (figure~\ref{sfig:F07-I-charges}). To draw a brane tiling with a gauge theory interpretation, we need to deform the NS5 brane boundaries as in~\subref{sfig:F07-I-tiling} and \subref{sfig:F07-I-tiling-var}. Which deformation we choose affects which $\SU(2)_i \times \U(1)_{i \pm 1}$ subgroup of $\SU(2)_1 \times \SU(2)_2$ is manifest in the gauge theory description. Focusing on~\subref{sfig:F07-I-tiling}, we find that $\quadCFT_1=\quadSO$ and $\quadCFT_2=\quadSp$. The corresponding quiver is \subref{sfig:F07-I-quiver}. Note that, while we draw double arrowheads to indicate a doublet of fields, in this case the direction of the arrows is meaningless, as the  $\SO\times\Sp$ bifundamental representation is pseudoreal.}
  \label{fig:F07-I}
\end{figure}

We go through this procedure step by step.
A deformed brane tiling is shown in figure~\ref{sfig:F07-I-tiling}.
To determine which $\TO_2$ CFTs $\quadCFT_1$ and $\quadCFT_2$ appear, we cross reference the local configuration against figure~\ref{fig:quaddictionary}, using the local charges in figure~\ref{sfig:F07-I-charges}. For instance, the $A$--$D$ pair (using the web diagram labels from figure~\ref{sfig:F07-web}) enclose an O5$^-$ where they intersect at the upper-left fixed point. Comparing with figure~\ref{fig:quaddictionary}, we conclude that $\quadCFT_1 = \quadSO$. Picking any adjacent pair of NS5 branes at the lower-right fixed point, we get $\quadCFT_2 = \quadSp$ by the same method. We therefore obtain the quiver in figure~\ref{sfig:F07-I-quiver}, with the corresponding charge table
\begin{equation}
  \begin{array}{c|cc|cccc}
    & \SO(N-2) & \Sp(N+2) & \SU(2)_1 & \U(1)_2 & \U(1)_{B} & \U(1)_R \\ \hline
    \quadSO^{(1)} & \ast & \ast & \ast & 0 & \frac{1}{2} & -\frac{3}{2} \\
    A_i[\phi_1] & \fund & \fund & \fund & 1 & 0 & 1 \\ \hline
    \quadSp^{(2)} & \ast & \ast  & \ast &0 & -\frac{1}{2}  &  -\frac{3}{2}\\
    B_i[\phi_2] & \fund & \fund  & \fund & -1 & 0 & 1 
  \end{array}
\end{equation}
Notice that only $\U(1)_2 \subset \SU(2)_2$ is manifest in this description. To see $\SU(2)_2$, we integrate in a vector-like pair of elementary mesons $\tilde{A}_2$, $\tilde{B}_1$, with the superpotential\footnote{Here and elsewhere in the paper, we will not be careful about signs or coupling constants in the superpotential, as these can always be absorbed into the definitions of the fields in toric orientifolds.}
\be
W = A_1 B_2 - \tilde{A}_2 \tilde{B}_1 +\tilde{A}_2 A_2 + B_1 \tilde{B}_1 \,.
\ee
This breaks the $\SU(2)_1$ symmetry in the UV without affecting the infrared fixed point. We then use $\tilde{A}_2$ and $\tilde{B}_1$ to flip the $\TO_2$ mesons $A_2$ and $B_1$, which gives
\begin{equation}
  \begin{array}{c|cc|cccc}
    & \SO(N-2) & \Sp(N+2) & \U(1)_1 & \SU(2)_2 & \U(1)_{B} & \U(1)_R \\ \hline
    \quadSp^{(1)} & \ast & \ast & 0 & \ast & \frac{1}{2} & -\frac{3}{2} \\
    \tilde{A}_i[\phi_1] & \fund & \fund & 1 & \fund & 0 & 1 \\ \hline
    \quadSO^{(2)} & \ast & \ast  & 0 & \ast & -\frac{1}{2}  &  -\frac{3}{2}\\
    \tilde{B}_i[\phi_2] & \fund & \fund  & -1 & \fund & 0 & 1 
  \end{array}
\end{equation}
after relabeling $A_1$ and $B_2$, with the superpotential $W = \varepsilon^{i j} \tilde{A}_i \tilde{B}_j$. In this alternate description, which corresponds to the brane tiling shown in figure~\ref{sfig:F07-I-tiling-var}, $\SU(2)_2$ becomes manifest while $\SU(2)_1$ is hidden. Since the two descriptions are in the same universality class by construction, we see that the infrared fixed point has the full $\SU(2)_1 \times \SU(2)_2 \times \U(1)_B \times \U(1)_R$ symmetry, as expected.

To determine $[F]$, we read off $\phi_i = (-1)^{F_i}$ using (\ref{eqn:RRtorsionprescription}):
\be \label{eqn:F0-O7-I-Ftorsion}
\begin{split}
F_1 &\equiv [F] \cdot Y_{CD}=\alpha_F \pmod 2  \,,\\ F_2 &\equiv [F] \cdot Y_{BC} = \beta_F \pmod 2 \,,
\end{split}
\ee
hence $[F] = F_1 \vev{a} + F_2 \vev{b}$, where $N$ is constrained to be even. We denote this phase as $\I^{\phi_1 \phi_2}$ for future reference.

\subsubsection*{Duality predictions}

Using the torsion assignments derived above, we predict the following $\SL(2,\bZ)$ multiplets
\begin{align} \label{eqn:F0multiplets}
&(\I^{++})\,, & &(\I^{-+}, \II^+)\,, & &(\I^{+-}, \tII^+)\,, & &(\I^{--}, \III^+)\,, & &(\II^-, \tII^-, \III^-)\,.
\end{align}
These include the duality between phases \II\ and \III\ for odd $N$ that was noticed in~\cite{dualities1}, now revealed to be a triality once the action of $\SL(2,\bZ)$ on the $\SU(2)\times\SU(2)$ global symmetry is taken into account. The other dualities all involve the non-classical phase \I, explaining why they were not seen in~\cite{dualities1}. The rank relations between the duals are fixed by~(\ref{eqn:QD3}), which gives $Q_{\rm D3} = N/2$ for all of the phases.

In appendix~\ref{app:F0-SCIs}, we present evidence that the superconformal indices match within each multiplet in~(\ref{eqn:F0multiplets}), which is a powerful consistency check of these proposed dualities, and by extension of the discrete torsion dictionary developed in this paper.

\subsection{Complex cone over \alt{$\bF_0$}{F0}: O3 planes}
\label{sec:F0-O3}

\begin{figure}
  \centering
  \begin{subfigure}{0.25\textwidth}
    \centering
    \includegraphics[height=3.5cm]{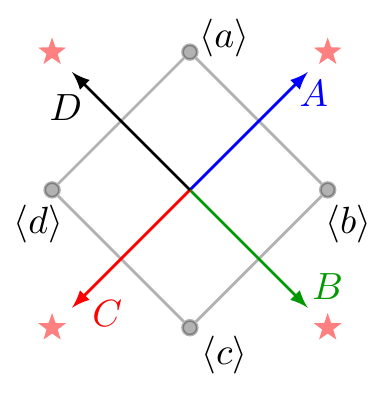}
    \caption{Toric structure}
    \label{sfig:F03-web}
  \end{subfigure}
  \hfill
    \begin{subfigure}{0.175\textwidth}
      \centering
      \includegraphics[height=2.75cm]{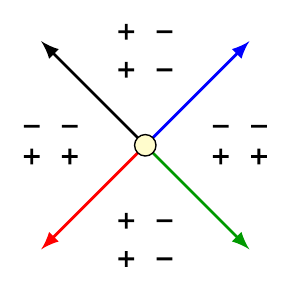}
      \caption{Phase $\I$}
      \label{sfig:F03-I-charges}
    \end{subfigure}
    \hfill
    \begin{subfigure}{0.175\textwidth}
      \centering
      \includegraphics[height=2.75cm]{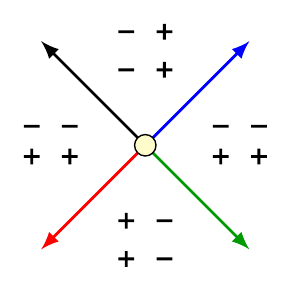}
      \caption{Phase $\II$}
      \label{sfig:F03-II-charges}
    \end{subfigure}
    \hfill
       \\    \begin{subfigure}{0.175\textwidth}
      \centering
      \includegraphics[height=2.75cm]{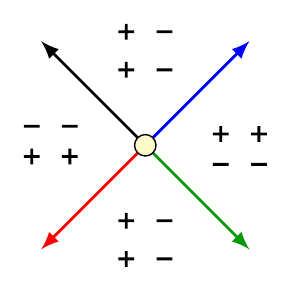}
      \caption{Phase \tII}
      \label{sfig:F03-tII-charges}
    \end{subfigure}
    \hfill
    \begin{subfigure}{0.175\textwidth}
      \centering
      \includegraphics[height=2.75cm]{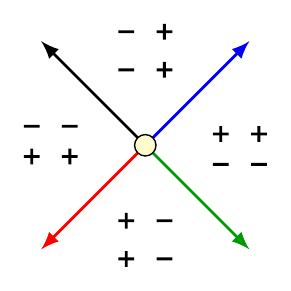}
      \caption{Phase \III}
      \label{sfig:F03-III-charges}
    \end{subfigure}
  \caption{\subref{sfig:F03-web} Toric data for the isolated O3
    orientifold of the complex cone over $\bF_0$. We have named the
    external legs and torsion generators for future reference.}
  \label{fig:F03-toric}
\end{figure}

As mentioned above, there is another toric orientifold of $\bF_0$ with an isolated fixed point, corresponding to the toric data shown in figure~\ref{sfig:F03-web}.
The details are closely analogous to \S\ref{sec:F0-O7}, so we merely summarize the results without going into too much detail.

\begin{figure}
\centering
\begin{subfigure}[b]{0.325\textwidth}
    \centering
    \includegraphics[height=4cm]{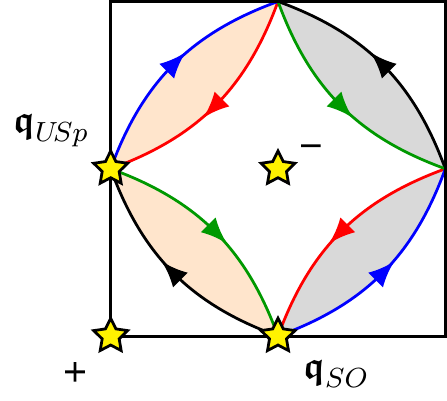}
    \caption{Phase I}
    \label{sfig:F03-I-tiling}
\end{subfigure}
\begin{subfigure}[b]{0.325\textwidth}
    \centering
    \includegraphics[height=3.5cm]{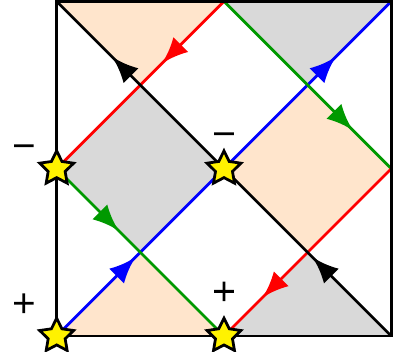}
    \vspace{0.5cm}
    \caption{Phase II}
    \label{sfig:F03-II-tiling}
\end{subfigure}
\begin{subfigure}[b]{0.325\textwidth}
    \centering
    \includegraphics[height=4cm]{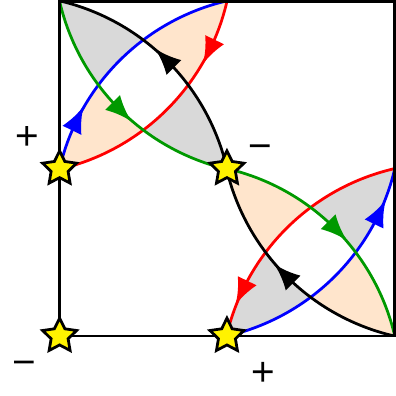}
    \caption{Phase III}
    \label{sfig:F03-III-tiling}
\end{subfigure}
\caption{Brane tilings for the various phases of the $\bF_0$ orientifold with four O3 planes on the exceptional divisor.}
\label{fig:F03-tiling}
\end{figure}

\begin{figure}
\centering
\begin{subfigure}[b]{0.45\textwidth}
    \centering
    \includegraphics[width=\textwidth]{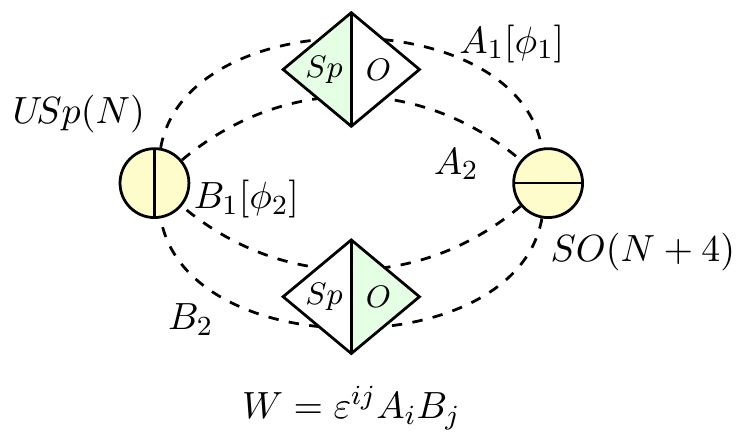}
    \caption{Phase \I}
    \label{sfig:F03-I-quiver}
\end{subfigure} \\[0.25cm]
\begin{subfigure}[b]{0.45\textwidth}
    \centering
    \includegraphics[width=\textwidth]{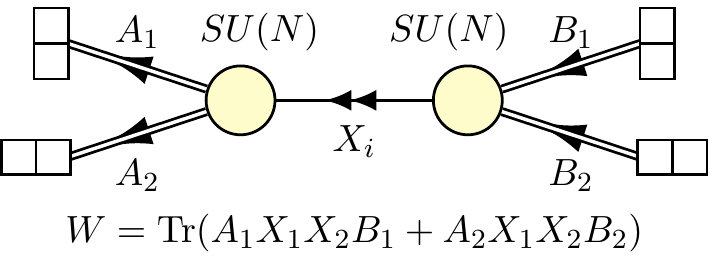}
    \caption{Phase \II}
    \label{sfig:F03-II-quiver}
\end{subfigure}
\hspace{0.1\textwidth}
\begin{subfigure}[b]{0.325\textwidth}
    \centering
    \includegraphics[width=\textwidth]{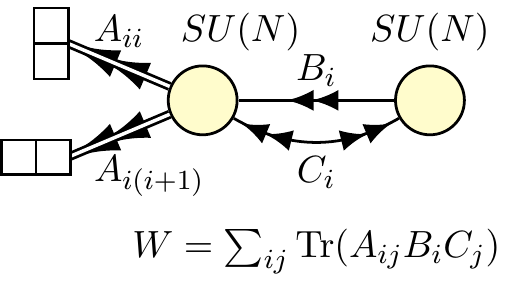}
    \caption{Phase \III}
    \label{sfig:F03-III-quiver}
\end{subfigure}
\caption{Quiver diagrams corresponding to the brane tilings in figure~\ref{fig:F03-tiling}.}
\label{fig:F03-quiver}
\end{figure}

The $\bF_0 \cong \bP^1 \times \bP^1$ exceptional divisor now contains four O3 planes, located at the poles of the $\bP^1$s. This breaks the isometry group $\SU(2)^2 \to \U(1)^2$, which will also be visible in the corresponding gauge theories. The divisor basis still satisfies $\vev{a} = \vev{c}$ and $\vev{b} = \vev{d}$, so we use the same torsion basis and phase labels as before. The local charges are shown in figures~\ref{sfig:F03-I-charges}--\subref{sfig:F03-III-charges} and the corresponding brane tilings and quiver diagrams are shown in figures~\ref{fig:F03-tiling} and \ref{fig:F03-quiver}.

The charge table for phase \I\ is now
\begin{equation}
  \begin{array}{c|cc|cccc}
    & \SO(N+2) & \Sp(N-2) & \U(1)_{B} & \U(1)_X & \U(1)_Y & \U(1)_R \\ \hline
    \quadSp^{(1)} & \ast & \ast & -\frac{1}{2} & 0 & 0 & -\frac{1}{2}  \\
    A_1[\phi_1] & \fund & \fund & 0 & 1 & 1 &  1 \\ 
    A_2 & \fund & \fund & 0 & -1 & 1 & 1 \\ \hline
    \quadSO^{(2)} & \ast & \ast & \frac{1}{2}  & 0 &0  &  -\frac{1}{2}\\
    B_1[\phi_2] & \fund & \fund  & 0 & 1 & -1 & 1 \\
    B_2 & \fund & \fund  & 0 & -1 & -1 & 1 
  \end{array}
\end{equation}
whereas for phase \II\ we have
\begin{equation}
  \label{eq:F03-II-charges}
    \begin{array}{c|cc|cccc}
    & SU(N) & SU(N) & U(1)_B & U(1)_X & U(1)_Y & U(1)_R\\
    \hline
    X_1 & \fund & \ov\fund & \frac{1}{N} & 1 & 0 & \frac{1}{2}\\
    X_2 & \fund & \ov\fund & \frac{1}{N} & -1 & 0 & \frac{1}{2}\\
    B_1 & \ov\asymm & {\bf 1} & -\frac{1}{N} & 0 & -1-\frac{2}{N} & \frac{1}{2}\\
    B_2 & \ov\symm & {\bf 1} & -\frac{1}{N} & 0 & 1-\frac{2}{N} & \frac{1}{2}\\
    A_1 & {\bf 1} & \asymm & -\frac{1}{N} & 0 & 1+\frac{2}{N} & \frac{1}{2}\\
    A_2 & {\bf 1} & \symm & -\frac{1}{N} & 0 & -1+\frac{2}{N} & \frac{1}{2}
  \end{array}
\end{equation}
and for phase \III
\begin{equation}
  \label{eqref:F03-III-charges}
    \begin{array}{c|cc|cccc}
    & SU(N) & SU(N) & U(1)_B & U(1)_X & U(1)_Y & U(1)_R\\
    \hline
    B_1 & \fund & \ov\fund & \frac{1}{N} & 1 & 0 & \frac{1}{2} \\
    B_2 & \fund & \ov\fund & \frac{1}{N} & -1 & 0 & \frac{1}{2} \\
    C_1 & \ov\fund & \ov\fund & -\frac{1}{N} & 0 & -1 & \frac{1}{2} \\
    C_2 & \ov\fund & \ov\fund & -\frac{1}{N} & 0 & 1 & \frac{1}{2} \\
    A_{11} & {\bf 1} & \asymm & 0 & -1 & 1 & 1 \\
    A_{22} & {\bf 1} & \asymm & 0 & 1 & -1 & 1 \\
    A_{12} & {\bf 1} & \symm & 0 & -1 & -1 & 1 \\
    A_{21} & {\bf 1} & \symm & 0 & 1 & 1 & 1
  \end{array}
\end{equation}

It is straightforward to compute $[F]$ in phases \II\ and \III; we obtain
\be
\begin{split}
\II: N &\equiv \beta_F \pmod 2 \,, \\
\III: N &\equiv \alpha_F + \beta_F \pmod 2 \,,
\end{split}
\ee
as above. In phase \I, however, there is a slight difference due to the change in the local charges at the upper-left and lower-right fixed points. Denoting the associated flavor parities as $\phi_1 = (-1)^{F_1}$ and $\phi_2 = (-1)^{F_2}$, respectively, we obtain
\be
\begin{aligned}
F_1 &\equiv \beta_F \pmod 2 \,, &
F_2 &\equiv \alpha_F \pmod 2 \,,
\end{aligned}
\ee
which is different from~(\ref{eqn:F0-O7-I-Ftorsion}). As a consequence, the predicted duality multiplets are now
\begin{align} \label{eqn:F0-O3-multiplets}
&(\I^{++})\,, & &(\I^{+-}, \II^+)\,, & &(\I^{-+}, \tII^+)\,, & &(\I^{--}, \III^+)\,, & &(\II^-, \tII^-, \III^-)\,,
\end{align}
in the same notation as before. These dualities have not been
previously discussed in the literature. In
appendix~\ref{app:F0-O3-SCIs} we present strong evidence that the
superconformal index matches between different phases in the same
$\SL(2,\bZ)$ multiplet, a necessary condition for these S-dualities to
exist, and a further consistency check on the discrete torsion
dictionary presented in this paper.

\subsection{Real cone over \alt{$Y^{4,0}$}{Y40}}
\label{sec:Y40}

There are infinitely many other toric orientifolds that we could consider. For instance, the real cones over $Y^{p,q}$~\cite{Gauntlett:2004yd, Martelli:2004wu} define a simple infinite class of toric singularities. Much like the infinite series of $\bC^3/\bZ_{2k+1}$ orientifolds pictured in figure~\ref{fig:Zk-toric}, there is an infinite series of $Y^{2p,2p-1}$ toric orientifolds which generalize the $dP_1$ (i.e., $Y^{2,1}$) toric orientifold considered in~\S\ref{sec:dP1}, with a corresponding set of S-dualities~\cite{Wrase:2013}. Likewise, there is an infinite class of $Y^{2p,0}$ toric orientifolds which generalize the $\bF_0$ (i.e., $Y^{2,0}$) orientifolds considered above.

The same methods discussed above can be applied to these examples, but the details of the discrete torsion calculations are largely analogous to what we have discussed already. As an illustration, we very briefly discuss $Y^{4,0}$ toric orientifolds. Since these were considered briefly in~\cite{dualities1}, we focus on confirming that the triality detected there is an $\SL(2,\bZ)$ sextet similar to figure~\ref{fig:sextet}.

\begin{figure}
  \centering
  \begin{subfigure}[b]{0.23\textwidth}
    \centering
    \includegraphics[height=5cm]{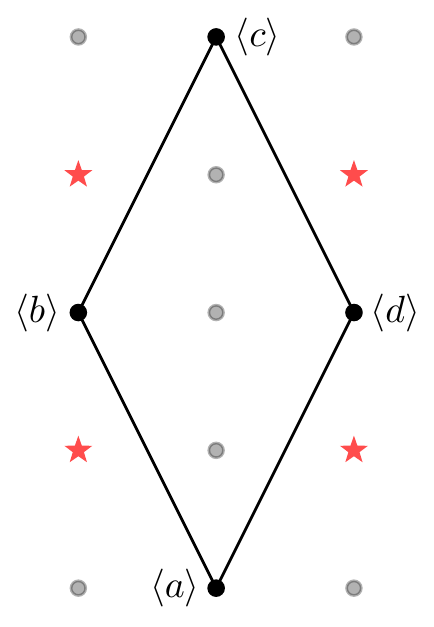}
    \caption{O3 planes}
    \label{sfig:Y40-3-toric}
  \end{subfigure}
  \begin{subfigure}[b]{0.23\textwidth}
    \centering
    \includegraphics[height=5cm]{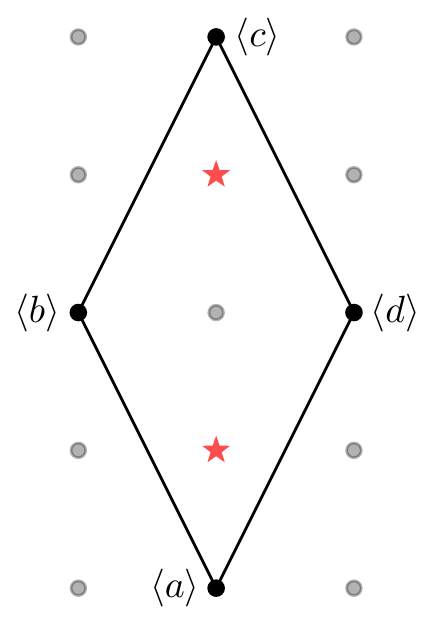}
    \caption{Isolated O7 planes}
    \label{sfig:Y40-7-toric}
  \end{subfigure}
  \begin{subfigure}[b]{0.5\textwidth}
   \begin{subfigure}[b]{0.48\textwidth}
    \centering
    \includegraphics[width=\textwidth]{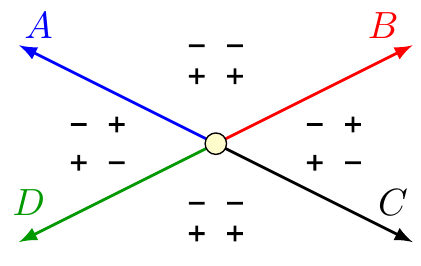}
    \caption{Phase I}
    \label{sfig:Y40-7-I}
  \end{subfigure}
  \hfill
  \begin{subfigure}[b]{0.48\textwidth}
    \centering
    \includegraphics[width=\textwidth]{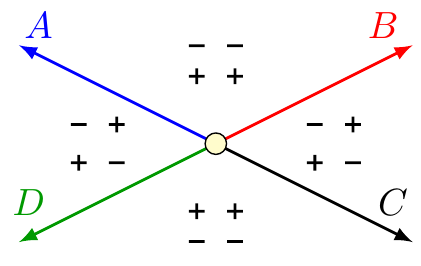}
    \caption{Phase II}
    \label{sfig:Y40-7-II}
  \end{subfigure}
  \\
  \begin{subfigure}[b]{0.48\textwidth}
    \centering
    \includegraphics[width=\textwidth]{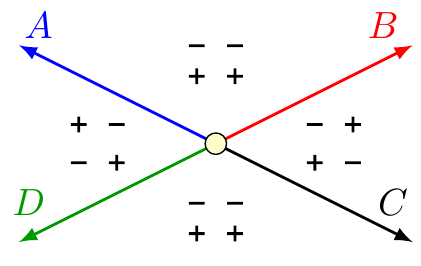}
    \caption{Phase III}
    \label{sfig:Y40-7-III}
  \end{subfigure}
  \hfill
  \begin{subfigure}[b]{0.48\textwidth}
    \centering
    \includegraphics[width=\textwidth]{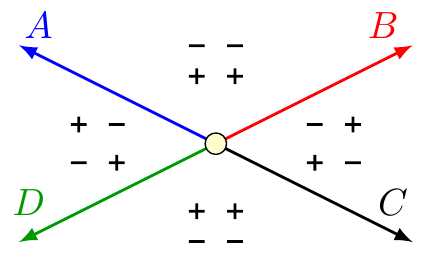}
    \caption{Phase IV}
    \label{sfig:Y40-7-IV}
  \end{subfigure}
  \end{subfigure}
  \caption{Toric data for the $Y^{4,0}$ toric orientifolds with~\subref{sfig:Y40-3-toric} O3 planes and~\subref{sfig:Y40-7-toric} O7 planes in the resolved geometry. \subref{sfig:Y40-7-I}--\subref{sfig:Y40-7-IV} Local charges for the four phases in the O7 case, following the conventions of table~\ref{table:Y40-7-phases}.}
  \label{fig:Y40-toric}
\end{figure}

As in the $\bF_0$ case, there are two possible toric involutions with an isolated fixed point, differing by whether O3 planes (figure~\ref{sfig:Y40-3-toric}) or O7 planes (figure~\ref{sfig:Y40-7-toric}) appear when the singularity is resolved.
We consider the latter case, as this will lead to the orientifold gauge theories considered in~\cite{dualities1}. The divisor basis once again satisfies $\vev{a} = \vev{c}$ and $\vev{b} = \vev{d}$, so we take the ansatz $[H] = \alpha \vev{a} + \beta \vev{b}$ and $[F] = \alpha_F \vev{a} + \beta_F \vev{b}$. There are four choices of $[H]$, labelled by the phase numbers shown in table~\ref{table:Y40-7-phases}.
\begin{table}
  \centering
  \begin{tabular}{r|cccc}
    Phase: & \I & \II & \III & \IV \\
    \hline
    $H_3$ torsion &  (00) & (10) & (01) & (11)
  \end{tabular}
  \caption{The four different choices of NSNS torsion in the $Y^{4,0}$ isolated toric orientifold with O7 planes, written in the form $(\alpha\beta)$ with $[H] = \alpha \vev{a}+\beta\vev{b}$.}
  \label{table:Y40-7-phases}
\end{table}
The corresponding local charges are shown in figures~\ref{sfig:Y40-7-I}--\subref{sfig:Y40-7-IV}.

\begin{figure}
  \centering
  \begin{subfigure}[b]{0.4\textwidth}
    \centering
    \includegraphics[height=4.5cm]{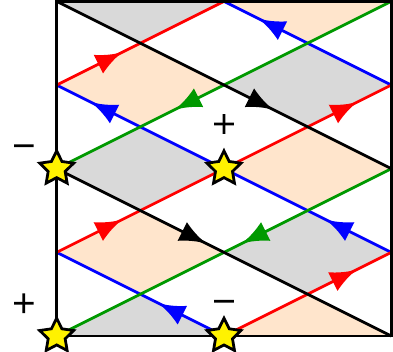}
    \caption{Tiling for phase \II}
    \label{sfig:Y40-7-II-tiling}
  \end{subfigure}
  \hspace{1cm}
  \begin{subfigure}[b]{0.5\textwidth}
    \centering
    \includegraphics[width=\textwidth]{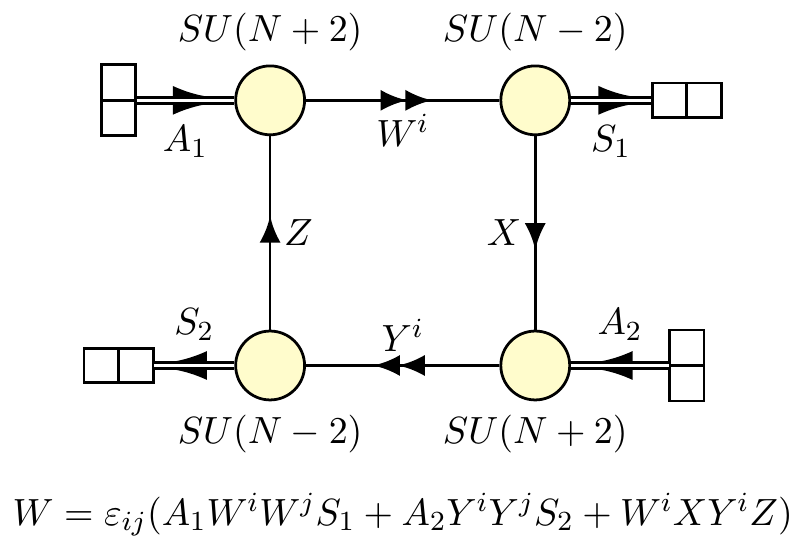}
    \caption{Quiver for phase \II}
    \label{sfig:Y40-7-II-quiver}
  \end{subfigure}
  \caption{Tiling and quiver for phase \II\ of the $Y^{4,0}$ toric orientifold in figure~\ref{sfig:Y40-7-toric}.}
  \label{fig:Y40-7-II-gauge}
\end{figure}

\begin{figure}
  \centering
  \begin{subfigure}[b]{0.4\textwidth}
    \centering
    \includegraphics[height=4cm]{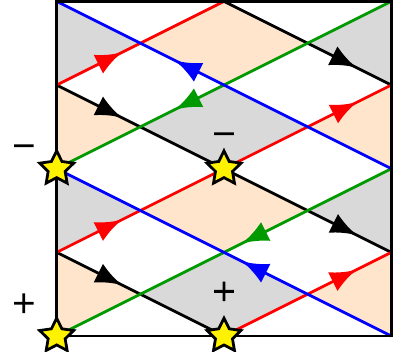}
    \caption{Tiling for phase \III}
    \label{sfig:Y40-7-III-tiling}
  \end{subfigure}
  \hspace{1cm}
  \begin{subfigure}[b]{0.5\textwidth}
    \centering
    \includegraphics[width=\textwidth]{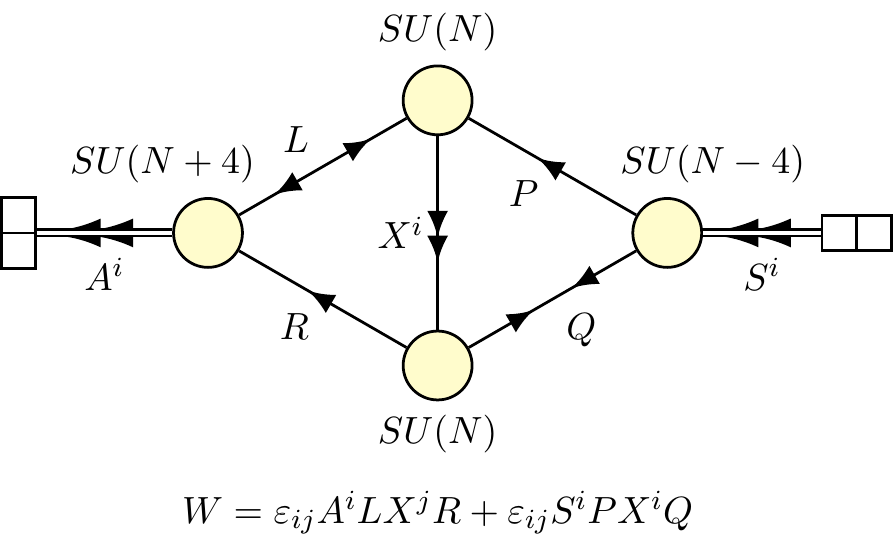}
    \caption{Quiver for phase \III}
    \label{sfig:Y40-7-III-quiver}
  \end{subfigure}
  \caption{Tiling and quiver for phase \III\ of the $Y^{4,0}$ toric orientifold in figure~\ref{sfig:Y40-7-toric}.}
  \label{fig:Y40-7-III-gauge}
\end{figure}

\begin{figure}
  \centering
  \begin{subfigure}[b]{0.4\textwidth}
    \centering
    \includegraphics[height=4cm]{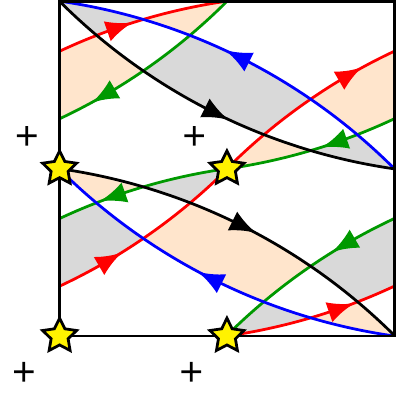}
    \caption{Tiling for phase \IV}
    \label{sfig:Y40-7-IV-tiling}
  \end{subfigure}
  \hspace{1cm}
  \begin{subfigure}[b]{0.5\textwidth}
    \centering
    \includegraphics[width=\textwidth]{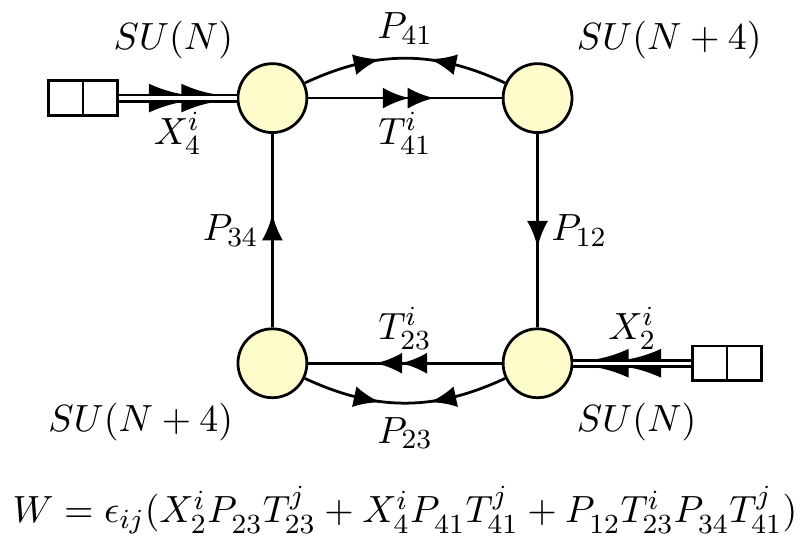}
    \caption{Quiver for phase \IV}
    \label{sfig:Y40-7-IV-quiver}
  \end{subfigure}
  \caption{Tiling and quiver for phase \IV\ of the $Y^{4,0}$ toric orientifold in figure~\ref{sfig:Y40-7-toric}.}
  \label{fig:Y40-7-IV-gauge}
\end{figure}

Using these, we construct the brane tilings and quivers for the ``classical'' phases \II, \III\ and \IV, shown in figures~\ref{fig:Y40-7-II-gauge}, \ref{fig:Y40-7-III-gauge} and \ref{fig:Y40-7-IV-gauge}, respectively.
The quivers in figures~\ref{sfig:Y40-7-II-quiver}, \ref{sfig:Y40-7-III-quiver} and \ref{sfig:Y40-7-IV-quiver} are exactly those appearing in~\cite{dualities1}, under the names $\II^{\rm (here)} = \Ib^{\rm (there)}$, $\III^{\rm (here)} = \Ia^{\rm (there)}$ and $\IV^{\rm (here)} = \II^{\rm (there)}$.

As usual in classical phases, the $[F]$ torsion is straightforward to read off
\be
\begin{split}
\II&: N \equiv \alpha_F \pmod 2 \,, \\
\III&: N \equiv \beta_F \pmod 2 \,, \\
\IV&: N \equiv \alpha_F+\beta_F \pmod 2 \,. \\
\end{split}
\ee
Denoting these phases by $\II^n$, $\III^n$, and $\IV^n$ with $n = (-1)^N$, we conclude that $(\II^-, \III^-, \IV^-)$ is indeed an $\SL(2,\bZ)$ sextet, matching the triality observed in~\cite{dualities1}.\footnote{Since $Q_{\rm D3} = N/2$ for all three phases, the relation between the ranks is trivial.} This is analogous to the $\bF_0$ sextet $(\II^-, \tII^-, \III^-)$, except that in this case $\II$ and $\III$ do not give isomorphic brane tilings, as the $D_4$ invariance of the $\bF_0$ toric diagram is reduced to the Klein four-group, $\bZ_2 \times \bZ_2$.

In the above discussion, we have deliberately omitted phase \I\ and the S-dualities involving it. This is done both to save space and to provide a tractable exercise for any readers wishing to try their hand at calculations.

\subsection{Complex cone over \alt{$dP_2$}{dP2}}
\label{sec:dP2}

Our final example, the toric orientifold of the Calabi-Yau cone over
$dP_2$ with an O7 plane wrapping the exceptional divisor, is one that
was entirely intractable before $\TO_2$ CFTs were understood
in~\cite{dP1paper}. As pointed out in that paper (see also
\cite{Retolaza:2016alb}), ``classical'' phases of toric orientifolds
do not exist when the toric diagram has at least five
sides.\footnote{In particular, since each NS5 brane boundary
  intersects two O5 planes, the toric diagram has $\sum_{i=1}^4 k_i$
  sides for configurations $\TO_{k_i}$ at the four fixed points. Thus,
  tilings with only the ``classical'' $\TO_{0,1}$ configurations
  correspond to toric diagrams with at most four sides.} One way to
think about this is as a consequence of the fact that
$[F] \cong [F]+[H]$ contains at least two bits of information in every
phase, whereas classical phases have only color parity and no
additional flavor parities. In any case, every phase of the $dP_2$
orientifold contains at least one $\TO_2$
configuration. 

Even with the results of~\cite{dP1paper} in hand, the study of the $dP_2$ orientifold is daunting, due to the large number of phases, torsion assignments, and possible S-dualities that need to be resolved. This task becomes much more tractable with the systematic techniques developed in this paper, and we present a complete description of all the phases and their conjectural S-dualities below (which we check by matching the superconformal index between members of the same $\SL(2,\bZ)$ multiplet).

\begin{figure}
  \centering
        \includegraphics[height=5cm]{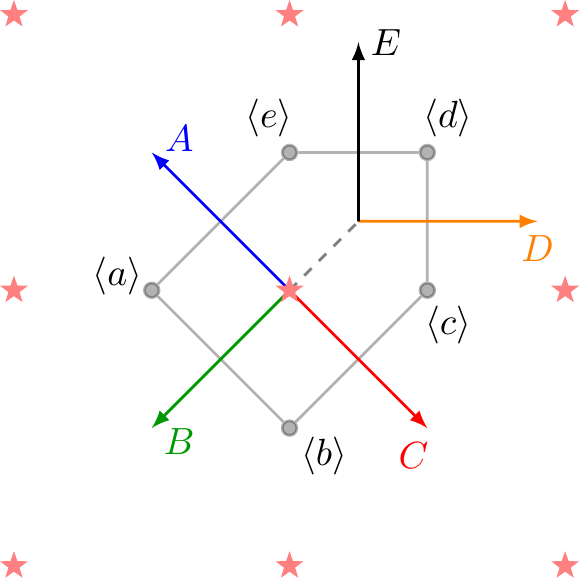}
       \caption{Toric data for the isolated
    orientifold of the complex Calabi-Yau cone over $dP_2$. We have
    named the external $(p,q)$ legs for later reference. For ease of
    depiction we have displaced slightly the $D$ and $E$ external
    legs, keeping their slopes.}
  \label{fig:dP2-toric}
\end{figure}

The toric data for the orientifold considered here are shown in figure~\ref{fig:dP2-toric}.
The divisor basis relations are $\vev{a}+\vev{c} = \vev{b}+\vev{e} = \vev{d}$. We choose the basis $\vev{e_i}=\{\vev{c}, \vev{e}, \vev{c}+\vev{d}+\vev{e}\}$, $i\in\{1,2,3\}$, to describe the discrete torsions, which has the convenient property that $\vev{e_i} \cdot \vev{e_j} = \delta_{i j}$. This basis is also convenient when considering the partial resolutions to the $dP_1$ orientifold singularity plus an O3 plane, or to the $\bC^3/\bZ_3$ orientifold singularity plus two O3 planes. In particular, in the latter case $\vev{e_{1,2}}$ generate the torsions of the two O3 planes and $\vev{e_3}$ generates the $\bC^3/\bZ_3$ orientifold torsion. For future reference, we note that
\be
\begin{aligned}
\vev{a} &= \vev{e_2} + \vev{e_3}\,, & \vev{b} &= \vev{e_1} + \vev{e_3}\,, & \vev{c} &= \vev{e_1}\,, \\
\vev{d} &=  \vev{e_1} + \vev{e_2} +  \vev{e_3}\,, & \vev{e} &= \vev{e_2}\,,
\end{aligned}
\ee
in this torsion basis.

Writing $[H] = h_1 \vev{e_1} + h_2 \vev{e_2} + h_3 \vev{e_3}$, we find the eight phases listed in table~\ref{table:dP2-phases}. 
\begin{table}
  \centering
  \begin{tabular}{r|cccccccc}
    Phase: & \IA & \IB & \IIA & \IIB & \IIIA & \IIIB & \tIIIA & \tIIIB \\
    \hline
              $H_3$ torsion: & (000) & (111) & (110) & (001) & (100) & (011) & (010) & (101)
  \end{tabular}
  \caption{Phase labels for the eight choices of $[H] =  h_1 \vev{e_1} + h_2 \vev{e_2} + h_3 \vev{e_3}$ in the $dP_2$ orientifold, specified in the form $(h_1 h_2 h_3)$.}
  \label{table:dP2-phases}
\end{table}
Note that the toric diagram has a $\bZ_2$ reflection symmetry which maps $\vev{e_1} \leftrightarrow \vev{e_2}$. Since the phases $\IIIA \longleftrightarrow \tIIIA$ and $\IIIB \longleftrightarrow \tIIIB$ are related by this reflection, the corresponding SCFTs will be isomorphic up to a relabeling of the global symmetries, hence we will not discuss $\tIIIA$ and $\tIIIB$ further until we organize our results into $\SL(2,\bZ)$ multiplets.

\subsubsection*{Phase \IA}

\begin{figure}
  \centering
  \begin{subfigure}[t]{0.23\textwidth}
    \centering
    \includegraphics[height=3.5cm]{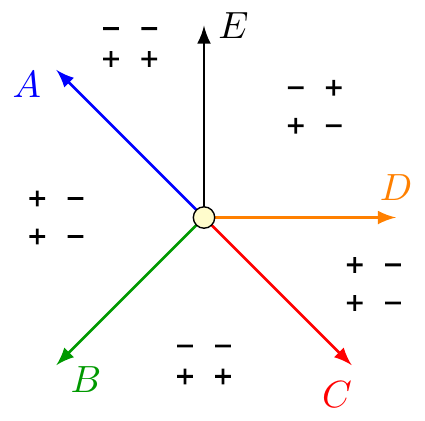}
    \caption{Local charges.}
    \label{sfig:dP2-IA-charges}
  \end{subfigure}
  \hfill
  \begin{subfigure}[t]{0.3\textwidth}
    \centering
    \includegraphics[height=4cm]{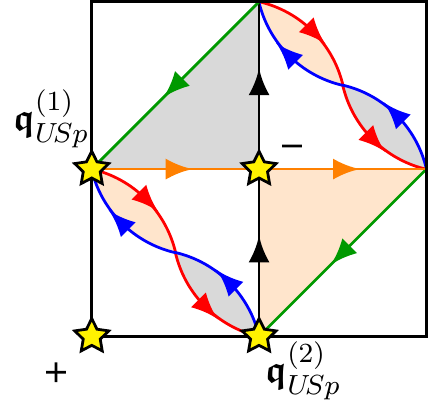}
    \caption{Brane tiling.}
    \label{sfig:dP2-IA-tiling}
  \end{subfigure}
  \hfill
  \begin{subfigure}[t]{0.45\textwidth}
    \centering
    \includegraphics[width=\textwidth]{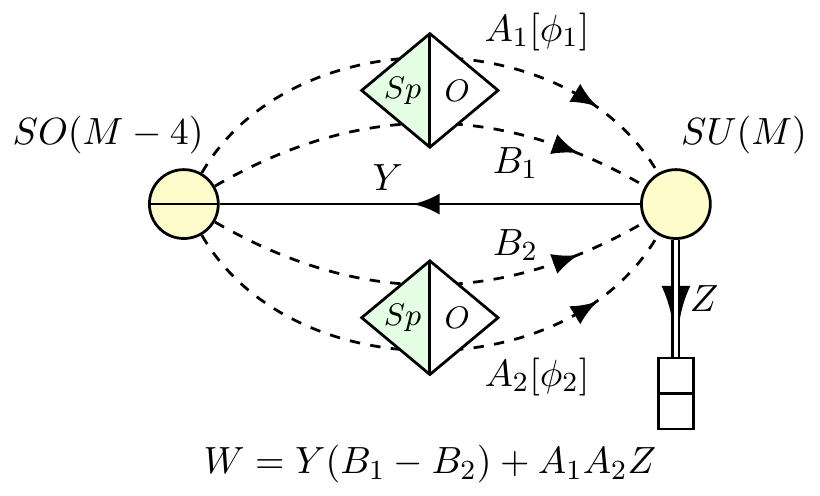}
    \caption{Quiver and superpotential.}
    \label{sfig:dP2-IA-quiver}
  \end{subfigure}
  \caption{Brane tiling and field theory data for phase {\IA} of the
    complex cone over $dP_2$. In~\subref{sfig:dP2-IA-tiling} we
    have slightly deformed the overlapping parallel NS5 branes to obtain a quiver description, as in~\S\ref{sec:F0-O7}.}
  \label{fig:dP2-IA}
\end{figure}

We start with the phase \IA\, for which $[H] = 0$. The local charges, brane tiling, quiver diagram and superpotential are shown in figure~\ref{fig:dP2-IA}.
 The global symmetries for this phase are
\begin{equation} \label{eqn:dP2-IA-charges}
  \begin{array}{c|cc|ccccc}
    & \SO(M-4) & \SU(M) & \U(1)_{B+} & \U(1)_{B-} & \U(1)_{X+} & \U(1)_{X-} & \U(1)_R \\ \hline
    Y & \ydiagram{1} & \ydiagram{1} & -\frac{2}{M} & 0 &  -\frac{2}{M} & 0 & - \frac{2}{M} \\
    Z & \mathbf{1} & \ydiagram{1,1} & -\frac{4}{M} & 0 & 2-\frac{4}{M} & 0 & 2- \frac{4}{M}  \\ \hline
    \quadSp^{(1)} & \ast & \ast & \frac{1}{2} & \frac{1}{2} & -\frac{M-2}{4} & \frac{M-2}{4} & -\frac{M+4}{4} \\
    A_1[\phi_1] & \ydiagram{1} & \overline{\ydiagram{1}} & \frac{2}{M} & 0 & -1+ \frac{2}{M} & 1 & \frac{2}{M} \\
    B_1 & \ydiagram{1} & \overline{\ydiagram{1}} & \frac{2}{M} & 0 & \frac{2}{M} & 0 & 2+ \frac{2}{M} \\ \hline
    \quadSp^{(2)} & \ast & \ast &  \frac{1}{2} & -\frac{1}{2} & -\frac{M-2}{4} & -\frac{M-2}{4}  &  -\frac{M+4}{4}\\
    A_2[\phi_2] & \ydiagram{1} & \overline{\ydiagram{1}} & \frac{2}{M} & 0 & -1+ \frac{2}{M} & -1 & \frac{2}{M} \\
    B_2 & \ydiagram{1} & \overline{\ydiagram{1}} & \frac{2}{M} & 0 &  \frac{2}{M} & 0 & 2+ \frac{2}{M}
  \end{array}
\end{equation}
 This basis of
generators for the global symmetries (reused in all subsequent phases) is chosen to ensure several
desirable properties: (i) there is a $\bZ_2$ reflection symmetry in
the toric diagram which manifests itself as the exchange
$\mathfrak{q}^{(1)} \leftrightarrow \mathfrak{q}^{(2)}$. The symmetries
$\U(1)_{B+}, \U(1)_{X+}$, and $\U(1)_R$ are invariant under this
exchange, whereas $\U(1)_{B-}$ and $\U(1)_{X-}$ change sign.\footnote{Making this reflection symmetry manifest is also the motivation for including the elementary meson $Y$ in the table, rather than integrating it out by flipping $B_1$ or $B_2$.} (ii)
Taking $M \to -M$ will produce the quiver diagram and charge table for phase $\I_B$ by negative rank duality,
see figure~\ref{fig:dP2-IB} and~(\ref{eqn:dP2-IB-charges}). This basis is chosen to ensure matching
anomalies up to $\U(1)_{B\pm} \to - \U(1)_{B\pm}$. (iii) The charges
satisfy the simple quantization conditions
$Q_{B+} \pm Q_{B-} \in \bZ$, $Q_{X+} \pm Q_{X-} \in \bZ$ and
$Q_R \in \bZ$.

The 't Hooft anomalies in this basis are as follows:
\begin{equation} \label{eqn:dP2anomalies}
\begin{array}{|c|c|c|c|}
\hline
\U(1)_{X+}^3 & 3 Q^2 - \frac{3}{4} & \U(1)_{X-}^2 \U(1)_{X+} & -Q^2+\frac{1}{4}\\
 \U(1)_{X+}^2 \U(1)_{B+} & -4Q & \U(1)_{X-}^2 \U(1)_{B+} & 4Q \\
 \U(1)_{B+}^2 \U(1)_{X+} & 4 &  \U(1)_{B+} \U(1)_{B-} \U(1)_{X-} & 2 \\ \hline
\U(1)_{X+}^2 \U(1)_R & Q^2-\frac{17}{4} & \U(1)_{X-}^2 \U(1)_R & -Q^2-\frac{15}{4} \\
 \U(1)_{B+}^2 \U(1)_R & - 6 & \U(1)_{B-}^2 \U(1)_R & - 2 \\
\U(1)_R^3 & -83 & \U(1)_{X+} \U(1)_R^2 & -16 \\
\U(1)_R & -11 & & \\
\hline
\end{array}
\end{equation}
with $Q \equiv M-5/2 = 2 Q_{\rm D3}$ and all other anomalies vanish. The $a$-maximizing R-charge---which we use later to compute the superconformal index---is given by
\begin{equation}
\U(1)_R^{\rm (sc)} = \U(1)_R + \frac{Q a_X^2}{2 a_X-3} \U(1)_{B+} + a_X \U(1)_{X+} \,,
\end{equation}
where $a_X$ is the root of the quartic equation
\begin{equation}
(Q^2-9/4) a_X (4 a_X^2 - 3 a_X - 18)+(a_X + 12)(a_X-2)(3 a_X^2 +4 a_X - 6) = 0\,,
\end{equation}
that lies in the range $-2 < a_X < -\frac{3}{8} (\sqrt{33}-1) \approx -1.779$ for $Q> 7/2$ ($M> 6$).\footnote{The case $M=6$ confines, in some cases with chiral symmetry breaking. A discussion of the confining dynamics is beyond the scope of this paper, though it is potentially of interest in the string theory dual.}

To fix $[F] = f_1 \vev{e_1} +  f_2 \vev{e_2} +  f_3 \vev{e_3}$, we read off $m = (-1)^M$ and $\phi_{1,2} = (-1)^{F_{1,2}}$ from the brane tiling using~(\ref{eqn:colorparity}) and (\ref{eqn:RRtorsionprescription}). We find
\be
\begin{split}
M &\equiv f_1 + f_2 + f_3 \pmod 2\,, \\
F_{1,2} &\equiv f_{1,2} \pmod 2\,,
\end{split}
\ee
where we use $Y_{CD} = \vev{c} = \vev{e_1}$ and $Y_{EA} = \vev{e} = \vev{e_2}$ to determine $F_1$ and $F_2$, respectively. We label this phase as $\IA^{m;\phi_1\phi_2}$ for future reference.

\subsubsection*{Phase \IB}

\begin{figure}
  \centering
  \begin{subfigure}[t]{0.25\textwidth}
    \centering
    \includegraphics[height=3.5cm]{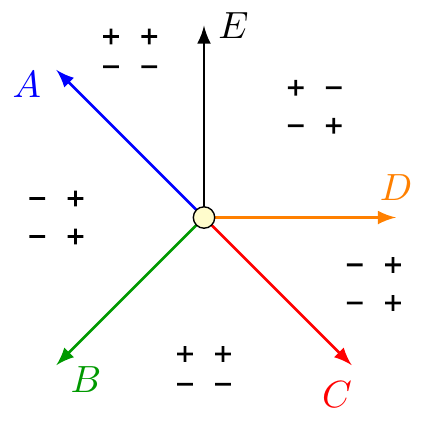}
    \caption{Local charges.}
    \label{sfig:dP2-IB-charges}
  \end{subfigure}
  \hfill
  \begin{subfigure}[t]{0.25\textwidth}
    \centering
    \includegraphics[height=3.5cm]{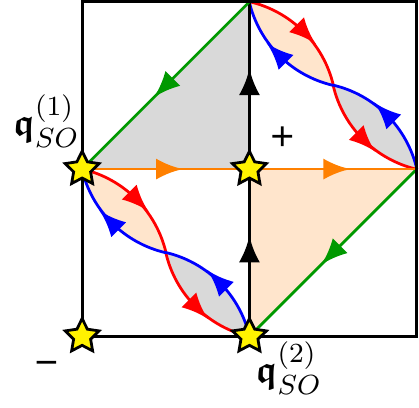}
    \caption{Brane tiling.}
    \label{sfig:dP2-IB-tiling}
  \end{subfigure}
  \hfill
  \begin{subfigure}[t]{0.45\textwidth}
    \centering
    \includegraphics[width=\textwidth]{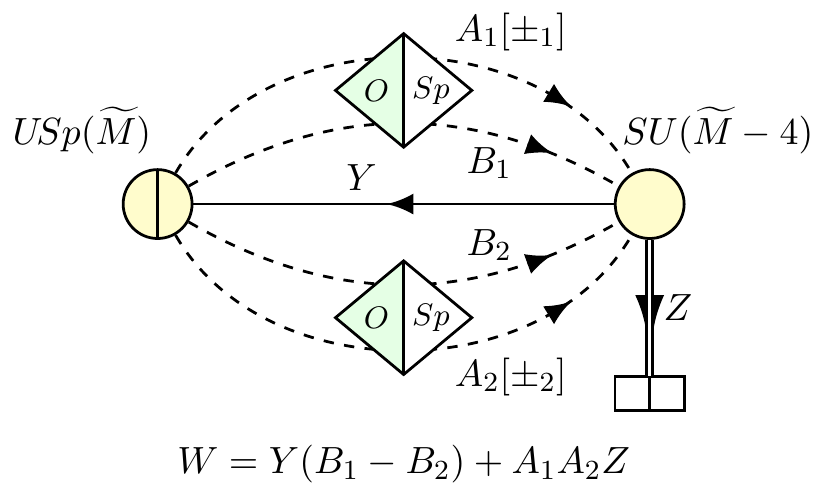}
    \caption{Quiver and superpotential.}
    \label{sfig:dP2-IB-quiver}
  \end{subfigure}
  \caption{Brane tiling and field theory data for phase {\IB} of the
    complex cone over $dP_2$, where we include a half-period translation on $T^2$ to make the negative rank duality with figure~\ref{fig:dP2-IA} manifest.}
  \label{fig:dP2-IB}
\end{figure}

We next consider phase $\IB$. As the negative rank dual of phase \IA, many of its superficial properties are analogous, and can be obtained from those above by simple procedures such as inverting all of the local charges in the brane tiling (see also \cite{dualities1} for a discussion of negative rank duality in the quiver and charge table). The local charges, brane tiling, quiver diagram, and superpotential are shown in figure~\ref{fig:dP2-IB}.
The global symmetries are described by the charge table
\begin{equation} \label{eqn:dP2-IB-charges}
  \begin{array}{c|cc|ccccc}
    & \Sp(\tM+4) & \SU(\tM) & \U(1)_{B+} & \U(1)_{B-} & \U(1)_{X+} & \U(1)_{X-} & \U(1)_R \\ \hline
    Y & \ydiagram{1} & \ydiagram{1} & -\frac{2}{\tM} & 0 &  \frac{2}{\tM} & 0 & \frac{2}{\tM} \\
    Z & \mathbf{1} & \ydiagram{2} & -\frac{4}{\tM} & 0 & 2+\frac{4}{\tM} & 0 & 2 + \frac{4}{\tM}  \\ \hline
    \mathfrak{q}_{\SO}^{(1)} & \ast & \ast & -\frac{1}{2} & -\frac{1}{2} & \frac{\tM+2}{4} & -\frac{\tM+2}{4} & \frac{\tM-4}{4} \\
    A_1[\phi_1] & \ydiagram{1} & \overline{\ydiagram{1}} & \frac{2}{\tM} & 0 & -1- \frac{2}{\tM} & 1 & -\frac{2}{\tM} \\
    B_1 & \ydiagram{1} & \overline{\ydiagram{1}} & \frac{2}{\tM} & 0 & -\frac{2}{\tM} & 0 & 2- \frac{2}{\tM} \\ \hline
    \mathfrak{q}_{\SO}^{(2)} & \ast & \ast &  -\frac{1}{2} & \frac{1}{2} & \frac{\tM+2}{4} & \frac{\tM+2}{4}  & \frac{\tM-4}{4}\\
    A_2[\phi_2] & \ydiagram{1} & \overline{\ydiagram{1}} & \frac{2}{\tM} & 0 & -1-\frac{2}{\tM} & -1 & -\frac{2}{\tM} \\
    B_2 & \ydiagram{1} & \overline{\ydiagram{1}} & \frac{2}{\tM} & 0 & - \frac{2}{\tM} & 0 & 2- \frac{2}{\tM}
  \end{array}
\end{equation}
The 't Hooft anomalies match with~(\ref{eqn:dP2anomalies}) for $Q = \tM+5/2$, which is the same relation one obtains by computing the D3 charge $Q_{\rm D3} = \tM/2 + 5/4$ using~(\ref{eqn:QD3}).

Computing the flavor parities $\phi_i = (-1)^{F_i}$ as above, we obtain
\be \label{eqn:dP2-IB-Ftorsion}
F_i \equiv f_i + f_3 \pmod 2\,,
\ee
where we use $Y_{BC} = \vev{b} = \vev{e_1} + \vev{e_3}$ and $Y_{A B} = \vev{a} = \vev{e_2} + \vev{e_3}$ to compute $F_1$ and $F_2$, respectively, and $\tM$ is constrained to be even. Notice that (\ref{eqn:dP2-IB-Ftorsion}) is invariant under $f_i \to f_i +1$, as expected from the $[F] \to [F]+[H]$, $\tau \to \tau+1$ equivalence. We denote this phase as $\IB^{\phi_1 \phi_2}$ for future reference.

\subsubsection*{Phase \IIA}

\begin{figure}
  \centering
  \begin{subfigure}[t]{0.25\textwidth}
    \centering
    \includegraphics[height=3.5cm]{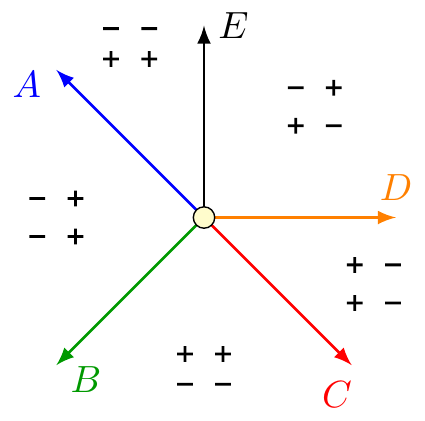}
    \caption{Local charges.}
    \label{sfig:dP2-IIA-charges}
  \end{subfigure}
  \hfill
  \begin{subfigure}[t]{0.25\textwidth}
    \centering
    \includegraphics[height=3.8cm]{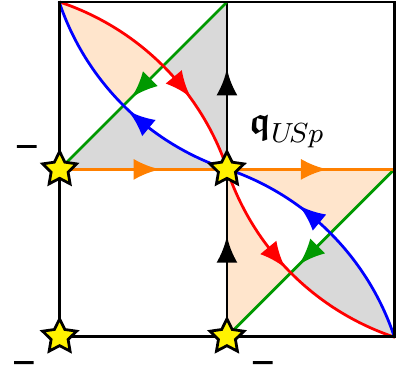}
    \caption{Brane tiling.}
    \label{sfig:dP2-IIA-tiling}
  \end{subfigure}
  \hfill
  \begin{subfigure}[t]{0.45\textwidth}
    \centering
    \includegraphics[width=\textwidth]{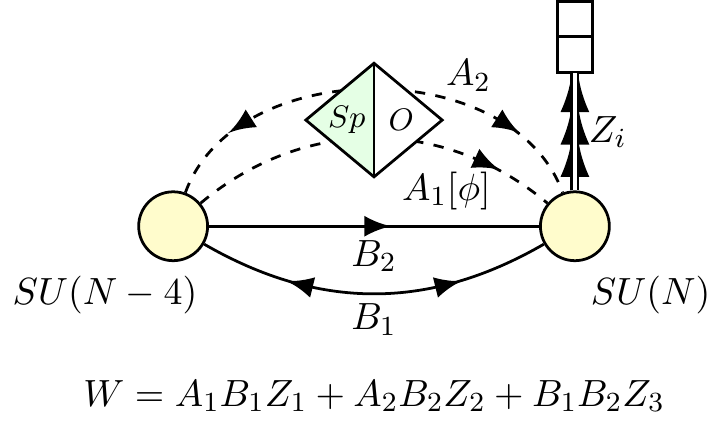}
    \caption{Quiver and superpotential.}
    \label{sfig:dP2-IIA-quiver}
  \end{subfigure}
  \caption{Brane tiling and field theory data for phase {\IIA} of the
    complex cone over $dP_2$. We include a half-period translation on $T^2$ to make the tiling easier to read.}
  \label{fig:dP2-IIA}
\end{figure}

This phase is depicted in figure~\ref{fig:dP2-IIA}. The resulting
charge table is
\begin{equation} \label{eqn:dP2-IIA-charges}
  \begin{array}{c|cc|ccccc}
    & \SU(N-4) & \SU(N) & \U(1)_{B+} & \U(1)_{B-} & \U(1)_{X+} & \U(1)_{X-} & \U(1)_R \\ \hline
    B_1 & \ov\fund & \ov\fund & -\frac{1}{N} & -\frac{1}{N-4} &  0 & -\frac{2}{N-4} & \frac{2}{N} \\
    B_2 & \fund & \ov\fund & -\frac{1}{N} & \frac{1}{N-4} & 0 & \frac{2}{N-4} & \frac{2}{N} \\
    Z_1 & \singlet & \asymm & \frac{2}{N} & 0 & -1 & -1 & -\frac{4}{N}  \\ 
    Z_2 & \singlet & \asymm & \frac{2}{N} & 0 & -1 & 1 & -\frac{4}{N}  \\ 
    Z_3 & \singlet & \asymm & \frac{2}{N} & 0 & 0 & 0 & 2-\frac{4}{N}  \\ \hline
    \mathfrak{q}_{\Sp} & \ast & \ast & -\frac{3}{2} & 0 & \frac{N}{2} & 0 & \frac{N-4}{4} \\
    A_1[\phi] & \fund & \ov\fund & -\frac{1}{N} & \frac{1}{N-4} & 1 & 1+\frac{2}{N-4} & 2+\frac{2}{N} \\
    A_2 & \ov\fund & \ov\fund & -\frac{1}{N} & -\frac{1}{N-4} & 1 & -1-\frac{2}{N-4} & 2+\frac{2}{N} 
  \end{array}
\end{equation}
The anomalies match~(\ref{eqn:dP2anomalies}) for $Q=N-1/2$, where one can check that $Q = 2 Q_{\rm D3}$ as before.

The color and flavor parities $n = (-1)^N$ and $\phi = (-1)^F$ are straightforward to compute as before
\be
\begin{split}
N &\equiv f_3 \pmod 2 \,,\\
F & \equiv f_1 + f_2 + f_3 \pmod 2\,,
\end{split}
\ee
where we use $Y_{DE} = \vev{d} = \vev{e_1}+\vev{e_2}+\vev{e_3}$ to read off the flavor parity. We label this phase as $\IIA^{n;\phi}$ for future reference.

\subsubsection*{Phase \IIB}

\begin{figure}
  \centering
  \begin{subfigure}[t]{0.25\textwidth}
    \centering
    \includegraphics[height=3.5cm]{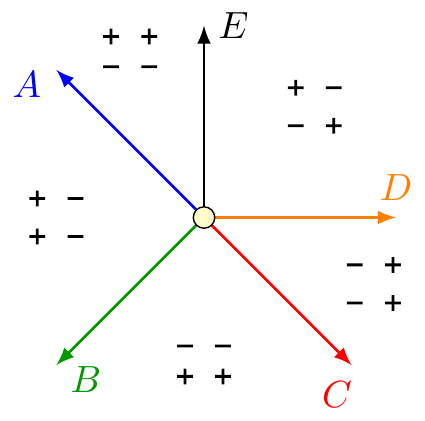}
    \caption{Local charges.}
    \label{sfig:dP2-IIB-charges}
  \end{subfigure}
  \hfill
  \begin{subfigure}[t]{0.25\textwidth}
    \centering
    \includegraphics[height=4cm]{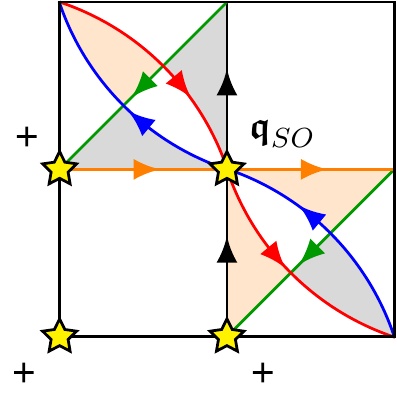}
    \caption{Brane tiling.}
    \label{sfig:dP2-IIB-tiling}
  \end{subfigure}
  \hfill
  \begin{subfigure}[t]{0.45\textwidth}
    \centering
    \includegraphics[width=\textwidth]{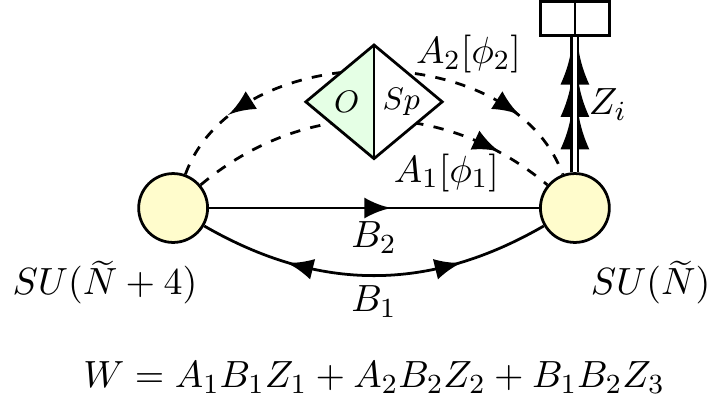}
    \caption{Quiver and superpotential.}
    \label{sfig:dP2-IIB-quiver}
  \end{subfigure}
  \caption{Brane tiling and field theory data for phase {\IIB} of the
    complex cone over $dP_2$.}
  \label{fig:dP2-IIB}
\end{figure}

As with phase \IB, many of the properties of phase \IIB, shown in figure~\ref{fig:dP2-IIB}, can be derived from those of \IIA using negative rank duality. 
We obtain the charge table
\begin{equation} \label{eqn:dP2-IIB-charges}
  \begin{array}{c|cc|ccccc}
    & \SU(\tN+4) & \SU(\tN) & \U(1)_{B+} & \U(1)_{B-} & \U(1)_{X+} & \U(1)_{X-} & \U(1)_R \\ \hline
    B_1 & \ov\fund & \ov\fund & -\frac{1}{\tN} & -\frac{1}{\tN+4} &  0 & \frac{2}{\tN+4} & -\frac{2}{\tN} \\
    B_2 & \fund & \ov\fund & -\frac{1}{\tN} & \frac{1}{\tN+4} & 0 & -\frac{2}{\tN+4} & -\frac{2}{\tN} \\
    Z_1 & \singlet & \symm & \frac{2}{\tN} & 0 & -1 & -1 & \frac{4}{\tN}  \\ 
    Z_2 & \singlet & \symm & \frac{2}{\tN} & 0 & -1 & 1 & \frac{4}{\tN}  \\ 
    Z_3 & \singlet & \symm & \frac{2}{\tN} & 0 & 0 & 0 & 2+\frac{4}{\tN}  \\ \hline
    \mathfrak{q}_{\SO} & \ast & \ast & \frac{3}{2} & 0 & -\frac{\tN}{2} & 0 & -\frac{\tN+4}{4} \\
    A_1[\phi_1] & \fund & \ov\fund & -\frac{1}{\tN} & \frac{1}{\tN+4} & 1 & 1-\frac{2}{\tN+4} & 2-\frac{2}{\tN} \\
    A_2[\phi_2] & \ov\fund & \ov\fund & -\frac{1}{\tN} & -\frac{1}{\tN+4} & 1 & -1+\frac{2}{\tN+4} & 2-\frac{2}{\tN}
  \end{array}
\end{equation}
where the 't Hooft anomalies match with~(\ref{eqn:dP2anomalies}) for $Q = \tN + 1/2$.

\begin{figure}
  \centering
  \begin{subfigure}[b]{0.3\textwidth}
    \centering
    \includegraphics[width=\textwidth]{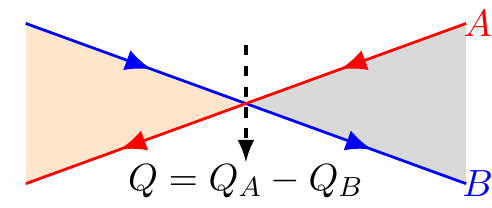}
    \vspace*{1cm}
    \caption{NS5 brane $\U(1)$ convention}
   \label{sfig:NS5U1convention}
  \end{subfigure}
  \hfill
  \begin{subfigure}[b]{0.3\textwidth}
    \centering
    \includegraphics[width=\textwidth]{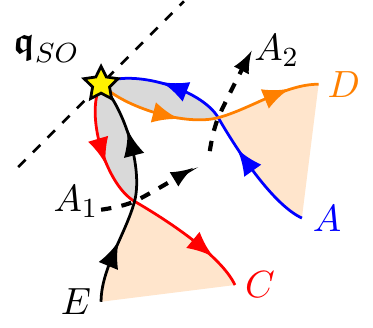}
     \vspace*{0cm}
    \caption{Making the mesons visible}
   \label{sfig:meson-charges}
  \end{subfigure}
  \hfill
  \begin{subfigure}[b]{0.3\textwidth}
    \centering
    \includegraphics[width=\textwidth]{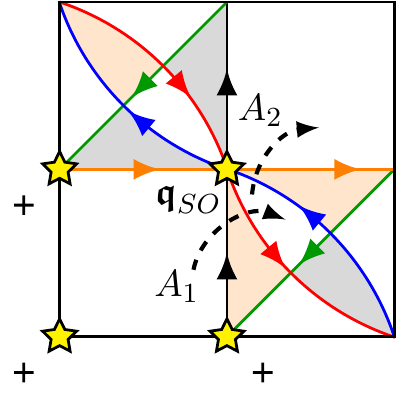}
    \caption{Mesons in the tiling}
   \label{sfig:dP2-IIB-mesons}
  \end{subfigure}
  \caption{\subref{sfig:NS5U1convention}~The $\U(1)$ global symmetries of the quiver theory originate from the $\U(1)$ worldvolume gauge theories on the NS5 branes, where the bifundamental fields carry opposite charges under the two intersecting NS5 branes where they are localized. We choose the sign convention for the NS5 brane $\U(1)$ charges shown here. \subref{sfig:meson-charges}~To read off the charges of $\TO_k$ mesons, it is helpful to recross the NS5 branes to make the mesons fundamental fields. In the example shown here, we consider the charges of the phase \IIB\ \quadSO\ mesons under $\U(1)_{X-}=Q_A+Q_B+Q_C$. The meson with charge $+1$ ($-1$) is labelled $A_1$ ($A_2$), following the conventions of the charge table~(\ref{eqn:dP2-IIB-charges}).~\subref{sfig:dP2-IIB-mesons} These same mesons, illustrated in the full tiling.}
    \end{figure}

However, reading off the flavor parities $\phi_i = (-1)^{F_i}$ in this case is somewhat subtle, because the meson parities $\phi_{1,2}$ associated to $A_1$ and $A_2$ differ by $\tilde{n} = (-1)^{\tilde{N}}$. To do so properly, we use the prescription shown in figure~\ref{subfig:MesonParity}. First, however, we need to distinguish the mesons $A_1$ and $A_2$ in the tiling. To do so, we note that $A_1$ carries charge $1 + O(1/\tN)$ under $\U(1)_{X-}$, whereas $A_2$ carries charge $-1 + O(1/\tN)$. Note that the $\U(1)$ global symmetries are associated to the $\U(1)$ world-volume gauge theories on the NS5 branes (up to contributions from the anomalous $\U(1) \subseteq \U(N)$ of each D5 brane face, which enter the charges at $O(1/N)$), see, e.g.,~\cite{Hanany:2005ss,Kennaway:2007tq}. In particular, $Q_{X-} = Q_A + Q_B + Q_C + O(1/N)$, where the charges $Q_{A,B,C}$ are defined by the conventions in figure~\ref{sfig:NS5U1convention}, and the correct linear combination can be found by cross referencing the charge tables~(\ref{eqn:dP2-IA-charges}) and (\ref{eqn:dP2-IIA-charges}) with their respective tilings. 
Based on this, we identify $A_{1,2}$ in the tiling, as shown in figure~\ref{sfig:dP2-IIB-mesons}.

We read off
\be
\begin{split}
\tN&\equiv f_1 + f_2 \pmod 2 \,,\\
F_i &\equiv f_i  \pmod 2\,,
\end{split}
\ee
where we use $Y_{CD} = \vev{c} = \vev{e_1}$ and $Y_{EA} = \vev{e} = \vev{e_2}$ to read off the meson parities associated to $A_1$ and $A_2$, respectively. As expected, $\tilde{n} = \phi_1 \phi_2$. We denote this phase $\IIB^{\phi_1 \phi_2}$ for future reference, where the color parity is fixed implicitly by $\tN \equiv F_1 + F_2 \pmod 2$.

\subsubsection*{Phase \IIIA}

\begin{figure}
  \centering
  \begin{subfigure}[b]{0.25\textwidth}
    \centering
    \includegraphics[height=3.5cm]{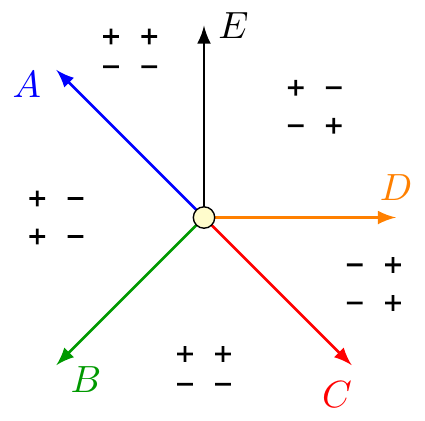}
    \caption{Local charges.}
    \label{sfig:dP2-IIIA-charges}
  \end{subfigure}
  \hfill
  \begin{subfigure}[b]{0.25\textwidth}
    \centering
    \includegraphics[height=3.8cm]{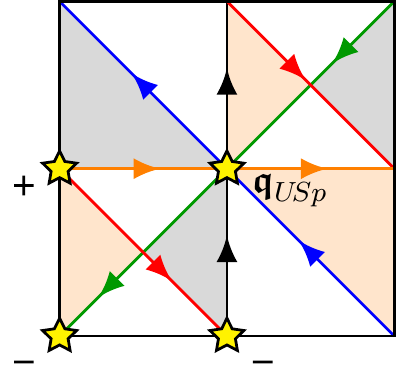}
    \caption{Brane tiling.}
    \label{sfig:dP2-IIIA-tiling}
  \end{subfigure}
  \hfill
  \begin{subfigure}[b]{0.4\textwidth}
    \centering
    \includegraphics[width=\textwidth]{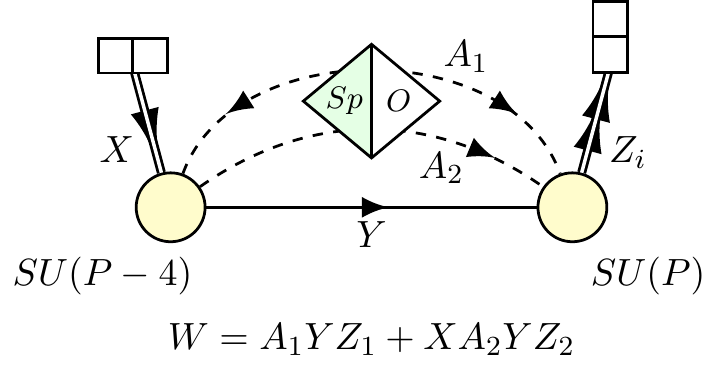}
    \caption{Quiver and superpotential.}
    \label{sfig:dP2-IIIA-quiver}
  \end{subfigure}
  \caption{Brane tiling and field theory data for phase {\IIIA} of the
    complex cone over $dP_2$.}
  \label{fig:dP2-IIIA}
\end{figure}

The local charges, brane tiling, quiver diagram and superpotential for this phase are shown in figure~\ref{fig:dP2-IIIA}.
The corresponding charge table is
{\small
\begin{equation}
\begin{array}{c|cc|ccccc}
& \SU(P-4) & \SU(P) & \U(1)_{B+} & \U(1)_{B-} & \U(1)_{X+} & \U(1)_{X-} & \U(1)_R \\ \hline
X & \ov\symm & \singlet & \frac{3}{P-4} & \frac{1}{P-4} & -1-\frac{3}{P-4} & 1+\frac{1}{P-4} & 0 \\
Y & \fund & \ov\fund & -\frac{3}{2(P-4)}+\frac{1}{2P} & -\frac{1}{2(P-4)}-\frac{1}{2P} & \frac{3}{2(P-4)}+\frac{1}{2P}  & -\frac{1}{2(P-4)}+\frac{1}{2P} & \frac{2}{P} \\
Z_1 & \singlet & \asymm & -\frac{1}{P} & \frac{1}{P} & 1-\frac{1}{P} & 1-\frac{1}{P}  & 2-\frac{4}{P}  \\  
Z_2 & \singlet & \asymm & -\frac{1}{P} & \frac{1}{P} & -\frac{1}{P} & -\frac{1}{P} & -\frac{4}{P}  \\ \hline
\quadSp & \ast & \ast & \frac{5}{4} & -\frac{1}{4} & -\frac{2P-3}{4} & -\frac{1}{4} & -\frac{P+4}{4} \\
A_1[\phi] & \ov\fund & \ov\fund & \frac{3}{2(P-4)}+\frac{1}{2P} & \frac{1}{2(P-4)}-\frac{1}{2P} & -1-\frac{3}{2(P-4)}+\frac{1}{2P} & -1+\frac{1}{2(P-4)}+\frac{1}{2P} & \frac{2}{P} \\
A_2 & \fund & \ov\fund & -\frac{3}{2(P-4)}+\frac{1}{2P} & -\frac{1}{2(P-4)}-\frac{1}{2P} & 1+\frac{3}{2(P-4)}+\frac{1}{2P} & -1-\frac{1}{2(P-4)}+\frac{1}{2P}  & 2+\frac{2}{P}
\end{array}
\end{equation}}
The anomalies match~(\ref{eqn:dP2anomalies}) for $Q = P-3/2$.

Reading off the color and flavor parities, $p = (-1)^P$ and $\phi = (-1)^F$, is by now a straightforward exercise. We find
\be
\begin{split}
P &\equiv f_2 + f_3 \pmod 2 \,, \\
F &\equiv f_2 \pmod 2 \,,
\end{split}
\ee
where we use $Y_{EA} = \vev{e} = \vev{e_2}$ to find the flavor parity. We denote this phase as $\IIIA^{p;\phi}$ for future reference.

The phase $\tIIIA^{p;\phi}$ is the same as the one described above, except that we replace $\U(1)_{X-} \to -\U(1)_{X-}$, $\U(1)_{B-} \to - \U(1)_{B-}$, and change the associated torsions by $\vev{e_1} \leftrightarrow \vev{e_2}$.

\subsubsection*{Phase \IIIB}

\begin{figure}
  \centering
  \begin{subfigure}[t]{0.25\textwidth}
    \centering
    \includegraphics[height=3.5cm]{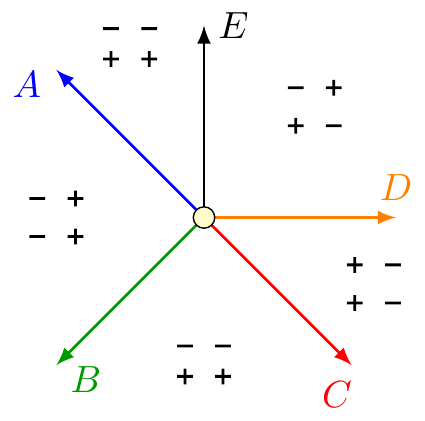}
    \caption{Local charges.}
    \label{sfig:dP2-IIIB-charges}
  \end{subfigure}
  \hfill
  \begin{subfigure}[t]{0.25\textwidth}
    \centering
    \includegraphics[height=3.5cm]{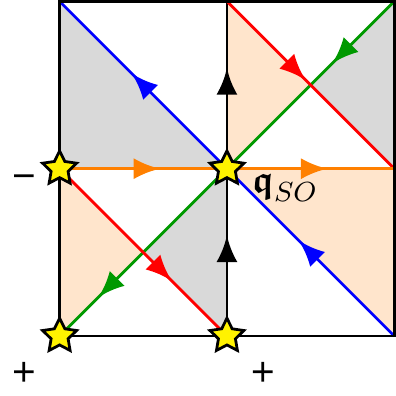}
    \caption{Brane tiling.}
    \label{sfig:dP2-IIIB-tiling}
  \end{subfigure}
  \hfill
  \begin{subfigure}[t]{0.4\textwidth}
    \centering
    \includegraphics[width=\textwidth]{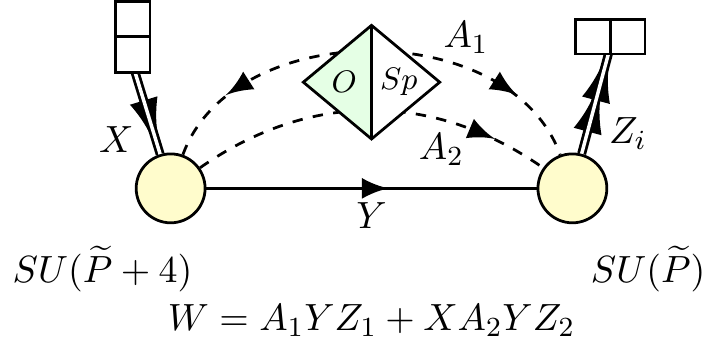}
    \caption{Quiver and superpotential.}
    \label{sfig:dP2-IIIB-quiver}
  \end{subfigure}
  \caption{Brane tiling and field theory data for phase {\IIIB} of the
    complex cone over $dP_2$.}
  \label{fig:dP2-IIIB}
\end{figure}

As before the local charges, tiling, quiver and superpotential (figure~\ref{fig:dP2-IIIB}) can be obtained by negative rank duality,
as can the charge table:
{\small
\begin{equation}
\begin{array}{c|cc|ccccc}
& \SU(\tP+4) & \SU(\tP) & \U(1)_{B+} & \U(1)_{B-} & \U(1)_{X+} & \U(1)_{X-} & \U(1)_R \\ \hline
X & \ov\asymm & \singlet & \frac{3}{\tP+4} & \frac{1}{\tP+4} & -1+\frac{3}{\tP+4} & 1-\frac{1}{\tP+4} & 0 \\
Y & \fund & \ov\fund & -\frac{3}{2(\tP+4)}+\frac{1}{2\tP} & -\frac{1}{2(\tP+4)}-\frac{1}{2\tP} & -\frac{3}{2(\tP+4)}-\frac{1}{2\tP}  & \frac{1}{2(\tP+4)}-\frac{1}{2\tP} & -\frac{2}{\tP} \\
Z_1 & \singlet & \symm & -\frac{1}{\tP} & \frac{1}{\tP} & 1+\frac{1}{\tP} & 1+\frac{1}{\tP}  & 2+\frac{4}{\tP}  \\  
Z_2 & \singlet & \symm & -\frac{1}{\tP} & \frac{1}{\tP} & \frac{1}{\tP} & \frac{1}{\tP} & \frac{4}{\tP}  \\ \hline
\quadSO & \ast & \ast & -\frac{5}{4} & \frac{1}{4} & \frac{2\tP+3}{4} & -\frac{1}{4} & \frac{\tP-4}{4} \\
A_1[\phi_1] & \ov\fund & \ov\fund & \frac{3}{2(\tP+4)}+\frac{1}{2\tP} & \frac{1}{2(\tP+4)}-\frac{1}{2\tP} & -1+\frac{3}{2(\tP+4)}-\frac{1}{2\tP} & -1-\frac{1}{2(\tP+4)}-\frac{1}{2\tP} & -\frac{2}{\tP} \\
A_2[\phi_2] & \fund & \ov\fund & -\frac{3}{2(\tP+4)}+\frac{1}{2\tP} & -\frac{1}{2(\tP+4)}-\frac{1}{2\tP} & 1-\frac{3}{2(\tP+4)}-\frac{1}{2\tP} & -1+\frac{1}{2(\tP+4)}-\frac{1}{2\tP}  & 2-\frac{2}{\tP}
\end{array}
\end{equation}}
where the anomalies match~(\ref{eqn:dP2anomalies}) for $Q = \tP+3/2$.

As in phase \IIB, the meson parities $\phi_i = (-1)^{F_i}$ associated to the two \quadSO\ mesons $A_1$ and $A_2$ differ by the color parity $\tilde{p} = (-1)^{\tP}$. However, as $A_1$ and $A_2$ appear in cubic and quartic superpotential terms, respectively, in this case there is no ambiguity in locating $A_1$ and $A_2$ in the tiling. Using the prescription shown in figure~\ref{subfig:MesonParity}, we find the color and flavor parities
\be
\begin{split}
\tP&\equiv f_1 \pmod 2 \,, \\
F_1 &\equiv f_2 + f_3 \pmod 2 \,,\\
F_2 &\equiv f_1 + f_2 + f_3 \pmod 2 \,,
\end{split}
\ee
where we use $Y_{AB} = \vev{a} = \vev{e_2} + \vev{e_3}$ and $Y_{DE} = \vev{d} = \vev{e_1} + \vev{e_2} + \vev{e_3}$ to read off the meson parities $F_1$ and $F_2$ associated to $A_1$ and $A_2$, respectively. As expected, $\phi_1 \phi_2 = \tilde{p}$. We label this phase as $\IIIB^{\tilde{p};\phi_1}$ for future convenience. The phase $\tIIIB^{\tilde{p};\phi_1}$ is again related by the replacements $\U(1)_{X-} \to -\U(1)_{X-}$, $\U(1)_{B-} \to - \U(1)_{B-}$ as well transforming the associated torsions by $\vev{e_1} \leftrightarrow \vev{e_2}$.

\subsubsection*{Duality predictions}

To classify the S-dualities expected to relate the $dP_2$ orientifold SCFTs, we begin by summarizing the relation between the phase labels and the discrete torsion, determined above:
\be
\begin{aligned}
\IA^{n;\phi_1 \phi_2}&: (\phi_1,\phi_2,n \phi_1\phi_2;+,+,+)\,, & \IB^{\phi_1 \phi_2}&: (\pm\phi_1,\pm\phi_2,\pm;-,-,-)  \\
\IIA^{n;\phi}&: (\pm n\phi,\pm,n;-,-,+)\,, & \IIB^{\phi_1 \phi_2}&: (\phi_1,\phi_2,\pm;+,+,-)  \\
\IIIA^{n;\phi}&: (\pm,\phi,n \phi;-,+,+)\,, & \IIIB^{n;\phi}&: (n,\pm\phi,\pm;+,-,-)  \\
\tIIIA^{n;\phi}&: (\phi,\pm,n \phi;+,-,+)\,, & \tIIIB^{n;\phi}&: (\pm\phi,n,\pm;-,+,-)  
\end{aligned}
\ee
where we write the torsion as $((-1)^{f_1},(-1)^{f_2},(-1)^{f_3};(-1)^{h_1},(-1)^{h_2},(-1)^{h_3})$, and the $\pm$ signs for phases other than \IA\ indicate the equivalence under $[F] \to [F]+[H]$, $\tau \to \tau+1$.

\begin{figure}
\begin{center}
  \includegraphics[height=6cm]{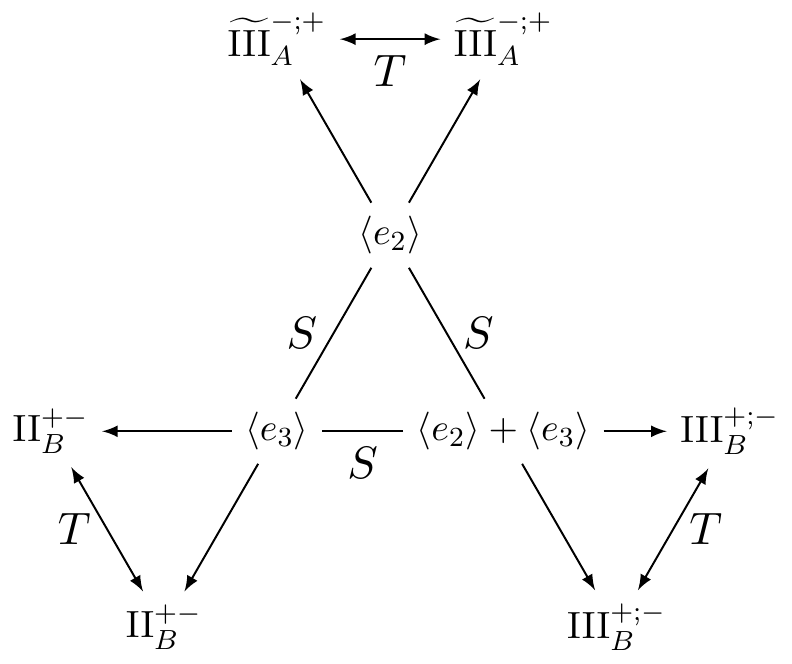}
  \caption{Mapping the triple $(\vev{e_2}, \vev{e_3}, \vev{e_2}+\vev{e_3})$ onto an $\SL(2,\bZ)$ sextet. The discrete torsion for the phase at each corner of the diagram is $([H],[F])$, read inwards from the corner along the diagonal line connected to it.}
  \label{fig:dP2sextet}
\end{center}
\end{figure}

To fill out the $\SL(2,\bZ)$ multiplets, we note that the generators $S: \tau \to -1/\tau$ and $T: \tau \to \tau+1$ act by permutations on the triple $([H],[F],[F]+[H])$: $S$ exchanges the first two entries, and $T$ exchanges the second two. For instance, consider the triple $(\vev{e_2}, \vev{e_3}, \vev{e_2}+\vev{e_3})$. Choosing all possible orderings we obtain the various phases filling out the sextet $(\IIB^{+-}, \tIIIA^{-;+}, \IIIB^{+;-})$ as shown in figure~\ref{fig:dP2sextet}.
We fill out all the $\SL(2,\bZ)$ multiplets in the same manner, where sextets, triplets, and singlets correspond to triples of the form $(\vev{A}, \vev{B}, \vev{C})$, $(\vev{A}, \vev{A}, 0)$, and $(0, 0, 0)$, respectively, for non-trivial torsion classes $\vev{A}, \vev{B}, \vev{C}$ satisfying $\vev{A}+\vev{B}+\vev{C}=0$. Our results are shown in table~\ref{table:dP2-orbits}.

\begin{table}
  \centering
   \begin{tabular}{c|p{0.75\textwidth}}
    Singlet & $\IA^{+;++}$ \\
    \hline
    Triplets & $(\IA^{-;--}, \IB^{++})$, $(\IA^{+;--}, \IIA^{+;+})$, $(\IA^{-;++}, \IIB^{++})$, \newline
   		 $(\IA^{+;+-}, \IIIB^{+;+}) \leftrightarrow (\IA^{+;-+}, \tIIIB^{+;+})$,
                 $(\IA^{-;+-}, \tIIIA^{+;+}) \leftrightarrow (\IA^{-;-+}, \IIIA^{+;+})$ \\
    \hline
    Sextets & $(\IIA^{+;-}, \IIIA^{-;-}, \tIIIA^{-;-})$, $(\IIA^{-;+}, \IIIB^{-;-}, \tIIIB^{-;-})$, $(\IIA^{-;-}, \IB^{--}, \IIB^{--})$, \newline
                $(\IIB^{+-}, \tIIIA^{-;+}, \IIIB^{+;-}) \leftrightarrow (\IIB^{-+}, \IIIA^{-;+}, \tIIIB^{+;-})$,\newline
                $(\IB^{+-}, \tIIIA^{+;-}, \tIIIB^{-;+}) \leftrightarrow (\IB^{-+}, \IIIA^{+;-}, \IIIB^{-;+})$
  \end{tabular}
  \caption{$SL(2,\bZ)$ multiplets for the $\mathcal{N}=1$ SCFTs on D3 branes probing the $dP_2$ toric orientifold considered in the text. In addition to the singlet, there are seven triplets and seven sextets, in agreement the counting given in~\cite{dP1paper}. When two multiplets are exchanged by the $\bZ_2$ automorphism of the toric diagram, we indicate this by $(\ldots) \leftrightarrow (\ldots)$. The other multiplets are mapped to themselves.}
  \label{table:dP2-orbits}
\end{table}

We check the predictions of table~\ref{table:dP2-orbits} in detail by computing the superconformal index for each phase, expanding to fixed order in the fugacity $t = \sqrt{p q}$ for a fixed rank (specified by $Q=2 Q_{\rm D3}$).\footnote{When two phases are related by the $\bZ_2$ automorphism of the toric diagram their indices will be related by $x_- \to x_-^{-1}$, $b_- \to b_-^{-1}$, hence it is not necessary to compute the index separately for each.} Our results, presented in appendix~\ref{app:dP2-SCIs}, show that theories in different multiplets have distinct indices, whereas theories in the same multiplet have the same index up to the order in $t$ that we were able to check. This is strong evidence that the superconformal indices of two $dP_2$ SCFTs match if and only if they lie in the same S-duality multiplet, a necessary condition for the conjectured S-dualities to exist, and a very non-trivial check that they do.

\section{Conclusions}
\label{sec:conclusions}

In this paper we have presented a prescription for relating the
discrete data in an orientifolded brane tiling to the discrete fluxes
present in the dual orientifolded singularity. For the NSNS torsion,
under the assumption that NS5 brane and O5 charges are determined by
NSNS data only, we have proven that our proposal is correct. We have
no proof that the proposal for the RR torsion is correct, but we have
verified its validity in a large number of examples. Furthermore, when
formulated in the toric language which is natural of the problem,
equations~\eqref{eqn:Htorsion} and \eqref{eqn:RRtorsionprescription}
are extremely simple, which gives us further confidence in their
validity. The counting arguments in \S\ref{sec:RR-torsion} further
support the validity of our proposal for the RR torsion.

Assuming that the dictionary in \eqref{eqn:Htorsion} and
\eqref{eqn:RRtorsionprescription} is indeed correct, we
have obtained a most remarkable result: we now have a systematic way
of producing an infinite number of $SL(2,\bZ)$ duality multiplets of
$\cN=1$ theories, completing the program initiated in
\cite{dualities1,dualities2}. More abstractly, in this paper we have
completely solved the problem of describing the effective theory at
cusps of the conformal manifold for a large class of $\cN=1$ SCFTs.

There are various directions for further research in immediately related areas. A more physical explanation for our conjectured RR torsion dictionary~\eqref{eqn:RRtorsionprescription} would be very useful, ideally leading to a full proof of the dictionary. Checking at least some examples involving $\TO_k$ CFTs beyond the $k=2$ case considered in this paper would also be prudent, for instance to understand the $dP_3$ orientifold discussed in~\cite{dP1paper}.
An important technical assumption in this paper is that of having an
isolated toric orientifold of a toric singularity. It would be
interesting to expand our analysis to remove the requirement for an isolated singularity; we expect that the same basic
picture will hold after slightly resolving the singularity. More non-trivial and physically interesting would be to
drop the requirement that there are no O7 planes extending away from the singularity. Doing so should naturally lead to S-dualities with flavor symmetries as D7 branes are needed to cancel the tadpole generated by the non-compact O7 planes, ensuring that $\tau_{\rm IIB}$ remains a modulus. In both cases, removing these requirements would allow us to catalog and understand a larger class of $\mathcal{N}=1$ S-dualities.

More broadly, it would be very interesting and informative to connect this class of theories to other systematic treatments of S-duality in the literature, such as \cite{Gaiotto:2009we,Gaiotto:2015usa}.

\acknowledgments

We would like to thank D.~Gaiotto and M.~Yamazaki for illuminating
discussions. BH was supported in part by the Fundamental Laws Initiative of the Harvard Center for the Fundamental Laws of Nature and in part by Perimeter Institute for Theoretical Physics. Research at Perimeter Institute is supported by the Government of Canada through Industry Canada and by the Province of Ontario through the Ministry of Economic Development and Innovation. I.G.-E. would
like to thank the Aspen Center for Physics, which is supported by
National Science Foundation grant PHY-1066293, for hospitality while
this work was being completed.

\appendix

\addtocontents{toc}{\protect\setcounter{tocdepth}{1}}

\section{Toric geometry} \label{app:toric}

In this appendix, we briefly review the basics of toric geometry, following the homogenous coordinate approach pioneered in~\cite{Cox:1993fz}. Some statements are presented without proof, and some definitions are omitted or abbreviated. More comprehensive treatments include~\cite{Cox:2003tn, Cox:2005tn, Closset:2009sv, Garcia:2010tn, Cox:2011tv}. 
\subsection{Toric varieties, homogenous coordinates, and fans}  \label{subapp:fans}

A toric variety is a $d$-dimensional abstract variety with a $(\bC^\star)^d$ dense subset whose natural $(\bC^\star)^d \cong (\U(1)_{\bC})^d$ group action extends to the whole space.\footnote{There are additional technical restrictions in the definition which we do not review here. In what follows, ``toric variety'' will refer to what is known as a ``normal toric variety'' in the math literature.} A $d$-dimensional toric variety is completely described by its fan, $\Sigma$, which is a collection of cones in $\bR^d$ subject to certain conditions. Specifically, the cones must be \emph{strongly convex rational polyhedral cones}, in that they take the form:
\be
\sigma = \Cone(\{\vec{u}_i\}) \equiv \left\{\left.\sum_{i=1}^n r_i \vec{u}_i \right\vert r_i \ge 0\right\}
\ee
for some generators $\vec{u}_{i=1, \ldots n} \in \bZ^d$ such that $\Cone(\{\vec{u}_i\}) \cap \Cone(\{-\vec{u}_i\}) = \{0\}$. We assume without loss of generality that for each $\vec{u}_i$
\be
\gcd(u^1_i,u^2_i,\ldots,u^d_i)=1\,, \qquad \vec{u}_i \notin \Cone(\{\vec{u}_{j \ne i}\})
\ee
since otherwise $\vec{u}_i$ can be rescaled or eliminated from the list of generators without changing $\sigma$. The set of generators $\{\vec{u}_i\}$ is then in one-to-one correspondence with $\sigma=\mathrm{Cone}(\{\vec{u}_i\})$. The cone is said to be \emph{simplicial} if $\{\vec{u}_i\}$ forms part of a basis of $\bR^d$, i.e.\ if the $\{\vec{u}_i\}$ are linearly independent. A simplicial cone is \emph{smooth} if $\{\vec{u}_i\}$ forms part of a basis of $\bZ^d$.

A subset $\{\vec{v}_j\} \subseteq \{\vec{u}_i\}$ generates a \emph{face} of the cone $\sigma$ if $\Cone(\{\vec{v}_j\}) = \sigma \cap H$ for some half-space $H\subset \bR^d$. The one-dimensional faces are \emph{rays} corresponding to each generator. Each cone is its own face, and the origin is a face of any cone. A cone is simplicial if and only if any subset of $\{\vec{u}_i\}$ generates a face of the cone.

By construction, each face of a cone is itself a cone. One can show that the intersection of two faces $\sigma$ is a face of $\sigma$, as is any face of a face of $\sigma$. This motivates a more general construct: an \emph{fan}, $\Sigma$, is a set of (strongly convex rational polyhedral) cones in $\bR^d$ such that for any $\sigma \in \Sigma$, all the faces of $\sigma$ are in $\Sigma$, and for any $\sigma_1, \sigma_2 \in \Sigma$, $\sigma_1 \cap \sigma_2$ is a face of both $\sigma_1$ and $\sigma_2$. The set of faces of a cone $\sigma$ (including $\sigma$ itself) forms a fan, denoted $\Fan(\sigma)$. A simplicial (smooth) fan is a fan whose cones are all simplicial (smooth). A \emph{complete} fan is a fan which covers $\bR^d$.

The $d$-dimensional toric variety corresponding to a fan $\Sigma$ can
be constructed as follows:\footnote{We assume for simplicity that the
  rays in the fan span $\bR^d$. Otherwise, the toric variety will be a
  direct product $X\times \bC^{\star}$ for some $(d-1)$-dimensional
  toric variety $X$.} to each ray in $\Sigma$, generated by
$\vec{u}_i$, we associate a homogenous coordinate $z_i \in \bC$, where
$i=1,\ldots n\ge d$ indexes the generators. We construct a subset
$V_\Sigma \subseteq \bC^n$ consisting of all points
$p=(z_1,\ldots,z_n)$ for which the vanishing coordinate components, if any, correspond to rays which all lie on the same cone in $\Sigma$. We define a group $G_\Sigma \subseteq (\bC^\star)^n$ which acts on $V_\Sigma$ via $z_i \to \mu_i z_i$, where
\be \label{eqn:Gdef}
G_\Sigma \equiv \left\{(\mu_1,\ldots,\mu_n) \,\left| \,\forall \vec{q} \in \bZ^d, \prod_i \mu_i^{\vec{q}\cdot \vec{u}_i} = 1 \right.\right\}
\ee
The toric variety $X_\Sigma$ is then:
\be \label{eqn:GCquot}
X_{\Sigma} = V_\Sigma \sslash G_\Sigma
\ee
where the quotient space consists of $G_\Sigma$ orbits which are
closed in $V_\Sigma$. If the fan is simplicial, then all $G_\Sigma$
orbits are closed, and the generalized quotient $V_\Sigma \sslash
G_\Sigma$ reduces to the usual quotient $V_\Sigma / G_\Sigma$. In
either case, there is a corresponding $\mathcal{N}=1$ gauged linear
sigma model (GLSM) with chiral fields $z_i$ and gauge group $G_\Sigma$
such that $X_{\Sigma}$ is the classical moduli space. In this
description, the Fayet-Iliopoulos (FI) terms corresponding to each
$\U(1) \subseteq G_\Sigma$ specify which higher dimensional cones are
in $\Sigma$, and thus determine $V_\Sigma$, whereas the generalized
quotient $\sslash$ is related to GLSM gauge transformations, as in~\cite{Luty:1995sd}.

Thus, the toric variety $X_\Sigma$ is determined uniquely by its fan $\Sigma$. Likewise, the fan $\Sigma$ is uniquely determined by the toric variety $X_\Sigma$, up to $\GL(d,\bZ)$ transformations on the fan, which preserve the $\bZ^d$ lattice. $X_\Sigma$ is compact if and only if $\Sigma$ is complete.

One can show that the closed $G_\Sigma$ orbits in $V_\Sigma$ are precisely those for which the vanishing coordinates correspond to rays which \emph{generate} a cone in $\Sigma$. Thus, each $k$-dimensional cone in the fan corresponds to the $(d-k)$-dimensional subvariety where the homogenous coordinates corresponding to the generators of the cone vanish. The subvariety is a singular locus if the corresponding cone is not smooth, and is an nonorbifold singularity if the cone is not simplicial. Thus, the singularities of $X_\Sigma$ are directly encoded in the fan $\Sigma$, and $X_\Sigma$ is smooth if and only if $\Sigma$ is smooth.

Just as the fan $\Sigma$ determines the global structure of the toric variety, the cones $\sigma \in \Sigma$ determine its local structure. An \emph{affine} toric variety is a toric variety whose fan is $\Fan(\sigma)$ for some cone $\sigma$. Each point in an arbitrary toric variety $p\in X_\Sigma$ has a neighborhood $N_p$ which is itself an affine toric variety. In particular, if $\sigma_p \in \Sigma$ of dimension $k$ corresponds to the vanishing homogenous coordinates at $p$, then $N_p \cong \bC^{d-k} \times X_{\Fan(\sigma_p)}$ where $X_{\Fan(\sigma_p)}$ is the $k$-dimensional affine toric variety whose fan is $\Fan(\sigma_p)$.

\subsection{Calabi-Yau toric varieties and toric diagrams} \label{subapp:CYtoric}

A toric variety $X_\Sigma$ is Calabi-Yau if and only if~\cite{Closset:2009sv}
\be \label{eqn:CYcondition}
\prod_i \mu_i = 1 \qquad \forall\; (\mu_1, \ldots, \mu_n) \in G_\Sigma\,.
\ee
This is ensured if the generators take the form $\vec{u}_i = (\vb{u}_i,1)$ after a $\GL(d,\bZ)$ transformation. In fact, this is a necessary and sufficient condition for $X_\Sigma$ to be Calabi-Yau. This implies that a toric Calabi-Yau manifold is always noncompact, since $\Sigma$ cannot be complete if the generators take this form. We now study the special properties of the fan of a toric Calabi-Yau manifold.

If the generators $\{\vec{u}_i\}$ of a cone $\sigma$ take the form $\vec{u}_i = (\vb{u}_i,1)$, then the cone admits an alternate description in terms of the \emph{lattice polytope}:
\be
P(\sigma) = \Conv(\{\vec{u}_i\}) \equiv \left\{\left.\sum_{i=1}^n r_i \vb{u}_i \right\vert r_i \ge 0, \sum_{i=1}^n r_i=1\right\}
\ee
where $P$, the \emph{convex hull} of $\{\vb{u}_i\}$, is the \emph{base} of the cone $\sigma$. A face of the polytope $P$ is the convex hull of a subset $\{\vb{v}_j\} \subseteq \{\vb{u}_i\}$ such that $\Conv(\{\vb{v}_j\}) = P \cap H$ for some half-space $H$ in $\bR^{d-1}$. As with cones, intersections of faces and faces of faces are again faces, and each face is itself a lattice polytope. The zero-dimensional faces are the \emph{vertices} $\vb{u}_i$. The \emph{cone over} $P$, $\sigma = \sigma(P)$, can be recovered by mapping the vertices to rays, $\vb{u}_i \to \vec{u}_i = (\vb{u}_i,1)$.  Each $k$ dimensional face of $P$ corresponds to an $k+1$ dimensional face of $\sigma(P)$

In complete analogy with the definition of a fan, we define a \emph{toric diagram}, $\Pi$, as a collection of lattice polytopes such that for any $P \in \Pi$, every face of $P$ is in $\Pi$, and for any $P_1, P_2 \in \Pi$, $P_1 \cap P_2$ is a face of both $P_1$ and $P_2$. The $(d-1)$-dimensional toric diagram $\Pi$ of a toric Calabi-Yau $d$-fold $X_\Sigma$ can be obtained from the fan $\Sigma$ by replacing each cone in $\Sigma$ by its base. Likewise the fan $\Sigma$ can be recovered from the toric diagram by taking the cone over each polytope in $\Pi$. The subgroup $\GL(d-1,\bZ) \ltimes \bZ^{d-1} \subset \GL(d,\bZ)$ of rotations and translations in $\bZ^{d-1}$ preserves the form $\vec{u}_i = (\vb{u}_i,1)$, and two toric diagrams represent the same Calabi-Yau variety if and only if they are related by a $\GL(d-1,\bZ) \ltimes \bZ^{d-1}$ transformation.

The vertices (zero-dimensional polytopes) of the toric diagram, $u_i \in \Pi$, correspond to a special class of divisors known as \emph{toric divisors}, of the form $z_i = 0$. Each such divisor is an $(d-1)$-dimensional subvariety whose fan can be read off from the toric diagram as follows. In the neighborhood of $\vb{u}_i$, the polytopes which contain $\vb{u}_i$ as a vertex appear locally as cones in $\bR^{d-1}$ with origin $\vb{u}_i$. These cones form an $(d-1)$-dimensional fan $\Sigma_i$, where $X_{\Sigma_i}$ is the divisor in question.

The holomorphic $d$-form for $X_\Sigma$ is given explicitly in terms of the homogenous coordinates as follows:\footnote{Since $X_\Sigma$ is noncompact, there may exist nontrivial holomorphic functions on $X_\Sigma$ with neither zeros nor poles, hence while this choice of holomorphic $d$-form is canonical, it is not in general unique.}
\be \label{eqn:Omega}
\Omega = \left(\prod_i z_i \right) \sum_{i_1, \ldots, i_d} u^1_{i_1} \ldots u^d_{i_d} \frac{\d z_{i_1}}{z_{i_1}} \wedge \ldots \wedge \frac{\d z_{i_d}}{z_{i_d}}\,,
\ee
where the factor in parentheses ensures that homogenous coordinates do not appear in the denominator, hence $\Omega$ is finite everywhere. We first prove that~(\ref{eqn:Omega}) is gauge invariant. The factor in parentheses is gauge invariant by~(\ref{eqn:CYcondition}). From~(\ref{eqn:Gdef}), we have:
\be
\prod_i (z_i)^{u^a_i} = \prod_i (z'_i)^{u^a_i}\,, \qquad a=1,\ldots,d\,,
\ee
whenever $z_i$ and $z_i'$ are gauge equivalent. Taking the exterior derivative of the log of each side of the equation, we obtain:
\be
\sum_i u^a_i \frac{\d z_i}{z_i} = \sum_i u^a_i \frac{\d z_i'}{z_i'} \,,
\ee
thus~(\ref{eqn:Omega}) is manifestly gauge invariant.

To show that $\Omega$ is nonvanishing at any smooth point in $X_\Sigma$, note that at any smooth point $p$ (or at an orbifold singularity) the vanishing homogenous coordinates generate a simplicial cone in $\Sigma$, hence the $k \le d$ generators $\{\vec{u}_i\}$ for these coordinates are linearly independent, and can be supplemented with $d-k$ other generators to form a basis, $B_p$. Since $z_i \ne 0$ for the remaining coordinates, we can gauge fix $z_i = 1$ for these coordinates, giving:
\be
\Omega = \det (\hat{u}) \d z^{\hat{1}} \wedge \ldots \wedge \d z^{\hat{d}}\,,
\ee
where $\hat{1},\ldots,\hat{d}$ index $B_p$ and $\hat{u}$ denotes the $d\times d$ matrix with components $u^a_i$ for $\vec{u}_i \in B_p$. Since $B_p$ is a basis, $\det (\hat{u}) \ne 0$, hence $\Omega$ is nonvanishing at $p$.

\subsection{Linear automorphisms of affine toric varieties} \label{subsec:automorph}

An automorphism of a toric variety $X_\Sigma$ is a biholomorphic map from $X_\Sigma$ to itself. The set of automorphisms of $X_\Sigma$ form a group $\Aut(X_\Sigma)$. For an affine toric variety (e.g.\ $\bC^d$), $\Aut(X_\Sigma)$ can be infinite dimensional, and is therefore difficult to construct. Moreover, in the case of Calabi-Yau toric varieties, most automorphisms will not preserve the Calabi-Yau metric, and are therefore not of direct physical interest. We instead consider a limited class of automorphisms of affine toric varieties.

Except in a few special cases, the origin, $z_i = 0$, is a special point in the geometry, and we only consider automorphisms which map the origin to itself. A simple subclass of automorphisms of this type consists of automorphisms which are \emph{linear} in the homogenous coordinates:
\be
z_i' = A_i^{\; j} z_j \,.
\ee
Linear automorphisms form a subgroup $\widehat{\Aut}(X_\Sigma) \subseteq \Aut(X_\Sigma)$. At least in the case of toric Calabi-Yau threefolds, $\widehat{\Aut}(X_\Sigma)$ will typically correspond to the complexified isometry group of the Calabi-Yau metric, and is therefore of direct physical interest.

To define a map from $X_\Sigma$ to itself, $A_i(z) = A_i^{\; j} z_j$ must preserve $G_\Sigma$ orbits, hence
\be
A^{-1} g A \in G_\Sigma \qquad \forall g \in G_\Sigma
\ee
and therefore $A \in N_{\GL(n,\bC)}(G_\Sigma)$ and $\widehat{\Aut}(X_\Sigma) \cong N_{\GL(n,\bC)}(G_\Sigma)/G_\Sigma$, where $N_G(H)$ is the normalizer of $H$ in $G$. We now construct $N_{\GL(n,\bC)}(G_\Sigma)$ and $\widehat{\Aut}(X_\Sigma)$.

The elements of $G_\Sigma$ are all diagonal, hence they share a common set of eigenvectors. For any $g \in G_\Sigma$, the $n$ eigenvectors of $g$ decompose into degenerate subspaces. Likewise $G_\Sigma$ induces a decomposition into degenerate subspaces of size $n_{\hat{i}}$ for $\hat{i} = 1,\ldots, \hat{n}$, where two eigenvectors are ``degenerate'' if their eigenvalues are equal for every element of $G_\Sigma$. Elements of $N_{\GL(n,\bC)}(G_\Sigma)$ must map eigenvectors to eigenvectors, hence an arbitrary element of $N_{\GL(n,\bC)}(G_\Sigma)$ can be decomposed as $A = P D$ where $P$ is a permutation matrix and $D \in \prod_{\hat{i}} \GL(n_{\hat{i}}, \bC)$ is a block-diagonal linear transformation within each degenerate subspace. Since $D$ normalizes $G_\Sigma$, $P$ must also do so. The group $G_\Sigma$ is uniquely encoded in the fan $\Sigma$, hence the permutations $P$ which normalize $G_\Sigma$ are precisely the automorphisms of the fan, $P \in \Aut(\Sigma)$. Multiplying by an element of $\Aut(\Sigma) \cap \prod_{\hat{i}} \GL(n_{\hat{i}}, \bC)$, we can restrict to $P\in \widetilde{\Aut}(\Sigma)$, permutations of the degenerate subspaces which preserve some canonical ordering of the elements within each subspace.

Thus, elements of $N_{\GL(n,\bC)}(G_\Sigma)$ can be decomposed as $A = P D$ with $P\in \widetilde{\Aut}(\Sigma)$ and $D \in \prod_{\hat{i}} \GL(n_{\hat{i}}, \bC)$. The decomposition is unique, hence $N_{\GL(n,\bC)}(G_\Sigma) \cong \prod_{\hat{i}} \GL(n_{\hat{i}}, \bC) \ltimes \widetilde{\Aut}(\Sigma)$. Quotienting by $G_\Sigma$, we find 
\be
\widehat{\Aut}(X_\Sigma) \cong \left\{\prod_{\hat{i}} \SL(n_{\hat{i}}, \bC)\times (\bC^\star)^{d-\sum_{\hat{i}} (n_{\hat{i}} -1)}\right\} \ltimes \widetilde{\Aut}(\Sigma)
\ee
corresponding to an isometry group $\left\{\prod_{\hat{i}} \SU(n_{\hat{i}})\times \U(1)^{d-\sum_{\hat{i}} (n_{\hat{i}} -1)}\right\} \ltimes \widetilde{\Aut}(\Sigma)$.

Elements $\cI \in \widehat{\Aut}(X_\Sigma)$ of finite order admit a simpler decomposition, up to conjugation $\cI\to A^{-1} \cI A$ for $A \in \widehat{\Aut}(X_\Sigma)$. Decomposing $\cI = P D$ as above, we write $P$ as a product of disjoint cycles permuting the degenerate subspaces. Suppose that $P = (12\ldots \hat{n})$. We fix each block $D_i = \id$ for $i=2,\ldots,\hat{n}$ using $A= D_i^{-1}$. Thus, $\cI^{\hat{n}}$ is block diagonal with $D_1$ on each block, and thus $D_1^p$ must be diagonal for some $p>0$, since $\cI$ has finite order. $D_1$ is therefore diagonalizable, and so $D$ can be diagonalized by conjugation. A similar argument applies when $P$ is any product of disjoint cycles, hence any element of finite order in $\widehat{\Aut}(X_\Sigma)$ is conjugate to an element $\hat{\cI}$ with the decomposition $\hat{\cI} = P T$ with $P \in \Aut(\Sigma)$ and $T \in (\bC^{\star})^n$.

\subsection{Affine varieties and horizons} \label{subsec:horizons}

In this subsection, we consider the geometry of an affine toric variety defined by a fan $\Sigma = \Fan(\sigma)$ in more detail. We work in the GLSM description, where the $z_i$ are treated as chiral superfields with charges $Q^i_a$, $a=1,\ldots,n-d$, under a gauge group $\U(1)^{n-d} \times \hat{G} \cong G_\Sigma \cap \U(n)$, where $Q^i_a$ forms a basis for the orthogonal complement of $\Span(\vec{u}_i)$ (so that $\sum_i Q^i_a \vec{u}_i = 0$) and $\hat{G}$ is a discrete group. The affine case corresponds to vanishing FI parameters, hence the D-term conditions are
\be
\sum_i Q^i_a |z_i|^2 = 0 \,.
\ee
By construction, $\vec{u}_i$ forms a basis for the orthogonal complement of $\Span(Q^i_a)$, so an arbitrary solution to the D-term conditions takes the form
\be
|z_i|^2 = \vec{u}_i\cdot \vec{r}
\ee
for some $\vec{r} \in \bR^d$. The positivity requirement $|z_i|^2 \ge 0$ is equivalent to $\vec{r} \in \sigma^\vee$, where the \emph{dual cone} $\sigma^\vee$ is defined by:
\be
\sigma^\vee \equiv \{\vec{r}\, \vert \,\forall \vec{u} \in \sigma,\, \vec{r}\cdot \vec{u} \ge 0 \}
\ee

If $\sigma$ is a strongly convex rational polyhedral cone of maximal dimension $d$, then so is $\sigma^\vee$. Moreover $(\sigma^\vee)^\vee = \sigma$, and there is a bijective map between the faces of $\sigma$ and $\sigma^\vee$, as follows (see e.g.~\cite{Cox:2011tv}). For any face $\tau \subset \sigma$, we define the dual face
\be
\tau^\ast \equiv \{\vec{r}\in \sigma^\vee \, \vert \, \forall \vec{u} \in \tau, \vec{r}\cdot\vec{u} = 0\} \,.
\ee
One can verify that $\tau^\ast$ is a face of $\sigma^\vee$ and that $(\tau^\ast)^\ast = \tau$, hence the map $\tau \to \tau^\ast$ is bijective. Moreover, if $\tau$ is dimension $k$, then $\tau^\ast$ is dimension $d-k$.

The dual cone provides a more intuitive picture of the affine toric variety than the original cone, as follows. A point in interior of the dual cone corresponds to a $T^d/G'$ subspace of the affine variety for some discrete group $G'$, as follows. By construction, specifying a point $\vec{r} \in \sigma^\vee$ fixes the modulus of the homogenous coordinates $z_i$. Choosing any subset $I\in \{1,\ldots,n\}$ of $d$ homogenous coordinates such that $\Span_\bR(\{\vec{u}_{i \in I}\}) \cong \bR^d$, we can gauge fix $\U(1)^{n-d}$ by setting $\arg(z_i) = 0$ for $i \notin I$. If $\Span_\bZ(\{\vec{u}_{i \in I}\}) \cong \bZ^d$, then this completely gauge fixes $G_\Sigma \cap \U(n)$, whereas otherwise there is some discrete remnant, $G'$, which acts on $T^d$ with coordinates $\arg z_i$, $i\in I$.

Thus, the toric variety is a $T^d/G'$ fibration over the dual cone. When $\vec{r}$ approaches a facet of $\sigma^\vee$, one cycle of $T^d$ shrinks to zero size, so that each point in the interior of the facet corresponds to an orbifold of $T^{d-1}$. Further one-cycles pinch off as $\vec{r}$ approaches faces of decreasing dimension. In particular, taking $\vec{r} \to 0$ shrinks $T^d/G'$ to a point, corresponding to the singular point at the origin of the affine variety.

This description is very convenient for describing the ``horizon'' of the affine variety, i.e.\ the $(2d-1)$-dimensional manifold $Y$ such that the affine variety is a real cone over $Y$. To recover $Y$ from the affine variety, we define a ``radial'' coordinate as the weighted average
\be
r \equiv \frac{1}{\sum_i \lambda_i} \sum_i \lambda_i |z_i|^2
\ee
for some choice of weights $\lambda_i > 0$. A surface of constant $r$ satisfies $\vec{u}_0 \cdot \vec{r} = r$, for
\be
\vec{u}_0 \equiv \frac{1}{\sum_i \lambda_i} \sum_i \lambda_i \vec{u}_i
\ee
where by construction $\vec{u}_0$ lies in the interior of $\sigma$. Consequently, $\vec{u}_0 \cdot \vec{r} > 0$ for all $\vec{r} \in \sigma^\vee$, and any point on the dual cone away from the origin $\vec{r}=0$ can be mapped to a point in the plane $\vec{u}_0 \cdot \vec{r} = 1$ by the rescaling $\vec{r} \to \frac{1}{\vec{u}_0\cdot\vec{r}}\, \vec{r}$. The intersection of $\sigma^\vee$ with this plane is a $(d-1)$-dimensional polytope whose faces are in one-to-one correspondence with the faces of $\sigma^\vee$, much like the relationship between a toric diagram and the corresponding fan. The horizon manifold $Y$ is a $T^d/G'$ fibration over this polytope.

Up to $\GL(d,\bR)$ transformations and rescaling of the generators, we can write $\vec{u}_i$ in the form $(\vb{u}_i,1)$, where $\Delta = \Conv(\vb{u}_i)$ is the ``base'' of the cone $\sigma$. By a further $\GL(d,\bR)$ transformation, we can set $\vec{u}_0 = (\vb{0},1)$.
In this case, the $(d-1)$-dimensional polytope defined by $r=1$ in the dual cone is the set $\{(\vb{r},1)\,\vert\, \vb{r} \in \Delta^\circ\}$, where the dual polytope $\Delta^\circ$ is defined as:
\be
\Delta^\circ \equiv \{\vb{r} \,\vert\, \forall \vb{u} \in \Delta,\, \vb{r}\cdot\vb{u} \ge -1\}
\ee
for full-dimensional $\Delta$ containing the origin in its interior. As above, $(\Delta^\circ)^\circ = \Delta$, where $\Delta^\circ$ is a full dimensional polytope containing the origin in its interior. Moreover, there is a bijective map between the proper faces of $\Delta$ and $\Delta^\circ$, taking $k$-dimensional faces to $(d-k-2)$-dimensional faces.

Thus, the horizon manifold is (an orbifold of) a torus fibration over the base of the dual cone, $\Delta^\circ$. The dimensions of this polytope vary depending on the choice of radial coordinate, i.e. the choice of origin within $\Delta$, but its topology is fixed by the bijective mapping between the faces of $\Delta$ and $\Delta^\circ$. In the Calabi-Yau case, $\Delta$ is the toric diagram, and $\Delta^\circ$ its dual. In particular, for $d=3$ the dual polytope is a polygon whose corners (edges) correspond to the edges (corners) of the toric diagram.

\section{Bilinear forms on \alt{$\bZ_2$}{Z2} vector spaces} \label{app:Z2bilinear}

In this appendix, we review a few properties of bilinear forms on a vector space $V$ over the field $\bZ_2$. A symmetric bilinear form on this space is a map $V\times V \to \bZ_2$ such that $A\cdot B = B\cdot A$. Equivalently, $A\cdot B = - B\cdot A$, hence there is no distinction between symmetric and antisymmetric bilinear forms.

The bilinear form is \emph{non-degenerate} if for every $A\in V$ there exists $B\in V$ such that $A\cdot B \ne 0$. There is no corresponding notion of positivity. Instead the norm $A \to A^2$ is a linear map $V \to \bZ_2$, because
\begin{equation}
(a A + B)^2 = a^2 A^2 +2 a A\cdot B+B^2 = a A^2 + B^2 \,,
\end{equation}
where we use the fact that $a^2 = a$ for $a \in \bZ_2$. We refer to
elements with norm 0 (1) as even (odd). If $A$ is odd then $B$ and
$B+A$ have opposite parity for any $B$. Thus, either all elements of
$V$ are even, or one-half are odd. We refer to non-degenerate inner products of the former type as ``symplectic'', and those of the latter as ``orthogonal''.

We now show that orthogonal (symplectic) inner product spaces admit a basis $\{e_i\}$ where $e_i \cdot e_j = \delta_{i j}$ ($e_i \cdot e_j = \Omega_{i j}$\footnote{Here $\Omega_{i j} = \diag\bigl(i \sigma_2 , i \sigma_2, \ldots \bigr) = -\Omega_{ji}$ is the standard symplectic form.}), justifying their names. We apply a modified Gramm-Schmidt procedure. Given a basis, we choose any odd element $e_1$ and replace $e_i \to e_i + (e_i \cdot e_1) e_1$ for $i>1$. If no odd element exists, we select another element of the basis $e_2$ such that $e_1 \cdot e_2 \ne 0$. We then replace $e_j \to e_j +(e_j\cdot e_1) e_2+(e_j\cdot e_2) e_1$ for $j>2$. In either case, the remaining elements of the basis are orthogonal to the chosen element(s), and we proceed recursively. The resulting basis may contain both even and odd components. Consider a subset such that $e_1$ is odd whereas $e_{2,3}$ are even with $e_2 \cdot e_3 \ne 0$, and replace $e_1 \to e_1 + e_2 + e_3$, $e_2 \to e_1 + e_2$, $e_3 \to e_1+ e_3$, so that $e_{1,2,3}$ are orthonormal. Applying this repeatedly, we obtain an orthonormal basis.

As a corollary, for any (not necessarily canonical) basis $\{e_i\}$ there exists a dual basis $\{e^j\}$ such that $e_i \cdot e^j = \delta_i^j$.\footnote{Let $\hat{e}_i$ be a canonical basis, with $\hat{e}^i = \sum_j (\hat{e}_i\cdot \hat{e}_j) \hat{e}_j$ its dual basis. If $\hat{e}^i = a^{i j} e_j$, then the dual basis is $e^j = \hat{e}_i a^{i j}$. To show this, we rewrite $e_i = b_{i j} \hat{e}^j$ and note that $b_{i j} a^{j k} = \delta_i^j$.} Thus, for any linear map $f: V \to \bZ_2$, there exists $\tilde{f} \in V$ such that $f(B) = B\cdot\tilde{f}$; explicitly, $\tilde{f} = f(e_i) e^i$. Similarly, $A\in V$ can be reconstructed from its components $A_i = e_i\cdot A$ via $A = A_i e^i$.

To illustrate these results, we note that there exists a unique
``norm'' element $\eta \in V$ such that $A^2 = \eta\cdot A$. For a
symplectic inner product, $\eta = 0$, whereas for an orthogonal one,
$\eta = \sum_i e_i$ in a canonical basis. It is possible to show that for any two
distinct non-zero elements $A,B \in V$ there is an automorphism
mapping $A$ to $B$ iff $A,B \ne \eta$ and $A\cdot \eta = B\cdot
\eta$.
Thus, $\eta$ measures the failure of the inner product space to be
isotropic. In general, $\eta$ is even (odd) for an even (odd)
dimensional space.

\section{Superconformal index checks for various examples} \label{app:SCI}

We use the same conventions for the fugacities $t$ and $J_n$ as in, e.g., \cite{dP1paper}. To save space, we display only a small number of terms in the index for a single choice of rank, sufficient to demonstrate that the indices match (do not match) for SCFTs in the same (different) $\SL(2,\bZ)$ multiplets. Computations to higher order in $t$ or for different ranks can be done using the computer program of~\cite{dP1paper}, and the indices of conjectural S-duals match in every example that we have checked.

\subsection{Complex cone over \alt{$\bF_0$}{F0} -- O7 plane}
\label{app:F0-SCIs}

We follow the notation and conventions described in \S\ref{sec:F0-O7}.
The shorthand $[\ldots]$ stands for the previous term(s) with the replacement $X \leftrightarrow Y$ and $b \to b^{-1}$.

\subsubsection*{For $N=7$:}

\noindent
$\II^- = \tII^- = \III^-$:
\begin{multline}
1-t^2\left(X_2+Y_2\right)\\
+t^3 \Bigl((1+t J_1)[(X_1+X_3+X_5)(Y_1+Y_3+Y_5)+(X_1+X_3)(Y_1+Y_3)] -J_1(1+X_2)(1+Y_2)\Bigr) \\
+t^4\left(-1+X_2Y_2+X_4Y_4 - X_1 Y_1 J_1-J_2\left(1+X_2\right)\left(1+Y_2\right)\right) \\
+t^{9/2}\bigg[\frac{X_5Y_4+X_3\left(1+3Y_2+3Y_4+2Y_6+Y_8\right)+X_1\left(2+3Y_2+4Y_4+2Y_6+Y_8\right)}{b}+[\ldots]\biggr] +\ldots
\end{multline}

\subsection*{For $N=8$:}

\noindent
$\I^{++}$:
\begin{multline}
1+t\left(\frac{Y_5}{b^{1/2}}+b^{1/2}X_5\right)\\
+t^2\left(\frac{Y_2+Y_6+Y_{10}}{b}+\frac{J_1Y_5}{b^{1/2}}+[\ldots]-X_2-Y_2+X_1Y_1+X_3Y_3+X_5Y_5\right)+\ldots
\end{multline}

\noindent
$\I^{-+} = \II^+$:
\begin{multline}
1+\frac{tY_5}{b^{1/2}}-t^2\left(\frac{-Y_2-Y_6-Y_{10}}{b}-\frac{J_1Y_5}{b^{1/2}}+\left(X_2+Y_2\right)\right)\\
+t^3\Biggl(\frac{Y_3+Y_5+Y_7+Y_9+Y_{11}+Y_{15}}{b^{3/2}}+\frac{J_1\left(1+Y_2+Y_4+Y_6+Y_8+Y_{10}\right)}{b} \\
+\frac{\left(-1+J_2\right)Y_5-X_2\left(Y_3+Y_5\right)-Y_7}{b^{1/2}} \\
+b^{1/2}\left(X_4Y_1+X_6Y_3+X_2\left(Y_1+Y_3\right)\right) 
-J_1\left(1+X_2\right)\left(1+Y_2\right)\Biggr)+\ldots
\end{multline}

\noindent
$\I^{+-} = \tII^+$: same as above with $X \leftrightarrow Y$, $b \to b^{-1}$. \\

\noindent
$\I^{--} = \III^+$:
\begin{multline}
1+t^2\left(1+ X_2 Y_2-(1+t J_1 + t^2 J_2)(X_2+Y_2-X_4Y_4)\right)\\
+t^4\Biggl(\frac{1+Y_4+X_2\left(Y_2+Y_6\right)}{b} + [\ldots] +X_8Y_8 +X_8Y_4 +X_4 Y_8+2X_6Y_6+X_6Y_2+X_2 Y_6  \\
+ 4 X_4 Y_4+5 X_2 Y_2+X_8+Y_8+2 X_4+2Y_4-X_2-Y_2+3 \Biggr)+\ldots
\end{multline}

\subsection{Complex cone over \alt{$\bF_0$}{F0} -- O3 planes}
\label{app:F0-O3-SCIs}

We follow the notation and conventions described in \S\ref{sec:F0-O3}. We also use the $O(2)$ characters $x_n \equiv x^n + x^{-n}$ and $y_n \equiv y^n + y^{-n}$; the shorthand $[\ldots]$ stands for the previous term(s) with the replacement $x \leftrightarrow y$ and $b \to b^{-1}$.

\subsubsection*{For $N=3$:}

\noindent
$\II^- = \tII^- = \III^-$:
\begin{multline}
1+t^{3/2}\left(\frac{y_1+y_3}{b}+b\left(x_1+x_3\right)\right)+t^2\left(x_2+y_2-2\right)+t^{5/2}\left(\frac{J_1y_3}{b}+bJ_1x_3\right) \\
+t^3\left(\frac{2+2y_2+y_4+y_6}{b^2} + [\ldots] -J_1(x_2-2)(y_2-2)+x_1 y_1+(x_1+x_3)(y_1+y_3)\right) \\
+t^{7/2}\left(\frac{-2y_1+\left(x_2+J_2-2\right)y_3+y_5}{b}+[\ldots]\right)+\ldots
\end{multline}

\subsubsection*{For $N=4$:}

\noindent
$\I^{++}$:
\begin{multline}
1+t\left(\frac{y_3}{b^{1/2}}+b^{1/2}x_3\right)\\
+t^2\left(\frac{1+2y_2+y_6}{b}-\frac{J_1\left(y_1-y_3\right)}{b^{1/2}}+[\ldots]+\left(-2+x_2+x_1y_1+y_2+x_3y_3\right)\right)+\ldots
\end{multline}

\noindent
$\I^{+-} = \II^+$:
\begin{multline}
1+\frac{ty_3}{b^{1/2}}+t^2\left(\frac{1+2y_2+y_6}{b}-\frac{J_1\left(y_1-y_3\right)}{b^{1/2}}+\left(-2+x_2+y_2\right)+b\left(1+x_2+x_4\right)\right)\\
-t^3\left(\frac{-2y_1-y_3-2y_5-y_9}{b^{3/2}}-\frac{J_1\left(y_2-y_4+y_6\right)}{b}+\frac{\left(-1+J_2\right)y_1-\left(-2+J_2+x_2\right)y_3-y_5}{b^{1/2}}\right.\\
\left.\vphantom{\frac{-2y_1-y_3-2y_5-y_9}{b^{3/2}}}+J_1\left(-2+x_2\right)\left(-2+y_2\right)+b^{1/2}\left(-\left(1+x_2\right)y_1-\left(1+x_4\right)y_3\right)-bJ_1x_4\right)+\ldots
\end{multline}

\noindent
$\I^{-+} = \tII^+$: same as above with $x \leftrightarrow y$, $b \to b^{-1}$. \\

\noindent
$\I^{--} = \III^+$: 
\begin{multline}
1+t^2\left(\frac{1+y_2+y_4}{b}+\left(y_2+x_2\left(1+y_2\right)\right)+b\left(1+x_2+x_4\right)\right)+t^3\left(\frac{J_1y_4}{b}-2J_1+bJ_1x_4\right)\\
+t^4\left(\frac{3+2y_2+2y_4+y_6+y_8}{b^2}+\frac{2-y_2+J_2y_4+y_6+x_2\left(1+2y_2+y_4+y_6\right)}{b} + [\ldots] \right.\\
\left.\vphantom{\frac{3+2y_2+2y_4+y_6+y_8}{b^2}}+7+2x_4+2y_4+3 x_4 y_4+x_4 y_2+x_2 y_4+4 x_2 y_2\right)+\ldots
\end{multline}

\subsection{Complex cone over \alt{$dP_2$}{dP2}}
\label{app:dP2-SCIs}

We follow the notation and conventions described in
\S\ref{sec:dP2}. In the following $\beta = 2 a_X - 3$.

\subsubsection*{For $Q=15/2$:}

\noindent
$\mathrm{I}_A^{+;++}$:
\begin{multline}
1+\frac{b_+^{3/2}}{x_+^{7/2}} \left(\frac{1}{b_-^{1/2}x_-^{9/2}}+b_-^{1/2}x_-^{9/2}\right)t_{\text{(0.729)}}^{\frac{93}{8}+\frac{405}{16\beta}+\frac{17\beta}{16}}
+\frac{1}{b_+^{1/2}x_+^{1/2}} \left(\frac{x_-^{1/2}}{b_-^{1/2}}+\frac{b_-^{1/2}}{x_-^{1/2}}\right)t_{\text{(0.8392)}}^{-\frac{67}{8}-\frac{135}{16\beta}-\frac{19\beta}{16}} \\
+\frac{x_+^8}{b_+^2} t_{\text{(0.8637)}}^{-\frac{5}{2}-\frac{135}{4\beta}+\frac{\beta}{4}}
+\frac{x_+^{1/2}}{b_+^{1/2}} \left(\frac{1}{b_-^{1/2}x_-^{1/2}}+b_-^{1/2}x_-^{1/2}\right)t_{\text{(0.9899)}}^{-\frac{39}{8}-\frac{135}{16\beta}-\frac{11\beta}{16}}+\ldots
\end{multline}

\noindent
$\mathrm{I}_A^{+;-+}=\tIIIB^{+;+}$:
\begin{multline}
1+\frac{b_+^{3/2}t_{\text{(0.729)}}^{\frac{93}{8}+\frac{405}{16\beta}+\frac{17\beta}{16}}}{b_-^{1/2}x_-^{9/2}x_+^{7/2}}+\frac{x_-^{1/2}t_{\text{(0.8392)}}^{-\frac{67}{8}-\frac{135}{16\beta}-\frac{19\beta}{16}}}{b_-^{1/2}b_+^{1/2}x_+^{1/2}}+\frac{x_+^8t_{\text{(0.8637)}}^{-\frac{5}{2}-\frac{135}{4\beta}+\frac{\beta}{4}}}{b_+^2}+\frac{x_+^{1/2}t_{\text{(0.9899)}}^{-\frac{39}{8}-\frac{135}{16\beta}-\frac{11\beta}{16}}}{b_-^{1/2}b_+^{1/2}x_-^{1/2}} \\
+\frac{x_+^{3/2}t_{\text{(1.141)}}^{-\frac{11}{8}-\frac{135}{16\beta}-\frac{3\beta}{16}}}{b_-^{1/2}b_+^{1/2}x_-^{3/2}}+\frac{x_+^{5/2}t_{\text{(1.291)}}^{\frac{17}{8}-\frac{135}{16\beta}+\frac{5\beta}{16}}}{b_-^{1/2}b_+^{1/2}x_-^{5/2}}+\frac{x_+^{7/2}t_{\text{(1.442)}}^{\frac{45}{8}-\frac{135}{16\beta}+\frac{13\beta}{16}}}{b_-^{1/2}b_+^{1/2}x_-^{7/2}}+\frac{b_+^3t_{\text{(1.458)}}^{\frac{93}{4}+\frac{405}{8\beta}+\frac{17\beta}{8}}}{b_-x_-^9x_+^7}+\ldots
\end{multline}

\noindent
$\mathrm{I}_A^{+;+-}=\IIIB^{+;+}$: same as above with $b_- \to b_-^{-1}$, $x_- \to x_-^{-1}$. \\

\noindent
$\mathrm{I}_A^{+;--}=\mathrm{II}_A^{+;+}$:
\begin{multline}
1+\frac{x_+^8}{b_+^2} (1+t J_1)\, t_{\text{(0.8637)}}^{-\frac{5}{2}-\frac{135}{4\beta}+\frac{\beta}{4}}+\frac{b_+}{x_+^4} \left(\frac{1}{x_-^4}+\frac{1}{x_-^2}+1+x_-^2+x_-^4 \right)t_{\text{(1.568)}}^{\frac{13}{4}+\frac{135}{8\beta}-\frac{\beta}{8}}\\
+\frac{b_+}{x_+^3} \left(\frac{1}{x_-^3}+\frac{1}{x_-}+x_-+x_-^3\right)t_{\text{(1.719)}}^{\frac{27}{4}+\frac{135}{8\beta}+\frac{3\beta}{8}}+\frac{x_+^{16}}{b_+^4} t_{\text{(1.727)}}^{-5-\frac{135}{2\beta}+\frac{\beta}{2}}+\frac{1}{b_+} t_{\text{(1.829)}}^{-\frac{53}{4}-\frac{135}{8\beta}-\frac{15\beta}{8}} \\
+\frac{b_+}{x_+^2} \left(\frac{1}{x_-^2}+1+x_-^2\right)t_{\text{(1.87)}}^{\frac{41}{4}+\frac{135}{8\beta}+\frac{7\beta}{8}}+\frac{x_+}{b_+} \left(\frac{1}{x_-}+x_-\right)t_{\text{(1.98)}}^{-\frac{39}{4}-\frac{135}{8\beta}-\frac{11\beta}{8}}-3t^2+\ldots
\end{multline}

\noindent
$\mathrm{II}_A^{+;-} = \tIIIA^{-;-} = \IIIA^{-;-}$:
\begin{multline}
1+\frac{b_+}{x_+^4} (1+J_1 t) \left(\frac{1}{x_-^4}+\frac{1}{x_-^2}+1+x_-^2+x_-^4\right)t_{\text{(1.568)}}^{\frac{13}{4}+\frac{135}{8\beta}-\frac{\beta}{8}} \\
+\frac{b_+}{x_+^3} (1+J_1 t) \left(\frac{1}{x_-^3}+\frac{1}{x_-}+x_-+x_-^3\right)t_{\text{(1.719)}}^{\frac{27}{4}+\frac{135}{8\beta}+\frac{3\beta}{8}} 
+\frac{b_+}{x_+^2} (1+J_1 t) \left(\frac{1}{x_-^2}+1+x_-^2\right)t_{\text{(1.87)}}^{\frac{41}{4}+\frac{135}{8\beta}+\frac{7\beta}{8}} \\
-3t^2+\frac{b_+}{x_+} \left(\frac{1}{x_-}+x_-\right)t_{\text{(2.02)}}^{\frac{55}{4}+\frac{135}{8\beta}+\frac{11\beta}{8}}-x_+ \left(\frac{1}{x_-}+x_-\right)t_{\text{(2.151)}}^{\frac{11}{2}+\frac{\beta}{2}} 
+b_+t_{\text{(2.171)}}^{\frac{69}{4}+\frac{135}{8\beta}+\frac{15\beta}{8}}\\
-\frac{J_1}{x_+} \left(\frac{1}{x_-}+x_-\right)t_{\text{(2.849)}}^{-\frac{1}{2}-\frac{\beta}{2}}
-J_1 \left(\frac{1}{x_-^2}+5+x_-^2\right) t^3 +\ldots
\end{multline}

\noindent
$\mathrm{II}_B^{-;-} = \tIIIA^{-;+} = \IIIB^{+;-}$:
\begin{multline}
1+\frac{b_-^{1/2}b_+^{3/2}x_-^{9/2} }{x_+^{7/2}} \biggl[1+J_1 t+\Bigl(J_2 - 3 - \frac{1}{x_-^2}\Bigr) t^2\biggr]\, t_{\text{(0.729)}}^{\frac{93}{8}+\frac{405}{16\beta}+\frac{17\beta}{16}}+\frac{b_-b_+^3x_-^9}{x_+^7} (1+J_1 t)\,t_{\text{(1.458)}}^{\frac{93}{4}+\frac{405}{8\beta}+\frac{17\beta}{8}} \\
-3t^2 
-x_+ \left(\frac{1}{x_-}+x_-\right)t_{\text{(2.151)}}^{\frac{11}{2}+\frac{\beta}{2}} 
+\frac{b_-^{3/2}b_+^{9/2}x_-^{27/2}t_{\text{(2.187)}}^{\frac{279}{8}+\frac{1215}{16\beta}+\frac{51\beta}{16}}}{x_+^{21/2}} \\
+\frac{b_+^{1/2}}{x_+^{7/2} b_-^{1/2}} \left(\frac{1}{x_-^{7/2}}+\frac{1}{x_-^{3/2}}+x_-^{1/2}+x_-^{5/2}\right) t_{\text{(2.558)}}^{-\frac{13}{8}+\frac{135}{16\beta}-\frac{13\beta}{16}} 
-\frac{b_-^{1/2}b_+^{3/2}x_-^{7/2}t_{\text{(2.578)}}^{\frac{81}{8}+\frac{405}{16\beta}+\frac{9\beta}{16}}}{x_+^{9/2}} \\
+\frac{b_+^{1/2}}{b_-^{1/2} x_+^{5/2}} \left(\frac{1}{x_-^{9/2}}+\frac{2}{x_-^{5/2}}+\frac{2}{x_-^{1/2}}+x_-^{3/2}\right) t_{\text{(2.709)}}^{\frac{15}{8}+\frac{135}{16\beta}-\frac{5\beta}{16}}
-\frac{J_1}{x_+} \left(\frac{1}{x_-}+x_-\right)t_{\text{(2.849)}}^{-\frac{1}{2}-\frac{\beta}{2}} +\ldots
\end{multline}

\noindent
$\mathrm{II}_B^{-;+} = \IIIA^{-;+} = \tIIIB^{+;-}$: same as above with $b_- \to b_-^{-1}$, $x_- \to x_-^{-1}$. \\

\subsubsection*{For $Q=17/2$:}

\noindent
$\mathrm{I}_A^{-;++} = \mathrm{II}_B^{+;+}$:
\begin{multline}
1+\frac{b_+^{3/2}}{x_+^4} (1+t J_1) \left(\frac{1}{b_-^{1/2}x_-^5}+b_-^{1/2}x_-^5\right)t_{\text{(0.9029)}}^{\frac{105}{8}+\frac{459}{16\beta}+\frac{19\beta}{16}}
+\frac{b_+^3}{x_+^8} \left(\frac{1}{b_-x_-^{10}}+1+b_- x_-^{10}\right)t_{\text{(1.806)}}^{\frac{105}{4}+\frac{459}{8\beta}+\frac{19\beta}{8}} \\
-3t^2+\ldots
\end{multline}

\noindent
$\mathrm{I}_A^{-;+-} = \tIIIA^{+;+}$:
\begin{multline}
1+\frac{b_-^{1/2}b_+^{3/2}x_-^5}{x_+^4} (1+t J_1)\, t_{\text{(0.9029)}}^{\frac{105}{8}+\frac{459}{16\beta}+\frac{19\beta}{16}}+\frac{x_-^{1/2}t_{\text{(1.064)}}^{-\frac{73}{8}-\frac{153}{16\beta}-\frac{21\beta}{16}}}{b_-^{1/2}b_+^{1/2}x_+^{1/2}}+\frac{x_+^{1/2}t_{\text{(1.229)}}^{-\frac{45}{8}-\frac{153}{16\beta}-\frac{13\beta}{16}}}{b_-^{1/2}b_+^{1/2}x_-^{1/2}} \\
+\frac{x_+^{3/2}t_{\text{(1.393)}}^{-\frac{17}{8}-\frac{153}{16\beta}-\frac{5\beta}{16}}}{b_-^{1/2}b_+^{1/2}x_-^{3/2}} 
+\frac{x_+^{5/2}t_{\text{(1.558)}}^{\frac{11}{8}-\frac{153}{16\beta}+\frac{3\beta}{16}}}{b_-^{1/2}b_+^{1/2}x_-^{5/2}}
+\frac{x_+^{7/2}t_{\text{(1.722)}}^{\frac{39}{8}-\frac{153}{16\beta}+\frac{11\beta}{16}}}{b_-^{1/2}b_+^{1/2}x_-^{7/2}}+\frac{b_-b_+^3x_-^{10}t_{\text{(1.806)}}^{\frac{105}{4}+\frac{459}{8\beta}+\frac{19\beta}{8}}}{x_+^8} \\
+\frac{x_+^{9/2}t_{\text{(1.887)}}^{\frac{67}{8}-\frac{153}{16\beta}+\frac{19\beta}{16}}}{b_-^{1/2}b_+^{1/2}x_-^{9/2}} 
+\frac{b_-^{1/2}b_+^{3/2}x_-^5J_1t_{\text{(1.903)}}^{\frac{113}{8}+\frac{459}{16\beta}+\frac{19\beta}{16}}}{x_+^4} \\
+\frac{b_+}{x_+^{9/2}} \left(\frac{1}{x_-^{9/2}}+\frac{1}{x_-^{5/2}}+\frac{1}{x_-^{1/2}}+x_-^{3/2}+x_-^{7/2}+x_-^{11/2}\right)t_{\text{(1.967)}}^{4+\frac{153}{8\beta}-\frac{\beta}{8}}-3t^2+\ldots
\end{multline}

\noindent
$\mathrm{I}_A^{-;-+} = \IIIA^{+;+}$:  same as above with $b_- \to b_-^{-1}$, $x_- \to x_-^{-1}$.\\

\noindent
$\mathrm{I}_A^{-;--} = \mathrm{I}_B^{++}$:
\begin{multline}
1+\frac{1}{b_+^{1/2}x_+^{1/2}} \left(\frac{x_-^{1/2}}{b_-^{1/2}}+\frac{b_-^{1/2}}{x_-^{1/2}}\right)t_{\text{(1.064)}}^{-\frac{73}{8}-\frac{153}{16\beta}-\frac{21\beta}{16}}
+\frac{x_+^{1/2}}{b_+^{1/2}} \left(\frac{1}{b_-^{1/2}x_-^{1/2}}+b_-^{1/2}x_-^{1/2}\right)t_{\text{(1.229)}}^{-\frac{45}{8}-\frac{153}{16\beta}-\frac{13\beta}{16}} \\
+\frac{x_+^{3/2}}{b_+^{1/2}} \left(\frac{1}{b_-^{1/2}x_-^{3/2}}+b_-^{1/2}x_-^{3/2}\right)t_{\text{(1.393)}}^{-\frac{17}{8}-\frac{153}{16\beta}-\frac{5\beta}{16}}
+\frac{x_+^{5/2}}{b_+^{1/2}} \left(\frac{1}{b_-^{1/2}x_-^{5/2}}+b_-^{1/2}x_-^{5/2}\right)t_{\text{(1.558)}}^{\frac{11}{8}-\frac{153}{16\beta}+\frac{3\beta}{16}} \\
+\frac{x_+^{7/2}}{b_+^{1/2}} \left(\frac{1}{b_-^{1/2}x_-^{7/2}}+b_-^{1/2}x_-^{7/2}\right)t_{\text{(1.722)}}^{\frac{39}{8}-\frac{153}{16\beta}+\frac{11\beta}{16}}
+\frac{x_+^{9/2}}{b_+^{1/2}} \left(\frac{1}{b_-^{1/2}x_-^{9/2}}+b_-^{1/2}x_-^{9/2}\right)t_{\text{(1.887)}}^{\frac{67}{8}-\frac{153}{16\beta}+\frac{19\beta}{16}} \\ -3t^2+\ldots
\end{multline}

\noindent
$\mathrm{I}_B^{+-} = \tIIIA^{+;-} = \tIIIB^{-;+}$:

\begin{multline}
1+\frac{x_-^{1/2}t_{\text{(1.064)}}^{-\frac{73}{8}-\frac{153}{16\beta}-\frac{21\beta}{16}}}{b_-^{1/2}b_+^{1/2}x_+^{1/2}}+\frac{x_+^{1/2}t_{\text{(1.229)}}^{-\frac{45}{8}-\frac{153}{16\beta}-\frac{13\beta}{16}}}{b_-^{1/2}b_+^{1/2}x_-^{1/2}}+\frac{x_+^{3/2}t_{\text{(1.393)}}^{-\frac{17}{8}-\frac{153}{16\beta}-\frac{5\beta}{16}}}{b_-^{1/2}b_+^{1/2}x_-^{3/2}}+\frac{x_+^{5/2}t_{\text{(1.558)}}^{\frac{11}{8}-\frac{153}{16\beta}+\frac{3\beta}{16}}}{b_-^{1/2}b_+^{1/2}x_-^{5/2}} \\
+\frac{x_+^{7/2}t_{\text{(1.722)}}^{\frac{39}{8}-\frac{153}{16\beta} +\frac{11\beta}{16}}}{b_-^{1/2}b_+^{1/2}x_-^{7/2}} 
+\frac{x_+^{9/2}t_{\text{(1.887)}}^{\frac{67}{8}-\frac{153}{16\beta}+\frac{19\beta}{16}}}{b_-^{1/2}b_+^{1/2}x_-^{9/2}}-3t^2+\ldots
\end{multline}

\noindent
$\mathrm{I}_B^{-+} = \IIIA^{+;-} = \IIIB^{-;+}$: same as above with $b_- \to b_-^{-1}$, $x_- \to x_-^{-1}$. \\

\noindent
$\mathrm{I}_B^{--} = \mathrm{II}_A^{-;-} =  \mathrm{II}_B^{+;-}$:
\begin{multline}
1-3t^2-x_+ \left(\frac{1}{x_-}+x_-\right)t_{\text{(2.165)}}^{\frac{11}{2}+\frac{\beta}{2}}+\frac{t_{\text{(2.293)}}^{-\frac{59}{4}-\frac{153}{8\beta}-\frac{17\beta}{8}}}{b_+}+\frac{x_+}{b_+} \left(\frac{1}{x_-}+x_-\right) t_{\text{(2.457)}}^{-\frac{45}{4}-\frac{153}{8\beta}-\frac{13\beta}{8}} \\
+\frac{x_+^2}{b_+} \left(\frac{1}{x_-^2}+2+x_-^2\right)t_{\text{(2.622)}}^{-\frac{31}{4}-\frac{153}{8\beta}-\frac{9\beta}{8}}
+\frac{x_+^3}{b_+} \left(\frac{1}{x_-^3}+\frac{2}{x_-}+2x_-+x_-^3\right)t_{\text{(2.786)}}^{-\frac{17}{4}-\frac{153}{8\beta}-\frac{5\beta}{8}} \\
-\frac{J_1}{x_+} \left(\frac{1}{x_-}+x_-\right)t_{\text{(2.835)}}^{-\frac{1}{2}-\frac{\beta}{2}}
+\frac{x_+^4}{b_+} \left(\frac{1}{x_-^4}+\frac{2}{x_-^2}+3+2x_-^2+x_-^4\right)t_{\text{(2.951)}}^{-\frac{3}{4}-\frac{153}{8\beta}-\frac{\beta}{8}} \\
-J_1 \left(\frac{1}{x_-^2}+5+x_-^2\right) t^3+\ldots
\end{multline}

\noindent
$\mathrm{II}_A^{-;+} = \mathrm{III}_B^{-;-} = \widetilde{\mathrm{III}}_B^{-;-}$:

\begin{multline}
1+\frac{x_+^9t_{\text{(1.066)}}^{-3-\frac{153}{4\beta}+\frac{\beta}{4}}}{b_+^2}-3t^2+\frac{x_+^9J_1t_{\text{(2.066)}}^{-2-\frac{153}{4\beta}+\frac{\beta}{4}}}{b_+^2}+\frac{x_+^{18}t_{\text{(2.132)}}^{-6-\frac{153}{2\beta}+\frac{\beta}{2}}}{b_+^4}-\left(\frac{x_+}{x_-}+x_-x_+\right)t_{\text{(2.165)}}^{\frac{11}{2}+\frac{\beta}{2}}\\
-\left(\frac{J_1}{x_-x_+}+\frac{x_-J_1}{x_+}\right)t_{\text{(2.835)}}^{-\frac{1}{2}-\frac{\beta}{2}}-\frac{\left(\frac{x_+^8}{x_-}+x_-x_+^8\right)t_{\text{(2.902)}}^{-\frac{9}{2}-\frac{153}{4\beta}-\frac{\beta}{4}}}{b_+^2}-t^3\left(\frac{J_1}{x_-^2}+5J_1+x_-^2J_1\right) \\
+\frac{x_+^9\left(-3+J_2\right)t_{\text{(3.066)}}^{-1-\frac{153}{4\beta}+\frac{\beta}{4}}}{b_+^2}+\frac{x_+^{18}J_1t_{\text{(3.132)}}^{-5-\frac{153}{2\beta}+\frac{\beta}{2}}}{b_+^4}-2\left(\frac{x_+J_1}{x_-}+x_-x_+J_1\right)t_{\text{(3.165)}}^{\frac{13}{2}+\frac{\beta}{2}}+\frac{x_+^{27}t_{\text{(3.198)}}^{-9-\frac{459}{4\beta}+\frac{3\beta}{4}}}{b_+^6} \\
-\frac{\left(\frac{x_+^{10}}{x_-}+x_-x_+^{10}\right)t_{\text{(3.231)}}^{\frac{5}{2}-\frac{153}{4\beta}+\frac{3\beta}{4}}}{b_+^2}-x_+^2J_1t_{\text{(3.329)}}^{10+\beta}+\left(\frac{b_+^3}{b_-x_-^7x_+^9}+\frac{b_-b_+^3x_-^7}{x_+^9}\right)t_{\text{(3.641)}}^{\frac{99}{4}+\frac{459}{8\beta}+\frac{15\beta}{8}} +\ldots
\end{multline}

\bibliographystyle{JHEP}
\bibliography{refs}

\end{document}